\documentclass[11pt]{article}
\usepackage[utf8]{inputenc}
\usepackage{comment}
\usepackage{cite}
\input epsf.sty

\usepackage{epsfig,amsfonts,amssymb}
\usepackage{hyperref}
\usepackage{cite}
\usepackage{cite}

\usepackage[margin=2cm]{geometry}
\usepackage{epsfig,amsfonts,amssymb,framed,amsmath,xcolor}
\usepackage{slashed,upgreek}
\usepackage{hyperref}
\usepackage{hhline}
\usepackage{bm}
\usepackage{array}
\usepackage{float}
\usepackage{mathtools}
\usepackage[makeroom]{cancel}
\usepackage{authblk}
\usepackage{doi}

\hypersetup{
  colorlinks   = true, %Colours links instead of ugly boxes
  urlcolor     = blue, %Colour for external hyperlinks
  linkcolor    = blue, %Colour of internal links
  citecolor   = red %Colour of citations
}
\newcommand{\be}{\begin{eqnarray}\displaystyle}
\newcommand{\ee}{\end{eqnarray}}
\newcommand{\bi}{\begin{itemize}}
\newcommand{\ei}{\end{itemize}}
\newcommand{\bse}{\begin{subequations}}
\newcommand{\ese}{\end{subequations}}

\newcommand{\f}{\frac}

\newcommand{\p}{\partial}

\newcommand{\non}{\nonumber}

\newcolumntype{P}[1]{>{\centering\arraybackslash}p{#1}}

\begin{document}
\numberwithin{equation}{section}
\baselineskip 24pt

\begin{center}

{\Large \textbf{Classical Sub-subleading Soft Photon and Soft Graviton Theorems in Four Spacetime Dimensions}}

\end{center}

\vskip .6cm
\medskip

\vspace*{4.0ex}

\baselineskip=18pt

\centerline{\large \textbf{   Biswajit Sahoo }}

\vspace*{4.0ex}

\centerline{\large \it Harish-Chandra Research Institute, HBNI}
\centerline{\large \it  Chhatnag Road, Jhunsi,
Prayagraj, 211019, India.}

\vspace*{1.0ex}
\centerline{\small E-mail: bisphysahoo@gmail.com}

\vspace*{5.0ex}

\centerline{\bf Abstract} \bigskip
Classical soft photon and soft graviton theorems determine long wavelength electromagnetic and gravitational waveforms for a general classical scattering process in terms of the electric charges and asymptotic momenta of the ingoing and outgoing macroscopic objects. Performing Fourier transformation of the electromagnetic and gravitational waveforms in the frequency variable one finds electromagnetic and gravitational waveforms at late and early retarded time. Here extending the formalism developed in \cite{1912.06413}, we derive sub-subleading electromagnetic and gravitational waveforms which behave like $u^{-2}(\ln u)$ at early and late retarded time $u$ in four spacetime dimensions. We also have derived the sub-subleading soft photon theorem analyzing two loop amplitudes in scalar QED. Finally, we conjectured the  structure of leading non-analytic contribution to (sub)$^{n}$-leading classical soft photon and graviton theorems which behave like $u^{-n}(\ln u)^{n-1}$ for  early and late retarded time $u$.

%write abstract

\vfill \eject

\baselineskip 18pt

\tableofcontents

\section{Introduction and summary}
In a theory of quantum gravity, soft graviton theorem gives an amplitude with a 
set of finite energy external particles (hard particles)
and
one or
more low energy external gravitons (soft gravitons), in terms of the amplitude without 
the low energy gravitons\cite{weinberg1,weinberg2,jackiw1,jackiw2,1103.2981,1401.7026,1404.4091,1405.3533,1406.6987,
1408.2228,1706.00759,1503.04816,1504.05558,1607.02700,1707.06803,
1808.03288,1809.01675,1406.6574}. 
 On the other hand, classical limit of multiple soft graviton theorem determines
the low frequency radiative mode of the 
gravitational waveform 
in terms of the momenta and spin of the macroscopic objects (scattering data) participating in the scattering process, without the detail knowledge of the interactions responsible for  the classical 
scattering process\cite{1801.07719,1804.09193,1906.08288}. This is also related to the classical gravitational memory\cite{mem1,mem2,mem3,mem4,christodoulou,thorne,bondi,1411.5745,1712.01204,1502.06120,1806.01872,2005.03613} after performing Fourier transformation in the frequency of the gravitational waveform. An analogous inter-connection has been established between soft photon theorem\cite{Gell-Mann,Low1,low,saito,burnett,bell,duca,weinberg1,weinberg2,jackiw1,jackiw2} and electromagnetic  memory\cite{1307.5098,1505.00716,1507.02584,1803.00738}. 

Since S-matrix for massless theory is IR divergent in four spacetime dimensions, analysis of soft theorem for loop amplitudes was known to be ambiguous\cite{1405.1015,1405.1410,1405.3413,Gervais}. But soft theorem relates two S-matrices, so one does not need to make individual S-matrices IR-finite, instead one can factor out the same IR divergent piece from both the S-matrices (if possible) and cancel the IR divergent piece from both sides in the soft theorem relation\footnote{In the analysis of single soft photon theorem it turns out that the S-matrices with soft photon and without soft photon have the same IR-divergent piece but for the analysis of soft graviton theorem, the S-matrix with soft graviton contains extra IR divergent factor relative to S-matrix without soft graviton. Therefore the IR divergent factors do not cancel entirely, but the extra IR divergent piece  can be analyzed using IR regulator\cite{1808.03288}.}. Using this prescription, soft photon and soft graviton theorems have been derived in \cite{1808.03288} up to subleading order in soft momentum expansion. The soft factor at subleading order becomes logarithmic in the energy of external soft photon/graviton. This logarithmic soft factor turns out to be universal and one loop exact. In \cite{1808.03288} the authors made an observation that in the Feynman diagrammatics of loop amplitude if one replaces the Feynman propagator for virtual photon/graviton by it's corresponding retarded propagator one gets only the classical soft factor at subleading order which is proportional to the classical electromagnetic/gravitational waveform\footnote{Recently in \cite{2007.02077}, the authors derived subleading  electromagnetic and gravitational waveform directly from the soft expansion of electromagnetic and gravitational S-matrices.}. 

A direct evaluation of long wavelength electromagnetic and gravitational waveform has been performed in \cite{1912.06413} for a general classical scattering process in the presence of long range electromagnetic and gravitational force. The strategy for deriving gravitational waveform followed in \cite{1912.06413}, is iteratively solving Einstein equation to find the metric fluctuation and geodesic equation to find the correction of asymptotic trajectories. Then performing Fourier transformation in the frequency of long wavelength gravitational waveform one finds the DC gravitational memory at leading order and $u^{-1}$ gravitational tail memory at subleading order for retarded time $u$. Here in this work, we have given a systematic extension of the formalism developed in \cite{1912.06413} for deriving classical soft photon and soft graviton  theorems up to arbitrary order in soft momentum expansion. Using this prescription we have explicitly derived classical soft photon and soft graviton theorem at sub-subleading order which in turn gives us $u^{-2}(\ln u)$ tail memory after Fourier transformation. We have also shown how the same result could also be derived from two loop amplitudes following \cite{1808.03288}. Finally we conjectured the structure of the leading non-analytic contribution of $(sub)^{n}$-leading classical soft photon and soft graviton theorems in terms of some undetermined functions which reproduce $u^{-n}(\ln u)^{n-1}$ tail memory. 

We shall begin by stating our result of classical soft photon theorem, ignoring gravitational interaction. Consider a scattering process in which $M$ number of objects come in with momenta $p'_{1},p'_{2},\cdots , p'_{M}$ and charges $q'_{1},q'_{2},\cdots ,q'_{M}$ undergo complicated interactions\footnote{This complicated interactions can have fusion, splitting, energy transfer or even quantum number transfer. Our result will be independent of all these.} in a large but finite region of spacetime and finally disperse to $N$ number of outgoing objects with momenta $p_{1},p_{2},\cdots ,p_{N}$ and charges $q_{1},q_{2},\cdots ,q_{N}$. Let $\mathcal{R}$ be the region of spacetime of linear size $L$ inside which all the complicated interactions happen and outside the region $\mathcal{R}$ only long range electromagnetic force acts between charged scattered objects. Now defining the scattering center to be some spacetime point well inside the region $\mathcal{R}$ and keeping the electromagnetic wave detector at distance $R$ from the scattering center, we want to derive the $\f{1}{R}$ component of the electromagnetic waveform at large retarded time $|u|=|t-R|>> L$. We are working in a unit where the speed of light $c=1$. In the large $u$ expansion the leading and subleading order electromagnetic waveforms are given in \cite{1307.5098, 1505.00716, 1507.02584, 1803.00738, 1804.09193, 1808.03288, 1912.06413} in four spacetime dimensions. At leading order the change of the electromagnetic waveform between $u=-\infty$ and $u=+\infty$ turns out to be constant (velocity kick memory for test charge) and at subleading order the early and late time electromagnetic waveforms go like $u^{-1}$ as $|u|\rightarrow\infty$ (acceleration memory for test charge). Here extending the analysis of  \cite{1912.06413}, we derive the leading non-analytic contribution of sub-subleading electromagnetic waveforms at early and late time. The results are as follows:
\begingroup
\allowdisplaybreaks
\be
\Delta_{(2)}A_{\mu}(t,R,\hat{n})\ &=&\ \f{1}{4\pi R}\ \f{\ln |u|}{u^{2}}\ \Bigg[\sum_{a=1}^{N}q_{a}\sum_{\substack{b=1\\ b\neq a}}^{N}\f{q_{a}q_{b}}{4\pi}\f{p_{a}^{2}p_{b}^{2}}{[(p_{a}.p_{b})^{2}-p_{a}^{2}p_{b}^{2}]^{3/2}}\ \f{\mathbf{n}^{\rho}}{p_{a}.\mathbf{n}}\non\\
&&\times\Big{\lbrace}p_{a\mu}p_{b\rho}-p_{a\rho}p_{b\mu}\Big{\rbrace}\sum_{\substack{c=1\\ c\neq a}}^{N}\f{q_{a}q_{c}}{4\pi}\f{p_{c}^{2}}{[(p_{a}.p_{c})^{2}-p_{a}^{2}p_{c}^{2}]^{3/2}}\Big{\lbrace}p_{a}^{2}p_{c}.\mathbf{n}-p_{a}.p_{c}p_{a}.\mathbf{n}\Big{\rbrace}\non\\
&&\ -2\sum_{a=1}^{N}\mathbf{n}^{\alpha}\mathcal{B}^{(2)}_{\alpha\mu}\big(q_{a},p_{a};\lbrace q_{b}\rbrace ,\lbrace p_{b}\rbrace\big)\Bigg]\ +\mathcal{O}(u^{-2})\hspace{2cm}\hbox{for\ $u\rightarrow +\infty$,}\label{sub2futureA}
\ee
\be
\Delta_{(2)}A_{\mu}(t,R,\hat{n})\ &=&\ \f{1}{4\pi R}\ \f{\ln |u|}{u^{2}}\ \Bigg[\sum_{a=1}^{M}q'_{a}\sum_{\substack{b=1\\ b\neq a}}^{M}\f{q'_{a}q'_{b}}{4\pi}\f{p_{a}^{\prime 2}p_{b}^{\prime 2}}{[(p'_{a}.p'_{b})^{2}-p_{a}^{\prime 2}p_{b}^{\prime 2}]^{3/2}}\ \f{\mathbf{n}^{\rho}}{p'_{a}.\mathbf{n}}\non\\
&&\times\Big{\lbrace}p'_{a\mu}p'_{b\rho}-p'_{a\rho}p'_{b\mu}\Big{\rbrace}\sum_{\substack{c=1\\ c\neq a}}^{M}\f{q'_{a}q'_{c}}{4\pi}\f{p_{c}^{\prime 2}}{[(p'_{a}.p'_{c})^{2}-p_{a}^{\prime 2}p_{c}^{\prime 2}]^{3/2}}\Big{\lbrace}p_{a}^{\prime 2}p'_{c}.\mathbf{n}-p'_{a}.p'_{c}p'_{a}.\mathbf{n}\Big{\rbrace}\non\\
&&\ -2\sum_{a=1}^{M}\mathbf{n}^{\alpha}\mathcal{B}^{(2)}_{\alpha\mu}\big(q'_{a},p'_{a};\lbrace q'_{b}\rbrace ,\lbrace p'_{b}\rbrace\big)\Bigg] \ +\ \mathcal{O}(u^{-2})\hspace{2cm}\hbox{for\ $u\rightarrow -\infty$},\label{sub2pastA}
\ee
\endgroup
where,
\be 
\vec{x}\ \equiv R\hat{n}\ ,\ \hspace{2cm}\ \mathbf{n}^{\mu}\ \equiv (1,\hat{n}).
\ee
Above the expression for $\mathcal{B}^{(2)}_{\alpha\mu}\big(q_{a},p_{a};\lbrace q_{b}\rbrace ,\lbrace p_{b}\rbrace\big)$ takes form,
\be
\mathcal{B}^{(2)}_{\alpha\mu}\big(q_{a},p_{a};\lbrace q_{b}\rbrace ,\lbrace p_{b}\rbrace\big)\ =\ q_{a}\ \Big[p_{a\mu}\mathcal{C}_{\alpha}\big(q_{a},p_{a};\lbrace q_{b}\rbrace ,\lbrace p_{b}\rbrace\big)-p_{a\alpha}\mathcal{C}_{\mu}\big(q_{a},p_{a};\lbrace q_{b}\rbrace ,\lbrace p_{b}\rbrace\big)\Big]
\ee
with the expression for $\mathcal{C}_{\alpha}\big(q_{a},p_{a};\lbrace q_{b}\rbrace ,\lbrace p_{b}\rbrace\big)$\footnote{In the expression of $\mathcal{C}_{\alpha}$ we can remove the terms proportional the $p_{a\alpha}$ as those terms ultimately cancel in the expression of $\mathcal{B}^{(2)}_{\alpha\mu}$, but we kept those explicitly to make direct connection with Feynman diagram analysis as discussed in \S\ref{FeynQED}.} :
\begingroup
\allowdisplaybreaks
\be
&&\mathcal{C}_{\alpha}\big(q_{a},p_{a};\lbrace q_{b}\rbrace ,\lbrace p_{b}\rbrace\big)\non\\
 &=& -\f{1}{4(4\pi)^{2}}\ \sum_{\substack{b=1\\ b\neq a}}^{N}\sum_{\substack{c=1\\ c\neq a}}^{N}q_{a}^{2}\ q_{b}q_{c}\f{1}{[(p_{a}.p_{b})^{2}-p_{a}^{2}p_{b}^{2}]^{5/2}}\ \f{1}{[(p_{a}.p_{c})^{2}-p_{a}^{2}p_{c}^{2}]^{3/2}}\ p_{b}^{2}p_{c}^{2}p_{a}.p_{b}\non\\
&&\times \Big[\lbrace p_{a}.p_{c}p_{a\alpha}-p_{a}^{2}p_{c\alpha}\rbrace\lbrace (p_{a}.p_{b})^{2}-p_{a}^{2}p_{b}^{2}\rbrace +3\lbrace p_{a}^{2}p_{b\alpha}-p_{a}.p_{b}p_{a\alpha}\rbrace\lbrace p_{a}.p_{b}p_{a}.p_{c}-p_{a}^{2}p_{b}.p_{c}\rbrace\Big]\non\\
&&\ - \f{1}{4(4\pi)^{2}}\ \sum_{\substack{b=1\\ b\neq a}}^{N}\sum_{\substack{c=1\\ c\neq b}}^{N}q_{a}q_{b}^{2}q_{c}\f{1}{[(p_{b}.p_{c})^{2}-p_{b}^{2}p_{c}^{2}]^{3/2}}\ \f{1}{[(p_{a}.p_{b})^{2}-p_{a}^{2}p_{b}^{2}]^{5/2}}\ (p_{b}^{2})^{2}p_{c}^{2}\non\\
&&\times\Big[ \lbrace (p_{a}.p_{b})^{2}-p_{a}^{2}p_{b}^{2}\rbrace \lbrace p_{a}.p_{b}p_{c\alpha}-p_{a}.p_{c}p_{b\alpha}\rbrace\ + 3\ \lbrace p_{b}^{2}p_{a}.p_{b}p_{a}.p_{c}p_{a\alpha}\ +p_{a}^{2}p_{b}.p_{c}p_{a}.p_{b}p_{b\alpha}\non\\
&&\ -(p_{a}.p_{b})^{2}p_{b}.p_{c}p_{a\alpha}-p_{a}^{2}p_{b}^{2}p_{a}.p_{c}p_{b\alpha}\rbrace\Big] .\label{Cexp}
\ee
\endgroup
The expression for $\mathcal{B}^{(2)}_{\alpha\mu}\big(q'_{a},p'_{a};\lbrace q'_{b}\rbrace ,\lbrace p'_{b}\rbrace\big)$ has the same functional form with the arguments having charges and momenta for incoming objects and the particle sums run over $M$ values. The above result is universal i.e. theory independent and determined in terms of four momenta and charges of scattered objects. On the other hand the $\mathcal{O}\Big(u^{-2}\Big)$ correction of the sub-subleading electromagnetic waveform turns out to be theory dependent. It also depends on the details of the scattering event and on the structures (via electromagnetic multipole moments) and spins of the objects participating in the scattering process.

The contribution of first two lines in the sub-subleading electromagnetic waveform of eq.\eqref{sub2futureA} and \eqref{sub2pastA} describes the effect of electromagnetic radiation due to late and early time acceleration of charged objects in the background leading electromagnetic field produced by  a pair of two charged objects due to their asymptotic straight line trajectories. On the other hand the contribution $\mathcal{B}^{(2)}_{\alpha\mu}\big(q_{a},p_{a};\lbrace q_{b}\rbrace ,\lbrace p_{b}\rbrace\big)$ or $\mathcal{B}^{(2)}_{\alpha\mu}\big(q'_{a},p'_{a};\lbrace q'_{b}\rbrace ,\lbrace p'_{b}\rbrace\big)$ describes the effect of electromagnetic radiation due to the subleading correction to the straight line trajectories of charged objects. For massless charged particle scattering process, sub-subleading electromagnetic waveforms in eq.\eqref{sub2futureA} and \eqref{sub2pastA} vanish.
% If we also take care of the long range gravitational force between the scattered objects there will be extra correction of the leading non-analytic contribution of (sub)$^{n}$-leading electromagnetic waveform. Taking care of both long range electromagnetic and gravitational forces between the scattered objects the late and early time electromagnetic waveforms are given in eq.\eqref{lateAemgr} and eq.\eqref{EarlyAemgr}.

Now consider the same scattering setup but take into account the effect of gravitational interaction. For simplicity we shall consider the scattering objects are charge neutral so that outside the region $\mathcal{R}$ only long range gravitational force acts. So we are interested in deriving the $\f{1}{R}$ component of the gravitational waveform at large retarded time $|u|>> L$. Let us define the deviation of metric from Minkowski metric as,
\be
h_{\mu\nu}(x)\ \equiv\ \f{1}{2}\Big(g_{\mu\nu}(x)-\eta_{\mu\nu}\Big)\ \hspace{1cm}\ \hbox{and}\hspace{1cm}\ e_{\mu\nu}(x)\ \equiv\ h_{\mu\nu}(x)-\f{1}{2}\eta_{\mu\nu}\ \eta^{\rho\sigma}h_{\rho\sigma}(x)
\ee
In the large $u$ expansion the leading and subleading order gravitational waveforms are given in \cite{mem1, mem2, mem3, mem4, christodoulou,thorne, bondi, Peters-GR, 1003.3486,  1312.6871, 1401.5831,  1611.03493, 1411.5745, 1804.09193, 1806.01872, 1912.06413,2005.03613} in four spacetime dimensions. At leading order the change of the gravitational waveform between $u=-\infty$ and $u=+\infty$ turns out to be constant and at subleading order the early and late time gravitational waveforms go like $u^{-1}$ as $|u|\rightarrow\infty$. At sub-subleading order the leading non-analytic contribution of gravitational waveform has been conjectured in \cite{1912.06413} which goes like $u^{-2}(\ln u)$ as $|u|\rightarrow \infty$ at late and early time. Here extending the analysis of  \cite{1912.06413}, we have given a proof of the conjectured sub-subleading gravitational waveforms at early and late retarded time $u=t-R+2G\ln R\sum\limits_{b=1}^{N}p_{b}\cdot \mathbf{n}$, which has the following form,
\begingroup
\allowdisplaybreaks
\be
&&\Delta_{(2)}e^{\mu\nu}(t,R,\hat{n})\non\\
&=&\ \f{2G}{R}\ \f{\ln |u|}{u^{2}}\ \Big{\lbrace}2G\sum_{b=1}^{N}p_{b}.\mathbf{n}\Big{\rbrace}^{2}\Bigg(\sum_{a=1}^{N}\f{p_{a}^{\mu}p_{a}^{\nu}}{p_{a}.\mathbf{n}}-\sum_{a=1}^{M}\f{p_{a}^{\prime\mu}p_{a}^{\prime\nu}}{p'_{a}.\mathbf{n}}\Bigg)\non\\
&&\ -\f{4G}{R}\ \f{\ln |u|}{u^{2}}\   \Big{\lbrace}2G\sum_{c=1}^{N}p_{c}.\mathbf{n}\Big{\rbrace}\  \sum_{a=1}^{N}\Bigg[(2G)\sum_{\substack{b=1\\ b\neq a}}^{N}\f{p_{a}.p_{b}}{[(p_{a}.p_{b})^{2}-p_{a}^{2}p_{b}^{2}]^{3/2}} \Big{\lbrace}\f{3}{2}p_{a}^{2}p_{b}^{2}-(p_{a}.p_{b})^{2}\Big{\rbrace}\ \f{p_{a}^{\mu}\mathbf{n}_{\rho}}{p_{a}.\mathbf{n}}\Big{\lbrace}p_{a}^{\nu}p_{b}^{\rho}-p_{a}^{\rho}p_{b}^{\nu}\Big{\rbrace}\Bigg]\non\\
&&\ -\f{2G}{R}\ \f{\ln |u|}{u^{2}}\  \Big{\lbrace}2G\sum_{c=1}^{N}p_{c}.\mathbf{n}\Big{\rbrace}\  \sum_{a=1}^{M}\Bigg[(2G)\sum_{\substack{b=1\\ b\neq a}}^{M}\f{p'_{a}.p'_{b}}{[(p'_{a}.p'_{b})^{2}-p_{a}^{\prime 2}p_{b}^{\prime 2}]^{3/2}} \Big{\lbrace}\f{3}{2}p_{a}^{\prime 2}p_{b}^{\prime 2}-(p'_{a}.p'_{b})^{2}\Big{\rbrace}\ \f{p_{a}^{\prime\mu}\mathbf{n}_{\rho}}{p'_{a}.\mathbf{n}}\Big{\lbrace}p_{a}^{\prime\nu}p_{b}^{\prime\rho}-p_{a}^{\prime\rho}p_{b}^{\prime\nu}\Big{\rbrace}\Bigg]\non\\
&&\ +\ \f{2G}{R}\ \f{\ln |u|}{u^{2}}   \sum_{a=1}^{N}\f{\mathbf{n}_{\rho}\mathbf{n}_{\sigma}}{p_{a}.\mathbf{n}}\Bigg[(2G)\sum_{\substack{b=1\\ b\neq a}}^{N}\f{p_{a}.p_{b}}{[(p_{a}.p_{b})^{2}-p_{a}^{2}p_{b}^{2}]^{3/2}}\Big{\lbrace}\f{3}{2}p_{a}^{2}p_{b}^{2}-(p_{a}.p_{b})^{2}\Big{\rbrace}\ \Big{\lbrace}p_{a}^{\mu}p_{b}^{\rho}-p_{a}^{\rho}p_{b}^{\mu}\Big{\rbrace}\Bigg]\non\\
&&\times \Bigg[(2G)\sum_{\substack{c=1\\ c\neq a}}^{N}\f{p_{a}.p_{c}}{[(p_{a}.p_{c})^{2}-p_{a}^{2}p_{c}^{2}]^{3/2}}\Big{\lbrace}\f{3}{2}p_{a}^{2}p_{c}^{2}-(p_{a}.p_{c})^{2}\Big{\rbrace}\ \Big{\lbrace}p_{a}^{\nu}p_{c}^{\sigma}-p_{a}^{\sigma}p_{c}^{\nu}\Big{\rbrace}\Bigg]\ +\mathcal{O}(u^{-2})\hspace{1cm}\hbox{for $u\rightarrow+\infty$},
\ee
\endgroup
and
\begingroup
\allowdisplaybreaks
\be
&&\Delta_{(2)}e^{\mu\nu}(t,R,\hat{n})\non\\
&=&\  \f{2G}{R}\ \f{\ln |u|}{u^{2}}\   \Big{\lbrace}2G\sum_{c=1}^{N}p_{c}.\mathbf{n}\Big{\rbrace}\  \sum_{a=1}^{M}\Bigg[(2G)\sum_{\substack{b=1\\ b\neq a}}^{M}\f{p'_{a}.p'_{b}}{[(p'_{a}.p'_{b})^{2}-p_{a}^{\prime 2}p_{b}^{\prime 2}]^{3/2}}\non\\
&&\times \Big{\lbrace}\f{3}{2}p_{a}^{\prime 2}p_{b}^{\prime 2}-(p'_{a}.p'_{b})^{2}\Big{\rbrace}\ \f{p_{a}^{\prime\mu}\mathbf{n}_{\rho}}{p'_{a}.\mathbf{n}}\Big{\lbrace}p_{a}^{\prime\nu}p_{b}^{\prime\rho}-p_{a}^{\prime\rho}p_{b}^{\prime\nu}\Big{\rbrace}\Bigg]\non\\
&&\ +\ \f{2G}{R}\ \f{\ln |u|}{u^{2}} \sum_{a=1}^{M}\f{\mathbf{n}_{\rho}\mathbf{n}_{\sigma}}{p'_{a}.\mathbf{n}}\Bigg[(2G)\sum_{\substack{b=1\\ b\neq a}}^{M}\f{p'_{a}.p'_{b}}{[(p'_{a}.p'_{b})^{2}-p_{a}^{\prime 2}p_{b}^{\prime 2}]^{3/2}}\Big{\lbrace}\f{3}{2}p_{a}^{\prime 2}p_{b}^{\prime 2}-(p'_{a}.p'_{b})^{2}\Big{\rbrace}\ \Big{\lbrace}p_{a}^{\prime\mu}p_{b}^{\prime\rho}-p_{a}^{\prime\rho}p_{b}^{\prime\mu}\Big{\rbrace}\Bigg]\non\\
&&\times \Bigg[(2G)\sum_{\substack{c=1\\ c\neq a}}^{M}\f{p'_{a}.p'_{c}}{[(p'_{a}.p'_{c})^{2}-p_{a}^{\prime2}p_{c}^{\prime2}]^{3/2}}\Big{\lbrace}\f{3}{2}p_{a}^{\prime2}p_{c}^{\prime2}-(p'_{a}.p'_{c})^{2}\Big{\rbrace}\ \Big{\lbrace}p_{a}^{\prime\nu}p_{c}^{\prime\sigma}-p_{a}^{\prime\sigma}p_{c}^{\prime\nu}\Big{\rbrace}\Bigg]+\mathcal{O}(u^{-2})\hspace{0.5cm}\hbox{for $u\rightarrow-\infty$}\non\\
\ee
\endgroup
The gravitational waveform given above at order $\mathcal{O}\Big(u^{-2}\ln|u|\Big)$  is universal i.e. theory independent and determined in terms of the four momenta of scattered objects. On the other hand the un-determined $\mathcal{O}(u^{-2})$ gravitational memory depends on the spin angular momenta of the scattered objects as well as on the details of the scattering region $\mathcal{R}$. As an observational prediction of our result, consider gravitational radiation from a binary black hole merger, which we can think of as a decay process with bound state of two black holes as the only initial state and one massive black hole and mass less gravitational radiation are the final states. For this process $M=1$ and one final state object is massive, and rest are massless. In \cite{1912.06413}, it is already discussed that with the order  $u^{-1}$ gravitational tail memory, the order $u^{-2}\ln|u|$ gravitational tail memory also vanishes for binary black hole merger problem. So the absence of both the tail memories for binary black hole merger is the prediction from Einstein gravity. But as discussed in \cite{1912.06413} for supernova explosion, neutron star merger and hyper-velocity star production we can have non-vanishing finite 
gravitational tail memory.

The organization of the paper is as follows: In \S\ref{Ssoftphotonthm} after reviewing the derivation of leading and subleading classical soft photon theorem, we derive the sub-subleading classical soft photon theorem. Then extending this formalism we give the structure of leading non-analytic contribution of (sub)$^n$-leading classical soft photon theorem. In \S\ref{proofsoftgr} after reviewing the derivation of leading and subleading classical soft graviton theorem, we derive the sub-subleading classical soft graviton theorem. In \S\ref{FeynQED} we derive sub-subleading soft photon theorem from the analysis of two loop amplitude and describe it's connection with sub-subleading classical soft photon theorem. In \S\ref{outlook}, first we give  a Feynman diagrammatic understanding of the derivation of sub-subleading soft graviton theorem with the expected result. Then we give the structure of (sub)$^n$-leading electromagnetic and gravitational waveform for a scattering process where both electromagnetic and gravitational long range forces act between the scattered objects. Finally we make some comments on the gravitational tail memory for spinning object scattering.

When this paper was largely complete, a paper \cite{2007.03627} appeared where a derivation of sub-subleading electromagnetic waveform has been attempted by directly using position space Green's function. But the results of eq.\eqref{sub2futureA} and eq.\eqref{sub2pastA} has not been reproduced completely in eq.(61) of \cite{2007.03627} after substituting the expression of $f_{i}^{\mu}$ from eq.(56) of the same paper. A direct comparison of $f_{i}^{\mu}$ expression in \cite{2007.03627} with eq.\eqref{Cexp} shows that the following contribution within the square bracket in the second line of RHS of eq.\eqref{Cexp} has been missed in \cite{2007.03627}: 
 $3\lbrace p_{a}^{2}p_{b\alpha}-p_{a}.p_{b}p_{a\alpha}\rbrace\lbrace p_{a}.p_{b}p_{a}.p_{c}-p_{a}^{2}p_{b}.p_{c}\rbrace$\footnote{I am thankful to Sayali Atul Bhatkar for confirming this disagreement.}. To understand the relation between the soft factors and Ward identities associated with conserved quantities, the reader can look at the reference \cite{2007.03627}.

\section{Proof of classical soft photon theorem}\label{Ssoftphotonthm}
Consider a scattering process where $M$ number of charged objects come in with masses $\lbrace m'_{a}\rbrace$, charges $\lbrace q'_{a}\rbrace$, four velocities $\lbrace v'_{a}\rbrace$ and four momenta $\lbrace p'_{a}=m'_{a}v'_{a}\rbrace$ for $a=1,2,\cdots M$, undergo complicated interactions in a finite region of spacetime $\mathcal{R}$ and disperse to $N$ number of charged objects with masses $\lbrace m_{a}\rbrace$, charges $\lbrace q_{a}\rbrace$, four velocities $\lbrace v_{a}\rbrace$ and four momenta $\lbrace p_{a}=m_{a}v_{a}\rbrace$ for $a=1,2,\cdots N$. Choose the spacetime region $\mathcal{R}$ such that outside this region only long range electromagnetic interaction acts between the charged objects. In this section we are ignoring the long range gravitational interaction between the charged objects. Here our goal will be to determine late and early time electromagnetic waveform emitted from this scattering event, at large distance from the scattering center. Let us consider the linear size of region $\mathcal{R}$ be of order $L$, choose the scattering center well inside the region $\mathcal{R}$ and place the detector of electromagnetic wave at distance $R$ from the scattering center. Then in this setup we want to determine the $\f{1}{R}$ component of the electromagnetic waveform for retarded time $u>>L$. This also translates to determining the radiative mode of gauge field with frequency $\omega$ in the range $R^{-1}<<\omega<<L^{-1}$. As discussed in \cite{1611.03493, 1801.07719, 1804.09193, 1906.08288, 1912.06413} the Fourier transformation in time variable of the radiative mode of the gauge field is related to the Fourier transform of current density determined in terms of the asymptotic trajectory of scattered objects in the following way,
\be
\widetilde{A}_{\mu}(\omega ,R,\hat{n})\ &\simeq &\ \f{1}{4\pi R}\ e^{i\omega R}\  \widehat{J}_{\mu}(k)\label{AJrelation}
\ee
where,
\be
\widetilde{A}_{\mu}(\omega ,\vec{x})\ &=&\ \int_{-\infty}^{\infty}dt\ e^{i\omega t}\ A_{\mu}(t,\vec{x})\\
\widehat{J}_{\mu}(k)\ &=&\ \int d^{4}x\ e^{-ik.x}\ J_{\mu}(t,\vec{x})
\ee
and $\vec{x}=R\hat{n}$ with $\hat{n}$ being the unit vector along the direction of detector from the scattering center.

Since $L$ must be bigger than the size of the objects involves in the scattering process and we are interested in determining electromagnetic waveform with wavelength larger than $L$, we can treat the objects involved in the scattering process as particles. This statement is true for determining the leading non-analytic contribution of (sub)$^{n}$-leading electromagnetic waveform which turns out to be of order $\mathcal{O}\big(\omega^{n-1}(\ln\omega)^{n}\big)$ in small $\omega$ expansion. But if we want to determine the order $\mathcal{O}\big(\omega^{n-1}(\ln\omega)^{k}\big)$ contribution of (sub)$^{n}$-leading electromagnetic waveform  for $k\leq n-1$, the point particle assumption breaks down.

\begin{center}
\begin{figure}
\includegraphics[scale=0.6]{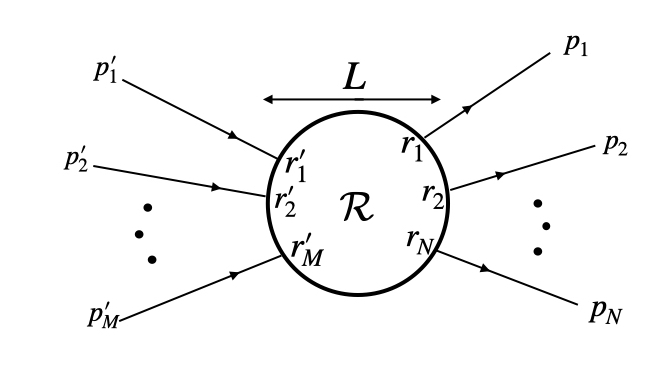}
\caption{A scattering process describing $M$ number of particles are coming into region $\mathcal{R}$, going through unspecified interaction inside the region $\mathcal{R}$ and disperse to $N$ number of particles. Outside the region $\mathcal{R}$ only long range electromagnetic and/or gravitational interaction is present. }\label{Scattering}
\end{figure}
\end{center}

\subsection{General setup}\label{GeneralsetupEM}
Consider the trajectories of outgoing particles with proper time $\sigma$ in the range $0\leq \sigma<\infty$ be $X_{a}(\sigma)$ for $a=1,2,\cdots , N$. Similarly the trajectories of ingoing particles with proper time $\sigma$ in the range $-\infty < \sigma\leq 0$ are denoted by $X'_{a}(\sigma)$ for $a=1,2,\cdots , M$.
For performing the analysis for both ingoing and outgoing objects at one go we use the following compact notations:
\be
q_{N+a}\ =\ -q'_{a},\hspace{5mm} m_{N+a}\ =\ m'_{a},\hspace{5mm}v_{N+a}\ =\ -v'_{a},\hspace{5mm}p_{N+a}\ =\ -p'_{a},\hspace{5mm}X_{N+a}(\sigma)\ =\ X'_{a}(-\sigma)
\ee
for $a=1,2,\cdots ,M$ and $0\leq \sigma <\infty$. Consider the asymptotic trajectory of a'th particle outside the region $\mathcal{R}$,
\be
X_{a}^{\mu}(\sigma)\ =\ r_{a}^{\mu}\ +\ v_{a}^{\mu}\sigma \ +\ Y_{a}^{\mu}(\sigma)
\ee
with boundary conditions,
\be
Y_{a}^{\mu}(\sigma)\Bigg{|}_{\sigma=0}\ =\ 0\ ,\hspace{1cm}\ \f{dY_{a}^{\mu}(\sigma)}{d\sigma}\Bigg{|}_{\sigma\rightarrow \infty}\ =\ 0\label{boundaryEM}
\ee
where $\lbrace r_{a}\rbrace$ denotes the set of points where particles' trajectory intersect the boundary of region $\mathcal{R}$ at $\sigma=0$, as indicated in Fig.\ref{Scattering}. Now to find the asymptotic trajectory of the particles and the electromagnetic gauge field we have to solve the following two equations,
\be
m_{a}\f{d^{2}X_{a}^{\mu}(\sigma)}{d\sigma^{2}}\ &=&\ q_{a}\ F^{\mu\nu}(X_{a}(\sigma))\ \f{dX_{a\nu}(\sigma)}{d\sigma}\\
\p^{\rho}\p_{\rho}A_{\mu}(x)\ &=&\ -J_{\mu}(x)
\ee
where in the last equation above we have used Lorenz gauge condition $\p_{\mu}A^{\mu}(x)=0$. The current density in terms of the asymptotic trajectories of the particles is given by,
\be
J^{\mu}(x)\ &=&\ \sum_{a=1}^{M+N}q_{a}\int_{0}^{\infty}d\sigma\  \delta^{(4)}(x-X_{a}(\sigma))\ \f{dX_{a}^{\mu}(\sigma)}{d\sigma}\label{PositionC}
\ee
Here we are using the unit where speed of light $c=1$ and the charges $\lbrace q_{a}\rbrace$ are in unit of $\sqrt{\alpha}$, where $\alpha$ is the fine structure constant. In this unit $\f{q^{2}\omega}{M}$ is a dimensionless parameter with $q$ being the parameter representing charges of scattered particles and $M$ being the parameter representing masses of scattered particles\footnote{Actually there could be some other parameters i.e. $L$, size of charged objects, spins etc. Now using some of this parameters and $\omega$ we can build up some dimensionless quantity but those will not affect the leading non-analytic contribution to $(sub)^{n}$-leading electromagnetic waveform in the small $\omega$ expansion. The reason behind this claim is that the leading non-analytic behaviour at each order in $\omega$ expansion will only be controlled by the minimal coupling of the worldline trajectory with the gauge field in the action. On the other hand if the scattered objects carry spin or have internal struture, they couple to one or more derivatives of gauge field causing more powers of soft momenta in the Fourier transform of current density relative to the contribution coming from minimal coupling.\label{footEM} }. So expansion of any quantity in power of $\omega$ is same as expansion of this quantity in powers of $q^{2}$. Hence let us expand correction of the straight line trajectory in powers of $q^{2}$ in the following way,
\be
Y_{a}^{\mu}(\sigma)\ &=&\ \ \Delta_{(1)}Y_{a}^{\mu}(\sigma)+\ \Delta_{(2)}Y_{a}^{\mu}(\sigma)+\cdots \label{Yexpansion}
\ee
where $\Delta_{(r)}Y_{a}^{\mu}(\sigma)$ is of order $\mathcal{O}(q^{2r})$. Now with this correction of trajectory the Fourier transform of the current density can be expanded as,
\be
\widehat{J}_{\mu}(k)\ &=&\ \Delta_{(0)}\widehat{J}_{\mu}(k)\ +\ \Delta_{(1)}\widehat{J}_{\mu}(k)\ +\ \Delta_{(2)}\widehat{J}_{\mu}(k)\ +\cdots
\ee
where $\Delta_{(r)}\widehat{J}_{\mu}(k)$ is of order $\mathcal{O}(q^{2r+1})$. Similarly the gauge field also has an expansion,
\be
A_{\mu}(x)\ &=&\ \Delta_{(0)}A_{\mu}(x)\ +\ \Delta_{(1)}A_{\mu}(x)\ +\ \Delta_{(2)}A_{\mu}(x)+\cdots
\ee
where
\be
\Delta_{(r)}A_{\mu}(x)\ &=&\ -\int \f{d^{4}\ell}{(2\pi)^{4}}\ G_{r}(\ell)\ e^{i\ell\cdot x}\ \Delta_{(r)}\widehat{J}_{\mu}(\ell)
\ee
for $r=0,1,2,\cdots$. Here $G_{r}(\ell)$ represents the momentum space retarded Greens function,
\be
G_{r}(\ell)\ =\ \f{1}{(\ell^{0}+i\epsilon)^{2}-|\vec{\ell}|^{2}}
\ee
Now from eq.\eqref{PositionC}, the expression of the Fourier transform of current density turns out to be,
\be
\widehat{J}_{\mu}(k)\ &=&\ \sum_{a=1}^{M+N}q_{a}\int_{0}^{\infty}d\sigma \ e^{-ik.X_{a}(\sigma)}\ \f{dX_{a\mu}(\sigma)}{d\sigma}\non\\
&=&\ \sum_{a=1}^{M+N}q_{a}\int_{0}^{\infty}d\sigma \ e^{-i\lbrace k.r_{a}+k.v_{a}\sigma +\sum\limits_{s=1}^{\infty}k\cdot \Delta_{(s)}Y_{a}(\sigma) \rbrace}\ \Big[v_{a\mu}+\sum_{t=1}^{\infty}\f{d\Delta_{(t)}Y_{a\mu}(\sigma)}{d\sigma} \Big]\non\\
&=&\ \sum_{a=1}^{M+N}q_{a}\int_{0}^{\infty}d\sigma \ e^{-i\lbrace k.r_{a}+k.v_{a}\sigma \rbrace }\sum_{u=0}^{\infty}\f{1}{u!}\Big[-ik\cdot \sum_{s=1}^{\infty}\Delta_{(s)}Y_{a}(\sigma)\Big]^{u}\ \Big[v_{a\mu}+\sum_{t=1}^{\infty}\f{d\Delta_{(t)}Y_{a\mu}(\sigma)}{d\sigma} \Big]\non\\
&\equiv & \ \sum_{r=0}^{\infty}\ \Delta_{(r)}\widehat{J}_{\mu}(k)\label{FourierJk}
\ee
Here we want to extract the leading non-analytic contribution of $ \Delta_{(r)}\widehat{J}_{\mu}(k)$ in the small $\omega$ expansion, which turns out to be of order $\mathcal{O}\big(\omega^{r-1}(ln\omega)^{r}\big)$. For the analysis of leading and subleading order we will closely follow \cite{1912.06413}, but here we will be brief and organize the results in a different way which will be useful for the generalization to (sub)$^{n}$-leading order.

\subsection{Derivation of leading order electromagnetic waveform}\label{subsectionleading}
The leading order current density follows form eq.\eqref{FourierJk},
\be
\Delta_{(0)}\widehat{J}_{\mu}(k)&=&\ \sum_{a=1}^{M+N}q_{a}\ e^{-ik.r_{a}}\ \int_{0}^{\infty}d\sigma e^{-ik.v_{a}\sigma}\ v_{a\mu}\non\\
&=&\ \sum_{a=1}^{M+N}q_{a}\f{p_{a\mu}}{i(p_{a}.k-i\epsilon)}\ e^{-ik.r_{a}}\label{leadingC}
\ee
where for the convergence of the integral at $\sigma=\infty$  we need to replace $\omega\rightarrow (\omega +i\eta_{a}\epsilon)$ where $\eta_{a}=\pm 1$ for particle-a being outgoing/ingoing. This implies the following replacement in the exponential $k\cdot v_{a}\rightarrow (k\cdot v_{a}-i\epsilon)$. To extract the leading order contribution in $\omega$ expansion, we have to expand $e^{-ik.r_{a}}=1-ik.r_{a}+\cdots$ and keep only the leading contribution. Now using the result of eq.\eqref{AJrelation} at leading order and performing Fourier transformation we recover the known leading electromagnetic memory,
\be
&&\Delta_{(0)}A_{\mu}(t,R,\hat{n})\Big{|}_{(t-R)\rightarrow +\infty}-\Delta_{(0)}A_{\mu}(t,R,\hat{n})\Big{|}_{(t-R)\rightarrow -\infty}\non\\
&&=\ \f{1}{4\pi R}\ \Bigg[-\sum_{a=1}^{N}\f{q_{a}p_{a\mu}}{p_{a}.\mathbf{n}}+\sum_{a=1}^{M}\f{q'_{a}p'_{a\mu}}{p'_{a}.\mathbf{n}}\Bigg]
\ee

\subsection{Derivation of subleading order electromagnetic waveform}
From the leading order current density in eq.\eqref{leadingC}, the leading gauge field expression becomes,
\be
\Delta_{(0)}A_{\mu}(x)\ &=&\ -\sum_{b=1}^{M+N}\ \int \f{d^{4}\ell}{(2\pi)^{4}}\ e^{i\ell\cdot (x-r_{b})}\ G_{r}(\ell)\ \f{q_{b}p_{b\mu}}{i(p_{b}.\ell -i\epsilon)}\label{leadingA}
\ee
The expression for the leading order  field strength produced due to the asymptotic straight line motion of particle-b is,
\be
\Delta_{(0)}F^{(b)}_{\nu\rho}(x) 
= 
- \int{d^4 \ell\over (2\pi)^4} e^{i\ell.(x-r_{b)}}\ G_{r}(\ell)  \ {q_b\over (\ell. p_{b}-i\epsilon)}
(\ell_\nu p_{b\rho}-\ell_\rho p_{b\nu})\, 
\ee
For the leading order gauge field in eq.\eqref{leadingA}, leading correction to the straight line trajectory of particle-a satisfy the following equation of motion,
\be
m_{a}\f{d^{2}\Delta_{(1)}Y_{a}^{\mu}(\sigma)}{d\sigma^{2}}\ =\ q_{a}\ \Delta_{(0)}F^{\mu\nu}(r_{a}+v_{a}\sigma)\ v_{a\nu}
\ee
where\footnote{The self force on the trajectory of charged particles can be neglected up to the order we are working in\cite{rohrlich}.},
\be 
\Delta_{(0)}F^{\mu\nu}(r_{a}+v_{a}\sigma)\ =\ \sum_{\substack{b=1\\ b\neq a}}^{M+N}\ \Delta_{(0)}F^{(b)\mu\nu}(r_{a}+v_{a}\sigma).
\ee
After using the boundary conditions following from eq.\eqref{boundaryEM}
\be
\Delta_{(1)}Y_{a}^{\mu}(\sigma)\Big{|}_{\sigma=0}\ =\ 0 \hspace{5mm} ,\ \hspace{5mm}\ {d\Delta_{(1)}Y_a^\mu(\sigma)\over d\sigma}\Bigg{|}_{\sigma\rightarrow \infty}\ =\ 0\label{boundaryEM1}
\ee
we get,
\be
{d\Delta_{(1)}Y_a^\mu(\sigma)\over d\sigma} = -{q_a\over m_a}\, 
\int_\sigma^\infty\, d\sigma' \, \Delta_{(0)}F^\mu_{~\nu}(v_a\, \sigma'+r_{a}) \, 
v_a^\nu\
\ee
\be
\Delta_{(1)}Y_a^\mu(\sigma) =-{q_a\over m_a}\, 
\int_0^\sigma\, d\sigma' \, \int_{\sigma'}^\infty\, d\sigma'' \, \Delta_{(0)}F^\mu_{~\nu}(v_a\, \sigma''+r_{a}) \, v_a^\nu\,
\ee
From eq.\eqref{FourierJk}, the expression of subleading order current density becomes,
\be
\Delta_{(1)}\widehat{J}_{\mu}(k)&=&\ \sum_{a=1}^{M+N}q_{a}\ e^{-ik.r_{a}}\ \int_{0}^{\infty}d\sigma e^{-i(k.v_{a}-i\epsilon)\sigma}\ \Big[ -ik.\Delta_{(1)}Y_{a}(\sigma)\ v_{a\mu}\ +\ \f{d\Delta_{(1)}Y_{a\mu}(\sigma)}{d\sigma}\Big]\label{subleadingC}
\ee
Now we will analyze each term within the square bracket above separately to extract contribution of order $\mathcal{O}(\ln\omega)$ following \cite{1912.06413}. The first term of subleading current density in eq.\eqref{subleadingC} is,
 \begingroup
\allowdisplaybreaks
\be
\Delta^{(1)}_{(1)}\widehat{J}_{\mu}(k)&=&\ \sum_{a=1}^{M+N}q_{a} e^{-ik.r_{a}}\ \int_{0}^{\infty}d\sigma e^{-i(k.v_{a}-i\epsilon)\sigma}\ \Big[ -ik.\Delta_{(1)}Y_{a}(\sigma)\ v_{a\mu}\Big]\non\\
&=&\  \sum_{a=1}^{M+N}q_{a}\ \f{v_{a\mu}}{i(k.v_{a}-i\epsilon)}\ e^{-ik.r_{a}}\int_{0}^{\infty}d\sigma e^{-i(k.v_{a}-i\epsilon)\sigma}\ \Big[ -ik\cdot \f{d\Delta_{(1)}Y_{a}(\sigma)}{d\sigma}\Big]\non\\
&=&\ \sum_{a=1}^{M+N}\sum_{\substack{b=1\\ b\neq a}}^{M+N}q_{a}^{2}q_{b}\f{p_{a\mu}}{p_{a}.k}\ e^{-ik.r_{a}}\int \f{d^{4}\ell}{(2\pi)^{4}}G_{r}(\ell)e^{i\ell .(r_{a}-r_{b})}\ \f{1}{\ell.p_{b}-i\epsilon}\ \f{1}{\ell.p_{a}+i\epsilon}\ \f{1}{(\ell-k).p_{a}+i\epsilon}\non\\
&&\hspace{4cm} \times \big[\ell.kp_{a}.p_{b}-\ell.p_{a}p_{b}.k\big]\non\\
%&\simeq &\ \f{1}{4\pi}\sum_{\substack{a,b\\ a\neq b\\ \eta_{a}\eta_{b=1}}}q_{a}^{2}q_{b}\ln\lbrace L(\omega+i\epsilon\eta_{a})\rbrace \ \f{p_{a\mu}}{p_{a}.n}\ \f{p_{b}^{2}}{[(p_{a}.p_{b})^{2}-p_{a}^{2}p_{b}^{2}]^{3/2}}\ \Big[ p_{a}^{2}p_{b}.n-p_{a}.p_{b}p_{a}.n\Big]\non\\
&\simeq &\ - i\sum_{a=1}^{M+N}q_{a}\f{p_{a\mu}}{p_{a}\cdot k}\ k_{\rho}\f{\p}{\p p_{a\rho}}\  K_{em}^{cl}  \label{Jhat11}
\ee
\endgroup
where, 
 \begingroup
\allowdisplaybreaks
\be
K_{em}^{cl}\ &=&\ -\f{i}{2}\sum_{\substack{b,c\\b\neq c}} q_{b}q_{c}\int_{\omega}^{L^{-1}}\ \f{d^{4}\ell}{(2\pi)^{4}}\ G_{r}(\ell)\ \f{1}{p_{b}.\ell +i\epsilon}\ \f{1}{p_{c}.\ell -i\epsilon}\ p_{b}\cdot p_{c}\non\\
&=&\ -\f{i}{2}\  \sum_{\substack{b,c\\b\neq c\\ \eta_{b}\eta_{c}=1}}\f{q_{b}q_{c}}{4\pi}\ \ln\Big{\lbrace}L(\omega +i\epsilon\eta_{b})\Big{\rbrace}\ \f{p_{b}\cdot p_{c}}{\sqrt{(p_{b}.p_{c})^{2}-p_{b}^{2}p_{c}^{2}}}\label{Kemcl}
\ee
\endgroup
Here  in eq.\eqref{Jhat11}, to get the second line of RHS from the first line we have first written $e^{-i(k.v_{a}-i\epsilon)\sigma}=\big{\lbrace}-i(k.v_{a}-i\epsilon)\big{\rbrace}^{-1}\f{d}{d\sigma}e^{-i(k.v_{a}-i\epsilon)\sigma}$, then performed integration by parts to move the $\sigma$ derivative to the rest of the integrand after using the boundary conditions of eq.\eqref{boundaryEM1}. To get the last line of RHS from the second last expression, we approximated the integrand in the integration region $|r^{\mu}_{a}-r_{b}^{\mu}|^{-1}\sim L^{-1}>>|\ell^{\mu}|>>\omega$ and the corresponding result is written using $\simeq $ sign ignoring the $\mathcal{O}(\omega^{0})$ contribution\footnote{Under $\simeq $ sign we are also neglecting terms of order $\mathcal{O}(\omega^{k}\ln\omega)$ for $k=1,2,\cdots$ which we can get by expanding $exp(-ik.r_{a})=1-ik.r_{a}+\cdots$ from the second last line of eq.\eqref{Jhat11}. But these terms will not affect the leading non-analytic contribution of $\Delta_{(r)}\widehat{J}_{\mu}(k)$ for $r=2,3,\cdots.$}. The integration of eq.\eqref{Kemcl} is explicitly evaluated in \cite{1912.06413} and the result is non-vanishing only if both the particles $b$ and $c$ are incoming/outgoing. We are using the convention that $\eta_{a}=+1$ if particle-$a$ is outgoing and $\eta_{a}=-1$ if particle-$a$ is incoming.

Similarly the second term of subleading current density in eq.\eqref{subleadingC} is evaluated in the following way,
\be
\Delta^{(2)}_{(1)}\widehat{J}_{\mu}(k)&=&\ \sum_{a=1}^{M+N}q_{a}e^{-ik.r_{a}}\ \int_{0}^{\infty}d\sigma e^{-ik.v_{a}\sigma}\ \f{d\Delta_{(1)}Y_{a\mu}(\sigma)}{d\sigma}\non\\
&=&\ -\sum_{a=1}^{M+N}\sum_{\substack{b=1\\ b\neq a}}^{M+N}q_{a}^{2}q_{b}e^{-ik.r_{a}}\ \int \f{d^{4}\ell}{(2\pi)^{4}}G_{r}(\ell)\ e^{i\ell.(r_{a}-r_{b})}\ \f{1}{\ell.p_{b}-i\epsilon}\ \f{1}{\ell.p_{a}+i\epsilon}\ \f{1}{(\ell-k).p_{a}+i\epsilon}\non\\
&&\hspace{4cm}\times \Big[ \ell_{\mu}p_{a}.p_{b}-p_{a}.\ell p_{b\mu}\Big]\non\\
%&\simeq &\ -\f{1}{4\pi}\sum_{\substack{a,b\\ a\neq b\\ \eta_{a}\eta_{b=1}}}q_{a}^{2}q_{b}\ln\lbrace L(\omega+i\epsilon\eta_{a})\rbrace \ \ \f{p_{b}^{2}}{[(p_{a}.p_{b})^{2}-p_{a}^{2}p_{b}^{2}]^{3/2}}\Big[p_{a}^{2}p_{b\mu}-p_{a}.p_{b}p_{a\mu}\Big]\non\\
&\simeq &\ +i\sum_{a=1}^{M+N}q_{a}\ \f{\p}{\p p_{a}^{\mu}}\ K_{em}^{cl}\label{Jhat12}
\ee
 Here also to get the last line of RHS from the second last expression, we approximated the integrand in the momentum region $L^{-1}>>|\ell^{\mu}|>>\omega$ and the corresponding result is written using $\simeq $ sign ignoring the $\mathcal{O}(\omega^{0})$ contribution.
Now summing contributions from both the terms, total subleading current density at order $\mathcal{O}(\ln\omega)$ becomes,
\be\label{subJ}
\Delta_{(1)}\widehat{J}_{\mu}(k)\
&\simeq &\  - i\sum_{a=1}^{M+N}q_{a}\f{k^{\rho}}{p_{a}\cdot k}\ \Bigg(p_{a\mu}\f{\p}{\p p_{a}^{\rho}}\ -\ p_{a\rho}\f{\p}{\p p_{a}^{\mu}} \Bigg)K_{em}^{cl}  \non\\
 &=&\ \f{1}{4\pi}\sum_{a=1}^{M+N}\sum_{\substack{b=1\\ b\neq a\\ \eta_{a}\eta_{b}=1}}^{M+N}\ q_{a}^{2}q_{b}\ln\lbrace L(\omega+i\epsilon\eta_{a})\rbrace \ \ \f{p_{a}^{2}p_{b}^{2}}{[(p_{a}.p_{b})^{2}-p_{a}^{2}p_{b}^{2}]^{3/2}}\ \Bigg[ \f{p_{b}.\mathbf{n}}{p_{a}.\mathbf{n}}p_{a\mu}-p_{b\mu}\Bigg]
\ee
where $k^{\mu}=\omega\mathbf{n}^{\mu}$ with $\mathbf{n}=(1,\hat{n})$. Now using the relation \eqref{AJrelation} at subleading order and performing Fourier transformation in $\omega$ variable we get the following early and late time electromagnetic waveforms \cite{1912.06413} after using the results of the integrations from eq.\eqref{FourierFn0} and eq.\eqref{FourierF0n} for $n=1$,
\be
&&\Delta_{(1)}A_{\mu}(t,R,\hat{n})\non\\
&=&\ -\f{1}{4\pi R}\ \f{1}{u}\ \sum_{a=1}^{N}\sum_{\substack{b=1\\ b\neq a}}^{N}\ \f{q_{a}^{2}q_{b}}{4\pi} \ \f{p_{a}^{2}p_{b}^{2}}{[(p_{a}.p_{b})^{2}-p_{a}^{2}p_{b}^{2}]^{3/2}}\ \Bigg[ \f{p_{b}.\mathbf{n}}{p_{a}.\mathbf{n}}p_{a\mu}-p_{b\mu}\Bigg]\hspace{1cm}\hbox{for $u\rightarrow +\infty$},
\ee
\be
&&\Delta_{(1)}A_{\mu}(t,R,\hat{n})\non\\
&=&\ \f{1}{4\pi R}\ \f{1}{u}\ \sum_{a=1}^{M}\sum_{\substack{b=1\\ b\neq a}}^{M}\ \f{q_{a}^{\prime 2}q'_{b}}{4\pi} \ \f{p_{a}^{\prime 2}p_{b}^{\prime 2}}{[(p'_{a}.p'_{b})^{2}-p_{a}^{\prime 2}p_{b}^{\prime 2}]^{3/2}}\ \Bigg[ \f{p'_{b}.\mathbf{n}}{p'_{a}.\mathbf{n}}p'_{a\mu}-p'_{b\mu}\Bigg]\hspace{1cm}\hbox{for $u\rightarrow -\infty$}.
\ee

\subsection{Derivation of sub-subleading order electromagnetic waveform}\label{Subsecsub2ph}
Sub-subleading order current density from eq.\eqref{FourierJk} takes the following form,
\be
&&\Delta_{(2)}\widehat{J}_{\mu}(k)\non\\
&=&\ \sum_{a=1}^{M+N}q_{a}\ e^{-ik.r_{a}}\ \int_{0}^{\infty}d\sigma e^{-i(k.v_{a}-i\epsilon)\sigma}\ \Bigg[  -\ ik\cdot \Delta_{(1)}Y_{a}(\sigma)\f{d\Delta_{(1)}Y_{a\mu}(\sigma)}{d\sigma}\ +\ \f{1}{2}\Big(-ik\cdot \Delta_{(1)}Y_{a}(\sigma)\Big)^{2}v_{a\mu}\ \non\\
&&\ -ik\cdot \Delta_{(2)}Y_{a}(\sigma)v_{a\mu}\ +\ \f{d\Delta_{(2)}Y_{a\mu}(\sigma)}{d\sigma}\Bigg]\label{sub-subleadingC}
\ee
Now we will analyze each terms within the square bracket above separately to extract contribution of order $\mathcal{O}\big(\omega(\ln\omega)^{2}\big)$, which turns out to be the leading non-analytic contribution at this order in the small $\omega$ expansion. The first term of sub-subleading current density in eq.\eqref{sub-subleadingC} contributes to,
\begingroup
\allowdisplaybreaks
\be
\Delta^{(1)}_{(2)}\widehat{J}_{\mu}(k)\ &=& \ \sum_{a=1}^{M+N}q_{a}e^{-ik.r_{a}}\ \int_{0}^{\infty}d\sigma \ e^{-i(k.v_{a}-i\epsilon)\sigma}\ \Big[ -ik\cdot \Delta_{(1)}Y_{a}(\sigma)\ \f{d\Delta_{(1)}Y_{a\mu}(\sigma)}{d\sigma}\Big]\non\\
&=&-ik^{\nu}\sum_{a=1}^{M+N}\f{q_{a}^{3}}{m_{a}^{2}}e^{-ik.r_{a}}\ v_{a}^{\rho}v_{a}^{\alpha}\int_{0}^{\infty}d\sigma \ e^{-i(k.v_{a}-i\epsilon)\sigma}\ \int_{0}^{\sigma}d\sigma' \int_{\sigma'}^{\infty}d\sigma''\non\\
&&\times \sum_{\substack{b=1\\ b\neq a}}^{M+N}\int \f{d^{4}\ell_{1}}{(2\pi)^{4}}\ e^{i\ell_{1}.(r_{a}-r_{b}+v_{a}\sigma'')}G_{r}(\ell_{1})\ \f{q_{b}}{\ell_{1}.p_{b}-i\epsilon}\ \Big{\lbrace}\ell_{1\nu}p_{b\rho}-\ell_{1\rho}p_{b\nu}\Big{\rbrace}\non\\
&&\times \int_{\sigma}^{\infty} d\sigma'''\sum_{\substack{c=1\\ c\neq a}}^{M+N}\int\f{d^{4}\ell_{2}}{(2\pi)^{4}}\ e^{i\ell_{2}.(r_{a}-r_{c}+v_{a}\sigma''')}G_{r}(\ell_{2})\ \f{q_{c}}{\ell_{2}.p_{c}-i\epsilon}\ \Big{\lbrace} \ell_{2\mu}p_{c\alpha}-\ell_{2\alpha}p_{c\mu}\Big{\rbrace}\label{Jhat21}
\ee
\endgroup
Now after using the following integration result in the above expression
\be
&&\int_{0}^{\infty}d\sigma e^{-i(k.v_{a}-i\epsilon)\sigma}\int_{0}^{\sigma}d\sigma' \int_{\sigma'}^{\infty}d\sigma''\ e^{i\ell_{1}.v_{a}\sigma''}\int_{\sigma}^{\infty}d\sigma''' e^{i\ell_{2}.v_{a}\sigma'''}\non\\
&=&\ \f{1}{\ell_{1}.v_{a}+i\epsilon}\ \f{1}{\ell_{2}.v_{a}+i\epsilon}\ \f{1}{(\ell_{2}-k).v_{a}+i\epsilon}\ \f{1}{(\ell_{1}+\ell_{2}-k).v_{a}+i\epsilon}\label{integration}
\ee
we get,
\begingroup
\allowdisplaybreaks
\be
\Delta^{(1)}_{(2)}\widehat{J}_{\mu}(k)
&=&\ -i\sum_{a=1}^{M+N}\ \sum_{\substack{b=1\\ b\neq a}}^{M+N}\ \sum_{\substack{c=1\\ c\neq a}}^{M+N}q_{a}^{3}q_{b}q_{c} e^{-ik.r_{a}}\int \f{d^{4}\ell_{1}}{(2\pi)^{4}}\ e^{i\ell_{1}.(r_{a}-r_{b})}G_{r}(\ell_{1})\ \f{1}{\ell_{1}.p_{b}-i\epsilon}\ \f{1}{\ell_{1}.p_{a}+i\epsilon}\non\\
&&\times \big[\ell_{1}.kp_{a}.p_{b}-\ell_{1}.p_{a}p_{b}.k\big]\ \int\f{d^{4}\ell_{2}}{(2\pi)^{4}}\ e^{i\ell_{2}.(r_{a}-r_{c})}G_{r}(\ell_{2})\ \f{1}{\ell_{2}.p_{c}-i\epsilon}\ \f{1}{\ell_{2}.p_{a}+i\epsilon}\ \non\\
&&\times \f{1}{(\ell_{2}-k).p_{a}+i\epsilon}\ \big[\ell_{2\mu}p_{a}.p_{c}-\ell_{2}.p_{a}p_{c\mu}\big]\ \f{1}{(\ell_{1}+\ell_{2}-k).p_{a}+i\epsilon}\label{hatJ21}
\ee
\endgroup
Now the above expression contributes to order $\mathcal{O}\big(\omega(\ln\omega)^{2}\big)$ in the integration region $L^{-1}>>|\ell_{1}^{\mu}|>>|\ell_{2}^{\mu}|>>\omega$. In this integration range approximating $\lbrace (\ell_{1}+\ell_{2}-k).p_{a}+i\epsilon\rbrace^{-1}\simeq \lbrace\ell_{1}.p_{a}+i\epsilon\rbrace^{-1}$ and $\lbrace (\ell_{2}-k).p_{a}+i\epsilon\rbrace^{-1}\simeq \lbrace\ell_{2}.p_{a}+i\epsilon\rbrace^{-1}$ we get,
\be
\Delta^{(1)}_{(2)}\widehat{J}_{\mu}(k)\ &\simeq &\ - i\ \sum_{a=1}^{M+N}\ \sum_{\substack{b=1\\ b\neq a}}^{M+N}\sum_{\substack{c=1\\ c\neq a}}^{M+N}q_{a}^{3}q_{b}q_{c} \int_{\omega}^{L^{-1}} \f{d^{4}\ell_{1}}{(2\pi)^{4}}\ G_{r}(\ell_{1})\ \f{1}{\ell_{1}.p_{b}-i\epsilon}\ \f{1}{[\ell_{1}.p_{a}+i\epsilon]^{2}}\ \big[\ell_{1}.kp_{a}.p_{b}-\ell_{1}.p_{a}p_{b}.k\big]\non\\
&&\times \int_{\omega}^{|\vec{\ell}_{1}|}\f{d^{4}\ell_{2}}{(2\pi)^{4}}\ G_{r}(\ell_{2})\ \f{1}{\ell_{2}.p_{c}-i\epsilon}\ \f{1}{[\ell_{2}.p_{a}+i\epsilon]^{2}}\ \big[\ell_{2\mu}p_{a}.p_{c}-\ell_{2}.p_{a}p_{c\mu}\big]\label{step}
\ee
Now in the above expression the $\int d^{4}\ell_{2}$ integral can be evaluated first by performing $\ell_{2}^{0}$ integration using contour prescription and then by performing the angular integral and finally doing the $\int_{\omega}^{|\vec{\ell}_{1}|}\f{d|\vec{\ell}_{2}|}{|\vec{\ell}_{2}|}$ integral. In the result of the $\int d^{4}\ell_{2}$ integration, the $\vec{\ell}_{1}$ dependence turns out $\ln(|\vec{\ell}_{1}| \omega^{-1})$. Then following the same steps for $\int d^{4}\ell_{1}$, the omega dependence appears as $\omega\times\int_{\omega}^{L^{-1}} \f{d |\vec{\ell}_{1}|}{|\vec{\ell}_{1}|} \ln(|\vec{\ell}_{1}|\omega^{-1}) = \f{1}{2}\omega\big\lbrace\ln(\omega L)\big\rbrace^{2}$. Clearly the origin of the half factor is due to the upper limit of $\int d^{4}\ell_{2}$ integration being $|\vec{\ell}_{1}|$, which comes from the momenta ordering $|\ell_{1}^{\mu}|>>|\ell_{2}^{\mu}|$. This suggests that we can also express the contribution of $\Delta^{(1)}_{(2)}\widehat{J}_{\mu}(k)$, approximated in the integration region $L^{-1}>>|\ell_{1}^{\mu}|>>|\ell_{2}^{\mu}|>>\omega$, by both $\int d^{4}\ell_{1}$ and $\int d^{4}\ell_{2}$ integrals having lower limit $\omega$ and upper limit $L^{-1}$ with including an overall multiplicative factor of $1/2$. Following this equivalence prescription of organizing the two momentum integrations we get,
\begingroup
\allowdisplaybreaks
\be
\Delta^{(1)}_{(2)}\widehat{J}_{\mu}(k)\ &\simeq &\ - \f{i}{2}\ \sum_{a=1}^{M+N}\ \sum_{\substack{b=1\\ b\neq a}}^{M+N}\sum_{\substack{c=1\\ c\neq a}}^{M+N}q_{a}^{3}q_{b}q_{c} \int_{\omega}^{L^{-1}} \f{d^{4}\ell_{1}}{(2\pi)^{4}}\ G_{r}(\ell_{1})\ \f{1}{\ell_{1}.p_{b}-i\epsilon}\ \f{1}{[\ell_{1}.p_{a}+i\epsilon]^{2}}\ \big[\ell_{1}.kp_{a}.p_{b}-\ell_{1}.p_{a}p_{b}.k\big]\non\\
&&\times \int_{\omega}^{L^{-1}}\f{d^{4}\ell_{2}}{(2\pi)^{4}}\ G_{r}(\ell_{2})\ \f{1}{\ell_{2}.p_{c}-i\epsilon}\ \f{1}{[\ell_{2}.p_{a}+i\epsilon]^{2}}\ \big[\ell_{2\mu}p_{a}.p_{c}-\ell_{2}.p_{a}p_{c\mu}\big]\non\\
&=&\ \f{i}{2}\ \sum_{a=1}^{M+N}q_{a}\ \Bigg[\f{\p}{\p p_{a}^{\mu}}\ K_{em}^{cl}\Bigg]\ \Bigg[ k_{\sigma}\f{\p}{\p p_{a\sigma}}K_{em}^{cl}\Bigg]\label{deltaJ21}
%&=&\ -\ \f{i}{2}\ \f{1}{(4\pi)^{2}}\sum_{a=1}^{m+n}\ \omega\big[ln\lbrace L(\omega+i\epsilon\eta_{a})\rbrace\big]^{2}\ q_{a}^{3}\sum_{\substack{b\neq a\\ \eta_{a}\eta_{b}=1}}\f{q_{b}p_{b}^{2}}{[(p_{a}.p_{b})^{2}-p_{a}^{2}p_{b}^{2}]^{3/2}}\Big{\lbrace} p_{a}^{2}p_{b}.n -p_{a}.n p_{a}.p_{b}\Big{\rbrace}\non\\
%&&\times \sum_{\substack{c\neq a\\ \eta_{a}\eta_{c}=1}} \f{q_{c}p_{c}^{2}}{[(p_{a}.p_{c})^{2}-p_{a}^{2}p_{c}^{2}]^{3/2}}\ \Big{\lbrace}p_{a}^{2}p_{c\mu}-p_{a}.p_{c}p_{a\mu}\Big{\rbrace}
\ee
\endgroup
Similarly the second term in the sub-subleading current density expression \eqref{sub-subleadingC} contributes to,
\begingroup
\allowdisplaybreaks
\be
\Delta^{(2)}_{(2)}\widehat{J}_{\mu}(k)\ &=&\ \f{1}{2}\sum_{a=1}^{M+N}q_{a}\ e^{-ik.r_{a}}\ \int_{0}^{\infty}d\sigma e^{-i(k.v_{a}-i\epsilon)\sigma}\ \Big(-ik\cdot \Delta_{(1)}Y_{a}(\sigma)\Big)^{2}\ v_{a\mu}\non\\
&=&\ \sum_{a=1}^{M+N}q_{a}e^{-ik.r_{a}} \f{1}{i(k.v_{a}-i\epsilon)}\ \int_{0}^{\infty}d\sigma e^{-i(k.v_{a}-i\epsilon)\sigma}\ \Big{\lbrace}-ik\cdot \Delta_{(1)}Y_{a}(\sigma)\Big{\rbrace}\ \Big{\lbrace}-ik\cdot \f{d\Delta_{(1)}Y_{a}(\sigma)}{d\sigma}\Big{\rbrace}\ v_{a\mu}\non\\
&=&\ i\sum_{a=1}^{M+N}\sum_{\substack{b=1\\ b\neq a}}^{M+N}\sum_{\substack{c=1\\ c\neq a}}^{M+N}\ q_{a}^{3}q_{b}q_{c}\ \f{p_{a\mu}}{p_{a}\cdot k}e^{-ik.r_{a}}\ \int\f{d^{4}\ell_{1}}{(2\pi)^{4}}\ e^{i\ell_{1}.(r_{a}-r_{b})}\ G_{r}(\ell_{1})\ \f{1}{\ell_{1}.p_{a}+i\epsilon}\ \f{1}{\ell_{1}.p_{b}-i\epsilon}\non\\
&&\times \Big{\lbrace} k.\ell_{1}\ p_{a}.p_{b}-p_{a}.\ell_{1}\ p_{b}.k\Big{\rbrace}\ \int\f{d^{4}\ell_{2}}{(2\pi)^{4}}\ e^{i\ell_{2}.(r_{a}-r_{c})}\ G_{r}(\ell_{2})\ \f{1}{\ell_{2}.p_{a}+i\epsilon}\ \f{1}{\ell_{2}.p_{c}-i\epsilon}\non\\
&&\times \Big{\lbrace} k.\ell_{2}\ p_{a}.p_{c}-p_{a}.\ell_{2}\ p_{c}.k\Big{\rbrace}\ \times \f{1}{(\ell_{2}-k).p_{a}+i\epsilon}\ \f{1}{(\ell_{1}+\ell_{2}-k).p_{a}+i\epsilon}
\ee
\endgroup
Above to get second line of RHS from the first line we have first written $e^{-i(k.v_{a}-i\epsilon)\sigma}=\big{\lbrace}-i(k.v_{a}-i\epsilon)\big{\rbrace}^{-1}\f{d}{d\sigma}e^{-i(k.v_{a}-i\epsilon)\sigma}$, then performed integration by parts to move the $\sigma$ derivative to the rest of the integrand after using the boundary conditions of eq.\eqref{boundaryEM1}. To get the last three lines of RHS from the second line we have used the result of the integration \eqref{integration}. Here also the order $\mathcal{O}\big(\omega(\ln\omega)^{2}\big)$ contribution comes from the integration region $L^{-1}>>|\ell_{1}^{\mu}|>>|\ell_{2}^{\mu}|>>\omega$. So using the same trick as described below eq.\eqref{step} we get,
\begingroup
\allowdisplaybreaks
\be
\Delta^{(2)}_{(2)}\widehat{J}_{\mu}(k)\ &\simeq &\ \f{i}{2}\sum_{a=1}^{M+N}\ \sum_{\substack{b=1\\ b\neq a}}^{M+N}\sum_{\substack{c=1\\ c\neq a}}^{M+N}\ q_{a}^{3}q_{b}q_{c}\ \f{p_{a\mu}}{p_{a}\cdot k}\ \int_{\omega}^{L^{-1}}\f{d^{4}\ell_{1}}{(2\pi)^{4}}\  G_{r}(\ell_{1})\ \f{1}{\big[\ell_{1}.p_{a}+i\epsilon\big]^{2}}\ \f{1}{\ell_{1}.p_{b}-i\epsilon}\non\\
&&\times \Big{\lbrace} k.\ell_{1}\ p_{a}.p_{b}-p_{a}.\ell_{1}\ p_{b}.k\Big{\rbrace}\ \int_{\omega}^{L^{-1}}\f{d^{4}\ell_{2}}{(2\pi)^{4}}\ G_{r}(\ell_{2})\ \f{1}{\big[\ell_{2}.p_{a}+i\epsilon\big]^{2}}\ \f{1}{\ell_{2}.p_{c}-i\epsilon}\non\\
&&\times \Big{\lbrace} k.\ell_{2}\ p_{a}.p_{c}-p_{a}.\ell_{2}\ p_{c}.k\Big{\rbrace}\non\\
&=&\  -\f{i}{2}\ \sum_{a=1}^{M+N}q_{a}\ \f{p_{a\mu}}{p_{a}\cdot k}\ \Bigg[ k_{\rho}\f{\p}{\p p_{a\rho}}K_{em}^{cl}\Bigg]\ \Bigg[ k_{\sigma}\f{\p}{\p p_{a\sigma}}K_{em}^{cl}\Bigg]\label{deltaJ22}
\ee
\endgroup
The  third and fourth terms within the square bracket in eq.\eqref{sub-subleadingC} also  contribute to order $\mathcal{O}(\omega(\ln\omega)^{2})$ as explicitly evaluated in appendix-\ref{appA}. Summing the contributions of eq.\eqref{appAJ23} and eq.\eqref{appAJ24} we get,
\be
&&\Delta_{(2)}^{(3)}\widehat{J}_{\mu}(k)\ +\ \Delta_{(2)}^{(4)}\widehat{J}_{\mu}(k)\non\\
&\simeq &\ (-i)\ \sum_{a=1}^{M+N}\Big{\lbrace}\ln\Big((\omega+i\epsilon\eta_{a})L\Big)\Big{\rbrace}^{2}\ k^{\alpha}\mathcal{B}^{(2)}_{\alpha\mu}\big(q_{a},p_{a};\lbrace q_{b}\rbrace ,\lbrace p_{b}\rbrace\big)\label{deltaJ23+24}
\ee
where
\be
\mathcal{B}^{(2)}_{\alpha\mu}\big(q_{a},p_{a};\lbrace q_{b}\rbrace ,\lbrace p_{b}\rbrace\big)\ =\ q_{a}\ \Big[p_{a\mu}\mathcal{C}_{\alpha}\big(q_{a},p_{a};\lbrace q_{b}\rbrace ,\lbrace p_{b}\rbrace\big)-p_{a\alpha}\mathcal{C}_{\mu}\big(q_{a},p_{a};\lbrace q_{b}\rbrace ,\lbrace p_{b}\rbrace\big)\Big]\label{mathcalB}
\ee
with the expression for $\mathcal{C}_{\alpha}\big(q_{a},p_{a};\lbrace q_{b}\rbrace ,\lbrace p_{b}\rbrace\big)$:
\begingroup
\allowdisplaybreaks
\be
&&\mathcal{C}_{\alpha}\big(q_{a},p_{a};\lbrace q_{b}\rbrace ,\lbrace p_{b}\rbrace\big)\non\\
 &=& -\f{1}{4(4\pi)^{2}}\ \sum_{\substack{b=1\\ b\neq a\\ \eta_{a}\eta_{b}=1}}^{M+N}\sum_{\substack{c=1\\ c\neq a\\ \eta_{a}\eta_{c}=1}}^{M+N}q_{a}^{2}\ q_{b}q_{c}\f{1}{[(p_{a}.p_{b})^{2}-p_{a}^{2}p_{b}^{2}]^{5/2}}\ \f{1}{[(p_{a}.p_{c})^{2}-p_{a}^{2}p_{c}^{2}]^{3/2}}\ p_{b}^{2}p_{c}^{2}p_{a}.p_{b}\non\\
&&\times \Big[\lbrace p_{a}.p_{c}p_{a\alpha}-p_{a}^{2}p_{c\alpha}\rbrace\lbrace (p_{a}.p_{b})^{2}-p_{a}^{2}p_{b}^{2}\rbrace +3\lbrace p_{a}^{2}p_{b\alpha}-p_{a}.p_{b}p_{a\alpha}\rbrace\lbrace p_{a}.p_{b}p_{a}.p_{c}-p_{a}^{2}p_{b}.p_{c}\rbrace\Big]\non\\
&&\ - \f{1}{4(4\pi)^{2}}\ \sum_{\substack{b=1\\ b\neq a\\ \eta_{a}\eta_{b}=1}}^{M+N}\sum_{\substack{c=1\\ c\neq b\\ \eta_{b}\eta_{c}=1}}^{M+N}q_{a}q_{b}^{2}q_{c}\f{1}{[(p_{b}.p_{c})^{2}-p_{b}^{2}p_{c}^{2}]^{3/2}}\ \f{1}{[(p_{a}.p_{b})^{2}-p_{a}^{2}p_{b}^{2}]^{5/2}}\ (p_{b}^{2})^{2}p_{c}^{2}\non\\
&&\times\Big[ \lbrace (p_{a}.p_{b})^{2}-p_{a}^{2}p_{b}^{2}\rbrace \lbrace p_{a}.p_{b}p_{c\alpha}-p_{a}.p_{c}p_{b\alpha}\rbrace\ + 3\ \lbrace p_{b}^{2}p_{a}.p_{b}p_{a}.p_{c}p_{a\alpha}\ +p_{a}^{2}p_{b}.p_{c}p_{a}.p_{b}p_{b\alpha}\non\\
&&\ -(p_{a}.p_{b})^{2}p_{b}.p_{c}p_{a\alpha}-p_{a}^{2}p_{b}^{2}p_{a}.p_{c}p_{b\alpha}\rbrace\Big]
\ee
\endgroup
Now summing over the contributions of eq.\eqref{deltaJ21}, \eqref{deltaJ22} and \eqref{deltaJ23+24}, we get the following expression for the total sub-subleading current density at order $\mathcal{O}(\omega(\ln\omega)^{2})$,
\be
\Delta_{(2)}\widehat{J}_{\mu}(k)\ &= & \ \Delta^{(1)}_{(2)}\widehat{J}_{\mu}(k)\ +\ \Delta^{(2)}_{(2)}\widehat{J}_{\mu}(k)+\Delta_{(2)}^{(3)}\widehat{J}_{\mu}(k)\ +\ \Delta_{(2)}^{(4)}\widehat{J}_{\mu}(k)\non\\
&=&\ - \f{i}{2}\ \sum_{a=1}^{M+N}\Bigg[q_{a}\f{k^{\rho}}{p_{a}\cdot k}\ \Bigg(p_{a\mu}\f{\p}{\p p_{a}^{\rho}}\ -\ p_{a\rho}\f{\p}{\p p_{a}^{\mu}} \Bigg)K_{em}^{cl} \Bigg]\ \Bigg[  k_{\sigma}\f{\p}{\p p_{a\sigma}}K_{em}^{cl}\Bigg]\non\\
&&\ -i\ \sum_{a=1}^{M+N}\Big{\lbrace}\ln\Big((\omega+i\epsilon\eta_{a})L\Big)\Big{\rbrace}^{2}\ k^{\alpha}\mathcal{B}^{(2)}_{\alpha\mu}\big(q_{a},p_{a};\lbrace q_{b}\rbrace ,\lbrace p_{b}\rbrace\big)\ +\mathcal{O}(\omega\ln\omega)
\ee
Now using the relation \eqref{AJrelation}, the radiative component of the sub-subleading order electromagnetic waveform becomes,
\begingroup
\allowdisplaybreaks
\be
\Delta_{(2)}\widetilde{A}_{\mu}(\omega ,R,\hat{n}) &=&\ i\omega \lbrace\ln(\omega+i\epsilon)\rbrace^{2}\ \f{1}{4\pi R}\ e^{i\omega R}\Bigg[\f{1}{2}\sum_{a=1}^{N}q_{a}\sum_{\substack{b=1\\ b\neq a}}^{N}\f{q_{a}q_{b}}{4\pi}\f{p_{a}^{2}p_{b}^{2}}{[(p_{a}.p_{b})^{2}-p_{a}^{2}p_{b}^{2}]^{3/2}}\ \f{\mathbf{n}^{\rho}}{p_{a}.\mathbf{n}}\non\\
&&\times\Big{\lbrace}p_{a\mu}p_{b\rho}-p_{a\rho}p_{b\mu}\Big{\rbrace}\sum_{\substack{c=1\\ c\neq a}}^{N}\f{q_{a}q_{c}}{4\pi}\f{p_{c}^{2}}{[(p_{a}.p_{c})^{2}-p_{a}^{2}p_{c}^{2}]^{3/2}}\Big{\lbrace}p_{a}^{2}p_{c}.\mathbf{n}-p_{a}.p_{c}p_{a}.\mathbf{n}\Big{\rbrace}\non\\
&&\ -\sum_{a=1}^{N}\mathbf{n}^{\alpha}\mathcal{B}^{(2)}_{\alpha\mu}\big(q_{a},p_{a};\lbrace q_{b}\rbrace ,\lbrace p_{b}\rbrace\big)\Bigg]\non\\
&&\ -\  i\omega \lbrace\ln(\omega -i\epsilon)\rbrace^{2}\ \f{1}{4\pi R}\ e^{i\omega R}\Bigg[\f{1}{2}\sum_{a=1}^{M}q'_{a}\sum_{\substack{b=1\\ b\neq a}}^{M}\f{q'_{a}q'_{b}}{4\pi}\f{p_{a}^{\prime 2}p_{b}^{\prime 2}}{[(p'_{a}.p'_{b})^{2}-p_{a}^{\prime 2}p_{b}^{\prime 2}]^{3/2}}\ \f{\mathbf{n}^{\rho}}{p'_{a}.\mathbf{n}}\non\\
&&\times\Big{\lbrace}p'_{a\mu}p'_{b\rho}-p'_{a\rho}p'_{b\mu}\Big{\rbrace}\sum_{\substack{c=1\\ c\neq a}}^{M}\f{q'_{a}q'_{c}}{4\pi}\f{p_{c}^{\prime 2}}{[(p'_{a}.p'_{c})^{2}-p_{a}^{\prime 2}p_{c}^{\prime 2}]^{3/2}}\Big{\lbrace}p_{a}^{\prime 2}p'_{c}.\mathbf{n}-p'_{a}.p'_{c}p'_{a}.\mathbf{n}\Big{\rbrace}\non\\
&&\ -\sum_{a=1}^{M}\mathbf{n}^{\alpha}\mathcal{B}^{(2)}_{\alpha\mu}\big(q'_{a},p'_{a};\lbrace q'_{b}\rbrace ,\lbrace p'_{b}\rbrace\big)\Bigg]\hspace{1cm} +\ \mathcal{O}(\omega\ln\omega)
\ee
\endgroup
Above the expression of $\mathcal{B}^{(2)}_{\alpha\mu}\big(q_{a},p_{a};\lbrace q_{b}\rbrace ,\lbrace p_{b}\rbrace\big)$ is same as the one given in eq.\eqref{mathcalB} but with the particle sums running over only outgoing objects and the expression of $\mathcal{B}^{(2)}_{\alpha\mu}\big(q'_{a},p'_{a};\lbrace q'_{b}\rbrace ,\lbrace p'_{b}\rbrace\big)$ takes the same functional form as in eq.\eqref{mathcalB} with the sums running over only incoming objects. Now taking Fourier transformation in $\omega$ variable and using the results of integration \eqref{FourierFn0} and \eqref{FourierF0n} for $n=2$, we find the following late and early time sub-subleading electromagnetic waveform:
 \begingroup
\allowdisplaybreaks
\be
\Delta_{(2)}A_{\mu}(t,R,\hat{n})\ &=&\ \f{1}{4\pi R}\ \f{\ln |u|}{u^{2}}\ \Bigg[\sum_{a=1}^{N}q_{a}\sum_{\substack{b=1\\ b\neq a}}^{N}\f{q_{a}q_{b}}{4\pi}\f{p_{a}^{2}p_{b}^{2}}{[(p_{a}.p_{b})^{2}-p_{a}^{2}p_{b}^{2}]^{3/2}}\ \f{\mathbf{n}^{\rho}}{p_{a}.\mathbf{n}}\non\\
&&\times\Big{\lbrace}p_{a\mu}p_{b\rho}-p_{a\rho}p_{b\mu}\Big{\rbrace}\sum_{\substack{c=1\\ c\neq a}}^{N}\f{q_{a}q_{c}}{4\pi}\f{p_{c}^{2}}{[(p_{a}.p_{c})^{2}-p_{a}^{2}p_{c}^{2}]^{3/2}}\Big{\lbrace}p_{a}^{2}p_{c}.\mathbf{n}-p_{a}.p_{c}p_{a}.\mathbf{n}\Big{\rbrace}\non\\
&&\ -2\sum_{a=1}^{N}\mathbf{n}^{\alpha}\mathcal{B}^{(2)}_{\alpha\mu}\big(q_{a},p_{a};\lbrace q_{b}\rbrace ,\lbrace p_{b}\rbrace\big)\Bigg]\ +\ \mathcal{O}(u^{-2}) \hspace{1cm}\hbox{for\ $u\rightarrow +\infty$,}
\ee
\be
\Delta_{(2)}A_{\mu}(t,R,\hat{n})\ &=&\ \f{1}{4\pi R}\ \f{\ln |u|}{u^{2}}\ \Bigg[\sum_{a=1}^{M}q'_{a}\sum_{\substack{b=1\\ b\neq a}}^{M}\f{q'_{a}q'_{b}}{4\pi}\f{p_{a}^{\prime 2}p_{b}^{\prime 2}}{[(p'_{a}.p'_{b})^{2}-p_{a}^{\prime 2}p_{b}^{\prime 2}]^{3/2}}\ \f{\mathbf{n}^{\rho}}{p'_{a}.\mathbf{n}}\non\\
&&\times\Big{\lbrace}p'_{a\mu}p'_{b\rho}-p'_{a\rho}p'_{b\mu}\Big{\rbrace}\sum_{\substack{c=1\\ c\neq a}}^{M}\f{q'_{a}q'_{c}}{4\pi}\f{p_{c}^{\prime 2}}{[(p'_{a}.p'_{c})^{2}-p_{a}^{\prime 2}p_{c}^{\prime 2}]^{3/2}}\Big{\lbrace}p_{a}^{\prime 2}p'_{c}.\mathbf{n}-p'_{a}.p'_{c}p'_{a}.\mathbf{n}\Big{\rbrace}\non\\
&&\ -2\sum_{a=1}^{M}\mathbf{n}^{\alpha}\mathcal{B}^{(2)}_{\alpha\mu}\big(q'_{a},p'_{a};\lbrace q'_{b}\rbrace ,\lbrace p'_{b}\rbrace\big)\Bigg] \ +\mathcal{O}(u^{-2})\hspace{1cm}\hbox{for\ $u\rightarrow -\infty$},
\ee
\endgroup
with $\vec{x}=R\hat{n}$, $\mathbf{n}=(1,\hat{n})$ and $u=t-R$.

\subsection{Structure of (sub)$^{n}$-leading electromagnetic waveform}\label{SsubnEM}
In this subsection we derive the leading non-analytic contribution of (sub)$^{n}$-leading electromagnetic waveform  up to some un-determined gauge invariant function. Before analyzing the contribution of (sub)$^{n}$-leading current density, let us list some useful observations from the last three subsections.
\begin{enumerate}
\item If we compare the expression of leading current density in eq.\eqref{leadingC} to the first term within the square bracket of subleading current density in eq.\eqref{subleadingC}, we find an extra factor of $\Big{\lbrace} -ik\cdot \Delta_{(1)}Y_{a}(\sigma)\Big{\rbrace}$ in the integrand of the first term of eq.\eqref{subleadingC}  relative to eq.\eqref{leadingC}. The effect of this extra factor produces an extra contribution of $k^{\rho}\f{\p}{\p p_{a}^{\rho}}K_{em}^{cl}$ relative to the contribution of \eqref{leadingC} as we can see from the contribution of eq.\eqref{Jhat11}. Similarly eq.\eqref{Jhat21} contains an extra factor of $\Big{\lbrace} -ik\cdot \Delta_{(1)}Y_{a}(\sigma)\Big{\rbrace}$ relative to eq.\eqref{Jhat12} and this gives an extra contribution of $k^{\rho}\f{\p}{\p p_{a}^{\rho}}K_{em}^{cl}$ in eq.\eqref{hatJ21} relative to eq.\eqref{Jhat12} with a numerical factor $\f{1}{2}$ due to momentum ordering of two integrals.
 \item For the analysis of the terms containing $\Delta_{(2)}Y_{a}(\sigma)$ in the integrand of sub-subleading current density in eq.\eqref{sub-subleadingC}, we need to use the expression of subleading gauge field $\Delta_{(1)}A_{\mu}(x)$ as explicitly analyzed in appendix-\ref{appA}. We also found that sum of the contributions from those terms is gauge invariant by itself. So we expect that in the analysis of (sub)$^{n}$-leading current density the part of the integrand containing $\Delta_{(r)}Y_{a}(\sigma)$ for $r\geq 2$ similarly contributes to order $\mathcal{O}\big(\omega^{n-1}(\ln\omega)^{n}\big)$ and the contribution will be gauge invariant by itself. But we will not be able to analyze those terms since for that we need the expressions of the gauge fields $\Big{\lbrace}\Delta_{(r)}A_{\mu}(x)\Big{\rbrace}$ for $r=1,...,n-1$, which we don't have yet beyond $r=2$. So we write the contribution from the analysis of those terms in terms of some undetermined function of momenta and charges, which can be derived by the method of induction.
\end{enumerate}
The (sub)$^{n}$-leading current density from eq.\eqref{FourierJk} turns out to be,
\be
\Delta_{(n)}\widehat{J}_{\mu}(k)\ &= &\  \sum_{a=1}^{M+N}q_{a}\ e^{-ik.r_{a}}\ \int_{0}^{\infty}d\sigma e^{-i(k.v_{a}-i\epsilon)\sigma}\ \Bigg[ \f{1}{(n-1)!}\Big{\lbrace} -ik\cdot \Delta_{(1)}Y_{a}(\sigma)\Big{\rbrace}^{n-1}\ \f{d\Delta_{(1)}Y_{a\mu}(\sigma)}{d\sigma} \non\\
&&\ +\f{1}{n!}\Big{\lbrace}-ik\cdot \Delta_{(1)}Y_{a}(\sigma)\Big{\rbrace}^{n}\ v_{a\mu}\Bigg{]}\ +\ \Delta_{(n)}^{rest}\widehat{J}_{\mu}(k)\label{Jnfull}
\ee
where $\Delta_{(n)}^{rest}\widehat{J}_{\mu}(k)$ contains terms $\Delta_{(r)}Y_{a}(\sigma)$ for $r\geq 2$ within the $\sigma$ integral. Now for the second term within the square bracket above, if we first write $e^{-i(k.v_{a}-i\epsilon)\sigma}=\big{\lbrace}-i(k.v_{a}-i\epsilon)\big{\rbrace}^{-1}\f{d}{d\sigma}e^{-i(k.v_{a}-i\epsilon)\sigma}$ and  perform integration by parts to move the $\sigma$ derivative to the rest of the integrand after using the boundary conditions of eq.\eqref{boundaryEM1} we get,
\be
\Delta_{(n)}\widehat{J}_{\mu}(k)\ &\simeq &\  \f{1}{(n-1)!}\ \sum_{a=1}^{M+N}q_{a}\ e^{-ik.r_{a}}\ \int_{0}^{\infty}d\sigma e^{-i(k.v_{a}-i\epsilon)\sigma}\ \Big{\lbrace} -ik\cdot \Delta_{(1)}Y_{a}(\sigma)\Big{\rbrace}^{n-1}\ \Bigg[ \ \f{d\Delta_{(1)}Y_{a\mu}(\sigma)}{d\sigma}\non\\
&& +\f{1}{i(k.v_{a}-i\epsilon)}\ \Big{\lbrace}-ik\cdot \f{d\Delta_{(1)}Y_{a}(\sigma)}{d\sigma}\Big{\rbrace}\ v_{a\mu}\Bigg{]} +\ \Delta_{(n)}^{rest}\widehat{J}_{\mu}(k)\label{subncurrent}
\ee
The first term within the square bracket contributes to,
\be
&&\Delta^{(1)}_{(n)}\widehat{J}_{\mu}(k)\non\\
 &= &\ \f{1}{(n-1)!}\sum_{a=1}^{M+N}q_{a}\ e^{-ik.r_{a}}\ \int_{0}^{\infty}d\sigma e^{-i(k.v_{a}-i\epsilon)\sigma}\ \prod_{i=1}^{n-1}\Bigg[(-i)\ \f{q_{a}q_{b_{i}}}{m_{a}}\sum_{\substack{b_{i}=1\\ b_{i}\neq a}}^{M+N}\int \f{d^{4}\ell_{i}}{(2\pi)^{4}}\non\\
&&\times e^{i\ell_{i}.(r_{a}-r_{b_{i}})}\ G_{r}(\ell_{i})\ \f{1}{\ell_{i}.p_{b_{i}}-i\epsilon}\ \f{1}{(\ell_{i}.v_{a}+i\epsilon)^{2}}\ \Big{\lbrace}\ell_{i}.k p_{b_{i}}.v_{a}-\ell_{i}.v_{a}p_{b_{i}}.k\Big{\rbrace}\ \Big[e^{i\ell_{i}.v_{a}\sigma}-1\Big]\ \Bigg]\non\\
&&\times \sum_{\substack{b_{n}=1\\ b_{n}\neq a}}^{M+N}\f{iq_{a}q_{b_{n}}}{m_{a}}\ \int\f{d^{4}\ell_{n}}{(2\pi)^{4}}\ e^{i\ell_{n}.(r_{a}-r_{b_{n}})}G_{r}(\ell_{n})\ \f{1}{\ell_{n}.p_{b_{n}}-i\epsilon}\ \f{1}{\ell_{n}.v_{a}+i\epsilon}\non\\
&&\ \times\Big{\lbrace}\ell_{n\mu}p_{b_{n}}.v_{a}-\ell_{n}.v_{a}p_{b_{n}\mu}\Big{\rbrace}\ e^{i\ell_{n}.v_{a}\sigma}\label{subnJfirst}
\ee
Now the above expression contributes to order $\mathcal{O}\big(\omega^{n-1}(\ln\omega)^{n}\big)$ in $(n-1)!$ numbers of integration region: $L^{-1}>> \mathcal{P} \big\lbrace \ell_{1}>>\ell_{2}>>\cdots >>\ell_{n-1}\big\rbrace>>\ell_{n}>>\omega$, where $\mathcal{P}$ denotes all possible permutations of $\ell_{i}$'s within the parentheses. In any of this $(n-1)!$ integration regions the leading contribution in $\omega$ expansion from the following integration becomes,
\be
\int_{0}^{\infty}d\sigma e^{-i(k.v_{a}-i\epsilon)\sigma}\prod_{i=1}^{n-1}\Big[e^{i(\ell_{i}.v_{a}+i\epsilon)\sigma}-1\Big]\ e^{i\ell_{n}.v_{a}\sigma}\simeq \ i(-1)^{n-1}\ \f{1}{\ell_{n}.v_{a}+i\epsilon}
\ee
Now substituting the result of the above integral in eq.\eqref{subnJfirst}, we find same contribution from all the integration regions. So the total contribution will be $(n-1)!$ times the contribution from one of the region, say $L^{-1}>> \ell_{1}>>\ell_{2}>>\cdots >>\ell_{n-1}>>\ell_{n}>>\omega$. Now using the procedure described below eq.\eqref{step}, the expression can also be written with all the $n$ number of integrating momenta run from $\omega$ to $L^{-1}$ by the cost of an overall $\f{1}{n!}$. Performing all these steps we find,
\be
\Delta^{(1)}_{(n)}\widehat{J}_{\mu}(k)\ &\simeq &\ \f{i}{n!}\sum_{a=1}^{M+N}q_{a}\   \prod_{i=1}^{n-1}\Bigg[i\ q_{a}q_{b_{i}}\sum_{\substack{b_{i}=1\\ b_{i}\neq a}}^{M+N}\int_{\omega}^{L^{-1}} \f{d^{4}\ell_{i}}{(2\pi)^{4}}\ G_{r}(\ell_{i})\ \f{1}{\ell_{i}.p_{b_{i}}-i\epsilon}\ \f{1}{(\ell_{i}.p_{a}+i\epsilon)^{2}}\non\\
&&\times \Big{\lbrace}\ell_{i}.k p_{b_{i}}.p_{a}-\ell_{i}.p_{a}p_{b_{i}}.k\Big{\rbrace}\Bigg] \ \sum_{\substack{b_{n}=1\\ b_{n}\neq a}}^{M+N}(iq_{a}q_{b_{n}})\ \int_{\omega}^{L^{-1}}\f{d^{4}\ell_{n}}{(2\pi)^{4}}\ G_{r}(\ell_{n})\ \f{1}{\ell_{n}.p_{b_{n}}-i\epsilon}\non\\
&&\times \f{1}{(\ell_{n}.p_{a}+i\epsilon)^{2}}\Big{\lbrace}\ell_{n\mu}p_{b_{n}}.p_{a}-\ell_{n}.p_{a}p_{b_{n}\mu}\Big{\rbrace}\non\\
&=&\ \f{i}{n!} \sum_{a=1}^{M+N}q_{a}\ \Bigg[\f{\p}{\p p_{a}^{\mu}}\ K_{em}^{cl}\Bigg]\ \Bigg[ k_{\sigma}\f{\p}{\p p_{a\sigma}}K_{em}^{cl}\Bigg]^{n-1}
\ee
Now if we compare the above expression with the second term in the sub-leading current density in eq.\eqref{Jhat12}, we find that in the above expression the integrand contains an extra multiplicative factor of $\f{1}{(n-1)!}\Big{\lbrace} -ik\cdot \Delta_{(1)}Y_{a}(\sigma)\Big{\rbrace}^{n-1}$ relative to the integrand of eq.\eqref{Jhat12}. This suggest to include an extra factor of $\f{1}{n!}\Bigg(k^{\sigma}\f{\p}{\p p_{a}^{\sigma}}K_{em}^{cl}\Bigg)^{n-1}$ in the contribution of $\Delta_{(n)}\widehat{J}_{\mu}(k)$ relative to the contribution of $\Delta_{(1)}^{(2)}\widehat{J}_{\mu}(k)$ following our first observation in the beginning of this subsection. So the above derivation gives a direct generalization of the first observation made in the beginning of this subsection. An analogous analysis for the second term within the square bracket of eq.\eqref{subncurrent}  gives the following contribution at order $\mathcal{O}\big(\omega^{n-1}(\ln\omega)^{n}\big)$,
\be
\Delta^{(2)}_{(n)}\widehat{J}_{\mu}(k)\ &\simeq & \ -i\ \f{1}{n!}\sum_{a=1}^{M+N}q_{a}\f{p_{a\mu}}{p_{a}.k}\ \Bigg[ k_{\sigma}\f{\p}{\p p_{a\sigma}}K_{em}^{cl}\Bigg]^{n}
\ee
The contribution of $\Delta_{(n)}^{rest}\widehat{J}_{\mu}(k)$ has not been evaluated explicitly due to the reason explained in the second observation in the beginning of this subsection but the general form follows from gauge invariance\cite{1801.05528,1802.03148,1810.04619}, 
\be
\Delta_{(n)}^{rest}\widehat{J}_{\mu}(k)&\simeq & (-i)\sum_{a=1}^{M+N}\omega^{n-1}\Big{\lbrace}\ln\Big((\omega+i\epsilon\eta_{a})L\Big)\Big{\rbrace}^{n}\ \mathbf{n}^{\alpha_{1}}\mathbf{n}^{\alpha_{2}}\cdots\mathbf{n}^{\alpha_{n-1}}\mathcal{B}^{(n)}_{\alpha_{1}\alpha_{2}\cdots\alpha_{n-1}\mu}\big(q_{a},p_{a};\lbrace q_{b}\rbrace ,\lbrace p_{b}\rbrace\big)\non\\
\ee
with the property that $\mathcal{B}^{(n)}_{\alpha_{1}\alpha_{2}\cdots\alpha_{n-1}\mu}\big(q_{a},p_{a};\lbrace q_{b}\rbrace ,\lbrace p_{b}\rbrace\big)$ is anti-symmetric in $\mu$ and any one of the $\alpha_{r}$ indices for $r=1,\cdots ,n-1$. We expect that $\mathcal{B}^{(n)}_{\alpha_{1}\alpha_{2}\cdots\alpha_{n-1}\mu}\big(q_{a},p_{a};\lbrace q_{b}\rbrace ,\lbrace p_{b}\rbrace\big)$ will be independent of the details of the scattering event as well as will be independent of the nature of short range forces acting between the scattered objects inside region $\mathcal{R}$. After summing all the contributions, the $\mathcal{O}\big(\omega^{n-1}(\ln\omega)^{n}\big)$ contribution of $\Delta_{(n)}\widehat{J}_{\mu}(k)$ becomes,
 \begingroup
\allowdisplaybreaks
\be
\Delta_{(n)}\widehat{J}_{\mu}(k)&\simeq &-i\sum_{a=1}^{M+N}\Bigg[q_{a}\f{k^{\rho}}{p_{a}\cdot k}\ \Bigg(p_{a\mu}\f{\p}{\p p_{a}^{\rho}}\ -\ p_{a\rho}\f{\p}{\p p_{a}^{\mu}} \Bigg)K_{em}^{cl}\Bigg] \times \f{1}{n!}\Bigg{[}k_{\sigma}\f{\p}{\p p_{a\sigma}}K_{em}^{cl}\Bigg{]}^{n-1} \non\\
&&-i\sum_{a=1}^{M+N}\omega^{n-1}\Big{\lbrace}\ln\Big((\omega+i\epsilon\eta_{a})L\Big)\Big{\rbrace}^{n}\ \mathbf{n}^{\alpha_{1}}\mathbf{n}^{\alpha_{2}}\cdots\mathbf{n}^{\alpha_{n-1}}\mathcal{B}^{(n)}_{\alpha_{1}\alpha_{2}\cdots\alpha_{n-1}\mu}\big(q_{a},p_{a};\lbrace q_{b}\rbrace ,\lbrace p_{b}\rbrace\big)\non\\
&=&\ (-i)\ \sum_{a=1}^{M+N}\omega^{n-1}\Big(\ln\Big{\lbrace}L(\omega +i\epsilon\eta_{a})\Big{\rbrace}\Big)^{n}\Bigg[\f{i}{4\pi}\sum_{\substack{b=1\\b\neq a\\ \eta_{a}\eta_{b}=1}}^{M+N}q_{a}^{2}q_{b}\ \f{p_{a}^{2}p_{b}^{2}}{[(p_{a}.p_{b})^{2}-p_{a}^{2}p_{b}^{2}]^{3/2}}\non\\
 &&\ \times \f{\mathbf{n}^{\rho}}{p_{a}.\mathbf{n}}\Big{\lbrace}p_{a\mu}p_{b\rho}-p_{a\rho}p_{b\mu}\Big{\rbrace}\Bigg]\times \f{1}{n!}\Bigg[ \f{i}{4\pi}\sum_{\substack{c=1\\c\neq a\\ \eta_{a}\eta_{c}=1}}^{M+N}q_{a}q_{c}\f{p_{c}^{2}}{[(p_{a}.p_{c})^{2}-p_{a}^{2}p_{c}^{2}]^{3/2}}\Big{\lbrace}p_{a}^{2}p_{c}.\mathbf{n}-p_{a}.p_{c}p_{a}.\mathbf{n}\Big{\rbrace}\Bigg]^{n-1}\non\\
 &&-i\sum_{a=1}^{M+N}\omega^{n-1}\Big{\lbrace}\ln\Big((\omega+i\epsilon\eta_{a})L\Big)\Big{\rbrace}^{n}\ \mathbf{n}^{\alpha_{1}}\mathbf{n}^{\alpha_{2}}\cdots\mathbf{n}^{\alpha_{n-1}}\mathcal{B}^{(n)}_{\alpha_{1}\alpha_{2}\cdots\alpha_{n-1}\mu}\big(q_{a},p_{a};\lbrace q_{b}\rbrace ,\lbrace p_{b}\rbrace\big)\label{SubnJmu}
\ee
\endgroup
Now using the relation of eq.\eqref{AJrelation} at (sub)$^{n}$-leading order, performing Fourier transformation in $\omega$ variable and using the results of Fourier transformations \eqref{FourierFn0},\eqref{FourierF0n} we get the following late and early time electromagnetic waveform,
 \begingroup
\allowdisplaybreaks
\be
&&\Delta_{(n)}A_{\mu}(t,R,\hat{n})\non\\
&=&\ \f{(-1)^{n}}{4\pi R}\ \f{\big(\ln|u|\big)^{n-1}}{u^{n}}\sum_{a=1}^{N}\ \Bigg[\f{1}{4\pi}\sum_{\substack{b=1\\b\neq a}}^{N}q_{a}^{2}q_{b}\ \f{p_{a}^{2}p_{b}^{2}}{[(p_{a}.p_{b})^{2}-p_{a}^{2}p_{b}^{2}]^{3/2}}\ \f{\mathbf{n}^{\rho}}{p_{a}.\mathbf{n}}\Big{\lbrace}p_{a\mu}p_{b\rho}-p_{a\rho}p_{b\mu}\Big{\rbrace}\Bigg]\non\\
&&\times \Bigg[ \f{1}{4\pi}\sum_{\substack{c=1\\c\neq a}}^{N}q_{a}q_{c}\f{p_{c}^{2}}{[(p_{a}.p_{c})^{2}-p_{a}^{2}p_{c}^{2}]^{3/2}}\Big{\lbrace}p_{a}^{2}p_{c}.\mathbf{n}-p_{a}.p_{c}p_{a}.\mathbf{n}\Big{\rbrace}\Bigg]^{n-1}\non\\
&&\ +\f{(i)^{n}}{4\pi R}\f{\big(\ln|u|\big)^{n-1}}{u^{n}}\ n!\sum_{a=1}^{N} \mathbf{n}^{\alpha_{1}}\mathbf{n}^{\alpha_{2}}\cdots\mathbf{n}^{\alpha_{n-1}}\mathcal{B}^{(n)}_{\alpha_{1}\alpha_{2}\cdots\alpha_{n-1}\mu}\big(q_{a},p_{a};\lbrace q_{b}\rbrace ,\lbrace p_{b}\rbrace\big)\non\\
&& +\ \mathcal{O}\Big(u^{-n}(\ln|u|)^{n-2}\Big)\hspace{4cm}\hbox{for}\ u\rightarrow+\infty
\ee
and
\be
&&\Delta_{(n)}A_{\mu}(t,R,\hat{n})\non\\
&=&\ \f{1}{4\pi R}\ \f{\big(\ln|u|\big)^{n-1}}{u^{n}}\sum_{a=1}^{M}\ \Bigg[\f{1}{4\pi}\sum_{\substack{b=1\\b\neq a}}^{M}q_{a}^{\prime 2}q'_{b}\ \f{p_{a}^{\prime 2}p_{b}^{\prime 2}}{[(p'_{a}.p'_{b})^{2}-p_{a}^{\prime 2}p_{b}^{\prime 2}]^{3/2}}\ \f{\mathbf{n}^{\rho}}{p'_{a}.\mathbf{n}}\Big{\lbrace}p'_{a\mu}p'_{b\rho}-p'_{a\rho}p'_{b\mu}\Big{\rbrace}\Bigg]\non\\
&&\times \Bigg[ \f{1}{4\pi}\sum_{\substack{c=1\\c\neq a}}^{M}q'_{a}q'_{c}\f{p_{c}^{\prime 2}}{[(p'_{a}.p'_{c})^{2}-p_{a}^{\prime 2}p_{c}^{\prime 2}]^{3/2}}\Big{\lbrace}p_{a}^{\prime 2}p'_{c}.\mathbf{n}-p'_{a}.p'_{c}p'_{a}.\mathbf{n}\Big{\rbrace}\Bigg]^{n-1}\non\\
&&\ +(-1)^{n}\f{(i)^{n}}{4\pi R}\f{\big(\ln|u|\big)^{n-1}}{u^{n}}\ n!\sum_{a=1}^{M} \mathbf{n}^{\alpha_{1}}\mathbf{n}^{\alpha_{2}}\cdots\mathbf{n}^{\alpha_{n-1}}\mathcal{B}^{(n)}_{\alpha_{1}\alpha_{2}\cdots\alpha_{n-1}\mu}\big(q'_{a},p'_{a};\lbrace q'_{b}\rbrace ,\lbrace p'_{b}\rbrace\big)\non\\
&& +\ \mathcal{O}\Big(u^{-n}(\ln|u|)^{n-2}\Big)\hspace{4cm}\hbox{for}\ u\rightarrow -\infty
\ee
\endgroup
From the reality of electromagnetic waveform, we expect that $(i)^{n}\mathcal{B}^{(n)}$ is real and from momentum and charge power counting $\mathcal{B}^{(n)}\big(-q_{a},-p_{a};\lbrace -q_{b}\rbrace ,\lbrace -p_{b}\rbrace\big)=(-1)^{n-1}\mathcal{B}^{(n)}\big(q_{a},p_{a};\lbrace q_{b}\rbrace ,\lbrace p_{b}\rbrace\big)$.

\section{Proof of classical soft graviton theorem}\label{proofsoftgr}
Consider the same scattering process as described in \S\ref{Ssoftphotonthm} but for simplicity consider the scattered objects are charge neutral. Now the region $\mathcal{R}$ is chosen such that all complicated interactions take place inside spacetime region $\mathcal{R}$ and outside only long range gravitational interaction is present as described in Fig.\ref{Scattering}. We define the scattering center be a spacetime point well inside the region $\mathcal{R}$ of linear size $L$ and place the detector of gravitational wave at distance $R$ from the scattering center. Then our goal is to determine the $\f{1}{R}$ component of gravitational waveform for retarded time $u>>L$. This translates to determining the radiative mode gravitational waveform with frequency $\omega$ in the range $R^{-1}<<\omega <<L^{-1}$.  Let us define the deviation of metric from Minkowski metric as,
\be
h_{\mu\nu}(x)\ \equiv\ \f{1}{2}\Big(g_{\mu\nu}(x)-\eta_{\mu\nu}\Big)\ \hspace{1cm}\ \hbox{and}\hspace{1cm}\ e_{\mu\nu}(x)\ \equiv\ h_{\mu\nu}(x)-\f{1}{2}\eta_{\mu\nu}\ \eta^{\rho\sigma}h_{\rho\sigma}(x)
\ee
 Now following \cite{1611.03493, 1801.07719, 1804.09193, 1906.08288, 1912.06413}, the radiative component of the trace reversed metric fluctuation is related to the Fourier transform of the total energy-momentum tensor $T^{\mu\nu}(x)$  in the following way,
\be
\widetilde{e}^{\mu\nu}(\omega ,R,\hat{n})\ &\simeq &\ \f{2G}{R}\ e^{i\omega R}\  \widehat{T}^{\mu\nu}(k)\label{eTrelation}
\ee
where,
\be
\widetilde{e}^{\mu\nu}(\omega ,\vec{x})\ &=&\ \int_{-\infty}^{\infty}dt\ e^{i\omega t}\ e^{\mu\nu}(t,\vec{x})\\
\widehat{T}^{\mu\nu}(k)\ &=&\ \int d^{4}x\ e^{-ik.x}\ T^{\mu\nu}(t,\vec{x})
\ee
and $\vec{x}=R\hat{n}$ , $k^{\mu}=\omega \mathbf{n}^{\mu}=\omega(1,\hat{n})$ with $\hat{n}$ being the unit vector along the direction of the detector from the scattering center.

\subsection{General setup}
We use the following compact notations for analyzing ingoing and outgoing particles' trajectories as done in \S\ref{GeneralsetupEM},
\be
m_{N+a}\ =\ m'_{a},\hspace{5mm}v_{N+a}\ =\ -v'_{a},\hspace{5mm}p_{N+a}\ =\ -p'_{a},\hspace{5mm}X_{N+a}(\sigma)\ =\ X'_{a}(-\sigma)
\ee
for $a=1,2,\cdots ,M$ and $0\leq \sigma <\infty$. Consider the asymptotic trajectory of a'th object outside the region $\mathcal{R}$,
\be
X_{a}^{\mu}(\sigma)\ =\ r_{a}^{\mu}\ +\ v_{a}^{\mu}\sigma \ +\ Y_{a}^{\mu}(\sigma)
\ee
where $r_{a}^{\mu}$ denotes the point on the boundary of the region $\mathcal{R}$ where particle-a enters or exits region $\mathcal{R}$ at $\sigma=0$. $Y_{a}^{\mu}(\sigma)$ satisfies the following boundary conditions,
\be
Y_{a}^{\mu}(\sigma)\Bigg{|}_{\sigma=0}\ =\ 0\ ,\hspace{1cm}\ \f{dY_{a}^{\mu}(\sigma)}{d\sigma}\Bigg{|}_{\sigma\rightarrow \infty}\ =\ 0\label{BoundaryGR}
\ee
Now we have to solve Einstein equation and geodesic equation\footnote{If the scattered objects have some internal structure and spin the geodesic equation will modify accordingly, but as we discuss in \S\ref{speculation}, they do not affect our result to the order we are working in.} iteratively to determine gravitational waveform in terms of the scattering data.
\be
\sqrt{-\det g}\ \Bigg(R^{\mu\nu}-\f{1}{2}R\ g^{\mu\nu}\Bigg)\ &=&\ 8\pi G\ T^{X\mu\nu}\label{EEq}\\
\f{d^{2}X_{a}^{\mu}(\sigma)}{d\sigma^{2}}\ +\ \Gamma^{\mu}_{\nu\rho}(X_{a}(\sigma))\ \f{dX_{a}^{\nu}(\sigma)}{d\sigma}\ \f{dX_{a}^{\rho}(\sigma)}{d\sigma}\ &=&\ 0\label{geoX}
\ee
where $T^{X\mu\nu}$ is the matter energy-momentum tensor given by\footnote{The matter energy-momentum tensor written here is for non-spinning point like objects. Presence of spin and internal structure will modify the $T^{X}$, but as we  discuss in \S\ref{speculation}, internal structure and spin don't affect the leading non-analytic behaviour of gravitational waveform in small $\omega$ expansion.},
\be
T^{X\mu\nu}(x)\ &=&\ \sum_{a=1}^{M+N}m_{a}\int_{0}^{\infty}d\sigma \ \f{dX_{a}^{\mu}(\sigma)}{d\sigma}\f{dX_{a}^{\nu}(\sigma)}{d\sigma}\ \delta^{(4)}(x-X_{a}(\sigma))\label{TX}
\ee
In de Donder gauge $\eta^{\rho\mu}\p_{\rho}e_{\mu\nu}=0$, the Einstein eq.\eqref{EEq} takes the following form,
\be
\eta^{\rho\sigma}\p_{\rho}\p_{\sigma}e^{\mu\nu}(x)\ &=&\ -8\pi G\ \Big(T^{X\mu\nu}(x)\ +\ T^{h\mu\nu}(x)\Big)\ \equiv \ -8\pi G\ T^{\mu\nu}(x)\label{LinearEE}
\ee
where $T^{h\mu\nu}(x)$ is the gravitational energy-momentum tensor defined as,
\be
T^{h\mu\nu}\ &\equiv &\ -\f{1}{8\pi G}\Bigg[\sqrt{-\det g}\ \Bigg(R^{\mu\nu}-\f{1}{2}R\ g^{\mu\nu}\Bigg)\  +\ \eta^{\rho\sigma}\p_{\rho}\p_{\sigma}e^{\mu\nu}\Bigg]\label{gravEMtensor}
\ee

Here we are following the unit convention $c=1$ and  in this unit $GM\omega$ is a dimensionless quantity, where $\omega$ is frequency of soft gravitational radiation and $M$ is the parameter representing masses of scattered objects\footnote{Here also we don't have to consider the other  parameters e.g. impact parameter $\sim |r_{a}-r_{b}|\sim L$ , Spin $\Sigma_{a}$, internal structure of the objects in terms of multipole moments etc. along with $\omega$ to construct dimensionless expansion parameters due to the analogous reasoning given in footnote-\ref{footEM} for charged particle's scattering.  }. So expansion in power of $\omega$ is same as expansion in power of gravitational constant $G$. Let us expand the Fourier transform of total energy momentum tensor in a power series expansion of $G$,
\be
\widehat{T}^{\mu\nu}(k)\ &=&\ \Delta_{(0)}\widehat{T}^{\mu\nu}(k)\ +\  \Delta_{(1)}\widehat{T}^{\mu\nu}(k)\ +\  \Delta_{(2)}\widehat{T}^{\mu\nu}(k)\ +\cdots
\ee
where $\Delta_{(r)}\widehat{T}^{\mu\nu}(k)$ is of order $(G)^{r}$ in gravitational constant expansion. Similarly the correction to the straight line trajectory has an expansion in power of $G$,
\be
Y_{a}^{\mu}(\sigma)\ &=&\ \ \Delta_{(1)}Y_{a}^{\mu}(\sigma)+\ \Delta_{(2)}Y_{a}^{\mu}(\sigma)\ +\ \Delta_{(3)}Y_{a}^{\mu}(\sigma)\ +\cdots
\ee
where $\Delta_{(r)}Y_{a}^{\mu}(\sigma)$ is of order $(G)^{ r}$ in gravitational constant expansion. This implies that the trace reversed metric fluctuation has the following expansion in power of $G$ if we use \eqref{LinearEE},
\be
e_{\mu\nu}(x)\ &=&\ \Delta_{(0)}e_{\mu\nu}(x)\ +\ \Delta_{(1)}e_{\mu\nu}(x)\ +\ \Delta_{(2)}e_{\mu\nu}(x)\ +\ \Delta_{(3)}e_{\mu\nu}(x)+\cdots
\ee
where
\be
\Delta_{(r)}e_{\mu\nu}(x)\ =\ -8\pi G\int \f{d^{4}\ell}{(2\pi)^{4}}\ G_{r}(\ell)\ e^{i\ell.x}\ \Delta_{(r)}\widehat{T}_{\mu\nu}(\ell) .
\ee
This implies $\Delta_{(r)}e_{\mu\nu}(x)$ is of order $(G)^{ r+1}$ in gravitational constant expansion. Analogous to the analysis of electromagnetic waveform case here we expect that the leading non-analytic contribution of $\Delta_{(n)}\widehat{T}^{\mu\nu}(k)$ should be of order $\omega^{n-1}(\ln\omega)^{n}$ in small $\omega$ expansion and for extracting gravitational waveform at this order we can treat the scattered objects as non-spinning point particles\footnote{This is along the same line of thought as discussed in \cite{1611.03493,Dixon,Tulczyjew,0511061,0511133,0604099,0804.0260,1709.06016,
1712.09250,1812.06895,0409156,0912.4254,
1203.2962,1705.09263,1806.07388}.
 We justify this claim in \S\ref{speculation}}.

Fourier transform of matter energy momentum tensor in terms of  corrected trajectory takes the following form,
\be
\widehat{T}^{X\mu\nu}(k)\ &=&\ \int d^{4}x\ e^{-ik.x}\ T^{X\mu\nu}(x)\non\\
&=&\ \sum_{a=1}^{M+N}m_{a}\int_{0}^{\infty}d\sigma \ e^{-ik\cdot X_{a}(\sigma)}\ \f{dX_{a}^{\mu}(\sigma)}{d\sigma}\ \f{dX_{a}^{\nu}(\sigma)}{d\sigma}\ \non\\
&=&\ \sum_{a=1}^{M+N}m_{a}e^{-ik.r_{a}}\ \int_{0}^{\infty}d\sigma \ e^{-ik\cdot v_{a}\sigma}\ \sum_{w=0}^{\infty}\f{1}{w!}\Bigg{\lbrace}-ik\cdot \sum_{s=1}^{\infty}\Delta_{(s)}Y_{a}(\sigma)\Bigg{\rbrace}^{w}\non\\
&&\ \times \ \Bigg{\lbrace} v_{a}^{\mu}\ +\ \sum_{t=1}^{\infty}\f{d\Delta_{(t)}Y_{a}^{\mu}(\sigma)}{d\sigma}\Bigg{\rbrace}\ \Bigg{\lbrace} v_{a}^{\nu}\ +\ \sum_{u=1}^{\infty}\f{d\Delta_{(u)}Y_{a}^{\nu}(\sigma)}{d\sigma}\Bigg{\rbrace}\non\\
&\equiv &\ \sum_{r=0}^{\infty}\ \Delta_{(r)}\widehat{T}^{X\mu\nu}(k)\label{FourierTX}
\ee
The gravitational energy-momentum tensor at order $(G)^{r}$ have the following kind of dependence on $\lbrace\Delta_{(s)}e_{\mu\nu}\rbrace$,
\be
\Delta_{(r)}T^{h\mu\nu}(x)\ \sim \ \f{1}{8\pi G}\Big[ \p\p(\Delta_{(0)}e)^{r+1}\ +\ \p\p \lbrace(\Delta_{(0)}e)^{r-1}\Delta_{(1)}e\rbrace\ +\cdots +\ \p\p \lbrace(\Delta_{(0)}e)(\Delta_{(r-1)}e)\rbrace\Big]\label{Thsketch}
\ee
where $\p\p$ above is just showing that each term in gravitational energy-momentum tensor has two derivatives on metric fluctuation\footnote{Here we are considering the theory of gravity is general relativity without any higher derivative corrections. If one includes higher derivative corrections to the Einstein-Hilbert action, those contribute to terms in the gravitational energy-momentum tensor having four or more derivatives acting on the metric fluctuations. One can explicitly show that these higher derivative terms will not contribute to the order $\mathcal{O}(\omega^{n-1}(\ln\omega)^{n})$ in the analysis of $\Delta_{(n)}\widehat{T}^{h\mu\nu}(k)$. } and the various $\lbrace\Delta_{(s)}e\rbrace$ dependence is fixed from the requirement of $\Delta_{(r)}T^{h\mu\nu}(x)$ should be of order $(G)^{r}$. For $r=1$ and $r=2$ the expressions for energy-momentum tensor are explicitly given in appendix-\ref{appC}.

Though our final goal is to determine the sub-subleading gravitational waveform at order $\mathcal{O}(\omega(\ln\omega)^{2})$, to set up the stage we will go through the derivation of subleading order gravitational waveform briefly following \cite{1912.06413} as some of the subleading order results are needed for the analysis of sub-subleading order energy-momentum tensor.

\subsection{Derivation of leading order gravitational waveform}
Fourier transform of the leading order matter energy-momentum tensor from eq.\eqref{FourierTX} takes form,
\be
\Delta_{(0)}\widehat{T}^{X\mu\nu}(k)\ &=&\  \sum_{a=1}^{M+N}m_{a}e^{-ik.r_{a}}\ \int_{0}^{\infty}d\sigma \ e^{-ik\cdot v_{a}\sigma}\ v_{a}^{\mu}v_{a}^{\nu}\non\\
&= &\ \sum_{a=1}^{M+N}\ \f{p_{a}^{\mu}p_{a}^{\nu}}{i(p_{a}.k -i\epsilon)}\ e^{-ik.r_{a}}\label{leadingT}
\ee
Similar to \S\ref{subsectionleading}, from the convergence of the integral at $\sigma=\infty$ for both ingoing and outgoing particles we need to replace $\omega\rightarrow (\omega +i\eta_{a}\epsilon)$ where $\eta_{a}=\pm 1$ for particle-a being outgoing/ingoing. This implies the following replacement in the exponential $k\cdot v_{a}\rightarrow (k\cdot v_{a}-i\epsilon)$ which we will use from now on. The gravitational energy momentum tensor at this order vanishes. Now  using the relation \eqref{eTrelation} for the leading energy-momentum tensor and performing Fourier transformation in $\omega$ variable we get the expected DC gravitational memory\cite{mem1,mem2,mem3,mem4,christodoulou,thorne,bondi,1411.5745,1712.01204,1502.06120,1806.01872,1912.06413},
\be
\Delta_{(0)}e^{\mu\nu}(t,R,\hat{n})\Big{|}_{(t-R)\rightarrow\infty}-\Delta_{(0)}e^{\mu\nu}(t,R,\hat{n})\Big{|}_{(t-R)\rightarrow -\infty}
&=& \f{2G}{R}\ \Bigg[-\sum_{a=1}^{N}\f{p_{a}^{\mu}p_{a}^{\nu}}{p_{a}.\mathbf{n}}+\sum_{a=1}^{M}\f{p_{a}^{\prime\mu}p_{a}^{\prime\nu}}{p_{a}^{\prime}.\mathbf{n}}\Bigg]
\ee

Now from the leading order energy momentum tensor in eq.\eqref{leadingT}, the leading order metric fluctuation takes the form,
\be
\Delta_{(0)}e_{\mu\nu}(x)\ &=&\ -8\pi G\ \sum_{a=1}^{M+N}\ \int \f{d^{4}\ell}{(2\pi)^{4}}\ G_{r}(\ell)\ e^{i\ell .(x-r_{a})}\ \f{p_{a\mu}p_{a\nu}}{i(p_{a}.\ell -i\epsilon)}\label{Leadinge}
\ee
\be
\Delta_{(0)}h_{\mu\nu}(x)\ &=&\ -8\pi G\ \sum_{a=1}^{M+N}\ \int \f{d^{4}\ell}{(2\pi)^{4}}\ G_{r}(\ell)\ e^{i\ell .(x-r_{a})}\ \f{\Big{\lbrace}p_{a\mu}p_{a\nu}-\f{1}{2}p_{a}^{2}\eta_{\mu\nu}\Big{\rbrace}}{i(p_{a}.\ell -i\epsilon)}\label{Leadingh}
\ee
The leading order Christoffel connection produced due to the asymptotic straight line trajectory of particle $b$ is given by,
\be
\Delta_{(0)}\Gamma^{(b)\alpha} _{\ \beta\gamma}(x)\ &=&\ \p_{\beta}\Delta_{(0)}h_{\gamma}^{(b)\alpha}+\p_{\gamma}\Delta_{(0)}h_{\beta}^{(b)\alpha}-\p^{\alpha}\Delta_{(0)}h^{(b)}_{\beta\gamma}\non\\
&=&\ -8\pi G\ \int\f{d^{4}\ell}{(2\pi)^{4}}G_{r}(\ell)\ e^{i\ell.(x-r_{b})}\ \f{1}{p_{b}.\ell -i\epsilon}\ \Bigg[ \Big{\lbrace} \ell_{\beta}p_{b}^{\alpha}p_{b\gamma}+\ell_{\gamma}p_{b}^{\alpha}p_{b\beta}-\ell^{\alpha}p_{b\beta}p_{b\gamma}\Big{\rbrace}\non\\
&&\hspace{10mm} -\f{1}{2}p_{b}^{2}\Big{\lbrace}\ell_{\beta}\delta^{\alpha}_{\gamma}+\ell_{\gamma}\delta^{\alpha}_{\beta}-\ell^{\alpha}\eta_{\beta\gamma}\Big{\rbrace}\Bigg]
\ee

\subsection{Derivation of subleading order gravitational waveform}\label{Ssecsubgr}
For the leading order metric fluctuation in eq.\eqref{Leadingh}, leading correction to the straight line trajectory of particle-$a$ satisfies the following geodesic equation as follows from eq.\eqref{geoX},
\be
\f{d^{2}\Delta_{(1)}Y_{a}^{\mu}(\sigma)}{d\sigma^{2}}\ =\ -\Delta_{(0)}\Gamma^{\mu}_{\nu\rho}(r_{a}+v_{a}\sigma)\ v_{a}^{\nu}v_{a}^{\rho}\label{subGeo}
\ee
where\footnote{The effect of self force will not be important to the order we are working in\cite{0306052}.}
\be 
\Delta_{(0)}\Gamma^{\mu}_{\nu\rho}(r_{a}+v_{a}\sigma)\ =\ \sum_{\substack{b=1\\ b\neq a}}^{M+N}\ \Delta_{(0)}\Gamma^{(b)\mu}_{\nu\rho}(r_{a}+v_{a}\sigma).
\ee
Now after using the boundary conditions 
\be
\Delta_{(1)}Y_{a}^{\mu}(\sigma)\Big{|}_{\sigma=0}\ =\ 0 \hspace{5mm} ,\ \hspace{5mm}\ {d\Delta_{(1)}Y_a^\mu(\sigma)\over d\sigma}\Bigg{|}_{\sigma\rightarrow \infty}\ =\ 0\label{boundarycon} ,
\ee
in eq.\eqref{subGeo} and performing integration in $\sigma$ variable we get,
\be
{d\Delta_{(1)}Y_a^\mu(\sigma)\over d\sigma} &=& \, 
\int_\sigma^\infty\, d\sigma' \, \Delta_{(0)}\Gamma^\mu_{\nu\rho}(v_a\, \sigma'+r_{a}) \, 
v_a^\nu v_{a}^{\rho}\label{dDeltaY1}\\
\Delta_{(1)}Y_a^\mu(\sigma) &=&\  
\int_0^\sigma\, d\sigma' \, \int_{\sigma'}^\infty\, d\sigma'' \, \Delta_{(0)}\Gamma^\mu_{\nu\rho}(v_a\, \sigma''+r_{a}) \, v_a^\nu v_{a}^{\rho}\label{DeltaY1}
\ee

The expression for  the subleading order matter energy-momentum tensor from eq.\eqref{FourierTX} takes form,
\be
\Delta_{(1)}\widehat{T}^{X\mu\nu}(k)\ &=&\  \sum_{a=1}^{M+N}m_{a}e^{-ik.r_{a}}\ \int_{0}^{\infty}d\sigma \ e^{-i(k\cdot v_{a}-i\epsilon)\sigma}\ \Bigg[\Big{\lbrace}-ik\cdot \Delta_{(1)}Y_{a}(\sigma)\Big{\rbrace}\ v_{a}^{\mu}v_{a}^{\nu}\non\\
&&\ +\ v_{a}^{\mu}\f{d\Delta_{(1)}Y_{a}^{\nu}(\sigma)}{d\sigma}\ +\ v_{a}^{\nu}\f{d\Delta_{(1)}Y_{a}^{\mu}(\sigma)}{d\sigma}\Bigg]\label{subleadingTX}
\ee
For the analysis of this matter energy-momentum tensor we will be very brief and follow ref.\cite{1912.06413} mostly but analyze the terms within the square bracket separately and organize the results in a different way which will help us to get the structure to higher order matter energy-momentum tensor contribution in small $\omega$ expansion. In the first term within the square bracket of eq.\eqref{subleadingTX} if we first replace $e^{-i(k.v_{a}-i\epsilon)\sigma}=\big{\lbrace}-i(k.v_{a}-i\epsilon)\big{\rbrace}^{-1}\f{d}{d\sigma}e^{-i(k.v_{a}-i\epsilon)\sigma}$ and  then perform integration by parts to move the $\sigma$ derivative to the rest of the integrand after using the boundary conditions from eq.\eqref{boundarycon}, we get the following contribution,
\be
\Delta_{(1)}^{(1)}\widehat{T}^{X\mu\nu}(k)\ 
&=&\ \sum_{a=1}^{M+N}m_{a}e^{-ik.r_{a}}\ \f{v_{a}^{\mu}v_{a}^{\nu}}{i(k.v_{a}-i\epsilon)}\int_{0}^{\infty}d\sigma \ e^{-i(k.v_{a}-i\epsilon)\sigma}\ \Big{\lbrace}-ik\cdot \f{d\Delta_{(1)}Y_{a}(\sigma)}{d\sigma}\Big{\rbrace}\non\\
%&=&\ (8\pi G)\ \sum_{a}\sum_{\substack{b\\b\neq a}}\  m_{a}e^{-ik.r_{a}}\ \f{v_{a}^{\mu}v_{a}^{\nu}}{v_{a}.k-i\epsilon}\int \f{d^{4}\ell}{(2\pi)^{4}}\ G_{r}(\ell)\ e^{i\ell .(r_{a}-r_{b})}\ \f{1}{p_{b}.\ell -i\epsilon}\non\\
%&&\ \bigg[\Big{\lbrace} 2v_{a}.\ell p_{b}.k v_{a}.p_{b}-k.\ell (v_{a}.p_{b})^{2}\Big{\rbrace}-\f{1}{2}p_{b}^{2}\Big{\lbrace}2v_{a}.\ell k.v_{a}-\ell.k v_{a}^{2}\Big{\rbrace}\Bigg]\non\\
%&&\ \times \int_{0}^{\infty}d\sigma e^{-i(k.v_{a}-i\epsilon)\sigma}\ \int_{\sigma}^{\infty}d\sigma'\ e^{i\ell.v_{a}\sigma'}\non\\
&=&\ -(8\pi G)\ \sum_{a=1}^{M+N}\sum_{\substack{b=1\\b\neq a}}^{M+N}\ e^{-ik.r_{a}}\ \f{p_{a}^{\mu}p_{a}^{\nu}}{p_{a}.k}\int\f{d^{4}\ell}{(2\pi)^{4}}\ e^{i\ell.(r_{a}-r_{b})}\ G_{r}(\ell)\ \f{1}{p_{b}.\ell -i\epsilon}\f{1}{p_{a}.\ell +i\epsilon}\non\\
&&\ \f{1}{p_{a}.(\ell -k)+i\epsilon}\ \Bigg[ \Big{\lbrace}2p_{a}.p_{b}p_{b}.kp_{a}.\ell - (p_{a}.p_{b})^{2}k.\ell\Big{\rbrace}-\f{1}{2}p_{b}^{2}\Big{\lbrace}2p_{a}.\ell p_{a}.k -p_{a}^{2}k.\ell\Big{\rbrace}\Bigg]\non\\
&\simeq &\ -i\ \sum_{a=1}^{M+N}\f{p_{a}^{\mu}p_{a}^{\nu}}{p_{a}.k}\ k^{\rho}\f{\p}{\p p_{a}^{\rho}}\ K_{gr}^{cl}\label{hatT11}
\ee
where in the last line above  we have approximated the integrand in the integration range $L^{-1}>>|\ell^{\mu}|>>\omega$  and written the approximated  result using $\simeq $ sign\footnote{Under $\simeq $ sign we are not only neglecting the order $\mathcal{O}(\omega^{0})$ contribution but also neglecting terms of order $\mathcal{O}(\omega^{k}\ln\omega)$ for $k=1,2,\cdots$ which we can get by expanding $exp(-ik.r_{a})=1-ik.r_{a}+\cdots$ from the second and third lines of the RHS of the above expression. But these terms will not affect the leading non-analytic contribution of $\Delta_{(r)}\widehat{T}^{\mu\nu}(k)$ for $r=2,3,\cdots.$}. The expression for $K_{gr}^{cl}$ is given below and the integration is explicitly evaluated in \cite{1912.06413},
\be
K_{gr}^{cl}\ &=&\ -\f{i}{2}\ (8\pi G)\ \sum_{\substack{b,c\\ c\neq b}}\int_{\omega}^{L^{-1}}\ \f{d^{4}\ell}{(2\pi)^{4}}\ G_{r}(\ell)\ \f{1}{p_{b}.\ell +i\epsilon}\ \f{1}{p_{c}.\ell-i\epsilon}\ \Big{\lbrace}(p_{b}.p_{c})^{2}-\f{1}{2}p_{b}^{2}p_{c}^{2}\Big{\rbrace}\non\\
&=&\ -\f{i}{2}\ (2G)\ \sum_{\substack{b,c\\ c\neq b\\ \eta_{b}\eta_{c}=1}}\ \ln\Big{\lbrace}L(\omega +i\epsilon \eta_{b})\Big{\rbrace}\ \f{(p_{b}.p_{c})^{2}-\f{1}{2}p_{b}^{2}p_{c}^{2}}{\sqrt{(p_{b}.p_{c})^{2}-p_{b}^{2}p_{c}^{2}}}\label{Kgrcl}
\ee
Performing an analogous analysis for the second term within the square bracket of eq.\eqref{subleadingTX} we get,
\be
\Delta_{(1)}^{(2)}\widehat{T}^{X\mu\nu}(k)\ &=&\ \sum_{a=1}^{M+N}m_{a}e^{-ik.r_{a}}\int_{0}^{\infty}d\sigma \ e^{-i(k.v_{a}-i\epsilon)\sigma}\ v_{a}^{\mu}\f{d\Delta_{(1)}Y_{a}^{\nu}(\sigma)}{d\sigma}\non\\
&=&\ (8\pi G)\ \sum_{a=1}^{M+N}\sum_{\substack{b=1\\b\neq a}}^{M+N}e^{-ik.r_{a}}p_{a}^{\mu}\ \int \f{d^{4}\ell}{(2\pi)^{4}}\ e^{i\ell.(r_{a}-r_{b})}\ G_{r}(\ell)\ \f{1}{p_{b}.\ell-i\epsilon}\ \f{1}{p_{a}.\ell +i\epsilon}\non\\
&&\ \f{1}{p_{a}.(\ell -k)+i\epsilon}\ \Bigg[ \Big{\lbrace} 2p_{a}.p_{b}p_{a}.\ell p_{b}^{\nu}-(p_{a}.p_{b})^{2}\ell^{\nu}\Big{\rbrace}-\f{1}{2}p_{b}^{2}\Big{\lbrace} 2p_{a}.\ell p_{a}^{\nu}-p_{a}^{2}\ell^{\nu}\Big{\rbrace}\Bigg]\non\\
&\simeq &\ i\sum_{a=1}^{M+N}p_{a}^{\mu}\f{\p}{\p p_{a\nu}}\ K_{gr}^{cl}\label{hatT12}
\ee
where in the last line above we have approximated the integrand in the integration range $L^{-1}>>|\ell^{\mu}|>>\omega$ under the sign $\simeq$. Similarly in this integration range the third term within the square bracket of eq.\eqref{subleadingTX} contributes,
\be
\Delta_{(1)}^{(3)}\widehat{T}^{X\mu\nu}(k)\ &=&\ \sum_{a=1}^{M+N}m_{a}e^{-ik.r_{a}}\int_{0}^{\infty}d\sigma \ e^{-i(k.v_{a}-i\epsilon)\sigma}\ v_{a}^{\nu}\f{d\Delta_{(1)}Y_{a}^{\mu}(\sigma)}{d\sigma}\non\\
&\simeq &\ i\sum_{a=1}^{M+N}p_{a}^{\nu}\f{\p}{\p p_{a\mu}}\ K_{gr}^{cl}\label{hatT13}
\ee
Hence the order $\mathcal{O}(\ln\omega)$ contribution from subleading order matter energy-momentum tensor becomes,
\be
\Delta_{(1)}\widehat{T}^{X\mu\nu}(k)\ &=&\ (-i)\ \sum_{a=1}^{M+N}\ \f{p_{a}^{\mu}k_{\rho}}{p_{a}.k}\ \Bigg(p_{a}^{\nu}\f{\p}{\p p_{a\rho}}-p_{a}^{\rho}\f{\p}{\p p_{a\nu}}\bigg)\ K_{gr}^{cl}\ +\  i\sum_{a=1}^{M+N}p_{a}^{\nu}\f{\p}{\p p_{a\mu}}\ K_{gr}^{cl}\label{subTXtotal}
\ee
Fourier transform of subleading order gravitational energy-momentum tensor follows from \eqref{T2h},
\be
\Delta_{(1)}\widehat{T}^{h\mu\nu}(k) &=& -(8\pi G)\sum_{a,b=1}^{M+N}e^{-ik.r_{b}}\int \f{d^{4}\ell}{(2\pi)^{4}}\ e^{i\ell.(r_{b}-r_{a})}\ G_{r}(k-\ell)G_{r}(\ell)\ \f{1}{p_{a}.\ell-i\epsilon}\ \f{1}{p_{b}.(k-\ell)-i\epsilon}\non\\
&&\times \Big{\lbrace} p_{a\alpha}p_{a\beta}-\f{1}{2}p_{a}^{2}\eta_{\alpha\beta}\Big{\rbrace}\ \mathcal{F}^{\mu\nu, \alpha\beta , \rho\sigma}(k,\ell)\ \Big{\lbrace}p_{b\rho}p_{b\sigma}-\f{1}{2}p_{b}^{2}\eta_{\rho\sigma}\Big{\rbrace}\label{Thath}
\ee
where,
\begingroup
\allowdisplaybreaks
\be
&&\mathcal{F}^{\mu\nu , \alpha\beta , \rho\sigma }(k,\ell)\non\\
 &=&\ 2\ \Big[\ \f{1}{2}\ell^{\mu}(k-\ell)^{\nu}\eta^{\rho\alpha}\eta^{\sigma\beta}+(k-\ell)^{\mu}(k-\ell)^{\nu}\eta^{\rho\alpha}\eta^{\sigma\beta}-(k-\ell)^{\nu}(k-\ell)^{\beta}\eta^{\rho\alpha}\eta^{\sigma\mu}\non\\
 &&\ -(k-\ell)^{\mu}(k-\ell)^{\beta}\eta^{\rho\alpha}\eta^{\sigma\nu}
\ +(k-\ell)^{\alpha}(k-\ell)^{\beta}\eta^{\rho\mu}\eta^{\sigma\nu}+(k-\ell).\ell\eta^{\beta\nu}\eta^{\alpha\rho}\eta^{\sigma\mu}-\ell^{\rho}(k-\ell)^{\alpha}\eta^{\beta\nu}\eta^{\sigma\mu}\non\\
&&\ -\f{1}{2}(k-\ell)^{2}\eta^{\alpha\mu}\eta^{\beta\nu}\eta^{\rho\sigma}\ +\eta^{\alpha\mu}\eta^{\beta\rho}\eta^{\nu\sigma}(k-\ell)^{2}\ +\ \eta^{\alpha\nu}\eta^{\beta\rho}\eta^{\mu\sigma}(k-\ell)^{2}
\ \Big]\non\\
&&\ -\eta^{\mu\nu}\ \Big[\ \f{3}{2}(k-\ell).\ell\eta^{\rho\alpha}\eta^{\sigma\beta}+2(k-\ell)^{2}\eta^{\rho\alpha}\eta^{\sigma\beta}-\ell^{\sigma}(k-\ell)^{\alpha}\eta^{\rho\beta}\ \Big]\non\\
&&\ -\eta^{\alpha\beta}(k-\ell)^{2}\eta^{\rho\mu}\eta^{\sigma\nu}+\f{1}{2}\eta^{\alpha\beta}(k-\ell)^{2}\eta^{\rho\sigma}\eta^{\mu\nu} .\label{F}
\ee
\endgroup
Now following \cite{1912.06413}, $\Delta_{(1)}\widehat{T}^{h\mu\nu}(k)$ contributes to $\ln\omega$ in two integration regions. In the integration region $R^{-1}<<|\ell^{\mu}|<<\omega$ with $R$ being the distance of the detector from the scattering center, we can approximate $\mathcal{F}^{\mu\nu , \alpha\beta , \rho\sigma }(k,\ell)$ up to gauge equivalence as,
\be
 \mathcal{F}^{\mu\nu , \alpha\beta , \rho\sigma }(k,\ell)\ &\approx &\ -2k^{\sigma}k^{\beta}\eta^{\alpha\rho}\eta^{\mu\nu}+2k^{\alpha}k^{\beta}\eta^{\mu\rho}\eta^{\nu\sigma}
\ee
Then the contribution from this integration region turns out to be,
\be
\Delta_{(1)}^{(1)}\widehat{T}^{h\mu\nu}(k)\ &=&\ -2G\ \ln\Big{\lbrace}(\omega+i\epsilon)R\Big{\rbrace}\sum_{b=1}^{N}p_{b}\cdot k\ \sum_{a=1}^{M+N}\f{p_{a}^{\mu}p_{a}^{\nu}}{p_{a}.k}\label{backscattering}
\ee
In the integration region $\omega<<|\ell^{\mu}|<<L^{-1}$, $\mathcal{F}^{\mu\nu , \alpha\beta , \rho\sigma }(k,\ell)$ can be approximated with the terms quadratic in $\ell$. While evaluating this integral one will encounter a linearly divergent term in this region which can be identified with the leading order energy-momentum tensor for soft radiation from finite energy gravitational radiation. Since this kind of contribution is already taken care of inside the finite energy particle's sum we need to subtract this contribution by hand. The expression for this soft radiation energy-momentum tensor is
\be
\Delta_{(0)}\widehat{T}^{R\mu\nu}(k)\ &=&\ \f{G}{\pi^{2}}\ \int _{\omega}^{L^{-1}}d^{4}\ell\ \delta(\ell^{2})\theta(\ell^{0})\ \sum_{a,b=1}^{M+N}\f{\Big{\lbrace}(p_{a}.p_{b})^{2}-\f{1}{2}p_{a}^{2}p_{b}^{2}\Big{\rbrace}}{(p_{a}.\ell-i\epsilon)\ (p_{b}.\ell +i\epsilon)}\ \f{\ell^{\mu}\ell^{\nu}}{i(k.\ell -i\epsilon)}\label{realrad}
\ee
Finally subtracting the contribution $\Delta_{(0)}\widehat{T}^{R\mu\nu}(k)$ from $\Delta_{(1)}^{(2)}\widehat{T}^{h\mu\nu}(k)$, which is the contribution from $\Delta_{(1)}\widehat{T}^{h\mu\nu}(k)$ in the integration region $\omega <<|\ell^{\mu}|<<L^{-1}$, we get,
\be
&&\ \Delta^{(2)}_{(1)}\widehat{T}^{h\mu\nu}(k)-\Delta_{(0)}\widehat{T}^{R\mu\nu}\non\\
&\simeq &\ -2G\ \sum_{a=1}^{M+N}\sum_{\substack{b=1\\b\neq a\\ \eta
_{a}\eta_{b}=1}}^{M+N}\ \ln\Big{\lbrace}L(\omega +i\epsilon\eta_{a})\Big{\rbrace}\ \f{1}{[(p_{a}.p_{b})^{2}-p_{a}^{2}p_{b}^{2}]^{3/2}}\non\\
&&\ \times \Bigg[ (p_{a}.p_{b})p_{a}^{\nu}p_{b}^{\mu}\Big{\lbrace}(p_{a}.p_{b})^{2}-\f{3}{2}p_{a}^{2}p_{b}^{2}\Big{\rbrace}\ +\f{1}{2}p_{a}^{2}(p_{b}^{2})^{2}p_{a}^{\mu}p_{a}^{\nu}\Bigg]\non\\
&=&\ -\  i\sum_{a=1}^{M+N}p_{a}^{\nu}\f{\p}{\p p_{a\mu}}\ K_{gr}^{cl}\label{subThtotal}
\ee
Now summing over eq.\eqref{subTXtotal}, eq.\eqref{subThtotal} and eq.\eqref{backscattering} we get the total subleading energy-momentum tensor at order $\mathcal{O}(\ln\omega)$,
\be
\Delta_{(1)}\widehat{T}^{\mu\nu}\ &=&\   (-i)\ \sum_{a=1}^{M+N}\ \f{p_{a}^{\mu}k_{\rho}}{p_{a}.k}\ \Bigg(p_{a}^{\nu}\f{\p}{\p p_{a\rho}}-p_{a}^{\rho}\f{\p}{\p p_{a\nu}}\bigg)\ K_{gr}^{cl}\non\\
&&\  -2G\ \ln\Big{\lbrace}(\omega+i\epsilon)R\Big{\rbrace}\sum_{b=1}^{N}p_{b}\cdot k\ \sum_{a=1}^{M+N}\f{p_{a}^{\mu}p_{a}^{\nu}}{p_{a}.k}\ +\ \mathcal{O}(\omega^{0})\label{subleadingTtotal}
\ee
Now to determine the radiative mode of subleading order gravitational waveform at late and early time we use the relation \eqref{eTrelation} at subleading order and then perform Fourier transformation in $\omega$ variable. Finally using the results of integrations from eq.\eqref{FourierFn0} and \eqref{FourierF0n} for $n=1$ we get \cite{1912.06413},
\be
\Delta_{(1)}e^{\mu\nu}(t,R,\hat{n})&=&\ \f{2G}{R}\ \f{1}{u}\ \Big{\lbrace}2G\sum_{b=1}^{N}p_{b}\cdot\mathbf{n}\Big{\rbrace}\Bigg(\sum_{a=1}^{N}\f{p_{a}^{\mu}p_{a}^{\nu}}{p_{a}.\mathbf{n}}-\sum_{a=1}^{M}\f{p_{a}^{\prime\mu}p_{a}^{\prime\nu}}{p'_{a}.\mathbf{n}}\bigg)\non\\
&&\ -\f{4G^{2}}{R}\ \f{1}{u}\sum_{a=1}^{N}\sum_{\substack{b=1\\ b\neq a}}^{N}\f{p_{a}.p_{b}}{[(p_{a}.p_{b})^{2}-p_{a}^{2}p_{b}^{2}]^{3/2}}\Big{\lbrace}\f{3}{2}p_{a}^{2}p_{b}^{2}-(p_{a}.p_{b})^{2}\Big{\rbrace}\non\\
&&\ \times \f{p_{a}^{\mu}\mathbf{n}_{\rho}}{p_{a}.\mathbf{n}}\Big{\lbrace}p_{a}^{\nu}p_{b}^{\rho}-p_{a}^{\rho}p_{b}^{\nu}\Big{\rbrace}\ \hspace{2cm}\hbox{for $u\rightarrow +\infty$}
\ee
\be
\Delta_{(1)}e^{\mu\nu}(t,R,\hat{n})&=&\ \f{4G^{2}}{R}\ \f{1}{u}\sum_{a=1}^{M}\sum_{\substack{b=1\\ b\neq a}}^{M}\f{p'_{a}.p'_{b}}{[(p'_{a}.p'_{b})^{2}-p_{a}^{\prime 2}p_{b}^{\prime 2}]^{3/2}}\Big{\lbrace}\f{3}{2}p_{a}^{\prime 2}p_{b}^{\prime 2}-(p'_{a}.p'_{b})^{2}\Big{\rbrace}\non\\
&&\ \times \f{p_{a}^{\prime\mu}\mathbf{n}_{\rho}}{p'_{a}.\mathbf{n}}\Big{\lbrace}p_{a}^{\prime\nu}p_{b}^{\prime\rho}-p_{a}^{\prime\rho}p_{b}^{\prime\nu}\Big{\rbrace}\ \hspace{2cm}\hbox{for $u\rightarrow -\infty$}
\ee
where retarded time $u=t-R+2G\ln R\sum\limits_{b=1}^{N}p_{b}\cdot \mathbf{n}$.

\subsection{Derivation of sub-subleading order gravitational waveform}\label{secsub2leadinggr}
Here in this section we derive the order $\mathcal{O}(\omega(\ln\omega)^{2})$ coefficient of the sub-subleading gravitational waveform. This turns out to be the leading non-analytic contribution at this order in small $\omega$ expansion.
\subsubsection{Analysis of sub-subleading matter energy-momentum tensor}
The expression for sub-subleading order energy-momentum tensor follows from eq.\eqref{FourierTX},
\be
\Delta_{(2)}\widehat{T}^{X\mu\nu}(k)\ &=&\ \  \sum_{a=1}^{M+N}m_{a}e^{-ik.r_{a}}\ \int_{0}^{\infty}d\sigma \ e^{-i(k\cdot v_{a}-i\epsilon)\sigma}\non\\
&&\ \times \Bigg[ \f{1}{2}\Big{\lbrace}-ik\cdot \Delta_{(1)}Y_{a}(\sigma)\Big{\rbrace}^{2}v_{a}^{\mu}v_{a}^{\nu}\ +\ \Big{\lbrace}-ik\cdot \Delta_{(1)}Y_{a}(\sigma)\Big{\rbrace}\ v_{a}^{\mu}\f{d\Delta_{(1)}Y_{a}^{\nu}(\sigma)}{d\sigma}\non\\
&&\  +\ \Big{\lbrace}-ik\cdot \Delta_{(1)}Y_{a}(\sigma)\Big{\rbrace}\ v_{a}^{\nu}\f{d\Delta_{(1)}Y_{a}^{\mu}(\sigma)}{d\sigma}\ +\ \f{d\Delta_{(1)}Y_{a}^{\mu}(\sigma)}{d\sigma}\ \f{d\Delta_{(1)}Y_{a}^{\nu}(\sigma)}{d\sigma}\non\\
&&\ +\ \Big{\lbrace}-ik\cdot \Delta_{(2)}Y_{a}(\sigma)\Big{\rbrace}\ v_{a}^{\mu}v_{a}^{\nu}\ +\ v_{a}^{\mu}\f{d\Delta_{(2)}Y_{a}^{\nu}(\sigma)}{d\sigma}\ +\ v_{a}^{\nu}\f{d\Delta_{(2)}Y_{a}^{\mu}(\sigma)}{d\sigma}\Bigg]\label{sub2leadingTX}
\ee
Contribution from the first term within the square bracket of eq.\eqref{sub2leadingTX} takes form,
\begingroup
\allowdisplaybreaks
\be
\Delta_{(2)}^{(1)}\widehat{T}^{X\mu\nu}(k)\ &=&\ \f{1}{2} \sum_{a=1}^{M+N}m_{a}e^{-ik.r_{a}}\int_{0}^{\infty}d\sigma\ e^{-i(k.v_{a}-i\epsilon)\sigma}\ \Big{\lbrace}-ik\cdot \Delta_{(1)}Y_{a}(\sigma)\Big{\rbrace}^{2}\ v_{a}^{\mu}v_{a}^{\nu}\non\\
&=&\ \sum_{a=1}^{M+N}m_{a}e^{-ik.r_{a}}\int_{0}^{\infty}d\sigma\ e^{-i(k.v_{a}-i\epsilon)\sigma}\ \f{v_{a}^{\mu}v_{a}^{\nu}}{i(k.v_{a}-i\epsilon)}\Big{\lbrace}-ik\cdot \Delta_{(1)}Y_{a}(\sigma)\Big{\rbrace}\ \Big{\lbrace}-ik\cdot \f{d\Delta_{(1)}Y_{a}(\sigma)}{d\sigma}\Big{\rbrace}\non\\
&=& \ i(8\pi G)^{2}\ \sum_{a=1}^{M+N}e^{-ik.r_{a}}\ \f{p_{a}^{\mu}p_{a}^{\nu}}{p_{a}.k}\ \sum_{\substack{b=1\\ b\neq a}}^{M+N}\ \sum_{\substack{c=1\\ c\neq a}}^{M+N}\ \int \f{d^{4}\ell_{1}}{(2\pi)^{4}}\ e^{i\ell_{1}.(r_{a}-r_{b})}\ G_{r}(\ell_{1})\ \f{1}{p_{a}.\ell_{1}+i\epsilon}\non\\
&&\times \f{1}{p_{b}.\ell_{1}-i\epsilon} \Bigg[\Big{\lbrace}2p_{a}.\ell_{1}p_{a}.p_{b}p_{b}.k -\ell_{1}.k(p_{a}.p_{b})^{2}\Big{\rbrace}-\f{1}{2}p_{b}^{2}\Big{\lbrace}2\ell_{1}.p_{a}p_{a}.k -\ell_{1}.k p_{a}^{2}\Big{\rbrace}\Bigg]\non\\
&&\times \int\f{d^{4}\ell_{2}}{(2\pi)^{4}}\ e^{i\ell_{2}.(r_{a}-r_{c})} G_{r}(\ell_{2})\ \f{1}{p_{a}.\ell_{2}+i\epsilon}\ \f{1}{p_{c}.\ell_{2}-i\epsilon}\ \Bigg[\Big{\lbrace}2p_{a}.\ell_{2}p_{a}.p_{c}p_{c}.k -\ell_{2}.k(p_{a}.p_{c})^{2}\Big{\rbrace}\non\\
&&\ -\f{1}{2}p_{c}^{2}\Big{\lbrace}2\ell_{2}.p_{a}p_{a}.k -\ell_{2}.k p_{a}^{2}\Big{\rbrace}\Bigg] \ \f{1}{(\ell_{2}-k).p_{a}+i\epsilon}\ \f{1}{(\ell_{1}+\ell_{2}-k).p_{a}+i\epsilon}\label{TX21}
\ee
\endgroup
Here to get second line of RHS from the first line we have first used $e^{-i(k.v_{a}-i\epsilon)\sigma}=\big{\lbrace}-i(k.v_{a}-i\epsilon)\big{\rbrace}^{-1}\f{d}{d\sigma}e^{-i(k.v_{a}-i\epsilon)\sigma}$ and then performed integration by parts to move the $\sigma$ derivative to the rest of the integrand after using the boundary conditions of eq.\eqref{boundarycon}. To get the last four lines from the second line of RHS we substituted the expressions of eq.\eqref{dDeltaY1} and eq.\eqref{DeltaY1} and then used the integration result of  eq.\eqref{integration}. The above expression contributes to order $\mathcal{O}\big(\omega(\ln\omega)^{2}\big)$ in the integration region $L^{-1}>>|\ell_{1}^{\mu}|>>|\ell_{2}^{\mu}|>>\omega$. Now following the procedure described below eq.\eqref{step} we can also get the same contribution approximating the integrand in the integration region $\Big{\lbrace}L^{-1}>>|\ell_{1}^{\mu}|>>\omega \ ,\ L^{-1}>>|\ell_{2}^{\mu}|>>\omega\Big{\rbrace}$ but with an overall multiplicative factor of $\f{1}{2}$ coming from the momentum ordering. So the $\mathcal{O}\big(\omega(\ln\omega)^{2}\big)$ contribution becomes,
\be
\Delta_{(2)}^{(1)}\widehat{T}^{X\mu\nu}(k)\ &\simeq &\ \f{i}{2}\ (8\pi G)^{2}\ \sum_{a=1}^{M+N} \f{p_{a}^{\mu}p_{a}^{\nu}}{p_{a}.k}\ \sum_{\substack{b=1\\ b\neq a}}^{M+N}\sum_{\substack{c=1\\c\neq a}}^{M+N}\ \int_{\omega}^{L^{-1}} \f{d^{4}\ell_{1}}{(2\pi)^{4}}\ G_{r}(\ell_{1})\ \f{1}{(p_{a}.\ell_{1}+i\epsilon)^{2}}\ \f{1}{p_{b}.\ell_{1}-i\epsilon}\non\\
&&\ \Bigg[\Big{\lbrace}2p_{a}.\ell_{1}p_{a}.p_{b}p_{b}.k -\ell_{1}.k(p_{a}.p_{b})^{2}\Big{\rbrace}-\f{1}{2}p_{b}^{2}\Big{\lbrace}2\ell_{1}.p_{a}p_{a}.k -\ell_{1}.k p_{a}^{2}\Big{\rbrace}\Bigg]\ \int_{\omega}^{L^{-1}}\f{d^{4}\ell_{2}}{(2\pi)^{4}}\  G_{r}(\ell_{2})\non\\
&& \f{1}{(p_{a}.\ell_{2}+i\epsilon)^{2}}\ \f{1}{p_{c}.\ell_{2}-i\epsilon}\ \Bigg[\Big{\lbrace}2p_{a}.\ell_{2}p_{a}.p_{c}p_{c}.k -\ell_{2}.k(p_{a}.p_{c})^{2}\Big{\rbrace}-\f{1}{2}p_{c}^{2}\Big{\lbrace}2\ell_{2}.p_{a}p_{a}.k -\ell_{2}.k p_{a}^{2}\Big{\rbrace}\Bigg]\non\\
&=& (-i)\ \f{1}{2}\sum_{a=1}^{M+N}\f{p_{a}^{\mu}p_{a}^{\nu}}{p_{a}.k}\ \Bigg(k^{\rho}\f{\p}{\p p_{a}^{\rho}}K_{gr}^{cl}\Bigg)\ \Bigg(k^{\sigma}\f{\p}{\p p_{a}^{\sigma}}K_{gr}^{cl}\Bigg)\label{hatT21}
\ee
Similarly the contribution from the second term within the square bracket of eq.\eqref{sub2leadingTX} turns out to be,
\begingroup
\allowdisplaybreaks
\be
\Delta_{(2)}^{(2)}\widehat{T}^{X\mu\nu}(k)\ &=&\ \sum_{a=1}^{M+N}m_{a}e^{-ik.r_{a}}\int_{0}^{\infty}d\sigma\ e^{-i(k.v_{a}-i\epsilon)\sigma}\ \Big{\lbrace}-ik\cdot \Delta_{(1)}Y_{a}(\sigma)\Big{\rbrace}\ v_{a}^{\mu}\f{d\Delta_{(1)}Y_{a}^{\nu}(\sigma)}{d\sigma}\non\\
&=&\ -i\ (8\pi G)^{2}\ \sum_{a=1}^{M+N}e^{-ik.r_{a}}\ p_{a}^{\mu}\  \sum_{\substack{b=1\\ b\neq a}}^{M+N}\sum_{\substack{c=1\\ c\neq a}}^{M+N}\ \int \f{d^{4}\ell_{1}}{(2\pi)^{4}}\ e^{i\ell_{1}.(r_{a}-r_{b})}\ G_{r}(\ell_{1})\ \f{1}{p_{a}.\ell_{1}+i\epsilon}\non\\
&&\times \f{1}{p_{b}.\ell_{1}-i\epsilon}\ \Bigg[\Big{\lbrace}2p_{a}.\ell_{1}p_{a}.p_{b}p_{b}.k -\ell_{1}.k(p_{a}.p_{b})^{2}\Big{\rbrace}-\f{1}{2}p_{b}^{2}\Big{\lbrace}2\ell_{1}.p_{a}p_{a}.k -\ell_{1}.k p_{a}^{2}\Big{\rbrace}\Bigg]\non\\
&&\times \int\f{d^{4}\ell_{2}}{(2\pi)^{4}}\ e^{i\ell_{2}.(r_{a}-r_{c})}\ G_{r}(\ell_{2})\ \f{1}{p_{a}.\ell_{2}+i\epsilon}\ \f{1}{p_{c}.\ell_{2}-i\epsilon}\ \Bigg[\Big{\lbrace}2p_{a}.\ell_{2}p_{a}.p_{c}p_{c}^{\nu} -\ell_{2}^{\nu}(p_{a}.p_{c})^{2}\Big{\rbrace}\non\\
&&\ -\f{1}{2}p_{c}^{2}\Big{\lbrace}2\ell_{2}.p_{a}p_{a}^{\nu} -\ell_{2}^{\nu} p_{a}^{2}\Big{\rbrace}\Bigg]\ \times \f{1}{(\ell_{2}-k).p_{a}+i\epsilon}\ \f{1}{(\ell_{1}+\ell_{2}-k).p_{a}+i\epsilon}
\ee
\endgroup
Similar to $\Delta_{(2)}^{(1)}\widehat{T}^{X\mu\nu}(k)$, the above expression also contributes to order $\mathcal{O}\big(\omega(\ln\omega)^{2}\big)$ in the integration region $L^{-1}>>|\ell_{1}^{\mu}|>>|\ell_{2}^{\mu}|>>\omega$. So following the procedure described below eq.\eqref{step} we can analyze the two integrals in the integration region $\Big{\lbrace}L^{-1}>>|\ell_{1}^{\mu}|>>\omega \ ,\ L^{-1}>>|\ell_{2}^{\mu}|>>\omega\Big{\rbrace}$ but with an overall multiplicative factor of $\f{1}{2}$ coming from the momentum ordering. So the order $\mathcal{O}\big(\omega(\ln\omega)^{2}\big)$ contribution becomes,
\be
\Delta_{(2)}^{(2)}\widehat{T}^{X\mu\nu}(k)\ &\simeq &  -\f{i}{2}\ (8\pi G)^{2}\ \sum_{a=1}^{M+N}\ p_{a}^{\mu}\  \sum_{\substack{b=1\\ b\neq a}}^{M+N}\sum_{\substack{c=1\\ c\neq a}}^{M+N}\ \int_{\omega}^{L^{-1}} \f{d^{4}\ell_{1}}{(2\pi)^{4}}\ G_{r}(\ell_{1})\ \f{1}{(p_{a}.\ell_{1}+i\epsilon)^{2}}\ \f{1}{p_{b}.\ell_{1}-i\epsilon}\non\\
&&\ \ \Bigg[\Big{\lbrace}2p_{a}.\ell_{1}p_{a}.p_{b}p_{b}.k -\ell_{1}.k(p_{a}.p_{b})^{2}\Big{\rbrace}-\f{1}{2}p_{b}^{2}\Big{\lbrace}2\ell_{1}.p_{a}p_{a}.k -\ell_{1}.k p_{a}^{2}\Big{\rbrace}\Bigg]\non\\
&&\ \times \int_{\omega}^{L^{-1}}\f{d^{4}\ell_{2}}{(2\pi)^{4}}\  G_{r}(\ell_{2})\ \f{1}{(p_{a}.\ell_{2}+i\epsilon)^{2}}\ \f{1}{p_{c}.\ell_{2}-i\epsilon}\ \non\\
&&\  \Bigg[\Big{\lbrace}2p_{a}.\ell_{2}p_{a}.p_{c}p_{c}^{\nu} -\ell_{2}^{\nu}(p_{a}.p_{c})^{2}\Big{\rbrace}\ -\f{1}{2}p_{c}^{2}\Big{\lbrace}2\ell_{2}.p_{a}p_{a}^{\nu} -\ell_{2}^{\nu} p_{a}^{2}\Big{\rbrace}\Bigg] \non\\
&=&\ \f{i}{2}\ \sum_{a=1}^{M+N}p_{a}^{\mu}\ \Bigg(\f{\p}{\p p_{a\nu}}K_{gr}^{cl}\Bigg)\ \Bigg(k^{\rho}\f{\p}{\p p_{a}^{\rho}}K_{gr}^{cl}\Bigg)\label{hatT22}
\ee
The contribution from the third term within the square bracket of eq.\eqref{sub2leadingTX} can be easily read out from eq.\eqref{hatT22} after interchanging $\mu\leftrightarrow\nu$ and the order $\mathcal{O}\big(\omega(\ln\omega)^{2}\big)$ contribution is,
\be
\Delta_{(2)}^{(3)}\widehat{T}^{X\mu\nu}(k)\ &=&\ \sum_{a=1}^{M+N}m_{a}e^{-ik.r_{a}}\int_{0}^{\infty}d\sigma\ e^{-i(k.v_{a}-i\epsilon)\sigma}\ \Big{\lbrace}-ik\cdot \Delta_{(1)}Y_{a}(\sigma)\Big{\rbrace}\ v_{a}^{\nu}\f{d\Delta_{(1)}Y_{a}^{\mu}(\sigma)}{d\sigma}\non\\
 &\simeq &\ \f{i}{2}\ \sum_{a=1}^{M+N}p_{a}^{\nu}\ \Bigg(\f{\p}{\p p_{a\mu}}K_{gr}^{cl}\Bigg)\ \Bigg(k^{\rho}\f{\p}{\p p_{a}^{\rho}}K_{gr}^{cl}\Bigg)
\ee
Contribution from the fourth term within the square bracket of eq.\eqref{FourierTX} takes the following form after performing the $\sigma$ integrations,
\begingroup
\allowdisplaybreaks
\be
\Delta_{(2)}^{(4)}\widehat{T}^{X\mu\nu}(k)\ &=&\ \sum_{a=1}^{M+N}m_{a}e^{-ik.r_{a}}\int_{0}^{\infty}d\sigma\ e^{-i(k.v_{a}-i\epsilon)\sigma}\ \f{d\Delta_{(1)}Y_{a}^{\mu}(\sigma)}{d\sigma}\  \f{d\Delta_{(1)}Y_{a}^{\nu}(\sigma)}{d\sigma}\non\\
&=&\ (-i)(8\pi G)^{2}\sum_{a=1}^{M+N}e^{-ik.r_{a}}\sum_{\substack{b=1\\ b\neq a}}^{M+N}\sum_{\substack{c=1\\ c\neq a}}^{M+N}\int \f{d^{4}\ell_{1}}{(2\pi)^{4}}G_{r}(\ell_{1})\ e^{i\ell_{1}.(r_{a}-r_{b})}\ \f{1}{p_{a}.\ell_{1}+i\epsilon}\ \f{1}{p_{b}.\ell_{1}-i\epsilon}\non\\
&&\ \Bigg[\Big{\lbrace}2p_{a}.\ell_{1}p_{a}.p_{b}p_{b}^{\mu} -\ell_{1}^{\mu}(p_{a}.p_{b})^{2}\Big{\rbrace}-\f{1}{2}p_{b}^{2}\Big{\lbrace}2\ell_{1}.p_{a}p_{a}^{\mu} -\ell_{1}^{\mu} p_{a}^{2}\Big{\rbrace}\Bigg]\ \int\f{d^{4}\ell_{2}}{(2\pi)^{4}}\ e^{i\ell_{2}.(r_{a}-r_{c})}\ G_{r}(\ell_{2})\non\\
&&\ \f{1}{p_{a}.\ell_{2}+i\epsilon}\ \f{1}{p_{c}.\ell_{2}-i\epsilon}\Bigg[\Big{\lbrace}2p_{a}.\ell_{2}p_{a}.p_{c}p_{c}^{\nu} -\ell_{2}^{\nu}(p_{a}.p_{c})^{2}\Big{\rbrace}-\f{1}{2}p_{c}^{2}\Big{\lbrace}2\ell_{2}.p_{a}p_{a}^{\nu} -\ell_{2}^{\nu} p_{a}^{2}\Big{\rbrace}\Bigg]\non\\
&&\ \times \f{1}{(\ell_{1}+\ell_{2}-k).p_{a}+i\epsilon}
\ee
\endgroup
Analyzing all possible integration regions we find that the above expression does not contribute at order $\mathcal{O}(\omega(\ln\omega)^{2})$. The expression $\Delta_{(2)}^{(4)}\widehat{T}^{X\mu\nu}(k)$ starts contributing at order $\mathcal{O}(\omega\ln\omega)$ in the integration region $L^{-1}>>|\ell_{1}^{\mu}|>>|\ell_{2}^{\mu}|>>\omega$ or $L^{-1}>>|\ell_{2}^{\mu}|>>|\ell_{1}^{\mu}|>>\omega$ in small $\omega$ expansion. It also contributes to order $\mathcal{O}(\omega^{0})$ but since order $\mathcal{O}(\omega^{0})$ is already undetermined from the analysis of subleading order and does not contribute to memory, we are ingoing that contribution\footnote{One point to note that order $\mathcal{O}(\omega^{0})$ contribution always dominates over order $\mathcal{O}(\omega(\ln\omega)^{2})$ contribution. But if we are only interested to extract $\mathcal{O}(\omega(\ln\omega)^{2})$ coefficient unambiguously, we can do so by operating $\p_{\omega}$ on the integral expression and extract the coefficient of $(\ln\omega)^{2}$. }.

Analysis of fifth, sixth and seventh terms within the square bracket of eq.\eqref{sub2leadingTX} is little involved as we need to find out subleading correction $\Delta_{(2)}Y_{a}^{\mu}(\sigma)$ to the straight line trajectory after knowing subleading order gravitational waveform $\Delta_{(1)}e_{\mu\nu}(x)$. A complete analysis has been performed in appendix-\ref{appB} and those terms also contribute to order $\mathcal{O}(\omega(\ln\omega)^{2})$. 

 Hence the contribution from the sub-subleading order matter energy-momentum tensor at order $\mathcal{O}(\omega(\ln\omega)^{2})$ becomes,
\be
\Delta_{(2)}\widehat{T}^{X\mu\nu}(k) &\simeq & (-i)\ \f{1}{2}\sum_{a=1}^{M+N}\f{k_{\rho}k_{\sigma}}{p_{a}.k}\ \Bigg[\Bigg(p_{a}^{\mu}\f{\p}{\p p_{a\rho}}-p_{a}^{\rho}\f{\p}{\p p_{a\mu}}\Bigg)K_{gr}^{cl}\Bigg]\ \Bigg[\Bigg(p_{a}^{\nu}\f{\p}{\p p_{a\sigma}}-p_{a}^{\sigma}\f{\p}{\p p_{a\nu}}\Bigg)K_{gr}^{cl}\Bigg]\non\\
&&\ +\f{i}{2}\ \sum_{a=1}^{M+N}(p_{a}.k)\ \Bigg(\f{\p}{\p p_{a\mu}}K_{gr}^{cl}\Bigg)\ \Bigg(\f{\p}{\p p_{a\nu}}K_{gr}^{cl}\Bigg)+\Delta_{(2)}^{(5)}\widehat{T}^{X\mu\nu}(k)+\Delta_{(2)}^{(6)}\widehat{T}^{X\mu\nu}(k)\non\\
&&\ +\Delta_{(2)}^{(7)}\widehat{T}^{X\mu\nu}(k)\label{sub2matterfinal}
\ee
Above we have added and subtracted a term $\f{i}{2}\ \sum\limits_{a=1}^{M+N}(p_{a}.k)\ \Big(\f{\p}{\p p_{a\mu}}K_{gr}^{cl}\Big)\ \Big(\f{\p}{\p p_{a\nu}}K_{gr}^{cl}\Big)$ to write the total contribution in a compact form. The expression for $\Delta_{(2)}^{(5)}\widehat{T}^{X\mu\nu}(k) , \Delta_{(2)}^{(6)}\widehat{T}^{X\mu\nu}(k)$ and $\Delta_{(2)}^{(7)}\widehat{T}^{X\mu\nu}(k)$ are given in eq.\eqref{TX25},\eqref{TX26} and \eqref{TX27}.

\subsubsection{Analysis of sub-subleading gravitational energy-momentum tensor}
Now we want to analyze the sub-subleading order gravitational energy-momentum tensor to extract $\mathcal{O}(\omega(\ln\omega)^{2})$ contribution. From eq.\eqref{Thsketch} if we substitute  $r=2$, we find that there are two kinds of contributions to sub-subleading order gravitational energy-momentum tensor. One is cubic in $\Delta_{(0)}h_{\mu\nu}$ which we denote by $\Delta_{(2)}T^{h\mu\nu}_{1}(x)$ and the other contribution is quadratic in metric fluctuation with one of them is $\Delta_{(0)}h_{\mu\nu}$ and the other is $\Delta_{(1)}h_{\mu\nu}$ which we denote by $\Delta_{(2)}T^{h\mu\nu}_{2}(x)$. The Fourier transform of the first part of the sub-subleading gravitational energy-momentum tensor follows from eq.\eqref{T3},
\begingroup
\allowdisplaybreaks
\be
\Delta_{(2)}\widehat{T}_{1}^{h\mu\nu}(k)&=& -i(8\pi G)^{2}\sum_{a,b,c=1}^{M+N}\ \int\f{d^{4}\ell_{1}}{(2\pi)^{4}}\f{d^{4}\ell_{2}}{(2\pi)^{4}}G_{r}(\ell_{1})G_{r}(\ell_{2})G_{r}(k-\ell_{1}-\ell_{2})\ e^{-i\ell_{1}.(r_{a}-r_{c})}e^{-i\ell_{2}.(r_{b}-r_{c})}e^{-ik.r_{c}}\non\\
&&\times \f{1}{p_{a}.\ell_{1}-i\epsilon}\f{1}{p_{b}.\ell_{2}-i\epsilon}\ \f{1}{p_{c}.(k-\ell_{1}-\ell_{2})-i\epsilon}\ \Big{\lbrace}p_{a}^{\alpha}p_{a}^{\beta}-\f{1}{2}p_{a}^{2}\eta^{\alpha\beta}\Big{\rbrace}\ \Big{\lbrace}p_{b}^{\rho}p_{b}^{\sigma}-\f{1}{2}p_{b}^{2}\eta^{\rho\sigma}\Big{\rbrace}\non\\
&&\ \Big{\lbrace}p_{c}^{\gamma}p_{c}^{\delta}-\f{1}{2}p_{c}^{2}\eta^{\gamma\delta}\Big{\rbrace}\ G^{\mu\nu}_{\alpha\beta ,\rho\sigma ,\gamma\delta}(\ell_{1},\ell_{2},k)\label{cubicTh}
\ee
\endgroup
where $ G^{\mu\nu}_{\alpha\beta ,\rho\sigma ,\gamma\delta}(\ell_{1},\ell_{2},k)$ can be read off from eq.\eqref{T3h} and the expression is quadratic in momenta. Since $T^{(3)h\mu\nu}(x)$ of eq.\eqref{T3h} contains two derivatives on one or two different $\lbrace h_{\tau\kappa}\rbrace$, we can always choose $h^{\gamma\delta}$ to be the metric fluctuation where no derivative operates, as explained in appendix-\ref{appC}. With this choice the momentum dependence of $ G^{\mu\nu}_{\alpha\beta ,\rho\sigma ,\gamma\delta}(\ell_{1},\ell_{2},k)$ turns out to be,
\be
 G^{\mu\nu}_{\alpha\beta ,\rho\sigma ,\gamma\delta}(\ell_{1},\ell_{2},k)\sim \# \ell_{1}\ell_{1}+\# \ell_{1}\ell_{2}+\# \ell_{2}\ell_{2} 
\ee
A detail power counting analysis shows that $\Delta_{(2)}\widehat{T}_{1}^{h\mu\nu}(k)$ does not contribute to order $\mathcal{O}\big(\omega(\ln\omega)^{2}\big)$ in any integration region. In small $\omega$ limit the leading non-analytic contribution comes from $\Delta_{(2)}\widehat{T}_{1}^{h\mu\nu}(k)$ at order $\mathcal{O}(\omega\ln\omega)$. Since we are not interested in order $\mathcal{O}(\omega\ln\omega)$ terms, we can ignore the contribution of $\Delta_{(2)}\widehat{T}_{1}^{h\mu\nu}(k)$ for the purpose of this paper.

The Fourier transformation of $\Delta_{(2)}T^{h\mu\nu}_{2}(x)$ takes the following form,
\be
\Delta_{(2)}\widehat{T}_{2}^{h\mu\nu}(k)\ &=&\ \f{1}{8\pi G}\ \int\f{d^{4}\ell_{1}}{(2\pi)^{4}}\ \Bigg[ \Delta_{(0)}\widehat{h}^{\alpha\beta}(\ell_{1})\mathcal{F}^{\mu\nu }\ _{\alpha\beta ,\rho\sigma }(k,\ell_{1})\Delta_{(1)}\widehat{h}^{\rho\sigma}(k-\ell_{1})+\Delta_{(1)}\widehat{h}^{\alpha\beta}(\ell_{1})\mathcal{F}^{\mu\nu}\ _{ \alpha\beta ,\rho\sigma }(k,\ell_{1})\non\\
&&\ \Delta_{(0)}\widehat{h}^{\rho\sigma}(k-\ell_{1})\Bigg]\label{Delta2Th2}
\ee
where
\be
\Delta_{(0)}\widehat{h}^{\mu\nu}(\ell_{1})&=&\ -8\pi G\ G_{r}(\ell_{1})\sum_{a=1}^{M+N}\ e^{-i\ell_{1}.r_{a}}\ \f{p_{a}^{\mu}p_{a}^{\nu}-\f{1}{2}p_{a}^{2}\eta^{\mu\nu}}{i(p_{a}.\ell_{1}-i\epsilon)}\\
\Delta_{(1)}\widehat{h}^{\mu\nu}(\ell_{1})&=&\ -8\pi G\ G_{r}(\ell_{1})\Big[\Delta_{(1)}\widehat{T}^{\mu\nu}(\ell_{1})-\f{1}{2}\eta^{\mu\nu}\Delta_{(1)}\widehat{T}^{\rho}_{\rho}(\ell_{1})\Big]
\ee
with the expression for $\Delta_{(1)}\widehat{T}^{\mu\nu}(\ell_{1})$ follows from eq.\eqref{subleadingTX} and eq.\eqref{Thath} :
\be
\Delta_{(1)}\widehat{T}^{\mu\nu}(\ell_{1})\ &=&\ -8\pi G \ \sum_{b=1}^{M+N}\sum_{\substack{c=1\\ c\neq b}}^{M+N}\ e^{-i\ell_{1}.r_{b}}\ 
\int\f{d^{4}\ell_{2}}{(2\pi)^{4}}\ e^{i\ell_{2}.(r_{b}-r_{c})}G_{r}(\ell_{2})\ \f{1}{p_{b}.\ell_{2}+i\epsilon}\f{1}{p_{c}.\ell_{2}-i\epsilon}\f{1}{p_{b}.(\ell_{2}-\ell_{1})+i\epsilon}\non\\
&&\ \times \mathcal{E}^{\mu\nu}(p_{b},p_{c},\ell_{1},\ell_{2})\non\\
&&\ -8\pi G\ \sum_{b=1}^{M+N}\sum_{c=1}^{M+N}e^{-i\ell_{1}.r_{c}}\int\f{d^{4}\ell_{2}}{(2\pi)^{4}}\ e^{i\ell_{2}.(r_{c}-r_{b})}G_{r}(\ell_{1}-\ell_{2})G_{r}(\ell_{2})\ \f{1}{p_{b}.\ell_{2}-i\epsilon}\f{1}{p_{c}.(\ell_{1}-\ell_{2})-i\epsilon}\non\\
&&\times \Big{\lbrace}p_{b\alpha}p_{b\beta}-\f{1}{2}p_{b}^{2}\eta_{\alpha\beta}\Big{\rbrace}\mathcal{F}^{\mu\nu,\alpha\beta,\rho\sigma}(\ell_{1},\ell_{2})\Big{\lbrace}p_{c\rho}p_{c\sigma}-\f{1}{2}p_{c}^{2}\eta_{\rho\sigma}\Big{\rbrace}
\ee
where the expression for $\mathcal{F}^{\mu\nu,\alpha\beta,\rho\sigma}$ is given in eq.\eqref{F} and the expression for for $\mathcal{E}^{\mu\nu}$ is the following,
\be
\mathcal{E}^{\mu\nu}(p_{b},p_{c},\ell_{1},\ell_{2})\ &=&\ \f{p_{b}^{\mu}p_{b}^{\nu}}{p_{b}.\ell_{1}-i\epsilon}\Big{\lbrace} 2p_{b}.p_{c}p_{b}.\ell_{2}p_{c}.\ell_{1}-\ell_{1}.\ell_{2}(p_{b}.p_{c})^{2}-p_{c}^{2}p_{b}.\ell_{1}p_{b}.\ell_{2}+\f{1}{2}\ell_{1}.\ell_{2}p_{b}^{2}p_{c}^{2}\Big{\rbrace}\non\\
&&\ -\Big{\lbrace} 2p_{b}.p_{c}\ell_{2}.p_{b}(p_{b}^{\mu}p_{c}^{\nu}+p_{b}^{\nu}p_{c}^{\mu})-(p_{b}.p_{c})^{2}(\ell_{2}^{\mu}p_{b}^{\nu}+\ell_{2}^{\nu}p_{b}^{\mu})-2p_{c}^{2}p_{b}.\ell_{2}p_{b}^{\mu}p_{b}^{\nu}\non\\
&&\ +\f{1}{2}p_{b}^{2}p_{c}^{2}(\ell_{2}^{\mu}p_{b}^{\nu}+\ell_{2}^{\nu}p_{b}^{\mu})\Big{\rbrace}
\ee
Here our goal will be to analyze the expression \eqref{Delta2Th2} in all possible integration regions after substituting the expressions of $\lbrace\Delta_{(1)}h_{\alpha\beta}\rbrace$ and $\lbrace\Delta_{(0)}h_{\alpha\beta}\rbrace$ to extract the order $\mathcal{O}\big(\omega(\ln\omega)^{2}\big)$ contribution. We only found the following three integration regions where eq.\eqref{Delta2Th2} contributes to order $\mathcal{O}\big(\omega(\ln\omega)^{2}\big)$ :\ $L^{-1}>>|\ell^{\mu}_{2}|>>\omega>>|\ell^{\mu}_{1}|>>R^{-1}$ , $\omega>>|\ell^{\mu}_{2}|>>|\ell^{\mu}_{1}|>>R^{-1}$ and $L^{-1}>>|\ell^{\mu}_{2}|>>|\ell^{\mu}_{1}|>>\omega$.

In the integration region $L^{-1}>>|\ell^{\mu}_{2}|>>\omega>>|\ell^{\mu}_{1}|>>R^{-1}$, approximating the expression in eq.\eqref{Delta2Th2} and performing the integration over $\ell_{2}$ analogous to \S\ref{Ssecsubgr} from the first term within the square bracket we get,
\be
\mathcal{U}_{1}\ &\equiv &\ i(8\pi G)\ 2G\sum_{a=1}^{M+N}\sum_{b=1}^{M+N}\sum_{\substack{c=1\\c\neq b\\ \eta_{b}\eta_{c}=1}}^{M+N}\int_{R^{-1}}^{\omega}\f{d^{4}\ell_{1}}{(2\pi)^{4}}\ G_{r}(\ell_{1}) \f{1}{k.\ell_{1}+i\epsilon}\ \f{1}{p_{a}.\ell_{1} -i\epsilon}\ \non\\
&&\times \f{1}{p_{b}.k}\ \ln\Big{\lbrace}L(\omega +i\epsilon\eta_{b})\Big{\rbrace}\ \f{p_{b}.p_{c}}{[(p_{b}.p_{c})^{2}-p_{b}^{2}p_{c}^{2}]^{3/2}}\ \Big{\lbrace}\f{3}{2}p_{b}^{2}p_{c}^{2}-(p_{b}.p_{c})^{2}\Big{\rbrace}\non\\
&&\ \times\Big{\lbrace}-(p_{a}.k)^{2}p_{c}.kp_{b}^{\mu}p_{b}^{\nu}+(p_{a}.k)^{2}p_{b}.kp_{b}^{\mu}p_{c}^{\nu}\Big{\rbrace}
\ee
Now using the result of the integration:
\be
&&\ \int_{R^{-1}}^{\omega}\f{d^{4}\ell}{(2\pi)^{4}}\ G_{r}(\ell) \f{1}{k.\ell+i\epsilon}\ \f{1}{p_{a}.\ell -i\epsilon}\ =\ \f{1}{4\pi}\ \delta_{\eta_{a},1}\ \f{1}{p_{a}.k}\ \ln\Big{\lbrace}R(\omega+i\epsilon)\Big{\rbrace}+\mathcal{O}(\omega^{-1})
\ee
the contribution of $\mathcal{U}_{1}$ turns out to be,
\be
\mathcal{U}_{1}\ &=&\ -i\ (2G)^{2}\sum_{a=1}^{N}p_{a}.k \ln\Big{\lbrace}R(\omega+i\epsilon)\Big{\rbrace}\sum_{b=1}^{M+N}\sum_{\substack{c=1\\c\neq b\\ \eta_{b}\eta_{c}=1}}^{M+N}\ln\Big{\lbrace}L(\omega+i\epsilon\eta_{b})\Big{\rbrace}\non\\
&&\times \f{p_{b}.p_{c}}{\big[(p_{b}.p_{c})^{2}-p_{b}^{2}p_{c}^{2}\big]^{3/2}}\ \Big{\lbrace}\f{3}{2}p_{b}^{2}p_{c}^{2}-(p_{b}.p_{c})^{2}\Big{\rbrace}\ \f{1}{p_{b}.k}\Big[p_{c}.kp_{b}^{\mu}p_{b}^{\nu}-p_{b}.kp_{b}^{\mu}p_{c}^{\nu}\Big]\label{U1}
\ee

In the region of integration integration $\omega>>|\ell^{\mu}_{2}|>>|\ell^{\mu}_{1}|>>R^{-1}$, approximating the expression in eq.\eqref{Delta2Th2} and performing the integration over $\ell_{2}$ analogous to \S\ref{Ssecsubgr} we find,
\be
\mathcal{U}_{2}\ &\equiv &\ -i(8\pi G)\ 2G\ \sum_{a=1}^{M+N}\sum_{b=1}^{N}\sum_{c=1}^{M+N}\int_{R^{-1}}^{\omega}\f{d^{4}\ell_{1}}{(2\pi)^{4}}\ G_{r}(\ell_{1})\ \f{1}{k.\ell_{1}+i\epsilon}\ \f{1}{p_{a}.\ell_{1}-i\epsilon}\ \f{p_{b}.k}{p_{c}.k}\non\\
&&\ \ln\Big{\lbrace}|\vec{\ell}_{1}|^{-1}(\omega+i\epsilon)\Big{\rbrace}\Big[p_{a}.kp_{c}.kp_{a}.p_{c}\eta^{\mu\nu}-(p_{a}.k)^{2}p_{c}^{\mu}p_{c}^{\nu}-\f{1}{2}p_{a}^{2}(p_{c}.k)^{2}\eta^{\mu\nu} \Big]
\ee
Now using the result of the following integral:
\be
&&\int_{R^{-1}}^{\omega}\f{d^{4}\ell}{(2\pi)^{4}}\ G_{r}(\ell) \f{1}{k.\ell+i\epsilon}\ \f{1}{p_{a}.\ell -i\epsilon}\  \ln\Big{\lbrace}|\vec{\ell}|^{-1}(\omega+i\epsilon)\Big{\rbrace}\ \non\\
&=&\ \f{1}{8\pi}\ \delta_{\eta_{a},1}\ \f{1}{p_{a}.k}\ \Bigg(\ln\Big{\lbrace}R(\omega+i\epsilon)\Big{\rbrace}\Bigg)^{2}\ +\ \mathcal{O}(\omega^{-1}\ln\omega)
\ee
we get,
\be
\mathcal{U}_{2}\ &=&\  -\ \f{i}{2}\ \Big[-2iG\ \sum_{a=1}^{N}p_{a}.k\  \ln\Big{\lbrace}R(\omega+i\epsilon)\Big{\rbrace}\Big]^{2}\ \sum_{c=1}^{M+N}\f{p_{c}^{\mu}p_{c}^{\nu}}{p_{c}.k}\label{U2}
\ee
For analyzing the expression \eqref{Delta2Th2} in the integration region $L^{-1}>>|\ell^{\mu}_{2}|>>|\ell^{\mu}_{1}|>>\omega$, we need to substitute
\be
G_{r}(\ell_{1})G_{r}(k-\ell_{1})=G^{*}_{r}(\ell_{1})G_{r}(k-\ell_{1})\ -\ 2\pi i\ \delta(\ell_{1}^{2})\big[H(\ell_{1}^{0})-H(-\ell_{1}^{0})\big]\ G_{r}(k-\ell_{1}).\label{GG}
\ee
First let us analyze the contribution of eq.\eqref{Delta2Th2} with $G^{*}_{r}(\ell_{1})G_{r}(k-\ell_{1})=G_{r}(-\ell_{1})G_{r}(k-\ell_{1})$. It turns out that the following part of $\Delta_{(2)}\widehat{T}_{2}^{h\mu\nu}(k)$ contributes to order $\mathcal{O}\big(\omega(\ln\omega)^{2}\big)$ in the integration region $L^{-1}>>|\ell^{\mu}_{2}|>>|\ell^{\mu}_{1}|>>\omega$,
\begingroup
\allowdisplaybreaks
\be
\mathcal{U}_{3}\ &\equiv &\ -\ i(8\pi G)^{2}\sum_{a=1}^{M+N}\sum_{b=1}^{M+N}\sum_{\substack{c=1\\ c\neq b}}^{M+N}\int \f{d^{4}\ell_{1}}{(2\pi)^{4}}\ \big{\lbrace}G_{r}(-\ell_{1})\big{\rbrace}^{2}\Big[1-2k.\ell_{1}\ G_{r}(-\ell_{1})\Big]\ \f{1}{p_{a}.\ell_{1}-i\epsilon}\non\\
&&\ \f{1}{p_{b}.\ell_{1}+i\epsilon}\Big[1+\f{p_{b}.k}{p_{b}.\ell_{1}+i\epsilon}\Big]\ \Big{\lbrace}p_{a}^{\alpha}p_{a}^{\beta}-\f{1}{2}p_{a}^{2}\eta^{\alpha\beta}\Big{\rbrace}\Big{\lbrace}p_{b}^{\rho}p_{b}^{\sigma}-\f{1}{2}p_{b}^{2}\eta^{\rho\sigma}\Big{\rbrace}\non\\
&& \Big[\Delta_{(\ell_{1}\ell_{1})}\mathcal{F}^{\mu\nu}\ _{\alpha\beta,\rho\sigma}(k,\ell_{1})+\Delta_{(k\ell_{1})}\mathcal{F}^{\mu\nu}\ _{\alpha\beta,\rho\sigma}(k,\ell_{1}) \Big]\times \int\f{d^{4}\ell_{2}}{(2\pi)^{4}}\ G_{r}(\ell_{2})\f{1}{p_{c}.\ell_{2}-i\epsilon}\non\\
&&\ \f{1}{(p_{b}.\ell_{2}+i\epsilon)^{2}}\Bigg[1-\f{p_{b}.(\ell_{1}-k)}{p_{b}.\ell_{2}+i\epsilon}\Bigg]\ \Big[\big{\lbrace}-2p_{b}.p_{c}p_{b}.\ell_{2}p_{c}.\ell_{1}+\ell_{1}.\ell_{2}(p_{b}.p_{c})^{2}+p_{c}^{2}p_{b}.\ell_{1}p_{b}.\ell_{2}\non\\
&&\ -\f{1}{2}\ell_{1}.\ell_{2}p_{b}^{2}p_{c}^{2}\big{\rbrace}+\big{\lbrace}2p_{b}.p_{c}p_{b}.\ell_{2}p_{c}.k -k.\ell_{2}(p_{b}.p_{c})^{2}-p_{c}^{2}p_{b}.kp_{b}.\ell_{2}+\f{1}{2}k.\ell_{2}p_{b}^{2}p_{c}^{2}\big{\rbrace}\Big]\non\\
&&\ -\ i(8\pi G)^{2}\ \sum_{a=1}^{M+N}\sum_{b=1}^{M+N}\sum_{\substack{c=1\\ c\neq b}}^{M+N}\int \f{d^{4}\ell_{1}}{(2\pi)^{4}}\ \big{\lbrace}G_{r}(-\ell_{1})\big{\rbrace}^{2}\Big[1-2k.\ell_{1}\ G_{r}(-\ell_{1})\Big]\ \f{1}{p_{b}.\ell_{1}-i\epsilon}\non\\
&&\f{1}{p_{a}.\ell_{1}+i\epsilon}\Big[1+\f{p_{a}.k}{p_{a}.\ell_{1}+i\epsilon}\Big]\ \Big{\lbrace}p_{b}^{\alpha}p_{b}^{\beta}-\f{1}{2}p_{b}^{2}\eta^{\alpha\beta}\Big{\rbrace}\Big{\lbrace}p_{a}^{\rho}p_{a}^{\sigma}-\f{1}{2}p_{a}^{2}\eta^{\rho\sigma}\Big{\rbrace}\non\\
&& \Big[\Delta_{(\ell_{1}\ell_{1})}\mathcal{F}^{\mu\nu}\ _{\alpha\beta,\rho\sigma}(k,\ell_{1})+\Delta_{(k\ell_{1})}\mathcal{F}^{\mu\nu}\ _{\alpha\beta,\rho\sigma}(k,\ell_{1}) \Big]\times \int\f{d^{4}\ell_{2}}{(2\pi)^{4}}\ G_{r}(\ell_{2})\f{1}{p_{c}.\ell_{2}-i\epsilon}\non\\
&&\f{1}{(p_{b}.\ell_{2}+i\epsilon)^{2}}\Bigg[1+\f{p_{b}.\ell_{1}}{p_{b}.\ell_{2}+i\epsilon}\Bigg]\ \Big{\lbrace} 2p_{b}.p_{c}p_{b}.\ell_{2}p_{c}.\ell_{1}-\ell_{1}.\ell_{2}(p_{b}.p_{c})^{2}-p_{c}^{2}p_{b}.\ell_{1}p_{b}.\ell_{2}+\f{1}{2}\ell_{1}.\ell_{2}p_{b}^{2}p_{c}^{2}\Big{\rbrace}\non\\ \label{U3}
\ee
\endgroup
In the above expression $\Delta_{(\ell_{1}\ell_{1})}\mathcal{F}^{\mu\nu}\ _{\alpha\beta,\rho\sigma}(k,\ell_{1})$ represents the order $\mathcal{O}(\ell_{1}\ell_{1})$ contribution of  $\mathcal{F}^{\mu\nu}\ _{\alpha\beta,\rho\sigma}(k,\ell_{1})$ and $\Delta_{(k\ell_{1})}\mathcal{F}^{\mu\nu}\ _{\alpha\beta,\rho\sigma}(k,\ell_{1})$ represents the order $\mathcal{O}(k\ell_{1})$ contribution of  $\mathcal{F}^{\mu\nu}\ _{\alpha\beta,\rho\sigma}(k,\ell_{1})$, which one can read out from the expression \eqref{F}. Now it turns out that if we pick up the leading contributions from all the square brackets above in the integration region $L^{-1}>>|\ell^{\mu}_{2}|>>|\ell^{\mu}_{1}|>>\omega$, the resulting expression does not contribute to order $\mathcal{O}\big(\omega(\ln\omega)^{2}\big)$. On the other hand if we take subleading contribution from one of the square bracket above and leading contributions from the rest of the square brackets that can contribute to order $\mathcal{O}\big(\omega(\ln\omega)^{2}\big)$. But for analyzing those contributions we have to use the following results which we get using integration by parts,
\be
&&\ \int \f{d^{4}\ell_{1}}{(2\pi)^{4}}\ \big[G_{r}(-\ell_{1})\big]^{2}\ \f{1}{p_{a}.\ell_{1}-i\epsilon}\f{1}{p_{b}.\ell_{1}+i\epsilon}\ \ell_{1}^{\alpha}\ell_{1}^{\beta}\non\\
&=&\ -\f{1}{2}\Big[p_{a}^{\alpha}\f{\p}{\p p_{a\beta}}+p_{b}^{\alpha}\f{\p}{\p p_{b\beta}}+\eta^{\alpha\beta}\Big]\ \int\f{d^{4}\ell_{1}}{(2\pi)^{4}}\ G_{r}(-\ell_{1})\ \f{1}{p_{a}.\ell_{1}-i\epsilon}\f{1}{p_{b}.\ell_{1}+i\epsilon} \ ,
\ee
\be
&&\int \f{d^{4}\ell_{1}}{(2\pi)^{4}}\ \big[G_{r}(-\ell_{1})\big]^{2}\ \f{1}{p_{a}.\ell_{1}-i\epsilon}\f{1}{(p_{b}.\ell_{1}+i\epsilon)^{2}}\ \ell_{1}^{\alpha}\ell_{1}^{\beta}\ell_{1}^{\tau}\non\\
&=&\ \f{1}{2}\Bigg[p_{a}^{\tau}\f{\p}{\p p_{a\alpha}}\f{\p}{\p p_{b\beta}}+p_{b}^{\tau}\f{\p}{\p p_{b\alpha}}\f{\p}{\p p_{b\beta}}+\Big(\eta^{\alpha\tau}\f{\p}{\p p_{b\beta}}+\eta^{\beta\tau}\f{\p}{\p p_{b\alpha}}\Big)\Bigg]\non\\
&&\ \times \ \int\f{d^{4}\ell_{1}}{(2\pi)^{4}}\ G_{r}(-\ell_{1})\ \f{1}{p_{a}.\ell_{1}-i\epsilon}\f{1}{p_{b}.\ell_{1}+i\epsilon}
\ee
Now after using this integration by parts results in $\mathcal{U}_{3}$ and evaluating the integrals  in the integration region $L^{-1}>>|\ell^{\mu}_{2}|>>|\ell^{\mu}_{1}|>>\omega$  we find,
\be
\mathcal{U}_{3}&\simeq &\ -\f{i}{2}\ \sum_{a=1}^{M+N}(p_{a}.k)\ \Bigg(\f{\p}{\p p_{a\mu}}K_{gr}^{cl}\Bigg)\ \Bigg(\f{\p}{\p p_{a\nu}}K_{gr}^{cl}\Bigg) -\Delta_{(2)}^{(5)}\widehat{T}^{X\mu\nu}(k) -\Delta_{(2)}^{(6)}\widehat{T}^{X\mu\nu}(k) -\Delta_{(2)}^{(7)}\widehat{T}^{X\mu\nu}(k)\non\\ \label{U3final}
\ee
We have to redefine the dummy indices for particle sums $a,b,c$ and organize the huge expression coming from eq.\eqref{U3} after performing the integrations, to bring the result in the compact form written above.

Now we also have to analyze the contribution of $\Delta_{(2)}\widehat{T}_{2}^{h\mu\nu}(k)$ in eq.\eqref{Delta2Th2} after substituting $-\ 2\pi i\ \delta(\ell_{1}^{2})\big[H(\ell_{1}^{0})-H(-\ell_{1}^{0})\big]\ G_{r}(k-\ell_{1})$ in place of $G_{r}(\ell_{1})G_{r}(k-\ell_{1})$ as follows from eq.\eqref{GG}. With this substituation $\Delta_{(2)}\widehat{T}_{2}^{h\mu\nu}(k)$ can contribute to some linearly divergent term at order $\mathcal{O}(\omega^{0}L^{-1})$ in the integration region where integrating momenta are large compare to $\omega$. We expect this contribution will represent  the subleading soft gravitational radiation from hard gravitational radiation which needs to be subtracted by hand as it's effect is already included in the sum over hard particles $a,b,c$\footnote{It would be interesting to independently compute subleading soft gravitational radiation from real hard gravitons at subleading order and compare with the $\mathcal{O}(\omega^{0}L^{-1})$ terms which we need to subtract here analogous to the leading order case as given in eq.\eqref{realrad}.}. After this subtraction we find the following integral structure from $\Delta_{(2)}\widehat{T}_{2}^{h\mu\nu}(k)$, which can potencially contribute to $\mathcal{O}(\omega(\ln\omega)^{2})$,
\be
&&\omega\ln\omega \int_{\omega}^{L^{-1}}\f{d^{4}\ell_{1}}{(2\pi)^{4}}\ \delta(\ell_{1}^{2})\ \big[H(\ell_{1}^{0})-H(-\ell_{1}^{0})\big] \f{1}{\mathbf{n}.\ell_{1}}\ \f{1}{p_{a}.\ell_{1}}\Bigg[f_{\alpha\beta\gamma}^{\mu\nu}(p_{a},p_{b},p_{c},\mathbf{n})\f{\ell_{1}^{\alpha}\ell_{1}^{\beta}\ell_{1}^{\gamma}}{(p_{b}.\ell_{1})^{3}}\non\\
&&+g^{\mu\nu}_{\alpha\beta}(p_{a},p_{b},p_{c},\mathbf{n})\f{\ell_{1}^{\alpha}\ell_{1}^{\beta}}{(p_{b}.\ell_{1})^{2}}+h^{\mu\nu}_{\alpha}(p_{a},p_{b},p_{c},\mathbf{n})\f{\ell_{1}^{\alpha}}{p_{b}.\ell_{1}}\Bigg]
\ee
where $f,g,h$ are functions of $p_{a},p_{b},p_{c},\mathbf{n}$ and the structure of the integral is written after performing integration over $\ell_{2}$. Now one can see that the part of the integrand inside the square bracket is symmetric under $\ell_{1}\leftrightarrow -\ell_{1}$ exchange. On the other hand the part of the integrand outside the square bracket is anti-symmetric under $\ell_{1}\leftrightarrow -\ell_{1}$ exchange. Hence from this the order $\mathcal{O}(\omega(\ln\omega)^{2})$ contribution vanishes.

Summing over the contributions of eq.\eqref{sub2matterfinal}, \eqref{U1}, \eqref{U2} and \eqref{U3final}, the order $\mathcal{O}\big(\omega(\ln\omega)^{2}\big)$ contribution of sub-subleading order energy-momentum tensor becomes,
\begingroup
\allowdisplaybreaks
\be
\Delta_{(2)}\widehat{T}^{\mu\nu}(k)\ &=&\ -\ \f{i}{2}\ \Big[-2iG\ \sum_{a=1}^{N}p_{a}.k\  \ln\Big{\lbrace}R(\omega+i\epsilon)\Big{\rbrace}\Big]^{2}\ \sum_{c=1}^{M+N}\f{p_{c}^{\mu}p_{c}^{\nu}}{p_{c}.k}\non\\
&&-i\ \Big[-2iG\ \sum_{a=1}^{N}p_{a}.k\ \ln\Big{\lbrace}R(\omega+i\epsilon)\Big{\rbrace}\Big]\sum_{b=1}^{M+N}\f{p_{b}^{\mu}k_{\rho}}{p_{b}.k}\Bigg[\Bigg(p_{b}^{\nu}\f{\p}{\p p_{b\rho}}-p_{b}^{\rho}\f{\p}{\p p_{b\nu}}\Bigg)K_{gr}^{cl}\Bigg]\non\\
&&- \f{i}{2}\sum_{a=1}^{M+N}\f{k_{\rho}k_{\sigma}}{p_{a}.k}\ \Bigg[\Bigg(p_{a}^{\mu}\f{\p}{\p p_{a\rho}}-p_{a}^{\rho}\f{\p}{\p p_{a\mu}}\Bigg)K_{gr}^{cl}\Bigg]\ \Bigg[\Bigg(p_{a}^{\nu}\f{\p}{\p p_{a\sigma}}-p_{a}^{\sigma}\f{\p}{\p p_{a\nu}}\Bigg)K_{gr}^{cl}\Bigg]\non\\
&&\ +\mathcal{O}\big(\omega\ln\omega\big)\label{delta2T}
\ee
\endgroup

\subsubsection{Sub-subleading order gravitational waveform}
Now using the relation in eq.\eqref{eTrelation}, the sub-subleading order gravitational waveform in frequency space becomes,
\begingroup
\allowdisplaybreaks
\be
&&\ \Bigg{\lbrace}\f{2G}{R}\Bigg{\rbrace}^{-1}\ \Bigg{\lbrace}exp\Big{\lbrace}i\omega R -2iG\sum_{a=1}^{N}p_{a}.k\ln R\Big{\rbrace}\Bigg{\rbrace}^{-1}\  \Delta_{(2)}\widetilde{e}^{\mu\nu}(\omega ,R,\hat{n})\non\\
&=&\ -i\ \omega\big{\lbrace}\ln(\omega+i\epsilon)\big{\rbrace}^{2}\times  \f{1}{2}\Big{\lbrace}-2iG\ \sum_{\substack{b=1}}^{N}p_{b}\cdot\mathbf{n}\Big{\rbrace}^{2}\ \sum_{a=1}^{M+N}\f{p_{a}^{\mu}p_{a}^{\nu}}{p_{a}.\mathbf{n}}\non\\
&&\ -i\ \omega \ln(\omega+i\epsilon)\ \Big{\lbrace}-2iG\ \sum_{\substack{c=1}}^{N}p_{c}\cdot \mathbf{n}\Big{\rbrace}\sum_{a=1}^{M+N}\ \ln(\omega +i\epsilon\eta_{a})\non\\
&&\times \Bigg[(2iG)\sum_{\substack{b=1\\ b\neq a\\ \eta_{a}\eta_{b}=1}}^{M+N}\f{p_{a}.p_{b}}{[(p_{a}.p_{b})^{2}-p_{a}^{2}p_{b}^{2}]^{3/2}}\Big{\lbrace}\f{3}{2}p_{a}^{2}p_{b}^{2}-(p_{a}.p_{b})^{2}\Big{\rbrace}\ \f{p_{a}^{\mu}\mathbf{n}_{\rho}}{p_{a}.\mathbf{n}}\Big{\lbrace}p_{a}^{\nu}p_{b}^{\rho}-p_{a}^{\rho}p_{b}^{\nu}\Big{\rbrace}\Bigg]\non\\
&&\ -\f{i}{2}\sum_{a=1}^{M+N}\omega\Big(\ln(\omega +i\epsilon\eta_{a})\Big)^{2} \f{\mathbf{n}_{\rho}\mathbf{n}_{\sigma}}{p_{a}.\mathbf{n}}\Bigg[(2iG)\sum_{\substack{b=1\\ b\neq a\\ \eta_{a}\eta_{b}=1}}^{M+N}\f{p_{a}.p_{b}}{[(p_{a}.p_{b})^{2}-p_{a}^{2}p_{b}^{2}]^{3/2}}\Big{\lbrace}\f{3}{2}p_{a}^{2}p_{b}^{2}-(p_{a}.p_{b})^{2}\Big{\rbrace}\ \Big{\lbrace}p_{a}^{\mu}p_{b}^{\rho}-p_{a}^{\rho}p_{b}^{\mu}\Big{\rbrace}\Bigg]\non\\
&&\times \Bigg[(2iG)\sum_{\substack{c=1\\ c\neq a\\ \eta_{a}\eta_{c}=1}}^{M+N}\f{p_{a}.p_{c}}{[(p_{a}.p_{c})^{2}-p_{a}^{2}p_{c}^{2}]^{3/2}}\Big{\lbrace}\f{3}{2}p_{a}^{2}p_{c}^{2}-(p_{a}.p_{c})^{2}\Big{\rbrace}\ \Big{\lbrace}p_{a}^{\nu}p_{c}^{\sigma}-p_{a}^{\sigma}p_{c}^{\nu}\Big{\rbrace}\Bigg]\non\\
&&\ +\ \mathcal{O}\big(\omega\ln\omega\big)
\ee
\endgroup
In the LHS of the above expression we have included a phase factor $exp\lbrace -2iG\sum\limits_{a=1}^{N}p_{a}.k\ln R\rbrace$ which corresponds to the time delay due to gravitational drag force on the emitted gravitational wave as derived in \cite{1808.03288}. The expansion of this extra phase factor up to order $\omega^{2}$ has already appeared in the expression \eqref{delta2T}. Now performing Fourier transformation in $\omega$ variable and using the integration results of eq.\eqref{FourierF0n}\eqref{FourierFn0}\eqref{FourierFn-rr} we get the following late and early time sub-subleading gravitational waveforms,
\begingroup
\allowdisplaybreaks
\be
&&\Delta_{(2)}e^{\mu\nu}(t,R,\hat{n})\non\\
&=&\ \f{2G}{R}\ \f{\ln |u|}{u^{2}}\ \Big{\lbrace}2G\sum_{b=1}^{N}p_{b}.\mathbf{n}\Big{\rbrace}^{2}\Bigg(\sum_{a=1}^{N}\f{p_{a}^{\mu}p_{a}^{\nu}}{p_{a}.\mathbf{n}}-\sum_{a=1}^{M}\f{p_{a}^{\prime\mu}p_{a}^{\prime\nu}}{p'_{a}.\mathbf{n}}\Bigg)\non\\
&&\ -\f{4G}{R}\ \f{\ln |u|}{u^{2}}\   \Big{\lbrace}2G\sum_{c=1}^{N}p_{c}.\mathbf{n}\Big{\rbrace}\  \sum_{a=1}^{N}\Bigg[(2G)\sum_{\substack{b=1\\ b\neq a}}^{N}\f{p_{a}.p_{b}}{[(p_{a}.p_{b})^{2}-p_{a}^{2}p_{b}^{2}]^{3/2}} \Big{\lbrace}\f{3}{2}p_{a}^{2}p_{b}^{2}-(p_{a}.p_{b})^{2}\Big{\rbrace}\ \f{p_{a}^{\mu}\mathbf{n}_{\rho}}{p_{a}.\mathbf{n}}\Big{\lbrace}p_{a}^{\nu}p_{b}^{\rho}-p_{a}^{\rho}p_{b}^{\nu}\Big{\rbrace}\Bigg]\non\\
&&\ -\f{2G}{R}\ \f{\ln |u|}{u^{2}}\  \Big{\lbrace}2G\sum_{c=1}^{N}p_{c}.\mathbf{n}\Big{\rbrace}\  \sum_{a=1}^{M}\Bigg[(2G)\sum_{\substack{b=1\\ b\neq a}}^{M}\f{p'_{a}.p'_{b}}{[(p'_{a}.p'_{b})^{2}-p_{a}^{\prime 2}p_{b}^{\prime 2}]^{3/2}} \Big{\lbrace}\f{3}{2}p_{a}^{\prime 2}p_{b}^{\prime 2}-(p'_{a}.p'_{b})^{2}\Big{\rbrace}\ \f{p_{a}^{\prime\mu}\mathbf{n}_{\rho}}{p'_{a}.\mathbf{n}}\Big{\lbrace}p_{a}^{\prime\nu}p_{b}^{\prime\rho}-p_{a}^{\prime\rho}p_{b}^{\prime\nu}\Big{\rbrace}\Bigg]\non\\
&&\ +\ \f{2G}{R}\ \f{\ln |u|}{u^{2}}   \sum_{a=1}^{N}\f{\mathbf{n}_{\rho}\mathbf{n}_{\sigma}}{p_{a}.\mathbf{n}}\Bigg[(2G)\sum_{\substack{b=1\\ b\neq a}}^{N}\f{p_{a}.p_{b}}{[(p_{a}.p_{b})^{2}-p_{a}^{2}p_{b}^{2}]^{3/2}}\Big{\lbrace}\f{3}{2}p_{a}^{2}p_{b}^{2}-(p_{a}.p_{b})^{2}\Big{\rbrace}\ \Big{\lbrace}p_{a}^{\mu}p_{b}^{\rho}-p_{a}^{\rho}p_{b}^{\mu}\Big{\rbrace}\Bigg]\non\\
&&\times \Bigg[(2G)\sum_{\substack{c=1\\ c\neq a}}^{N}\f{p_{a}.p_{c}}{[(p_{a}.p_{c})^{2}-p_{a}^{2}p_{c}^{2}]^{3/2}}\Big{\lbrace}\f{3}{2}p_{a}^{2}p_{c}^{2}-(p_{a}.p_{c})^{2}\Big{\rbrace}\ \Big{\lbrace}p_{a}^{\nu}p_{c}^{\sigma}-p_{a}^{\sigma}p_{c}^{\nu}\Big{\rbrace}\Bigg]\ +\mathcal{O}(u^{-2})\hspace{1cm}\hbox{for $u\rightarrow+\infty$},
\ee
\endgroup
and
\begingroup
\allowdisplaybreaks
\be
&&\Delta_{(2)}e^{\mu\nu}(t,R,\hat{n})\non\\
&=&\  \f{2G}{R}\ \f{\ln |u|}{u^{2}}\   \Big{\lbrace}2G\sum_{c=1}^{N}p_{c}.\mathbf{n}\Big{\rbrace}\  \sum_{a=1}^{M}\Bigg[(2G)\sum_{\substack{b=1\\ b\neq a}}^{M}\f{p'_{a}.p'_{b}}{[(p'_{a}.p'_{b})^{2}-p_{a}^{\prime 2}p_{b}^{\prime 2}]^{3/2}}\non\\
&&\times \Big{\lbrace}\f{3}{2}p_{a}^{\prime 2}p_{b}^{\prime 2}-(p'_{a}.p'_{b})^{2}\Big{\rbrace}\ \f{p_{a}^{\prime\mu}\mathbf{n}_{\rho}}{p'_{a}.\mathbf{n}}\Big{\lbrace}p_{a}^{\prime\nu}p_{b}^{\prime\rho}-p_{a}^{\prime\rho}p_{b}^{\prime\nu}\Big{\rbrace}\Bigg]\non\\
&&\ +\ \f{2G}{R}\ \f{\ln |u|}{u^{2}} \sum_{a=1}^{M}\f{\mathbf{n}_{\rho}\mathbf{n}_{\sigma}}{p'_{a}.\mathbf{n}}\Bigg[(2G)\sum_{\substack{b=1\\ b\neq a}}^{M}\f{p'_{a}.p'_{b}}{[(p'_{a}.p'_{b})^{2}-p_{a}^{\prime 2}p_{b}^{\prime 2}]^{3/2}}\Big{\lbrace}\f{3}{2}p_{a}^{\prime 2}p_{b}^{\prime 2}-(p'_{a}.p'_{b})^{2}\Big{\rbrace}\ \Big{\lbrace}p_{a}^{\prime\mu}p_{b}^{\prime\rho}-p_{a}^{\prime\rho}p_{b}^{\prime\mu}\Big{\rbrace}\Bigg]\non\\
&&\times \Bigg[(2G)\sum_{\substack{c=1\\ c\neq a}}^{M}\f{p'_{a}.p'_{c}}{[(p'_{a}.p'_{c})^{2}-p_{a}^{\prime2}p_{c}^{\prime2}]^{3/2}}\Big{\lbrace}\f{3}{2}p_{a}^{\prime2}p_{c}^{\prime2}-(p'_{a}.p'_{c})^{2}\Big{\rbrace}\ \Big{\lbrace}p_{a}^{\prime\nu}p_{c}^{\prime\sigma}-p_{a}^{\prime\sigma}p_{c}^{\prime\nu}\Big{\rbrace}\Bigg]+\mathcal{O}(u^{-2})\hspace{0.5cm}\hbox{for $u\rightarrow-\infty$}\non\\
\ee
\endgroup
where $u=t-R+2G\ln R\sum\limits_{b=1}^{N}p_{b}\cdot \mathbf{n}$ is the retarded time. This proves the conjecture made in \cite{1912.06413}. For a scattering event with massless external particles this result has been tested in \cite{1812.08137,1901.10986}.

\section{Sub-subleading soft photon theorem from two loop amplitudes}\label{FeynQED}
Here in this section we briefly discuss how we can derive sub-subleading soft photon theorem by analyzing two loop amplitudes in a theory of scalar QED following the analysis of \cite{1808.03288}. This will solve two purposes: one purpose is to show the connection between classical computation and Feynman diagrammatics and the other is to find out extra quantum contribution in sub-subleading soft photon factor. Consider a theory of $U(1)$ gauge filed $A_{\mu}(x)$ and $M+N$ number of complex scalar fields $\lbrace \phi_{a}\rbrace$ with masses $\lbrace m_{a}\rbrace$ and charges $\lbrace q_{a}\rbrace$ for $a=1,2,\cdots ,M+N$ satisfying $\sum\limits_{a=1}^{M+N}q_{a}=0$. The relevant part of the action needed for our analysis:
\be
&&\int d^{4}x\ \Big[ -\f{1}{4}F^{\mu\nu}F_{\mu\nu} -\sum_{a=1}^{M+N}\Big{\lbrace}(\p_{\mu}\phi^*_{a}+iq_{a}A_{\mu}\phi^*_{a})(\p^{\mu}\phi -iq_{a}A^{\mu}\phi_{a})\ +m_{a}^{2}\phi^*_{a}\phi_{a}\Big{\rbrace} \Big]\non\\
&&\ +\lambda \phi_{1}\phi_{2}\cdots\phi_{M+N}\ +\ \lambda \phi^*_{1}\phi^*_{2}\cdots\phi^*_{M+N}\Big]
\ee
Though in four spacetime dimensions S-matrix for this theory is IR divergent, still we can factor out the same IR-divergent piece from the S-matrix with  external photon and S-matrix without external photon and can derive the soft photon factor relating the IR finite parts of the two S-matrices \cite{1808.03288}. Let $\Gamma^{(M+N ,1)}$ denote the full perturbative scattering amplitude of $M+N$ number of external charged particles with momenta $\lbrace p_{a}\rbrace$, charges $\lbrace q_{a}\rbrace$ and one external outgoing photon with momentum $k$. Similarly $\Gamma^{(M+N )}$ denotes the full perturbative scattering amplitude with the same $M+N$ number of external charged particles but without any external photon. Following \cite{grammer}, let us decompose the photon propagator attached with two complex scalar lines with momenta $p_{a}$ and $p_{b}$ in the following way,
\be
-i\ \f{\eta^{\mu\nu}}{\ell^{2}-i\epsilon} \ =\ -i\ \f{1}{\ell^{2}-i\epsilon}\ \Bigg[ K_{(ab)}^{\mu\nu}(\ell,p_{a},p_{b})+G_{(ab)}^{\mu\nu}(\ell,p_{a},p_{b})\Bigg] 
\ee
where the photon momentum $\ell$ runs from particle-b to particle-a. We are using the convention that all the external particles are  outgoing and some of them can be made incoming by flipping the sign of their charges and momenta. The expression for $K_{(ab)}^{\mu\nu}(\ell,p_{a},p_{b})$ and $G_{(ab)}^{\mu\nu}(\ell,p_{a},p_{b})$ for $a\neq b$ are,
\be
K_{(ab)}^{\mu\nu}(\ell,p_{a},p_{b})\ &=&\ \ell^{\mu}\ell^{\nu}\f{(2p_{a}-\ell)\cdot(2p_{b}+\ell)}{(2p_{a}.\ell -\ell^{2}+i\epsilon)(2p_{b}.\ell +\ell^{2}-i\epsilon)}\non\\
G_{(ab)}^{\mu\nu}(\ell,p_{a},p_{b})\ &=&\ \eta^{\mu\nu}-K_{(ab)}^{\mu\nu}(\ell,p_{a},p_{b})
\ee
For $a=b$ we do not need to make any decomposition. The propagator with $K_{(ab)}$ numerator is named as K-photon propagator and the propagator with $G_{(ab)}$ numerator is named as G-photon propagator. It turns out that if we compute the amplitudes $\Gamma^{(M+N ,1)}$ and $\Gamma^{(M+N)}$ after the K-G decomposition of the photon propagators described above we get\cite{grammer},
\be
\Gamma^{(M+N ,1)}\ &=&\ exp[K_{em}]\ \Gamma^{(M+N ,1)}_{G}\ ,\hspace{1cm}\ \Gamma^{(M+N)}\ =\ exp[K_{em}]\ \Gamma^{(M+N)}_{G}\
\ee
where
\be
K_{em}\ &=&\ \f{i}{2}\sum_{a=1}^{M+N}\  \sum_{\substack{b=1\\b\neq a}}^{M+N}q_{a}q_{b}\ \int \f{d^{4}\ell}{(2\pi)^{4}} \ \f{1}{\ell^{2}-i\epsilon}\ \f{(2p_{a}-\ell)\cdot(2p_{b}+\ell)}{(2p_{a}.\ell-\ell^{2}+i\epsilon)\ (2p_{b}.\ell+\ell^{2}- i\epsilon)}\label{Kqmem}
\ee
Above $\Gamma^{(M+N ,1)}_{G}$ denotes the value of the IR-finite part of the amplitude $\Gamma^{(M+N ,1)}$ when all the photon propagators are replaced by G-photon propagators and $\Gamma^{(M+N)}_{G}$ denotes the value of the IR-finite part of the amplitude $\Gamma^{(M+N)}$ with all the photon propagators replaced by G-photon propagators. In \cite{1808.03288} analyzing one loop Feynman diagrams of $\Gamma^{(M+N ,1)}_{G}$ the subleading soft photon theorem is derived. Analogously analyzing two loop Feynman diagrams of $\Gamma^{(M+N ,1)}_{G}$ we can extract the leading non-analytic contribution of sub-subleading soft photon theorem. The algorithm for finding sub-subleading soft photon factor $S_{em}^{(2)}$ for photon polarization $\varepsilon$ turns out to be:
\be
\Gamma^{(M+N ,1)}_{G, 2-loop}\ &=&\ S_{em}^{(2)}\big(\varepsilon ,k,\lbrace q_{a},p_{a}\rbrace\big)\ \Gamma^{(M+N)}_{tree}\label{algo}
\ee
In principle, in the LHS of the above relation we should also include the contribution of the two loop Feynman diagrams which contain one or both virtual photons whose both ends are connected to a single scalar line. But a detailed analysis shows that those diagrams don't contribute to order $\mathcal{O}(\omega(\ln\omega)^{2})$.
%\begin{center}
\begin{figure}
\includegraphics[scale=0.25]{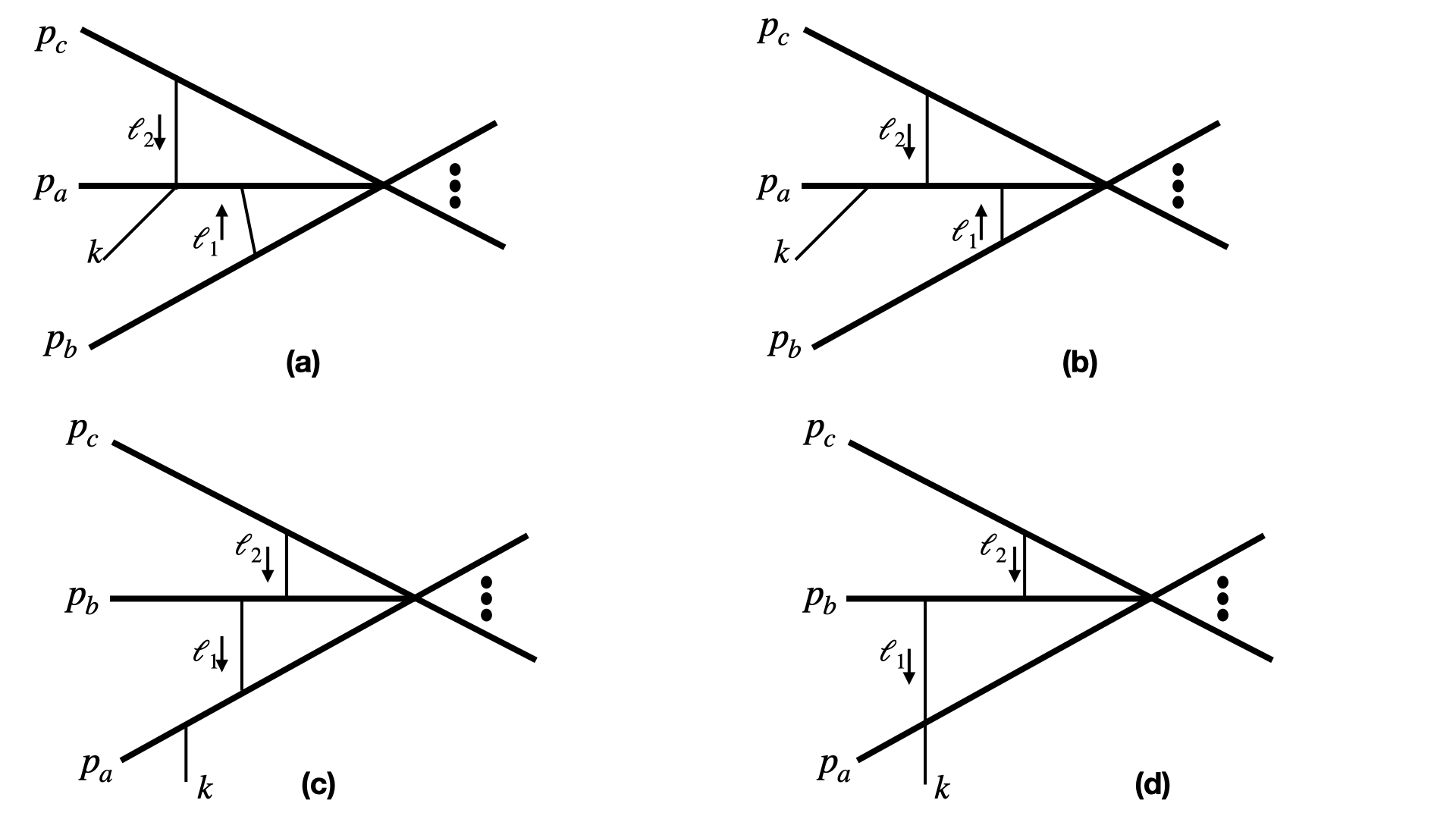}
\caption{Sets of two loop Feynman diagrams with two virtual photons connected to three different scalar lines which potentially can contribute to order $\mathcal{O}(\omega(\ln\omega)^{2})$. The thick lines represent massive complex scalars and the thin lines represent photons.}\label{Fsub2ph1}
\end{figure}
%\end{center}
%\begin{center}
\begin{figure}
\includegraphics[scale=0.25]{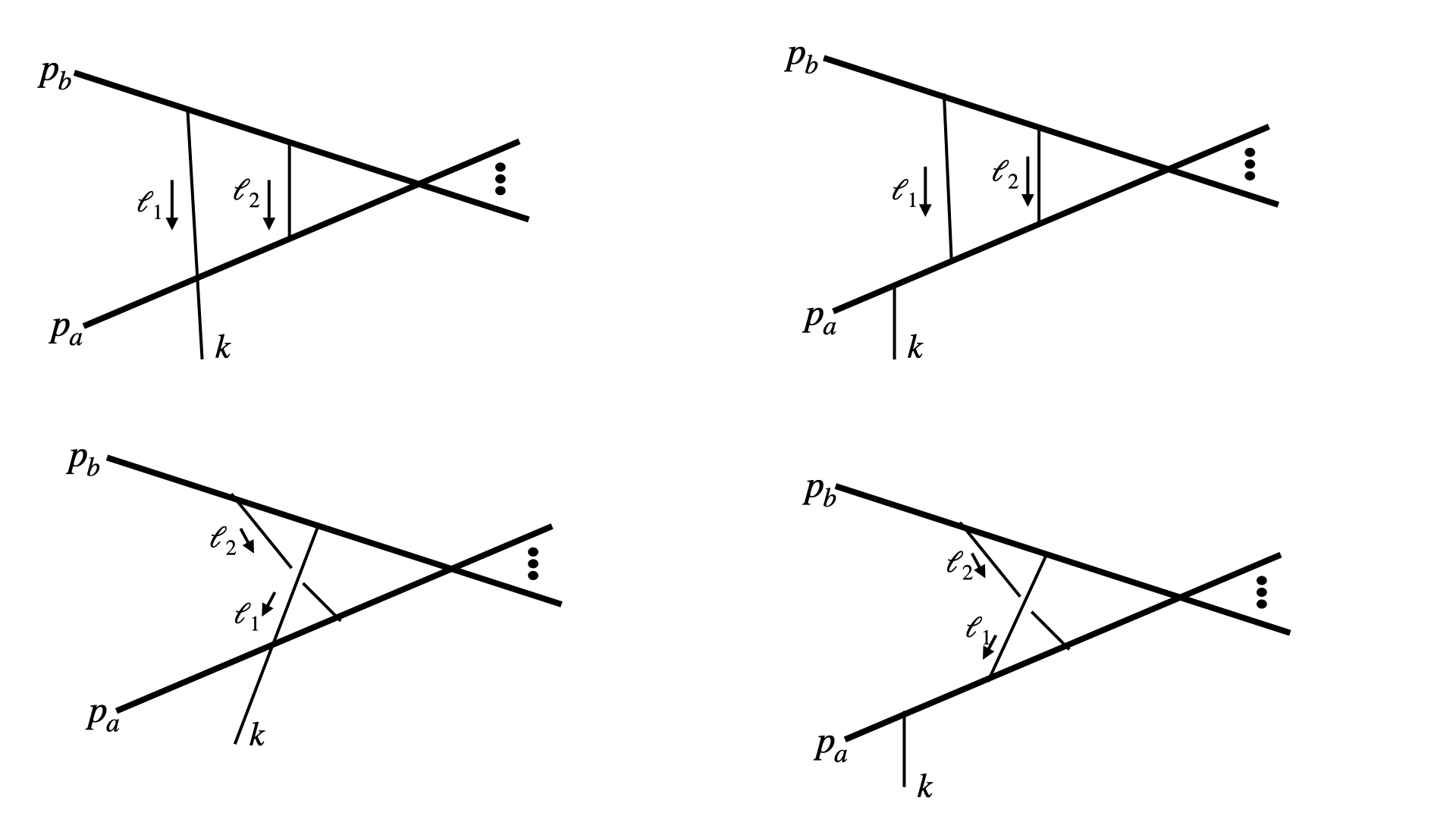}
\caption{Sets of two loop Feynman diagrams with two virtual photons connected to two different scalar lines which potentially can contribute to order $\mathcal{O}(\omega(\ln\omega)^{2})$. The thick lines represent massive complex scalars and the thin lines represent photons.}\label{Fsub2ph2}
\end{figure}
%\end{center}

Now one of the important observation made in \cite{1808.03288} is that if we replace the G-photon Feynman propagators by G-photon retarded propagators then analyzing loop amplitudes with retarded G-photon propagators one recovers only the classical soft factor. Analogously in the two loop case, if we replace $\lbrace \ell^{2}-i\epsilon\rbrace^{-1}$ by $ -G_{r}(\ell)$ in the analysis of Feynman diagrams of $\Gamma^{(M+N ,1)}_{G, 2-loop}$ and use the algorithm of eq.\eqref{algo} we expect the following relation will hold,
\be
\varepsilon^{\mu}\Delta_{(2)}\widehat{J}_{\mu}(k)\ &=&\ (-i)\ S_{em}^{(2)}\big(\varepsilon ,k,\lbrace q_{a},p_{a}\rbrace\big)\Big{|}_{classical}\label{JS}
\ee
So let us first re-derive our classical result of \S\ref{Subsecsub2ph} analyzing $\Gamma^{(M+N ,1)}_{G, 2-loop}$, under the replacement rule described above. 
\begingroup
\allowdisplaybreaks
\begin{itemize}
\item Contribution of diagram-(a) of Fig.\ref{Fsub2ph1} reproduces the expression of $\Delta_{(2)}^{(1)}\widehat{J}_{\mu}(k)$ in eq.\eqref{step} as well as the first three lines of the RHS of $\Delta_{(2)}^{(4)}\widehat{J}_{\mu}(k)$ in eq.\eqref{appAJ24} for $a\neq b\neq c$ via the relation \eqref{JS}.
\item Contribution of diagram-(b) of Fig.\ref{Fsub2ph1} reproduces the expression of $\Delta_{(2)}^{(2)}\widehat{J}_{\mu}(k)$ in eq.\eqref{deltaJ22} as well as the first three lines of the RHS of $\Delta_{(2)}^{(3)}\widehat{J}_{\mu}(k)$ in eq.\eqref{appAJ23} for $a\neq b\neq c$ via the relation \eqref{JS}.
\item Contribution of diagram-(c) of Fig.\ref{Fsub2ph1} reproduces the expression of the last three lines of the RHS of $\Delta_{(2)}^{(3)}\widehat{J}_{\mu}(k)$ in eq.\eqref{appAJ23} for $a\neq b\neq c$ via the relation \eqref{JS}.
\item Contribution of diagram-(d) of Fig.\ref{Fsub2ph1} reproduces the expression of the last three lines of the RHS of $\Delta_{(2)}^{(4)}\widehat{J}_{\mu}(k)$ in eq.\eqref{appAJ24} for $a\neq b\neq c$ via the relation \eqref{JS}.
\item On the other hand, sum over the diagrams of Fig.\ref{Fsub2ph2} reproduces sum over the contributions $\Delta_{(2)}^{(1)}\widehat{J}_{\mu}(k)+\Delta_{(2)}^{(2)}\widehat{J}_{\mu}(k)+\Delta_{(2)}^{(3)}\widehat{J}_{\mu}(k)+\Delta_{(2)}^{(4)}\widehat{J}_{\mu}(k)$ when $a=b$ or $b=c$ or $c=a$ via the relation \eqref{JS}.
\end{itemize}
\endgroup
The one to one mapping between Feynman diagrams and classical computation described above is only true when the integrals are approximated in the specified regions of integrations from where the $\mathcal{O}(\omega(\ln\omega)^{2})$ contribution comes.

If we analyze the diagrams in Fig.\ref{Fsub2ph1} and Fig.\ref{Fsub2ph2} using G-photon Feynman propagators then using the algorithm in eq.\eqref{algo} we find the full(classical +quantum) sub-subleading soft photon factor, which turn out to be:
\be
&&S_{em}^{(2)}\big(\varepsilon ,k,\lbrace q_{a},p_{a}\rbrace\big)\non\\
&=&\  \f{1}{2}\ \sum_{a=1}^{M+N}\Bigg[q_{a}\f{\varepsilon^{\mu}k^{\rho}}{p_{a}\cdot k}\ \Bigg(p_{a\mu}\f{\p}{\p p_{a}^{\rho}}\ -\ p_{a\rho}\f{\p}{\p p_{a}^{\mu}} \Bigg)K_{em}^{reg} \Bigg]\ \Bigg[  k_{\sigma}\f{\p}{\p p_{a\sigma}}K_{em}^{reg}\Bigg]\non\\
&&\ +\ (\ln \omega)^{2}\ \sum_{a=1}^{M+N} \varepsilon^{\mu} k^{\alpha}\mathcal{B}^{(2),reg}_{\alpha\mu}\big(q_{a},p_{a};\lbrace q_{b}\rbrace ,\lbrace p_{b}\rbrace\big)\ +\mathcal{O}(\omega\ln\omega)\label{2loopsoftthm}
\ee
where
\be
K_{em}^{reg}= -\f{i}{2}\sum_{b=1}^{M+N}\  \sum_{\substack{c=1\\c\neq b}}^{M+N}\f{q_{b}q_{c}}{4\pi}\ (\ln\omega) \ \f{p_{b}.p_{c}}{\sqrt{(p_{b}.p_{c})^{2}-p_{b}^{2}p_{c}^{2}}} \Bigg{\lbrace}\delta_{\eta_{b}\eta_{c},1}-\f{i}{2\pi}\ln\Bigg(\f{p_{b}.p_{c}+\sqrt{(p_{b}.p_{c})^{2}-p_{b}^{2}p_{c}^{2}}}{p_{b}.p_{c}-\sqrt{(p_{b}.p_{c})^{2}-p_{b}^{2}p_{c}^{2}}}\Bigg)\Bigg{\rbrace}
\ee
and
\be
\mathcal{B}^{(2),reg}_{\alpha\mu}\big(q_{a},p_{a};\lbrace q_{b}\rbrace ,\lbrace p_{b}\rbrace\big)\ =\ q_{a}\ \Big[p_{a\mu}\mathcal{C}^{reg}_{\alpha}\big(q_{a},p_{a};\lbrace q_{b}\rbrace ,\lbrace p_{b}\rbrace\big)-p_{a\alpha}\mathcal{C}^{reg}_{\mu}\big(q_{a},p_{a};\lbrace q_{b}\rbrace ,\lbrace p_{b}\rbrace\big)\Big] .
\ee
$\mathcal{C}^{reg}_{\alpha}\big(q_{a},p_{a};\lbrace q_{b}\rbrace ,\lbrace p_{b}\rbrace\big)$is given by:
\be
&&\mathcal{C}^{reg}_{\alpha}\big(q_{a},p_{a};\lbrace q_{b}\rbrace ,\lbrace p_{b}\rbrace\big)\non\\
 &=&-\sum_{\substack{b=1\\ b\neq a}}^{M+N}\sum_{\substack{c=1\\ c\neq a}}^{M+N}\f{q_{a}^{2}q_{b}q_{c}}{4}\ \lbrace p_{a}.p_{b}\delta_{\alpha}^{\gamma}-p_{a}^{\gamma}p_{b\alpha}\rbrace \f{\p}{\p p_{a\beta}}\Big{\lbrace}\mathcal{I}(p_{a},p_{c})\times p_{a}.p_{c}\Big{\rbrace} \f{\p}{\p p_{a}^{\gamma}}\f{\p}{\p p_{a}^{\beta}}\mathcal{I}(p_{a},p_{b})\non\\
 &&\ +\sum_{\substack{b=1\\ b\neq a}}^{M+N}\sum_{\substack{c=1\\ c\neq b}}^{M+N}\f{q_{a}q_{b}^{2}q_{c}}{4}\ \lbrace p_{a}.p_{b}\delta_{\alpha}^{\gamma}-p_{a}^{\gamma}p_{b\alpha}\rbrace \f{\p}{\p p_{b\beta}}\Big{\lbrace}\mathcal{I}(p_{b},p_{c})\times p_{b}.p_{c}\Big{\rbrace}\f{\p}{\p p_{a}^{\gamma}}\f{\p}{\p p_{a}^{\beta}}\mathcal{I}(p_{a},p_{b})
\ee
where,
\be
\mathcal{I}(p_{a},p_{b})&=&\ -\f{1}{4\pi}\ \f{1}{\sqrt{(p_{a}.p_{b})^{2}-p_{a}^{2}p_{b}^{2}}}\Bigg{\lbrace}\delta_{\eta_{a}\eta_{b},1}-\f{i}{2\pi}\ln\Bigg(\f{p_{a}.p_{b}+\sqrt{(p_{a}.p_{b})^{2}-p_{a}^{2}p_{b}^{2}}}{p_{a}.p_{b}-\sqrt{(p_{a}.p_{b})^{2}-p_{a}^{2}p_{b}^{2}}}\Bigg)\Bigg{\rbrace}\ .
\ee
The first and second lines of the RHS of sub-subleading soft photon factor expression \eqref{2loopsoftthm}, are gauge invariant separately. In principle the last line of eq.\eqref{2loopsoftthm} can also be written in the following form:
\be
(\ln\omega)^{2}\ [\varepsilon^{\mu}k^{\alpha}-\varepsilon^{\alpha}k^{\mu}]\sum_{a=1}^{M+N}q_{a}p_{a\mu}\mathcal{C}_{\alpha}^{reg}\big(q_{a},p_{a};\lbrace q_{b}\rbrace ,\lbrace p_{b}\rbrace\big) 
\ee
Hence, the origin of the above expression can be thought of as some effective coupling of electromagnetic field strength $F^{\mu\alpha}(k)=i(k^{\mu}\varepsilon^{\alpha}-k^{\alpha}\varepsilon^{\mu})$ to the finite energy charged particles which will appear in the quantum effective action of the theory. Though the leading non-analytic contribution to the sub-subleading soft photon factor given in eq.\eqref{2loopsoftthm} is derived by analyzing two loop amplitudes in scalar QED, we expect this result is universal and 2-loop exact. Similarly if we generalize this analysis to n-loop amplitudes, we can, in principle, derive the leading non-analytic contribution of (sub)$^{n}$-leading soft photon theorem whose classical limit will reproduce the result of eq.\eqref{SubnJmu}.

\section{Outlook and speculations: further understanding of electromagnetic and gravitational waveforms }\label{outlook}
In this section the results we are giving are not rigorously derived. But under some specified assumptions we expect that the results are correct, but need future investigation.

\subsection{Feynman diagrammatic understanding of the sub-subleading soft graviton theorem }
Emboldened by our understanding of classical sub-subleading soft photon factor from the analysis of two loop amplitude in \S\ref{FeynQED} with the replacement of Feynman propagators  by retarded propagators, we expect that a similar  analysis of two loop amplitudes for scalar coupled to gravity will reproduce classical sub-subleading soft graviton factor. But as described in \cite{1808.03288}, for the derivation of soft graviton theorem we need the following four extra ingredients relative to the derivation of soft photon theorem.
\begin{enumerate}
\item We have to perform K-G decomposition of the graviton propagator, whose both ends are attached to same scalar line.
\item We do not need to perform any K-G decomposition for the graviton propagators whose one or both ends are connected to any graviton line.
\item Consider the set of Feynman diagrams where a K-graviton is attached to the same scalar line where some other graviton is also attached. Sum over those kind of diagrams do not factorize completely in a sense that the contribution of the sum can not be written as sum of two kind of diagrams of which in one diagram the K-graviton is connected to the extreme left of the scalar line and in the other diagram the K-graviton is connected to the extreme right of the scalar line. After a partial factorization, we are left with some residual contribution which contributes to the IR finite terms in the loop amplitude. Also  these residual terms can contribute to order $\mathcal{O}(\omega(\ln\omega)^{2})$ soft factor in the two loop amplitude analysis.
\item Here $\Gamma^{(M+N,1)}$ contains some extra IR divergent factor relative to $\Gamma^{(M+N)}$. So the full cancellation of IR divergent factor from both the S-matrices does not happen in the soft theorem relation. The extra divergent piece in $\Gamma^{(M+N,1)}$ needs to be IR regulated using some IR cutoff.
\end{enumerate}
So analogous to eq.\eqref{algo}, here the algorithm for deriving sub-subleading soft graviton factor turns out to be,
\be
\Big[\Gamma^{(M+N,1)}_{G}+\Gamma^{(M+N,1)}_{G,self}+\Gamma^{(M+N,1)}_{residual}\ +\Gamma^{(M+N,1)}_{3,4-graviton}\Big]_{2-loop} &=&\ S^{(2)}_{gr}\big(\varepsilon ,k,\lbrace p_{a}\rbrace\big)\ \Gamma_{tree}^{(M+N)}\label{algo2}
\ee
where,
\begin{itemize}
\item Contribution of $\Gamma^{(M+N,1)}_{G}$ comes from the same sets of diagrams given in Fig.\ref{Fsub2ph1},\ref{Fsub2ph2} with the G-photon propagators replaced by G-graviton propagators.
\item For computing $\Gamma^{(M+N,1)}_{G,self}$ we have to analyze the set of two loop diagrams where at least for one of the G-graviton propagators, both ends are connected to the same scalar line.
\item  Contribution of $\Gamma^{(M+N,1)}_{residual}$ comes from the residual non-factorized contribution as discussed in the third point above.
\item Contribution of $\Gamma^{(M+N,1)}_{3,4-graviton}$ will come from two kinds of diagrams. One kind of diagrams contain one or two three graviton vertices which can contribute to order $\mathcal{O}(\omega(\ln\omega)^{2})$ as we have seen from classical analysis. On the other hand from the classical analysis it is suggestive that the other kind of diagrams containing four graviton vertex do not contribute to order $\mathcal{O}(\omega(\ln\omega)^{2})$.
\end{itemize}
We expect that with this understanding one can extract sub-subleading soft graviton factor using the algorithm \eqref{algo2}. Our expectation for the sub-subleading soft graviton factor is:
\be
&&S^{(2)}_{gr}\big(\varepsilon ,k,\lbrace p_{a}\rbrace\big)\non\\
 &=&\  \f{1}{2}\Big{\lbrace}K_{phase}^{reg}\Big{\rbrace}^{2}\ \sum_{a=1}^{M+N}\f{\varepsilon_{\mu\nu}p_{a}^{\mu}p_{a}^{\nu}}{p_{a}.k}\non\\
&& +\ \Big{\lbrace}K_{phase}^{reg}\Big{\rbrace}\sum_{a=1}^{M+N}\f{\varepsilon_{\mu\nu}p_{a}^{\mu}k_{\rho}}{p_{a}.k}\Bigg[\Bigg(p_{a}^{\nu}\f{\p}{\p p_{a\rho}}-p_{a}^{\rho}\f{\p}{\p p_{a\nu}}\Bigg)K_{gr}^{reg}\Bigg]\non\\
&&\ + \f{1}{2}\sum_{a=1}^{M+N}\f{\varepsilon_{\mu\nu}k_{\rho}k_{\sigma}}{p_{a}.k} \times \ \Bigg[\Bigg(p_{a}^{\mu}\f{\p}{\p p_{a\rho}}-p_{a}^{\rho}\f{\p}{\p p_{a\mu}}\Bigg)K_{gr}^{reg}\Bigg]\ \Bigg[\Bigg(p_{a}^{\nu}\f{\p}{\p p_{a\sigma}}-p_{a}^{\sigma}\f{\p}{\p p_{a\nu}}\Bigg)K_{gr}^{reg}\Bigg]\non\\
&&\ +\mathcal{O}(\omega\ln\omega)
\ee
where,
\begingroup
\allowdisplaybreaks
\be
K^{reg}_{gr}\ &=&\ \f{i}{2}\ (8\pi G)\sum_{a=1}^{M+N}\  \sum_{\substack{b=1\\b\neq a}}^{M+N}\Big{\lbrace} (p_{a}.p_{b})^{2}-\f{1}{2}p_{a}^{2}p_{b}^{2}\Big{\rbrace}\ \int_{\omega} \f{d^{4}\ell}{(2\pi)^{4}} \ \f{1}{\ell^{2}-i\epsilon}\ \f{1}{(p_{a}.\ell +i\epsilon)\ (p_{b}.\ell - i\epsilon)}\non\\
&\simeq &\ -\f{i}{2}(2G)\ \f{1}{4\pi}(\ln\omega)\sum_{a=1}^{M+N}\  \sum_{\substack{b=1\\b\neq a}}^{M+N} \ \f{\Big{\lbrace} (p_{a}.p_{b})^{2}-\f{1}{2}p_{a}^{2}p_{b}^{2}\Big{\rbrace}}{\sqrt{(p_{a}.p_{b})^{2}-p_{a}^{2}p_{b}^{2}}} \Bigg{\lbrace}\delta_{\eta_{a}\eta_{b},1}-\f{i}{2\pi}\ln\Bigg(\f{p_{a}.p_{b}+\sqrt{(p_{a}.p_{b})^{2}-p_{a}^{2}p_{b}^{2}}}{p_{a}.p_{b}\sqrt{(p_{a}.p_{b})^{2}-p_{a}^{2}p_{b}^{2}}}\Bigg)\Bigg{\rbrace}\non\\
\ee
\endgroup
and 
\be
K_{phase}^{reg}\ &=&\ i\ (8\pi G)\ \sum_{b=1}^{M+N}(p_{b}.k)^{2}\int_{R^{-1}}^{\omega}\f{d^{4}\ell}{(2\pi)^{4}}\ \f{1}{\ell^{2}-i\epsilon}\f{1}{k.\ell+i\epsilon}\f{1}{p_{b}.\ell -i\epsilon}\non\\
&=&\ -2iG\ \ln(\omega R)\ \Bigg[ \sum_{b=1}^{N}p_{b}.k\ -\ \f{i}{2\pi}\sum_{b=1}^{M+N}p_{b}.k\ \ln\Bigg(\f{p_{b}^{2}}{(p_{b}.\mathbf{n})^{2}}\Bigg)\Bigg]
\ee

\subsection{Structure of (sub)$^n$-leading electromagnetic and gravitational waveforms in presence of both long range electromagnetic and gravitational forces}\label{subngrem}
In this section we take care of both electromagnetic and gravitational long range forces between the charged scattered objects outside the region $\mathcal{R}$ to give the structure of electromagnetic and gravitational waveforms at late and early retarded time. The trajectories outside the region $\mathcal{R}$ satisfies,
\be
m_{a}\f{d^{2}X_{a}^{\mu}(\sigma)}{d\sigma^{2}}\ &=&\ q_{a}\ F^{\mu}\ _{ \nu}(X_{a}(\sigma))\ \f{dX_{a}^{\nu}(\sigma)}{d\sigma}\ -\ m_{a}\Gamma^{\mu}_{\nu\rho}(X_{a}(\sigma))\f{dX_{a}^{\nu}(\sigma)}{d\sigma}\f{dX_{a}^{\rho}(\sigma)}{d\sigma}\label{Eqofmot}
\ee
Here we have to expand the asymptotic trajectories of scattered objects in the same way as in eq.\eqref{Yexpansion}, considering $q^{2}$ and $G$ being the expansion parameters with equal weight.
\subsubsection*{Structure of (sub)$^n$-leading electromagnetic waveform}
To find the electromagnetic waveform we have to solve the following form of Maxwell equation in the background metric produces by other particles,
\be
\p_{\nu}\Big(\sqrt{-det (g)}g^{\nu\rho}g^{\mu\sigma}F_{\rho\sigma}\Big)\ &=&\ -J^{\mu}
\ee
Using Lorentz gauge condition $\eta^{\mu\nu}\p_{\mu}A_{\nu}=0$ this can also be written as,
\be
\eta^{\mu\alpha}\eta^{\rho\sigma}\p_{\rho}\p_{\sigma}A_{\alpha}=-J^{\mu}-J_{h}^{\mu}
\ee
where the gravitational contribution to current density is defined as,
\be
J_{h}^{\mu}\ &\equiv & \p_{\nu}\Big(\sqrt{-det (g)}g^{\nu\rho}g^{\mu\sigma}F_{\rho\sigma}\Big) -\ \eta^{\mu\alpha}\eta^{\rho\sigma}\p_{\rho}\p_{\sigma}A_{\alpha}\label{currentGR}
\ee
Now we have to expand the total current density in a power series of $q^{2}$ and $G$ with both having the same importance as expansion parameter. Analogous to \S\ref{SsubnEM}, here we expect the order $\mathcal{O}(\omega^{n-1}(\ln\omega)^{n})$ contribution to (sub)$^{n}$-leading current density will take the following form\footnote{A brief sketch of the derivation of this result is given in appendix-\ref{partialderivation}.},
\begingroup
\allowdisplaybreaks
\be
&&\Delta_{(n)}\widehat{J}^{\mu}(k)\ +\ \Delta_{(n)}\widehat{J}_{h}^{\mu}(k)\non\\
&\simeq &\ -i\ \omega^{n-1}\lbrace\ln(\omega+i\epsilon)\rbrace^{n}\times \f{1}{n!}\Big{\lbrace}-2iG\sum_{b=1}^{N}p_{b}\cdot \mathbf{n}\Big{\rbrace}^{n}\ \sum_{a=1}^{M+N}q_{a}\f{p_{a}^{\mu}}{p_{a}.\mathbf{n}}\non\\
&&\ -i\ \sum_{r=1}^{n} \omega^{n-1}\lbrace\ln(\omega+i\epsilon)\rbrace^{n-r} \f{1}{(n-r)!}\ \Big{\lbrace}-2iG\sum_{d=1}^{N}p_{d}\cdot \mathbf{n}\Big{\rbrace}^{n-r} \sum_{a=1}^{M+N}\Big{\lbrace}\ln(\omega +i\epsilon\eta_{a})\Big{\rbrace}^{r}\non\\
&&\times \Bigg[\sum_{\substack{b=1\\b\neq a\\ \eta_{a}\eta_{b}=1}}^{M+N} \f{1}{[(p_{a}.p_{b})^{2}-p_{a}^{2}p_{b}^{2}]^{3/2}}\ q_{a}\ \f{\mathbf{n}_{\rho}}{p_{a}.\mathbf{n}}\ \Big{\lbrace}p_{a}^{\mu}p_{b}^{\rho}-p_{a}^{\rho}p_{b}^{\mu}\Big{\rbrace}\ \Bigg(\f{i}{4\pi}\ q_{a}q_{b}\ p_{a}^{2}p_{b}^{2}\non\\
&&\ \  +\ 2iG\ (p_{a}.p_{b})\ \Big{\lbrace}\f{3}{2}p_{a}^{2}p_{b}^{2}-(p_{a}.p_{b})^{2}\Big{\rbrace}\Bigg)\Bigg]\times \f{1}{r!}\Bigg[ \sum_{\substack{c=1\\c\neq a\\ \eta_{a}\eta_{c}=1}}^{M+N}\f{1}{[(p_{a}.p_{c})^{2}-p_{a}^{2}p_{c}^{2}]^{3/2}}\non\\
&&\times \Bigg(\f{i}{4\pi}\ q_{a}q_{c}p_{c}^{2}\Big{\lbrace}p_{a}^{2}p_{c}.\mathbf{n}-p_{a}.p_{c}p_{a}.\mathbf{n}\Big{\rbrace} + 2iG\ \Big{\lbrace}\f{3}{2}p_{a}.p_{c}p_{a}^{2}p_{c}^{2}p_{c}.\mathbf{n}-(p_{a}.p_{c})^{3}p_{c}.\mathbf{n}-\f{1}{2}p_{a}^{2}(p_{c}^{2})^{2}p_{a}.\mathbf{n}\Big{\rbrace}\Bigg)\ \Bigg]^{r-1}\non\\
&& \ -i\sum_{r=2}^{n}\omega^{n-1}\Big{\lbrace}\ln(\omega+i\epsilon)\Big{\rbrace}^{n-r}\ \f{1}{(n-r)!}\ \Big{\lbrace}-2iG\sum_{b=1}^{N}p_{b}\cdot\mathbf{n}\Big{\rbrace}^{n-r}\non\\
&&\ \times \sum_{a=1}^{M+N}\Big{\lbrace}\ln(\omega+i\epsilon\eta_{a})\Big{\rbrace}^{r}\ \mathbf{n}_{\alpha_{1}}\mathbf{n}_{\alpha_{2}}\cdots\mathbf{n}_{\alpha_{r-1}}\ \mathbf{B}^{(r),\alpha_{1}\alpha_{2}\cdots\alpha_{r-1}\mu}\big(q_{a},p_{a}\big)\label{Jn+Jnh}
\ee
\endgroup
where the undetermined function $\mathbf{B}^{(r),\alpha_{1}\alpha_{2}\cdots\alpha_{r-1}\mu}\big(q_{a},p_{a}\big)$ is antisymmetric under $\mu$ and any $\alpha_{\ell}$ exchange for $\ell=1,2,...,r-1$. We expect that $\mathbf{B}^{(r)}$ will be independent of the details of the scattering event inside the region $\mathcal{R}$ and will only depend on the scattering data. Now using the following relation (a modification of \eqref{AJrelation})
\be
\widetilde{A}^{\mu}(\omega ,R,\hat{n})\ &\simeq &\ \f{1}{4\pi R}\ e^{i\omega R}\  \Big(\widehat{J}^{\mu}(k)\ +\ \widehat{J}^{\mu}_{h}(k)\Big)
\ee
and performing Fourier transformation in $\omega$ variable after using the results of eq.\eqref{FourierFn0},\eqref{FourierF0n} and \eqref{FourierFn-rr} we find the following late and early time (sub)$^{n}$-leading electromagnetic waveforms for $n\geq 2$,
\begingroup
\allowdisplaybreaks
\be
&&\Delta_{(n)}A^{\mu}(t ,R,\hat{n})\non\\
&=&\f{1}{4\pi R}\ \f{(\ln|u|)^{n-1}}{u^{n}}\ \Big{\lbrace}2G\sum_{b=1}^{N}p_{b}\cdot\mathbf{n}\Big{\rbrace}^{n}\ \Bigg(\sum_{a=1}^{N}q_{a}\f{p_{a}^{\mu}}{p_{a}.\mathbf{n}}-\sum_{a=1}^{M}q'_{a}\f{p_{a}^{\prime\mu}}{p'_{a}.\mathbf{n}}\Bigg)\non\\
&&\ +\f{1}{4\pi R}\ \f{(\ln|u|)^{n-1}}{u^{n}}\ \sum_{r=1}^{n}(-1)^{r}\ \f{n!}{r!(n-r)!}\ \Big{\lbrace}2G\sum_{d=1}^{N}p_{d}\cdot\mathbf{n}\Big{\rbrace}^{n-r}\ \sum_{a=1}^{N}\Bigg[\sum_{\substack{b=1\\b\neq a}}^{N} \f{1}{[(p_{a}.p_{b})^{2}-p_{a}^{2}p_{b}^{2}]^{3/2}}\non\\
&&\times q_{a}\ \f{\mathbf{n}_{\rho}}{p_{a}.\mathbf{n}}\ \Big{\lbrace}p_{a}^{\mu}p_{b}^{\rho}-p_{a}^{\rho}p_{b}^{\mu}\Big{\rbrace}\ \Bigg(\f{1}{4\pi}\ q_{a}q_{b}\ p_{a}^{2}p_{b}^{2}\ +\ 2G\ (p_{a}.p_{b})\ \Big{\lbrace}\f{3}{2}p_{a}^{2}p_{b}^{2}-(p_{a}.p_{b})^{2}\Big{\rbrace}\Bigg)\Bigg]\non\\
 &&\ \times \Bigg[ \sum_{\substack{c=1\\c\neq a }}^{N}\f{1}{[(p_{a}.p_{c})^{2}-p_{a}^{2}p_{c}^{2}]^{3/2}}\ \Bigg(\f{1}{4\pi}\ q_{a}q_{c}p_{c}^{2}\Big{\lbrace}p_{a}^{2}p_{c}.\mathbf{n}-p_{a}.p_{c}p_{a}.\mathbf{n}\Big{\rbrace}\non\\
 &&\ +\ 2G\ \Big{\lbrace}\f{3}{2}p_{a}.p_{c}p_{a}^{2}p_{c}^{2}p_{c}.\mathbf{n}-(p_{a}.p_{c})^{3}p_{c}.\mathbf{n}-\f{1}{2}p_{a}^{2}(p_{c}^{2})^{2}p_{a}.\mathbf{n}\Big{\rbrace}\Bigg)\ \Bigg]^{r-1}\non\\
 &&\ -\f{1}{4\pi R}\ \f{(\ln|u|)^{n-1}}{u^{n}}\ \sum_{r=1}^{n-1}\ \f{(n-1)!}{r!(n-r-1)!}\ \Big{\lbrace}2G\sum_{d=1}^{N}p_{d}\cdot\mathbf{n}\Big{\rbrace}^{n-r}\ \sum_{a=1}^{M}\Bigg[\sum_{\substack{b=1\\b\neq a}}^{M} \f{1}{[(p'_{a}.p'_{b})^{2}-p_{a}^{\prime 2}p_{b}^{\prime 2}]^{3/2}}\non\\
 &&\times q'_{a}\ \f{\mathbf{n}_{\rho}}{p'_{a}.\mathbf{n}}\ \Big{\lbrace}p_{a}^{\prime\mu}p_{b}^{\prime\rho}-p_{a}^{\prime\rho}p_{b}^{\prime\mu}\Big{\rbrace}\ \Bigg(\f{1}{4\pi}\ q'_{a}q'_{b}\ p_{a}^{\prime 2}p_{b}^{\prime 2}\ +\ 2G\ (p'_{a}.p'_{b})\ \Big{\lbrace}\f{3}{2}p_{a}^{\prime 2}p_{b}^{\prime 2}-(p'_{a}.p'_{b})^{2}\Big{\rbrace}\Bigg)\Bigg]\non\\
 &&\ \times \Bigg[ \sum_{\substack{c=1\\c\neq a }}^{M}\f{1}{[(p'_{a}.p'_{c})^{2}-p_{a}^{\prime 2}p_{c}^{\prime 2}]^{3/2}}\ \Bigg(\f{1}{4\pi}\ q'_{a}q'_{c}p_{c}^{\prime 2}\Big{\lbrace}p_{a}^{\prime 2}p'_{c}.\mathbf{n}-p'_{a}.p'_{c}p'_{a}.\mathbf{n}\Big{\rbrace}\non\\
 &&\ +\ 2G\ \Big{\lbrace}\f{3}{2}p'_{a}.p'_{c}p_{a}^{\prime 2}p_{c}^{\prime 2}p'_{c}.\mathbf{n}-(p'_{a}.p'_{c})^{3}p'_{c}.\mathbf{n}-\f{1}{2}p_{a}^{\prime 2}(p_{c}^{\prime 2})^{2}p'_{a}.\mathbf{n}\Big{\rbrace}\Bigg)\ \Bigg]^{r-1}\non\\
 &&\ +\f{1}{4\pi R}\ \f{(\ln|u|)^{n-1}}{u^{n}}\sum_{r=2}^{n}\f{i^{r}n!}{(n-r)!}\ \Big{\lbrace}2G\sum_{b=1}^{N}p_{b}\cdot\mathbf{n}\Big{\rbrace}^{n-r}\sum_{a=1}^{N}\mathbf{n}_{\alpha_{1}}\mathbf{n}_{\alpha_{2}}\cdots\mathbf{n}_{\alpha_{r-1}}\ \mathbf{B}^{(r),\alpha_{1}\alpha_{2}\cdots\alpha_{r-1}\mu}\big(q_{a},p_{a}\big)\non\\
 &&\ +\f{1}{4\pi R}\ \f{(\ln|u|)^{n-1}}{u^{n}}\sum_{r=2}^{n-1}\f{i^{r}(n-1)!}{(n-r-1)!}\ \Big{\lbrace}2G\sum_{b=1}^{N}p_{b}\cdot\mathbf{n}\Big{\rbrace}^{n-r}\sum_{a=1}^{M}\mathbf{n}_{\alpha_{1}}\mathbf{n}_{\alpha_{2}}\cdots\mathbf{n}_{\alpha_{r-1}}\ \mathbf{B}^{(r),\alpha_{1}\alpha_{2}\cdots\alpha_{r-1}\mu}\big(-q'_{a},-p'_{a}\big)\non\\
  &&\ +\ \mathcal{O}\Big(u^{-n}(\ln|u|)^{n-2}\Big)\ \hspace{5cm}\hbox{for}\ u\rightarrow+\infty\label{lateAemgr}
\ee
\endgroup
and
\begingroup
\allowdisplaybreaks
\be
&&\Delta_{(n)}A^{\mu}(t ,R,\hat{n})\non\\
 &=&\ \f{1}{4\pi R}\ \f{(\ln|u|)^{n-1}}{u^{n}}\ \sum_{r=1}^{n}\ \f{(n-1)!}{(r-1)!(n-r)!}\ \Big{\lbrace}2G\sum_{d=1}^{N}p_{d}\cdot\mathbf{n}\Big{\rbrace}^{n-r}\ \sum_{a=1}^{M}\Bigg[\sum_{\substack{b=1\\b\neq a}}^{M} \f{1}{[(p'_{a}.p'_{b})^{2}-p_{a}^{\prime 2}p_{b}^{\prime 2}]^{3/2}}\non\\
 &&\times q'_{a}\ \f{\mathbf{n}_{\rho}}{p'_{a}.\mathbf{n}}\ \Big{\lbrace}p_{a}^{\prime\mu}p_{b}^{\prime\rho}-p_{a}^{\prime\rho}p_{b}^{\prime\mu}\Big{\rbrace}\ \Bigg(\f{1}{4\pi}\ q'_{a}q'_{b}\ p_{a}^{\prime 2}p_{b}^{\prime 2}\ +\ 2G\ (p'_{a}.p'_{b})\ \Big{\lbrace}\f{3}{2}p_{a}^{\prime 2}p_{b}^{\prime 2}-(p'_{a}.p'_{b})^{2}\Big{\rbrace}\Bigg)\Bigg]\non\\
 &&\ \times \Bigg[ \sum_{\substack{c=1\\c\neq a }}^{M}\f{1}{[(p'_{a}.p'_{c})^{2}-p_{a}^{\prime 2}p_{c}^{\prime 2}]^{3/2}}\ \Bigg(\f{1}{4\pi}\ q'_{a}q'_{c}p_{c}^{\prime 2}\Big{\lbrace}p_{a}^{\prime 2}p'_{c}.\mathbf{n}-p'_{a}.p'_{c}p'_{a}.\mathbf{n}\Big{\rbrace}\non\\
 &&\ +\ 2G\ \Big{\lbrace}\f{3}{2}p'_{a}.p'_{c}p_{a}^{\prime 2}p_{c}^{\prime 2}p'_{c}.\mathbf{n}-(p'_{a}.p'_{c})^{3}p'_{c}.\mathbf{n}-\f{1}{2}p_{a}^{\prime 2}(p_{c}^{\prime 2})^{2}p'_{a}.\mathbf{n}\Big{\rbrace}\Bigg)\ \Bigg]^{r-1}\non\\
 &&-\f{1}{4\pi R}\ \f{(\ln|u|)^{n-1}}{u^{n}}\sum_{r=2}^{n}\f{i^{r}(n-1)!\ r}{(n-r)!}\ \Big{\lbrace}2G\sum_{b=1}^{N}p_{b}\cdot\mathbf{n}\Big{\rbrace}^{n-r}\sum_{a=1}^{M}\mathbf{n}_{\alpha_{1}}\mathbf{n}_{\alpha_{2}}\cdots\mathbf{n}_{\alpha_{r-1}}\ \mathbf{B}^{(r),\alpha_{1}\alpha_{2}\cdots\alpha_{r-1}\mu}\big(-q'_{a},-p'_{a}\big)\non\\
  &&\ +\ \mathcal{O}\Big(u^{-n}(\ln|u|)^{n-2}\Big)\ \hspace{5cm}\hbox{for}\ u\rightarrow -\infty\label{EarlyAemgr}
\ee
\endgroup
where the retarded time $u$ is given by $u=t-R+2G\ln R\sum\limits_{b=1}^{N}p_{b}\cdot \mathbf{n}$. It would be interesting to find out the expressions of $\lbrace \mathbf{B}^{(r)}\rbrace$ generalising our analysis to arbitrarily higher order.
\subsubsection*{Structure of (sub)$^n$-leading gravitational waveform}
 Analysis of (sub)$^{n}$-leading gravitational waveform for arbitrary value of $n$ will be very complicated as higher order gravitational energy-momentum tensor will carry lots of terms. But we can still give some structure of gravitational waveform compiling the results of \cite{1801.05528,1802.03148,1810.04619} and the observation made in \cite{1808.03288}. 
 
 In the classical analysis we expect the structure of order $\mathcal{O}(\omega^{n-1}(\ln\omega)^{n})$ contribution from the (sub)$^n$-leading  matter and gravitational energy-momentum tensor for $n\geq 3$ will take the following form\footnote{A brief sketch of the derivation of this result is given in appendix-\ref{partialderivation}.},
 \begingroup
\allowdisplaybreaks
 \be
&&\Delta_{(n)}\widehat{T}^{X\mu\nu}(k)+\Delta_{(n)}\widehat{T}^{h\mu\nu}(k)+\Delta_{(n)}\widehat{T}^{(em)\mu\nu}(k)\non\\
&\simeq &\ -i \f{1}{n!}\Big{\lbrace}-2iG\ln(\omega +i\epsilon)\ \sum_{\substack{b=1}}^{N}p_{b}\cdot k\Big{\rbrace}^{n}\ \sum_{a=1}^{M+N}\f{p_{a}^{\mu}p_{a}^{\nu}}{p_{a}.k}\non\\
&& -i\ \f{1}{(n-1)!}\Big{\lbrace}-2iG\ln(\omega +i\epsilon)\ \sum_{\substack{b=1}}^{N}p_{b}\cdot k\Big{\rbrace}^{n-1}\sum_{a=1}^{M+N}\f{p_{a}^{\mu}k_{\rho}}{p_{a}.k}\Bigg[\Bigg(p_{a}^{\nu}\f{\p}{\p p_{a\rho}}-p_{a}^{\rho}\f{\p}{\p p_{a\nu}}\Bigg)\Big{\lbrace}K_{gr}^{cl}+K_{em}^{cl}\Big{\rbrace}\Bigg]\non\\
&&\ -i\ \sum_{r=2}^{n}\f{1}{(n-r)!}\ \Big{\lbrace}-2iG\ln(\omega +i\epsilon)\ \sum_{\substack{b=1}}^{N}p_{b}\cdot k\Big{\rbrace}^{n-r}\ \sum_{a=1}^{M+N}\f{k_{\rho}k_{\sigma}}{p_{a}.k}\non\\
&&\ \times \ \Bigg[\Bigg(p_{a}^{\mu}\f{\p}{\p p_{a\rho}}-p_{a}^{\rho}\f{\p}{\p p_{a\mu}}\Bigg)\Big{\lbrace}K_{gr}^{cl}+K_{em}^{cl}\Big{\rbrace}\Bigg]\ \Bigg[\Bigg(p_{a}^{\nu}\f{\p}{\p p_{a\sigma}}-p_{a}^{\sigma}\f{\p}{\p p_{a\nu}}\Bigg)\Big{\lbrace}K_{gr}^{cl}+K_{em}^{cl}\Big{\rbrace}\Bigg]\non\\
&&\ \times \f{1}{r!}\Bigg[k^{\alpha}\f{\p}{\p p_{a}^{\alpha}}\Big{\lbrace}K_{gr}^{cl}+K_{em}^{cl}\Big{\rbrace}\Bigg]^{r-2}\non\\
&&\ -i\sum_{r=3}^{n}\omega^{n-1}\Big{\lbrace}\ln(\omega +i\epsilon)\Big{\rbrace}^{n-r}\f{1}{(n-r)!}\Big{\lbrace}-2iG\sum_{b=1}^{N}p_{b}\cdot \mathbf{n}\Big{\rbrace}^{n-r}\sum_{a=1}^{M+N}\Big{\lbrace}\ln(\omega+i\epsilon\eta_{a})\Big{\rbrace}^{r}\non\\
&&\ \times\ \mathbf{n}_{\alpha_{1}}\mathbf{n}_{\alpha_{2}}\cdots \mathbf{n}_{\alpha_{r-1}}\ \mathbf{C}^{(r),\ \mu\nu\alpha_{1}\cdots\alpha_{r-1}}(q_{a},p_{a})\label{TnX+Tnh}
\ee 
\endgroup
 where $\mathbf{C}^{(r),\ \mu\nu\alpha_{1}\cdots\alpha_{r-1}}(q_{a},p_{a})$ is anti-symmetric under $\mu \leftrightarrow\alpha_{i}$ exchange as well as $\nu \leftrightarrow\alpha_{j}$ for $i,j=1,2,\cdots ,r-1$. The functional behaviour of $\mathbf{C}^{(r)}(q_{a},p_{a})$ has to  determine by explicit computations following eq.\eqref{TXrest+Th2+Tem}, but we expect these functions will only depend on the scattering data. Using the relation of eq.\eqref{eTrelation} and performing Fourier transformation in $\omega$ variable after using the results of Fourier transformations given in eq.\eqref{FourierFn0},\eqref{FourierFn-rr} and \eqref{FourierF0n} we get the following late and early time (sub)$^{n}$-leading gravitational waveforms for $n\geq 3$,
 \begingroup
 \allowdisplaybreaks
\be
&&\Delta_{(n)}e^{\mu\nu}(t,R,\hat{n})\non\\
&=&\ \f{2G}{R}\ \f{\big(\ln |u|\big)^{n-1}}{u^{n}}\ \Big{\lbrace}2G\sum_{b=1}^{N}p_{b}.\mathbf{n}\Big{\rbrace}^{n}\Bigg(\sum_{a=1}^{N}\f{p_{a}^{\mu}p_{a}^{\nu}}{p_{a}.\mathbf{n}}-\sum_{a=1}^{M}\f{p_{a}^{\prime\mu}p_{a}^{\prime\nu}}{p'_{a}.\mathbf{n}}\Bigg)\non\\
&&\ -\f{2G}{R}\ \f{\big(\ln |u|\big)^{n-1}}{u^{n}}\ n\  \Big{\lbrace}2G\sum_{c=1}^{N}p_{c}.\mathbf{n}\Big{\rbrace}^{n-1}\  \sum_{a=1}^{N}\Bigg[\ \sum_{\substack{b=1\\ b\neq a}}^{N}\f{1}{[(p_{a}.p_{b})^{2}-p_{a}^{2}p_{b}^{2}]^{3/2}}\ \f{p_{a}^{\mu}\mathbf{n}_{\rho}}{p_{a}.\mathbf{n}}\Big{\lbrace}p_{a}^{\nu}p_{b}^{\rho}-p_{a}^{\rho}p_{b}^{\nu}\Big{\rbrace}\non\\
&&\times\Bigg(2G\ p_{a}.p_{b} \Big{\lbrace}\f{3}{2}p_{a}^{2}p_{b}^{2}-(p_{a}.p_{b})^{2}\Big{\rbrace}\ +\ \f{1}{4\pi}\ q_{a}q_{b}\ p_{a}^{2}p_{b}^{2}\Bigg)\ \Bigg]\non\\
&&\ -\f{2G}{R}\ \f{\big(\ln |u|\big)^{n-1}}{u^{n}}\ (n-1)\  \Big{\lbrace}2G\sum_{c=1}^{N}p_{c}.\mathbf{n}\Big{\rbrace}^{n-1}\  \sum_{a=1}^{M}\Bigg[\sum_{\substack{b=1\\ b\neq a}}^{M}\f{1}{[(p'_{a}.p'_{b})^{2}-p_{a}^{\prime 2}p_{b}^{\prime 2}]^{3/2}}\ \f{p_{a}^{\prime\mu}\mathbf{n}_{\rho}}{p'_{a}.\mathbf{n}}\Big{\lbrace}p_{a}^{\prime\nu}p_{b}^{\prime\rho}-p_{a}^{\prime\rho}p_{b}^{\prime\nu}\Big{\rbrace}\non\\
&&\times\Bigg(2G\ p'_{a}.p'_{b} \Big{\lbrace}\f{3}{2}p_{a}^{\prime 2}p_{b}^{\prime 2}-(p'_{a}.p'_{b})^{2}\Big{\rbrace}\ +\ \f{1}{4\pi}\ q'_{a}q'_{b}\ p_{a}^{\prime 2}p_{b}^{\prime 2}\Bigg)\ \Bigg]\non\\
&&\ +\ \f{2G}{R}\ \f{\big(\ln |u|\big)^{n-1}}{u^{n}}\sum_{r=2}^{n} (-1)^{r}\ \f{n!}{r!(n-r)!}\  \Big{\lbrace}2G\sum_{d=1}^{N}p_{d}.\mathbf{n}\Big{\rbrace}^{n-r}\ \times \sum_{a=1}^{N}\f{\mathbf{n}_{\rho}\mathbf{n}_{\sigma}}{p_{a}.\mathbf{n}}\non\\
&&\times\Bigg[\sum_{\substack{b=1\\ b\neq a}}^{N}\f{1}{[(p_{a}.p_{b})^{2}-p_{a}^{2}p_{b}^{2}]^{3/2}}\ \Big{\lbrace}p_{a}^{\mu}p_{b}^{\rho}-p_{a}^{\rho}p_{b}^{\mu}\Big{\rbrace}\Bigg(2G\ p_{a}.p_{b} \Big{\lbrace}\f{3}{2}p_{a}^{2}p_{b}^{2}-(p_{a}.p_{b})^{2}\Big{\rbrace}\ +\ \f{1}{4\pi}\ q_{a}q_{b}\ p_{a}^{2}p_{b}^{2}\Bigg)\ \Bigg]\non\\
&&\times\Bigg[\sum_{\substack{c=1\\ c\neq a}}^{N}\f{1}{[(p_{a}.p_{c})^{2}-p_{a}^{2}p_{c}^{2}]^{3/2}}\ \Big{\lbrace}p_{a}^{\nu}p_{c}^{\sigma}-p_{a}^{\sigma}p_{c}^{\nu}\Big{\rbrace}\Bigg(2G\ p_{a}.p_{c} \Big{\lbrace}\f{3}{2}p_{a}^{2}p_{c}^{2}-(p_{a}.p_{c})^{2}\Big{\rbrace}\ +\ \f{1}{4\pi}\ q_{a}q_{c}\ p_{a}^{2}p_{c}^{2}\Bigg)\ \Bigg]\non\\
&&\times \Bigg[\sum_{\substack{e=1\\ e\neq a}}^{N}\f{1}{[(p_{a}.p_{e})^{2}-p_{a}^{2}p_{e}^{2}]^{3/2}}\Bigg(\ (2G)\Big{\lbrace}\f{3}{2}p_{a}.p_{e}p_{a}^{2}p_{e}^{2}p_{e}.\mathbf{n}-(p_{a}.p_{e})^{3}p_{e}.\mathbf{n}-\f{1}{2}p_{a}^{2}(p_{e}^{2})^{2}p_{a}.\mathbf{n}\Big{\rbrace}\non\\
&&\ +\ \f{1}{4\pi}\ q_{a}q_{e}p_{e}^{2}\Big{\lbrace}p_{a}^{2}p_{e}.\mathbf{n}-p_{a}.p_{e}p_{a}.\mathbf{n}\Big{\rbrace}\Bigg)\Bigg]^{r-2} \non\\
&&\ -\ \f{2G}{R}\ \f{\big(\ln |u|\big)^{n-1}}{u^{n}}\ \sum_{r=2}^{n-1}\ \f{(n-1)! \ }{r!(n-r-1)!}\  \Big{\lbrace}2G\sum_{d=1}^{N}p_{d}.\mathbf{n}\Big{\rbrace}^{n-r}\ \times \sum_{a=1}^{M}\f{\mathbf{n}_{\rho}\mathbf{n}_{\sigma}}{p'_{a}.\mathbf{n}}\non\\
&&\times\Bigg[\sum_{\substack{b=1\\ b\neq a}}^{M}\f{1}{[(p'_{a}.p'_{b})^{2}-p_{a}^{\prime 2}p_{b}^{\prime 2}]^{3/2}}\ \Big{\lbrace}p_{a}^{\prime\mu}p_{b}^{\prime\rho}-p_{a}^{\prime\rho}p_{b}^{\prime\mu}\Big{\rbrace}\Bigg(2G\ p'_{a}.p'_{b} \Big{\lbrace}\f{3}{2}p_{a}^{\prime 2}p_{b}^{\prime 2}-(p'_{a}.p'_{b})^{2}\Big{\rbrace}\ +\ \f{1}{4\pi}\ q'_{a}q'_{b}\ p_{a}^{\prime 2}p_{b}^{\prime 2}\Bigg)\ \Bigg]\non\\
&&\times\Bigg[\sum_{\substack{c=1\\ c\neq a}}^{M}\f{1}{[(p'_{a}.p'_{c})^{2}-p_{a}^{\prime 2}p_{c}^{\prime 2}]^{3/2}}\ \Big{\lbrace}p_{a}^{\prime\nu}p_{c}^{\prime\sigma}-p_{a}^{\prime\sigma}p_{c}^{\prime\nu}\Big{\rbrace}\Bigg(2G\ p'_{a}.p'_{c} \Big{\lbrace}\f{3}{2}p_{a}^{\prime 2}p_{c}^{\prime 2}-(p'_{a}.p'_{c})^{2}\Big{\rbrace}\ +\ \f{1}{4\pi}\ q'_{a}q'_{c}\ p_{a}^{\prime 2}p_{c}^{\prime 2}\Bigg)\ \Bigg]\non\\
&&\times \Bigg[\sum_{\substack{e=1\\ e\neq a}}^{M}\f{1}{[(p'_{a}.p'_{e})^{2}-p_{a}^{\prime 2}p_{e}^{\prime 2}]^{3/2}}\Bigg(\ (2G)\Big{\lbrace}\f{3}{2}p'_{a}.p'_{e}p_{a}^{\prime 2}p_{e}^{\prime 2}p'_{e}.\mathbf{n}-(p'_{a}.p'_{e})^{3}p'_{e}.\mathbf{n}-\f{1}{2}p_{a}^{\prime 2}(p_{e}^{\prime 2})^{2}p'_{a}.\mathbf{n}\Big{\rbrace}\non\\
&&\ +\ \f{1}{4\pi}\ q'_{a}q'_{e}p_{e}^{\prime 2}\Big{\lbrace}p_{a}^{\prime 2}p'_{e}.\mathbf{n}-p'_{a}.p'_{e}p'_{a}.\mathbf{n}\Big{\rbrace}\Bigg)\Bigg]^{r-2}\non\\
&&\ +\ \f{2G}{R}\ \f{(\ln|u|)^{n-1}}{u^{n}}\sum_{r=3}^{n}\f{i^{r}n!}{(n-r)!}\Big{\lbrace}2G\sum_{b=1}^{N}p_{b}\cdot\mathbf{n}\Big{\rbrace}^{n-r}\sum_{a=1}^{N}\mathbf{n}_{\alpha_{1}}\mathbf{n}_{\alpha_{2}}\cdots \mathbf{n}_{\alpha_{r-1}}\ \mathbf{C}^{(r),\ \mu\nu\alpha_{1}\cdots\alpha_{r-1}}(q_{a},p_{a})\non\\
&&\ +\ \f{2G}{R}\f{(\ln|u|)^{n-1}}{u^{n}}\sum_{r=3}^{n-1}\f{i^{r}(n-1)!}{(n-r-1)!}\Big{\lbrace}2G\sum_{b=1}^{N}p_{b}\cdot\mathbf{n}\Big{\rbrace}^{n-r}\sum_{a=1}^{M}\mathbf{n}_{\alpha_{1}}\mathbf{n}_{\alpha_{2}}\cdots \mathbf{n}_{\alpha_{r-1}}\ \mathbf{C}^{(r),\ \mu\nu\alpha_{1}\cdots\alpha_{r-1}}(-q'_{a},-p'_{a})\non\\
&&\hspace{0.5cm} +\ \mathcal{O}\Big(u^{-n}(\ln|u|)^{n-2}\Big)\hspace{1.5cm}\hbox{for}\ u\rightarrow+\infty\label{Latewaveformgr}
\ee
\endgroup
and
\begingroup
\allowdisplaybreaks
\be
&&\Delta_{(n)}e^{\mu\nu}(t,R,\hat{n})\non\\
&=&\  \f{2G}{R}\ \f{\big(\ln |u|\big)^{n-1}}{u^{n}}\   \Big{\lbrace}2G\sum_{c=1}^{N}p_{c}.\mathbf{n}\Big{\rbrace}^{n-1}\  \sum_{a=1}^{M}\Bigg[\sum_{\substack{b=1\\ b\neq a}}^{M}\f{1}{[(p'_{a}.p'_{b})^{2}-p_{a}^{\prime 2}p_{b}^{\prime 2}]^{3/2}}\ \f{p_{a}^{\prime\mu}\mathbf{n}_{\rho}}{p'_{a}.\mathbf{n}}\Big{\lbrace}p_{a}^{\prime\nu}p_{b}^{\prime\rho}-p_{a}^{\prime\rho}p_{b}^{\prime\nu}\Big{\rbrace}\non\\
&&\times\Bigg(2G\ p'_{a}.p'_{b} \Big{\lbrace}\f{3}{2}p_{a}^{\prime 2}p_{b}^{\prime 2}-(p'_{a}.p'_{b})^{2}\Big{\rbrace}\ +\ \f{1}{4\pi}\ q'_{a}q'_{b}\ p_{a}^{\prime 2}p_{b}^{\prime 2}\Bigg)\ \Bigg]\non\\
&&\ +\ \f{2G}{R}\ \f{\big(\ln |u|\big)^{n-1}}{u^{n}}\ \sum_{r=2}^{n} \f{(n-1)! \ }{(r-1)!(n-r)!}\  \Big{\lbrace}2G\sum_{d=1}^{N}p_{d}.\mathbf{n}\Big{\rbrace}^{n-r}\times \sum_{a=1}^{M}\f{\mathbf{n}_{\rho}\mathbf{n}_{\sigma}}{p'_{a}.\mathbf{n}}\non\\
&&\times\Bigg[\sum_{\substack{b=1\\ b\neq a}}^{M}\f{1}{[(p'_{a}.p'_{b})^{2}-p_{a}^{\prime 2}p_{b}^{\prime 2}]^{3/2}}\ \Big{\lbrace}p_{a}^{\prime\mu}p_{b}^{\prime\rho}-p_{a}^{\prime\rho}p_{b}^{\prime\mu}\Big{\rbrace}\Bigg(2G\ p'_{a}.p'_{b} \Big{\lbrace}\f{3}{2}p_{a}^{\prime 2}p_{b}^{\prime 2}-(p'_{a}.p'_{b})^{2}\Big{\rbrace}\ +\ \f{1}{4\pi}\ q'_{a}q'_{b}\ p_{a}^{\prime 2}p_{b}^{\prime 2}\Bigg)\ \Bigg]\non\\
&&\times\Bigg[\sum_{\substack{c=1\\ c\neq a}}^{M}\f{1}{[(p'_{a}.p'_{c})^{2}-p_{a}^{\prime 2}p_{c}^{\prime 2}]^{3/2}}\ \Big{\lbrace}p_{a}^{\prime\nu}p_{c}^{\prime\sigma}-p_{a}^{\prime\sigma}p_{c}^{\prime\nu}\Big{\rbrace}\Bigg(2G\ p'_{a}.p'_{c} \Big{\lbrace}\f{3}{2}p_{a}^{\prime 2}p_{c}^{\prime 2}-(p'_{a}.p'_{c})^{2}\Big{\rbrace}\ +\ \f{1}{4\pi}\ q'_{a}q'_{c}\ p_{a}^{\prime 2}p_{c}^{\prime 2}\Bigg)\ \Bigg]\non\\
&&\times \Bigg[\sum_{\substack{e=1\\ e\neq a}}^{M}\f{1}{[(p'_{a}.p'_{e})^{2}-p_{a}^{\prime 2}p_{e}^{\prime 2}]^{3/2}}\Bigg(\ (2G)\Big{\lbrace}\f{3}{2}p'_{a}.p'_{e}p_{a}^{\prime 2}p_{e}^{\prime 2}p'_{e}.\mathbf{n}-(p'_{a}.p'_{e})^{3}p'_{e}.\mathbf{n}-\f{1}{2}p_{a}^{\prime 2}(p_{e}^{\prime 2})^{2}p'_{a}.\mathbf{n}\Big{\rbrace}\non\\
&&\ +\ \f{1}{4\pi}\ q'_{a}q'_{e}p_{e}^{\prime 2}\Big{\lbrace}p_{a}^{\prime 2}p'_{e}.\mathbf{n}-p'_{a}.p'_{e}p'_{a}.\mathbf{n}\Big{\rbrace}\Bigg)\Bigg]^{r-2}\non\\
&&\ -\ \f{2G}{R}\ \f{(\ln|u|)^{n-1}}{u^{n}}\sum_{r=3}^{n}\f{i^{r}(n-1)!\ r}{(n-r)!}\Big{\lbrace}2G\sum_{b=1}^{N}p_{b}\cdot\mathbf{n}\Big{\rbrace}^{n-r}\sum_{a=1}^{M}\mathbf{n}_{\alpha_{1}}\mathbf{n}_{\alpha_{2}}\cdots \mathbf{n}_{\alpha_{r-1}}\ \mathbf{C}^{(r),\ \mu\nu\alpha_{1}\cdots\alpha_{r-1}}(-q'_{a},-p'_{a})\non\\
&&\ \hspace{0.5cm} +\ \mathcal{O}\Big(u^{-n}(\ln|u|)^{n-2}\Big)\hspace{1.5cm}\hbox{for}\ u\rightarrow -\infty\label{earlywaveformgr}
\ee
\endgroup
where the retarded time $u$ is given by $u=t-R+2G\ln R\sum\limits_{b=1}^{N}p_{b}\cdot \mathbf{n}$.

\subsection{Comments on gravitational tail memory for spinning object scattering}\label{speculation}
In the analysis of \S\ref{proofsoftgr} we made a statement that internal structures or  spins of scattered objects will not affect the leading non-analytic contribution to (sub)$^n$-leading gravitational waveform. In this section we will try to present some arguments in support of this and also try to explain why determination of spin dependent gravitational memory is hard and ambiguous in our formalism.

We know that matter energy-momentum tensor has the following derivative expansion \cite{1803.02405,1712.09250,1906.08288},
\begingroup
\allowdisplaybreaks
\be
 T^{X\alpha\beta}(x)\ &=&\ \sum_{a=1}^{M+N}\int_{0}^{\infty} d\sigma\ \Bigg[ m_{a}\f{d X_{a}^{\alpha}(\sigma)}{d \sigma}\f{d X_{a}^{\beta}(\sigma)}{d \sigma}\ \delta^{(4)}\big(x-X_{a}(\sigma)\big)\non\\
 &&+\ \f{d X_{a}^{(\alpha}(\sigma)}{d \sigma}\ \Sigma_{a}^{\beta)\gamma}(\sigma)\ \p_{\gamma}\delta^{(4)}\big(x-X_{a}(\sigma)\big)\ +\ \cdots\Bigg]\label{TXspin}
\ee 
\endgroup
where $\Sigma_{a}$ represents the spin of object-a and "$\cdots$" contains terms having two or more derivatives operating on the delta function. The terms inside "$\cdots$" carry the information about the internal structure of the scattered objects.  For the spinning objects the geodesic equation is modified in the following way\cite{Papapetrou,Mathisson,Waldspin},
\be
\f{d^{2}X^{\mu}}{d\sigma^{2}}+\Gamma^{\mu}_{\alpha\beta}\f{dX^{\alpha}}{d\sigma}\f{dX^{\beta}}{d\sigma}=-\f{1}{m}\Big[\f{d^{2}\Sigma^{\mu}_{\ \nu}}{d\sigma^{2}}+\f{1}{2}R^{\mu}_{\ \nu\rho\sigma}\Sigma^{\rho\sigma}\Big]\ \f{dX^{\nu}}{d\sigma}
\ee
\be
\f{D\Sigma^{\mu\nu}}{d\sigma}-\f{dX^{\mu}}{d\sigma}\f{dX_{\rho}}{d\sigma}\f{D\Sigma^{\nu\rho}}{d\sigma}+\f{dX^{\nu}}{d\sigma}\f{dX_{\rho}}{d\sigma}\f{D\Sigma^{\mu\rho}}{d\sigma}=0
\ee
where $\f{D}{d\sigma}$ denotes covariant derivative along the world line. In the second equation above, we can substitute the leading large $\sigma$ behaviour for Christoffel connection which goes like $\Gamma^{\mu}_{\nu\rho}\sim \f{1}{\sigma^{2}}$ for large $\sigma$. In this background  solution of second equation gives $\Sigma^{\mu\nu}\sim \Sigma_{0}^{\mu\nu}+\mathcal{O}(\sigma^{-1})$ with $\Sigma_{0}^{\mu\nu}$ being the constant spin. But with this large $\sigma $ behaviour, the RHS of the first equation goes as $\mathcal{O}(\sigma^{-3})$. So to get the leading spin dependent behaviour, we can replace all the covariant derivatives in the second equation by ordinary derivatives. Then the second equation simplifies to,
\be
\f{d\Sigma^{\mu\nu}}{d\sigma}-v^{\mu}v_{\rho}\f{d\Sigma^{\nu\rho}}{d\sigma}+v^{\nu}v_{\rho}\f{d\Sigma^{\mu\rho}}{d\sigma}=0
\ee
This in some sense gives conservation relation between different spin components,
\be
\Sigma^{\mu\nu}-v^{\mu}v_{\rho}\Sigma^{\nu\rho}+v^{\nu}v_{\rho}\Sigma^{\mu\rho}=constant
\ee
So, to analyze the leading order spin dependence of matter energy-momentum tensor we can treat the spins of scattered objects $\lbrace \Sigma_{a}\rbrace$ to be constant. Now at the leading order considering the trajectory of object-a being  $X^{\mu}_{a}(\sigma)=r^{\mu}_{a}+v_{a}^{\mu}\sigma$, the Fourier transformation of energy-momentum tensor becomes,
 \be
 \Delta_{(0)}\widehat{T}^{X\alpha\beta}(k)\ &=&\ \sum_{a=1}^{M+N}e^{-ik.r_{a}}\f{1}{i(p_{a}.k-i\epsilon)}\ \Big[ p_{a}^{\alpha}p_{a}^{\beta}\ +\ ip_{a}^{(\alpha}\Sigma_{a}^{\beta)\gamma}k_{\gamma}\Big]
\ee 
Using the relation \eqref{eTrelation}, we find that the leading spin dependent contribution to $\widetilde{e}^{\mu\nu}(\omega, R,\hat{n})$ appear at order $\mathcal{O}(\omega^{0})$ , which does not contribute to memory. 

Solving linearized Einstein equation the leading order metric fluctuation takes form,
\be
\Delta_{(0)}h_{\alpha\beta}(x)\ &=&\ -8\pi G\ \sum_{a=1}^{M+N}\int \f{d^{4}\ell}{(2\pi)^{4}}\ e^{i\ell\cdot(x-r_{a})}\ G_{r}(\ell)\ \f{1}{i(p_{a}.\ell-i\epsilon)}\ \Big[ p_{a\alpha}p_{a\beta}-\f{1}{2}p_{a}^{2}\eta_{\alpha\beta}\ +\ ip_{a(\alpha}\Sigma_{a,\beta)\gamma}\ell^{\gamma}\Big]\non\\
\label{metric}
\ee
with constraint on the spin $p_{a}^{\alpha}\ \Sigma_{a,\alpha\beta}=0$.  Now we can find out the subleading correction to the straight line trajectories solving geodesic equation of scattered objects in the background of the above metric analogous to \S\ref{Ssecsubgr}.
Here the subleading contribution to the Fourier transform of matter energy-momentum tensor becomes,
\be
\Delta_{(1)}\widehat{T}^{X\mu\nu}(k)\ &=&\ \sum_{a=1}^{M+N}\int _{0}^{\infty}d\sigma e^{-ik.(v_{a}\sigma+r_{a})}\ \Bigg[-ik.\Delta_{(1)}Y_{a}(\sigma)\ m_{a}v_{a}^{\mu}v_{a}^{\nu}\ +m_{a}v_{a}^{\mu}\f{d\Delta_{(1)}Y_{a}^{\nu}(\sigma)}{d\sigma}\non\\
&&\ +m_{a}v_{a}^{\nu}\f{d\Delta_{(1)}Y_{a}^{\mu}(\sigma)}{d\sigma}+i\f{d\Delta_{(1)}Y_{a}^{(\mu}(\sigma)}{d\sigma}\Sigma_{a}^{\nu)\alpha}k_{\alpha} + k.\Delta_{(1)}Y_{a}(\sigma)\ v_{a}^{(\mu}\Sigma_{a}^{\nu)\alpha}k_{\alpha}\Bigg]
\ee
Now if we try to analyze the first term within the square bracket above, we find the following spin dependent contribution,
\be
\Delta^{(1)}_{(1)\Sigma}\widehat{T}^{X\mu\nu}(k)\ &=&\ i(8\pi G)\sum_{a,b}p_{a}^{\mu}p_{a}^{\nu}e^{-ik.r_{a}}\int \f{d^{4}\ell}{(2\pi)^{4}}\ e^{i\ell.(r_{a}-r_{b})}\ G_{r}(\ell)\f{1}{p_{b}.\ell-i\epsilon}\ \f{1}{p_{a}.\ell+i\epsilon}\ \f{1}{p_{a}.k}\non\\
&&\ \f{1}{p_{a}.(k-\ell)-i\epsilon} \Big[p_{a}.\ell\ p_{a}.p_{b}\ k.\Sigma_{b}.\ell+p_{a}.\ell\ p_{b}.k\ p_{a}.\Sigma_{b}.\ell-k.\ell\ p_{a}.p_{b}\ p_{a}.\Sigma_{b}.\ell\Big]
\ee
In the region of integration $L^{-1}>>|\ell^{\mu}|>>\omega$, the leading contribution turns out to be at order $\mathcal{O}(\omega^{0})$ with some $L$ dependent factor. On the other hand if we approximate $\lbrace p_{a}.(k-\ell)-i\epsilon\rbrace^{-1}\simeq -\lbrace p_{a}.\ell +i\epsilon\rbrace^{-1}-p_{a}.k\lbrace p_{a}.\ell +i\epsilon\rbrace^{-2}$ and analyze with the subleading term it contributes to $\mathcal{O}(\omega\ln\omega)$. Similar analysis of the other spin dependent terms of $\Delta_{(1)}\widehat{T}^{X\mu\nu}(k)$  contribute to $\mathcal{O}(\omega^{0})$ and $\mathcal{O}(\omega\ln\omega)$. We can also analyze the subleading gravitational energy-momentum tensor,
\be
\Delta_{(1)}\widehat{T}^{h\mu\nu}(k) &=& -(8\pi G)\sum_{a,b}e^{-ik.r_{b}}\int \f{d^{4}\ell}{(2\pi)^{4}}\ e^{i\ell.(r_{b}-r_{a})}\ G_{r}(k-\ell)G_{r}(\ell)\ \f{1}{p_{a}.\ell-i\epsilon}\ \f{1}{p_{b}.(k-\ell)-i\epsilon}\non\\
&&\times \Big{\lbrace} p_{a\alpha}p_{a\beta}-\f{1}{2}p_{a}^{2}\eta_{\alpha\beta}+ip_{a(\alpha}\Sigma_{a,\beta)\gamma}\ell^{\gamma}\Big{\rbrace}\ \mathcal{F}^{\mu\nu, \alpha\beta , \rho\sigma}(k,\ell)\ \Big{\lbrace}p_{b\rho}p_{b\sigma}-\f{1}{2}p_{b}^{2}\eta_{\rho\sigma}+ip_{b(\rho}\Sigma_{b,\sigma)\delta}(k-\ell)^{\delta}\Big{\rbrace}\non\\\label{Thathspin}
\ee
A detailed analysis shows that the spin dependent part of the  above expression  also contributes to $\mathcal{O}(\omega^{0})$ and $\mathcal{O}(\omega\ln\omega)$ similar to $\Delta_{(1)}\widehat{T}^{X\mu\nu}(k)$. So from this naive analysis we find that if the scattered objects carry spin then that will affect the order $\mathcal{O}(\omega\ln\omega)$ part of the energy-momentum tensor which in turn affect the order $\mathcal{O}(u^{-2})$ gravitational tail memory following eq.\eqref{eTrelation}. But as we have seen from the analysis of subleading and sub-subleading order energy-momentum tensor in \S\ref{Ssecsubgr} and \S\ref{secsub2leadinggr}, the order $\mathcal{O}(\omega\ln\omega)$ contribution is ambiguous as the contribution in this order depends on the details of scattering region $\mathcal{R}$ through $\lbrace r_{a}\rbrace$. This tells us that the order $\mathcal{O}(u^{-2})$ tail memory is not fully specified by the scattering data, it needs information about the scattering region. Hence it is not yet clear how to separate out this spin dependent gravitational memory at order $\mathcal{O}(u^{-2})$ from the other non-universal contributions at the same order.

Though we have not been able to extract spin dependent gravitational memory unambiguously in this subsection but the analysis shows that the spins of the scattered objects do not affect our sub-subleading order gravitational waveform at order $\mathcal{O}(\omega(\ln\omega)^{2})$. Analogous analysis of the (sub)$^{n}$-leading gravitational waveform shows that the spins of the scattered objects start contributing to gravitational tail memory at order $\mathcal{O}\big(u^{-n}(\ln u)^{n-2}\big)$. Since the internal structure dependence comes with more powers of momenta in the Fourier transform of matter energy-momentum tensor follows from eq.\eqref{TXspin}, those start affecting the (sub)$^{n}$-leading gravitational memory at order $\mathcal{O}\big(u^{-n}(\ln u)^{m}\big)$ for $m\leq n-3$.

\vspace{1.5cm}
{\bf{Acknowledgement:}} I am deeply thankful to Prof. Ashoke Sen for suggesting this problem, for many enlightening discussions and for correcting the initial version of the draft with immense patience. I would also like to thank Dileep Jatkar, Arnab Priya Saha, Alok Laddha and Sayali Atul Bhatkar for useful discussions and comments on the draft. I am also thankful to Abhishek, Kajal, Sachin and Vivek for helping me out surviving in this pandemic time.
I am grateful to the people of India for their continuous support to theoretical sciences.

\appendix
\section{Analysis of the terms containing $\Delta_{(2)}Y_{a}(\sigma)$ in the expression of $\Delta_{(2)}\widehat{J}(k)$}\label{appA}
Before we proceed for analyzing the third and fourth terms within the square bracket of eq.\eqref{sub-subleadingC}, let us find out the expression of $\f{d\Delta_{(2)}Y_{a\mu}(\sigma)}{d\sigma}$. 

Subleading current density for b'th particle motion from eq.\eqref{subleadingC},
\be
\Delta_{(1)}\widehat{J}^{(b)}_{\mu}(\ell_{1})\ &=&\ \sum_{\substack{c=1\\ c\neq b}}^{M+N}q_{b}^{2}q_{c}e^{-i\ell_{1}.r_{b}}\ \int\f{d^{4}\ell_{2}}{(2\pi)^{4}}G_{r}(\ell_{2})\ e^{i\ell_{2}.(r_{b}-r_{c})}\ \f{1}{p_{b}.\ell_{2}+i\epsilon}\ \f{1}{p_{b}.(\ell_{2}-\ell_{1})+i\epsilon}\ \f{1}{p_{c}.\ell_{2}-i\epsilon}\non\\
&&\times \Bigg[ \f{p_{b\mu}}{p_{b}.\ell_{1}-i\epsilon}\lbrace \ell_{1}.\ell_{2}p_{b}.p_{c}-p_{c}.\ell_{1}p_{b}.\ell_{2}\rbrace -\lbrace \ell_{2\mu}p_{b}.p_{c}-p_{c\mu}\ell_{2}.p_{b}\rbrace\Bigg]
\ee
With this form of subleading current density, the subleading field strength for b'th particle's motion takes the following form,
\be
\Delta_{(1)}F_{\mu\nu}^{(b)}(x)\ &=&\  -i\sum_{\substack{c=1\\ c\neq b}}^{M+N}q_{b}^{2}q_{c} \int\f{d^{4}\ell_{1}}{(2\pi)^{4}}\ e^{i\ell_{1}.(x-r_{b})}\ G_{r}(\ell_{1})\int \f{d^{4}\ell_{2}}{(2\pi)^{4}}\ e^{i\ell_{2}.(r_{b}-r_{c})}\ G_{r}(\ell_{2})\ \f{1}{p_{b}.\ell_{2}+i\epsilon}\non\\
&&\ \f{1}{p_{b}.(\ell_{2}-\ell_{1})+i\epsilon}\f{1}{p_{c}.\ell_{2}-i\epsilon}\Bigg[\f{1}{p_{b}.\ell_{1}-i\epsilon}\lbrace\ell_{1\mu}p_{b\nu}-\ell_{1\nu}p_{b\mu}\rbrace\lbrace \ell_{1}.\ell_{2}p_{b}.p_{c}-p_{c}.\ell_{1}p_{b}.\ell_{2}\rbrace\non\\
&&\ -\lbrace \ell_{1\mu}\ell_{2\nu}p_{b}.p_{c}-\ell_{1\nu}\ell_{2\mu}p_{b}.p_{c}-\ell_{1\mu}p_{c\nu}\ell_{2}.p_{b}+\ell_{1\nu}p_{c\mu}\ell_{2}.p_{b}\rbrace\Bigg]
\ee
The order $\mathcal{O}(q^{4})$ correction to the straight line trajectory satisfies the following eq. of motion,
\begingroup
\allowdisplaybreaks
\be
\f{d^{2}\Delta_{(2)}Y_{a\mu}(\sigma)}{d\sigma^{2}}\ &=&\ \f{q_{a}}{m_{a}}\sum_{\substack{b=1\\b\neq a}}^{M+N}\Big[\Delta_{(0)}F_{\mu\nu}^{(b)}(r_{a}+v_{a}\sigma +Y_{a}(\sigma))\ +\ \Delta_{(1)}F_{\mu\nu}^{(b)}(r_{a}+v_{a}\sigma)\Big]\ \Big{\lbrace}v_{a}^{\nu}+\f{d\Delta_{(1)}Y_{a}^{\nu}(\sigma)}{d\sigma}\Big{\rbrace}\ \Bigg{|}_{\mathcal{O}(q^{4})}\non\\
&=&\ \f{q_{a}}{m_{a}}\ \sum_{\substack{b=1\\ b\neq a}}^{M+N}\Delta_{(0)}F_{\mu\nu}^{(b)}(r_{a}+v_{a}\sigma)\ \f{d\Delta_{(1)}Y_{a}^{\nu}(\sigma)}{d\sigma}\non\\
&&-\f{q_{a}}{m_{a}}\ \sum_{\substack{b=1\\ b\neq a}}^{M+N}\int \f{d^{4}\ell_{1}}{(2\pi)^{4}}\ e^{i\ell_{1} .(r_{a}+v_{a}\sigma -r_{b})}\ \Big{\lbrace}i\ell_{1}\cdot \Delta_{(1)}Y_{a}(\sigma)\Big{\rbrace}\ G_{r}(\ell_{1})\ \f{q_{b}}{p_{b}.\ell_{1} -i\epsilon}(\ell_{1\mu}p_{b\nu}-\ell_{1\nu}p_{b\mu})v_{a}^{\nu}\non\\
&&+\f{q_{a}}{m_{a}}\  \sum_{\substack{b=1\\ b\neq a}}^{M+N}\Delta_{(1)}F_{\mu\nu}^{(b)}(r_{a}+v_{a}\sigma)\ v_{a}^{\nu}
\ee
\endgroup
Now using the boundary conditions follows from eq.\eqref{boundaryEM}:
\be
\Delta_{(2)}Y_{a}^{\mu}(\sigma)\Big{|}_{\sigma=0}\ =\ 0 \hspace{5mm} ,\ \hspace{5mm}\ {d\Delta_{(2)}Y_a^\mu(\sigma)\over d\sigma}\Bigg{|}_{\sigma\rightarrow \infty}\ =\ 0
\ee
we get,
\begingroup
\allowdisplaybreaks
\be
&&\f{d\Delta_{(2)}Y_{a\mu}(\sigma)}{d\sigma}\non\\
 &=&\ -\f{q_{a}}{m_{a}}\ \sum_{\substack{b=1\\ b\neq a}}^{M+N}\int_{\sigma}^{\infty}d\sigma'\ \Delta_{(0)}F_{\mu\nu}^{(b)}(r_{a}+v_{a}\sigma')\ \f{d\Delta_{(1)}Y_{a}^{\nu}(\sigma')}{d\sigma'}\non\\
&&+\f{q_{a}}{m_{a}}\ \sum_{\substack{b=1\\ b\neq a}}^{M+N}\int \f{d^{4}\ell_{1}}{(2\pi)^{4}}\ e^{i\ell_{1} .(r_{a}-r_{b})}G_{r}(\ell_{1})\ \f{q_{b}}{p_{b}.\ell_{1} -i\epsilon}(\ell_{1\mu}p_{b\nu}-\ell_{1\nu}p_{b\mu})v_{a}^{\nu}\int_{\sigma}^{\infty}d\sigma' e^{i\ell_{1} .v_{a}\sigma'}\ \Big{\lbrace}i\ell\cdot \Delta_{(1)}Y_{a}(\sigma')\Big{\rbrace}\non\\
&&-\f{q_{a}}{m_{a}}\  \sum_{\substack{b=1\\ b\neq a}}^{M+N}\int_{\sigma}^{\infty}d\sigma'\ \Delta_{(1)}F_{\mu\nu}^{(b)}(r_{a}+v_{a}\sigma')\ v_{a}^{\nu}
\ee
\endgroup
The contribution to sub-subleading current due to third term in the square bracket of eq.\eqref{sub-subleadingC},
\begingroup
\allowdisplaybreaks
\be
\Delta_{(2)}^{(3)}\widehat{J}_{\mu}(k)\ &=&\  \sum_{a=1}^{M+N}q_{a}\ e^{-ik.r_{a}}\ \int_{0}^{\infty}d\sigma e^{-i(k.v_{a}-i\epsilon)\sigma}\ \Big{\lbrace} -ik\cdot \Delta_{(2)}Y_{a}(\sigma)v_{a\mu}\ \Big{\rbrace}\non\\
&=&\ -\sum_{a=1}^{M+N}q_{a}e^{-ik\cdot r_{a}}\ \f{p_{a\mu}k^{\alpha}}{p_{a}\cdot k}\ \int_{0}^{\infty}d\sigma \ e^{-i(k.v_{a}-i\epsilon)\sigma}\ \f{d\Delta_{(2)}Y_{a\alpha}(\sigma)}{d\sigma}\non\\
&=&\ -\sum_{a=1}^{M+N}q_{a}e^{-ik\cdot r_{a}}\ \f{p_{a\mu}k^{\alpha}}{p_{a}\cdot k}\ \Bigg[-iq_{a}^{2}\sum_{\substack{b=1\\b\neq a}}^{M+N}\sum_{\substack{c=1\\c\neq a}}^{M+N}q_{b}q_{c}\ \int \f{d^{4}\ell_{1}}{(2\pi)^{4}}\ e^{i\ell_{1}.(r_{a}-r_{b})}G_{r}(\ell_{1})\non\\
&&\ \times\f{1}{p_{b}.\ell_{1}-i\epsilon}(\ell_{1\alpha}p_{b\nu}-\ell_{1\nu}p_{b\alpha})\ \int \f{d^{4}\ell_{2}}{(2\pi)^{4}}\ e^{i\ell_{2}.(r_{a}-r_{c})}\ G_{r}(\ell_{2})\ \f{1}{p_{c}.\ell_{2}-i\epsilon}\non\\
&&\ \times(\ell_{2}^{\nu}p_{c}.p_{a}-\ell_{2}.p_{a}p_{c}^{\nu})\ \f{1}{\ell_{2}.p_{a}+i\epsilon}\ \f{1}{(\ell_{1}+\ell_{2}).p_{a}+i\epsilon}\ \f{1}{(\ell_{1}+\ell_{2}-k).p_{a}+i\epsilon}\non\\
&&\ -iq_{a}^{2}\sum_{\substack{b=1\\b\neq a}}^{M+N}\sum_{\substack{c=1\\c\neq a}}^{M+N}q_{b}q_{c}\ \int \f{d^{4}\ell_{1}}{(2\pi)^{4}}\ e^{i\ell_{1}.(r_{a}-r_{b})}G_{r}(\ell_{1})\  \times\f{1}{p_{b}.\ell_{1}-i\epsilon}(\ell_{1\alpha}p_{b}.p_{a}-\ell_{1}.p_{a}p_{b\alpha})\non\\
&&\times \int \f{d^{4}\ell_{2}}{(2\pi)^{4}}\ e^{i\ell_{2}.(r_{a}-r_{c})}\ G_{r}(\ell_{2})\ \f{1}{p_{c}.\ell_{2}-i\epsilon}\ \f{1}{(\ell_{2}.p_{a}+i\epsilon)^{2}}\Bigg{\lbrace} \f{1}{(\ell_{1}+\ell_{2}).p_{a}+i\epsilon}\  \f{1}{(\ell_{1}+\ell_{2}-k).p_{a}+i\epsilon}\non\\
&&\ -\ \f{1}{\ell_{1}.p_{a}+i\epsilon}\  \f{1}{(\ell_{1}-k).p_{a}+i\epsilon}\Bigg{\rbrace}\times (\ell_{1}.\ell_{2}p_{a}.p_{c}-\ell_{1}.p_{c}\ell_{2}.p_{a})\non\\
&&\ -iq_{a}\ \sum_{\substack{b=1\\b\neq a}}^{M+N}\sum_{\substack{c=1\\c\neq b}}^{M+N}q_{b}^{2}q_{c}\int\f{d^{4}\ell_{1}}{(2\pi)^{4}}\ e^{i\ell_{1}.(r_{a}-r_{b})}G_{r}(\ell_{1})\ \f{1}{p_{a}.\ell_{1}+i\epsilon}\f{1}{p_{a}.(\ell_{1}-k)+i\epsilon} \int\f{d^{4}\ell_{2}}{(2\pi)^{4}} e^{i\ell_{2}.(r_{b}-r_{c})}\non\\
&&\times G_{r}(\ell_{2})\ \f{1}{p_{b}.\ell_{2}+i\epsilon}\f{1}{p_{b}.(\ell_{2}-\ell_{1})+i\epsilon}\f{1}{p_{c}.\ell_{2}-i\epsilon}\Bigg[\f{1}{p_{b}.\ell_{1}-i\epsilon}\lbrace\ell_{1\alpha}p_{b}.p_{a}-\ell_{1}.p_{a}p_{b\alpha}\rbrace\non\\
&&\ \lbrace \ell_{1}.\ell_{2}p_{b}.p_{c}-p_{c}.\ell_{1}p_{b}.\ell_{2}\rbrace \  -\lbrace \ell_{1\alpha}\ell_{2}.p_{a}p_{b}.p_{c}-\ell_{1}.p_{a}\ell_{2\alpha}p_{b}.p_{c}-\ell_{1\alpha}p_{c}.p_{a}\ell_{2}.p_{b}+\ell_{1}.p_{a}p_{c\alpha}\ell_{2}.p_{b}\rbrace\Bigg]\ \Bigg]\non\\
\ee
\endgroup
Analyzing the above expression in all the integration regions we find that only the last six lines above contribute to order $\mathcal{O}(\omega(\ln\omega)^{2})$ in the integration region $L^{-1}>>|\ell_{2}^{\mu}|>>|\ell_{1}^{\mu}|>>\omega$ \footnote{The expression above also contributes to the order $\mathcal{O}(\omega^{0})$ in some regions of integration and the contribution depends on $L$ , which we are neglecting as the order $\mathcal{O}(\omega^{0})$ contribution is ambiguous to determine as well as don't contribute to memory.}. So in this region of integration approximating $\lbrace p_{a}.(\ell_{1}-k)+i\epsilon\rbrace^{-1}\simeq \lbrace p_{a}.\ell_{1}+i\epsilon\rbrace^{-1}+p_{a}.k\lbrace p_{a}.\ell_{1}+i\epsilon\rbrace^{-2}$ and $\lbrace p_{b}.(\ell_{2}-\ell_{1})+i\epsilon\rbrace^{-1}\simeq \lbrace p_{b}.\ell_{2}+i\epsilon\rbrace^{-1}$ the order $\mathcal{O}(\omega(\ln\omega)^{2})$ contribution from the last six lines becomes,
\begingroup
\allowdisplaybreaks
\be
\Delta_{(2)}^{(3)}\widehat{J}_{\mu}(k)\ &\simeq &\ -i\sum_{a=1}^{M+N}q_{a}^{3}p_{a\mu}k^{\alpha}\sum_{\substack{b=1\\b\neq a}}^{M+N}\sum_{\substack{c=1\\c\neq a}}^{M+N}q_{b}q_{c}\int_{\omega}^{L^{-1}}\f{d^{4}\ell_{2}}{(2\pi)^{4}}\ G_{r}(\ell_{2})\ \f{1}{(p_{a}.\ell_{2}+i\epsilon)^{2}}\f{1}{p_{c}.\ell_{2}-i\epsilon}\non\\
&&\ \int_{\omega}^{|\vec{\ell}_{2}|}\f{d^{4}\ell_{1}}{(2\pi)^{4}}\ G_{r}(\ell_{1})\ \f{1}{(p_{a}.\ell_{1}+i\epsilon)^{3}}\f{1}{p_{b}.\ell_{1}-i\epsilon}\ \Big{\lbrace} \ell_{1\alpha}p_{a}.p_{b}-\ell_{1}.p_{a}p_{b\alpha}\Big{\rbrace}\non\\
&&\times \Big{\lbrace} \ell_{1}.\ell_{2}p_{a}.p_{c}-p_{c}.\ell_{1}p_{a}.\ell_{2}\Big{\rbrace}\non\\
&& +\  i\sum_{a=1}^{M+N}q_{a}^{2}\ p_{a\mu}k^{\alpha}\sum_{\substack{b=1\\b\neq a}}^{M+N}\sum_{\substack{c=1\\c\neq b}}^{M+N}q_{b}^{2}q_{c}\int_{\omega}^{L^{-1}}\f{d^{4}\ell_{2}}{(2\pi)^{4}}\ G_{r}(\ell_{2})\ \f{1}{(p_{b}.\ell_{2}+i\epsilon)^{2}}\f{1}{p_{c}.\ell_{2}-i\epsilon}\non\\
&&\ \int_{\omega}^{|\vec{\ell}_{2}|}\f{d^{4}\ell_{1}}{(2\pi)^{4}}\ G_{r}(\ell_{1})\ \f{1}{(p_{a}.\ell_{1}+i\epsilon)^{3}}\f{1}{p_{b}.\ell_{1}-i\epsilon}\ \Big{\lbrace} \ell_{1\alpha}p_{a}.p_{b}-\ell_{1}.p_{a}p_{b\alpha}\Big{\rbrace}\non\\
&&\times \Big{\lbrace} \ell_{1}.\ell_{2}p_{b}.p_{c}-p_{c}.\ell_{1}p_{b}.\ell_{2}\Big{\rbrace}\non\\
%&= &\ -\f{i}{2}\sum_{a=1}^{M+N}q_{a}^{2}p_{a\mu}k^{\alpha}\sum_{\substack{b=1\\ b\neq a}}^{M+N}\sum_{\substack{c=1\\ c\neq b}}^{M+N}q_{b}^{2}q_{c}\ \lbrace \delta_{\alpha}^{\rho}p_{a}.p_{b}-p_{a}^{\rho}p_{b\alpha}\rbrace\ \lbrace p_{b}.p_{c}\eta^{\sigma\gamma}-p_{b}^{\gamma}p_{c}^{\sigma}\rbrace\non\\
%&&\f{\p}{\p p_{b}^{\gamma}}\Bigg[ \int_{\omega}^{L^{-1}}\f{d^{4}\ell_{2}}{(2\pi)^{4}}G_{r}(\ell_{2})\ \f{1}{p_{b}.\ell_{2}+i\epsilon}\f{1}{p_{c}.\ell_{2}-i\epsilon}\Bigg]\f{\p}{\p p_{a}^{\rho}}\f{\p}{\p p_{a}^{\sigma}}\Bigg[\int_{\omega}^{|\vec{\ell}_{2}|}\f{d^{4}\ell_{1}}{(2\pi)^{4}}G_{r}(\ell_{1})\f{1}{p_{a}.\ell_{1}+i\epsilon}\f{1}{p_{b}.\ell_{1}-i\epsilon}\Bigg]\non\\
&=&\ \f{i}{4}\f{1}{(4\pi)^{2}}\sum_{a=1}^{M+N}\Big{\lbrace}\ln\Big((\omega+i\epsilon\eta_{a})L\Big)\Big{\rbrace}^{2}q_{a}^{3}p_{a\mu}k^{\alpha}\sum_{\substack{b=1\\ b\neq a\\ \eta_{a}\eta_{b}=1}}^{M+N}\sum_{\substack{c=1\\ c\neq a\\ \eta_{a}\eta_{c}=1}}^{M+N}q_{b}q_{c}\f{1}{[(p_{a}.p_{b})^{2}-p_{a}^{2}p_{b}^{2}]^{5/2}}\non\\
&&\times \f{1}{[(p_{a}.p_{c})^{2}-p_{a}^{2}p_{c}^{2}]^{3/2}}\ p_{b}^{2}p_{c}^{2}p_{a}.p_{b}\ \Big[\lbrace p_{a}.p_{c}p_{a\alpha}-p_{a}^{2}p_{c\alpha}\rbrace\lbrace (p_{a}.p_{b})^{2}-p_{a}^{2}p_{b}^{2}\rbrace\non\\
&&\ +3\lbrace p_{a}^{2}p_{b\alpha}-p_{a}.p_{b}p_{a\alpha}\rbrace\lbrace p_{a}.p_{b}p_{a}.p_{c}-p_{a}^{2}p_{b}.p_{c}\rbrace\Big]\non\\
&&+\f{i}{4}\ \f{1}{(4\pi)^{2}}\sum_{a=1}^{M+N}\Big{\lbrace}\ln\Big((\omega+i\epsilon\eta_{a})L\Big)\Big{\rbrace}^{2}q_{a}^{2}p_{a\mu}k^{\alpha}\sum_{\substack{b=1\\ b\neq a\\ \eta_{a}\eta_{b}=1}}^{M+N}\sum_{\substack{c=1\\ c\neq b\\ \eta_{b}\eta_{c}=1}}^{M+N}q_{b}^{2}q_{c}\f{1}{[(p_{b}.p_{c})^{2}-p_{b}^{2}p_{c}^{2}]^{3/2}}\non\\
&&\times \f{1}{[(p_{a}.p_{b})^{2}-p_{a}^{2}p_{b}^{2}]^{5/2}}\ (p_{b}^{2})^{2}p_{c}^{2}\Big[ \lbrace (p_{a}.p_{b})^{2}-p_{a}^{2}p_{b}^{2}\rbrace \lbrace p_{a}.p_{b}p_{c\alpha}-p_{a}.p_{c}p_{b\alpha}\rbrace +\ 3\ \lbrace p_{b}^{2}p_{a}.p_{b}p_{a}.p_{c}p_{a\alpha}\non\\
&&\ +p_{a}^{2}p_{b}.p_{c}p_{a}.p_{b}p_{b\alpha}-(p_{a}.p_{b})^{2}p_{b}.p_{c}p_{a\alpha}-p_{a}^{2}p_{b}^{2}p_{a}.p_{c}p_{b\alpha}\rbrace\Big]\label{appAJ23}
\ee
\endgroup

Similarly the contribution to sub-subleading current due to fourth term in the square bracket of eq.\eqref{sub-subleadingC} takes form,
\begingroup
\allowdisplaybreaks
\be
\Delta_{(2)}^{(4)}\widehat{J}_{\mu}(k)\ &=&\ \sum_{a=1}^{M+N}q_{a}\ e^{-ik.r_{a}}\ \int_{0}^{\infty}d\sigma e^{-i(k.v_{a}-i\epsilon)\sigma}\ \f{d\Delta_{(2)}Y_{a\mu}(\sigma)}{d\sigma}\non\\
&=&\  \sum_{a=1}^{M+N}q_{a}\ e^{-ik.r_{a}}\ \Bigg[-iq_{a}^{2}\sum_{\substack{b=1\\b\neq a}}^{M+N}\sum_{\substack{c=1\\c\neq a}}^{M+N}q_{b}q_{c}\ \int \f{d^{4}\ell_{1}}{(2\pi)^{4}}\ e^{i\ell_{1}.(r_{a}-r_{b})}G_{r}(\ell_{1})\non\\
&&\ \times\f{1}{p_{b}.\ell_{1}-i\epsilon}(\ell_{1\mu}p_{b\nu}-\ell_{1\nu}p_{b\mu})\ \int \f{d^{4}\ell_{2}}{(2\pi)^{4}}\ e^{i\ell_{2}.(r_{a}-r_{c})}\ G_{r}(\ell_{2})\ \f{1}{p_{c}.\ell_{2}-i\epsilon}\non\\
&&\ \times(\ell_{2}^{\nu}p_{c}.p_{a}-\ell_{2}.p_{a}p_{c}^{\nu})\ \f{1}{\ell_{2}.p_{a}+i\epsilon}\ \f{1}{(\ell_{1}+\ell_{2}).p_{a}+i\epsilon}\ \f{1}{(\ell_{1}+\ell_{2}-k).p_{a}+i\epsilon}\non\\
&&\ -iq_{a}^{2}\sum_{\substack{b=1\\b\neq a}}^{M+N}\sum_{\substack{c=1\\c\neq a}}^{M+N}q_{b}q_{c}\ \int \f{d^{4}\ell_{1}}{(2\pi)^{4}}\ e^{i\ell_{1}.(r_{a}-r_{b})}G_{r}(\ell_{1})\  \times\f{1}{p_{b}.\ell_{1}-i\epsilon}(\ell_{1\mu}p_{b}.p_{a}-\ell_{1}.p_{a}p_{b\mu})\non\\
&&\times \int \f{d^{4}\ell_{2}}{(2\pi)^{4}}\ e^{i\ell_{2}.(r_{a}-r_{c})}\ G_{r}(\ell_{2})\ \f{1}{p_{c}.\ell_{2}-i\epsilon}\ \f{1}{(\ell_{2}.p_{a}+i\epsilon)^{2}}\Bigg{\lbrace} \f{1}{(\ell_{1}+\ell_{2}).p_{a}+i\epsilon}\  \f{1}{(\ell_{1}+\ell_{2}-k).p_{a}+i\epsilon}\non\\
&&\ -\ \f{1}{\ell_{1}.p_{a}+i\epsilon}\  \f{1}{(\ell_{1}-k).p_{a}+i\epsilon}\Bigg{\rbrace}\times (\ell_{1}.\ell_{2}p_{a}.p_{c}-\ell_{1}.p_{c}\ell_{2}.p_{a})\non\\
&&\ -iq_{a}\ \sum_{\substack{b=1\\b\neq a}}^{M+N}\sum_{\substack{c=1\\c\neq b}}^{M+N}q_{b}^{2}q_{c}\int\f{d^{4}\ell_{1}}{(2\pi)^{4}}\ e^{i\ell_{1}.(r_{a}-r_{b})}G_{r}(\ell_{1})\ \f{1}{p_{a}.\ell_{1}+i\epsilon}\f{1}{p_{a}.(\ell_{1}-k)+i\epsilon} \int\f{d^{4}\ell_{2}}{(2\pi)^{4}} e^{i\ell_{2}.(r_{b}-r_{c})}\non\\
&&\times G_{r}(\ell_{2})\ \f{1}{p_{b}.\ell_{2}+i\epsilon}\f{1}{p_{b}.(\ell_{2}-\ell_{1})+i\epsilon}\f{1}{p_{c}.\ell_{2}-i\epsilon}\Bigg[\f{1}{p_{b}.\ell_{1}-i\epsilon}\lbrace\ell_{1\mu}p_{b}.p_{a}-\ell_{1}.p_{a}p_{b\mu}\rbrace\non\\
&&\ \lbrace \ell_{1}.\ell_{2}p_{b}.p_{c}-p_{c}.\ell_{1}p_{b}.\ell_{2}\rbrace \  -\lbrace \ell_{1\mu}\ell_{2}.p_{a}p_{b}.p_{c}-\ell_{1}.p_{a}\ell_{2\mu}p_{b}.p_{c}-\ell_{1\mu}p_{c}.p_{a}\ell_{2}.p_{b}+\ell_{1}.p_{a}p_{c\mu}\ell_{2}.p_{b}\rbrace\Bigg]\ \Bigg]\non\\
\ee
\endgroup
Analogously, analyzing the above expression in all the integration regions we find that only the last six lines contribute to order $\mathcal{O}(\omega(\ln\omega)^{2})$ in the integration region $L^{-1}>>|\ell_{2}^{\mu}|>>|\ell_{1}^{\mu}|>>\omega$. Now using the prescription described below eq.\eqref{step} we find the following order $\mathcal{O}(\omega(\ln\omega)^{2})$ contribution,
\begingroup
\allowdisplaybreaks
\be
\Delta_{(2)}^{(4)}\widehat{J}_{\mu}(k)\ &\simeq &\ + i\sum_{a=1}^{M+N}q_{a}^{3}\ (p_{a}.k)\sum_{\substack{b=1\\b\neq a}}^{M+N}\sum_{\substack{c=1\\c\neq a}}^{M+N}q_{b}q_{c}\int_{\omega}^{L^{-1}}\f{d^{4}\ell_{2}}{(2\pi)^{4}}\ G_{r}(\ell_{2})\ \f{1}{(p_{a}.\ell_{2}+i\epsilon)^{2}}\f{1}{p_{c}.\ell_{2}-i\epsilon}\non\\
&&\ \int_{\omega}^{|\vec{\ell}_{2}|}\f{d^{4}\ell_{1}}{(2\pi)^{4}}\ G_{r}(\ell_{1})\ \f{1}{(p_{a}.\ell_{1}+i\epsilon)^{3}}\f{1}{p_{b}.\ell_{1}-i\epsilon}\ \Big{\lbrace} \ell_{1\mu}p_{a}.p_{b}-\ell_{1}.p_{a}p_{b\mu}\Big{\rbrace}\non\\
&&\times \Big{\lbrace} \ell_{1}.\ell_{2}p_{a}.p_{c}-p_{c}.\ell_{1}p_{a}.\ell_{2}\Big{\rbrace}\non\\
&&\ - i\sum_{a=1}^{M+N}q_{a}^{2}\ (p_{a}.k)\sum_{\substack{b=1\\b\neq a}}^{M+N}\sum_{\substack{c=1\\c\neq b}}^{M+N}q_{b}^{2}q_{c}\int_{\omega}^{L^{-1}}\f{d^{4}\ell_{2}}{(2\pi)^{4}}\ G_{r}(\ell_{2})\ \f{1}{(p_{b}.\ell_{2}+i\epsilon)^{2}}\f{1}{p_{c}.\ell_{2}-i\epsilon}\non\\
&&\ \int_{\omega}^{|\vec{\ell}_{2}|}\f{d^{4}\ell_{1}}{(2\pi)^{4}}\ G_{r}(\ell_{1})\ \f{1}{(p_{a}.\ell_{1}+i\epsilon)^{3}}\f{1}{p_{b}.\ell_{1}-i\epsilon}\ \Big{\lbrace} \ell_{1\mu}p_{a}.p_{b}-\ell_{1}.p_{a}p_{b\mu}\Big{\rbrace}\non\\
&&\times \Big{\lbrace} \ell_{1}.\ell_{2}p_{b}.p_{c}-p_{c}.\ell_{1}p_{b}.\ell_{2}\Big{\rbrace}\non\\
&=&\ -\f{i}{4}\f{1}{(4\pi)^{2}}\sum_{a=1}^{M+N}\Big{\lbrace}\ln\Big((\omega+i\epsilon\eta_{a})L\Big)\Big{\rbrace}^{2}q_{a}^{3}\ (p_{a}.k)\sum_{\substack{b=1\\ b\neq a\\ \eta_{a}\eta_{b}=1}}^{M+N}\sum_{\substack{c=1\\ c\neq a\\ \eta_{a}\eta_{c}=1}}^{M+N}q_{b}q_{c}\f{1}{[(p_{a}.p_{b})^{2}-p_{a}^{2}p_{b}^{2}]^{5/2}}\non\\
&&\times \f{1}{[(p_{a}.p_{c})^{2}-p_{a}^{2}p_{c}^{2}]^{3/2}}\ p_{b}^{2}p_{c}^{2}p_{a}.p_{b}\ \Big[\lbrace p_{a}.p_{c}p_{a\mu}-p_{a}^{2}p_{c\mu}\rbrace\lbrace (p_{a}.p_{b})^{2}-p_{a}^{2}p_{b}^{2}\rbrace\non\\
&&\ +3\lbrace p_{a}^{2}p_{b\mu}-p_{a}.p_{b}p_{a\mu}\rbrace\lbrace p_{a}.p_{b}p_{a}.p_{c}-p_{a}^{2}p_{b}.p_{c}\rbrace\Big]\non\\
&&\ -\f{i}{4}\ \f{1}{(4\pi)^{2}}\sum_{a=1}^{M+N}\Big{\lbrace}\ln\Big((\omega+i\epsilon\eta_{a})L\Big)\Big{\rbrace}^{2}q_{a}^{2}\ (p_{a}.k)\ \sum_{\substack{b=1\\ b\neq a\\ \eta_{a}\eta_{b}=1}}^{M+N}\sum_{\substack{c=1\\ c\neq b\\ \eta_{b}\eta_{c}=1}}^{M+N}q_{b}^{2}q_{c}\f{1}{[(p_{b}.p_{c})^{2}-p_{b}^{2}p_{c}^{2}]^{3/2}}\non\\
&&\times \f{1}{[(p_{a}.p_{b})^{2}-p_{a}^{2}p_{b}^{2}]^{5/2}}\ (p_{b}^{2})^{2}p_{c}^{2}\Big[ \lbrace (p_{a}.p_{b})^{2}-p_{a}^{2}p_{b}^{2}\rbrace \lbrace p_{a}.p_{b}p_{c\mu}-p_{a}.p_{c}p_{b\mu}\rbrace +\ 3\ \lbrace p_{b}^{2}p_{a}.p_{b}p_{a}.p_{c}p_{a\mu}\non\\
&&\ +p_{a}^{2}p_{b}.p_{c}p_{a}.p_{b}p_{b\mu}-(p_{a}.p_{b})^{2}p_{b}.p_{c}p_{a\mu}-p_{a}^{2}p_{b}^{2}p_{a}.p_{c}p_{b\mu}\rbrace\Big]\label{appAJ24}
\ee
\endgroup

\section{Analysis of the terms containing $\Delta_{(2)}Y_{a}(\sigma)$ in the expression of $\Delta_{(2)}\widehat{T}^{X}(k)$}\label{appB}
Before analyzing the terms in the last line of the expression of $\Delta_{(2)}\widehat{T}^{X}(k)$ in eq.\eqref{sub2leadingTX}, first we need to evaluate the contribution of $\f{d\Delta_{(2)}Y_{a\mu}(\sigma)}{d\sigma}$. The subleading energy-momentum tensor for the motion of b'th particle as well as for the gravitational radiation from b'th particle has the following form as follows from eq.\eqref{subleadingTX} and eq.\eqref{Thath},
\be
\Delta_{(1)}\widehat{T}^{(b)\mu\nu}(\ell_{1})\ &=&\ -8\pi G \ e^{-i\ell_{1}.r_{b}}\ 
\sum_{\substack{c=1\\ c\neq b}}^{M+N}\int\f{d^{4}\ell_{2}}{(2\pi)^{4}}\ e^{i\ell_{2}.(r_{b}-r_{c})}G_{r}(\ell_{2})\ \f{1}{p_{b}.\ell_{2}+i\epsilon}\f{1}{p_{c}.\ell_{2}-i\epsilon}\f{1}{p_{b}.(\ell_{2}-\ell_{1})+i\epsilon}\non\\
&&\ \times \mathcal{E}^{\mu\nu}(p_{b},p_{c},\ell_{1},\ell_{2})\non\\
&&\ -8\pi G\ \sum_{c=1}^{M+N}e^{-i\ell_{1}.r_{c}}\int\f{d^{4}\ell_{2}}{(2\pi)^{4}}\ e^{i\ell_{2}.(r_{c}-r_{b})}G_{r}(\ell_{1}-\ell_{2})G_{r}(\ell_{2})\ \f{1}{p_{b}.\ell_{2}-i\epsilon}\f{1}{p_{c}.(\ell_{1}-\ell_{2})-i\epsilon}\non\\
&&\times \Big{\lbrace}p_{b\alpha}p_{b\beta}-\f{1}{2}p_{b}^{2}\eta_{\alpha\beta}\Big{\rbrace}\mathcal{F}^{\mu\nu,\alpha\beta,\rho\sigma}(\ell_{1},\ell_{2})\Big{\lbrace}p_{c\rho}p_{c\sigma}-\f{1}{2}p_{c}^{2}\eta_{\rho\sigma}\Big{\rbrace}
\ee
where the expression for $\mathcal{F}^{\mu\nu,\alpha\beta,\rho\sigma}$ is given in eq.\eqref{F} and the expression for for $\mathcal{E}^{\mu\nu}$ is given as,
\be
\mathcal{E}^{\mu\nu}(p_{b},p_{c},\ell_{1},\ell_{2})\ &=&\ \f{p_{b}^{\mu}p_{b}^{\nu}}{p_{b}.\ell_{1}-i\epsilon}\Big{\lbrace} 2p_{b}.p_{c}p_{b}.\ell_{2}p_{c}.\ell_{1}-\ell_{1}.\ell_{2}(p_{b}.p_{c})^{2}-p_{c}^{2}p_{b}.\ell_{1}p_{b}.\ell_{2}+\f{1}{2}\ell_{1}.\ell_{2}p_{b}^{2}p_{c}^{2}\Big{\rbrace}\non\\
&&\ -\Big{\lbrace} 2p_{b}.p_{c}\ell_{2}.p_{b}(p_{b}^{\mu}p_{c}^{\nu}+p_{b}^{\nu}p_{c}^{\mu})-(p_{b}.p_{c})^{2}(\ell_{2}^{\mu}p_{b}^{\nu}+\ell_{2}^{\nu}p_{b}^{\mu})-2p_{c}^{2}p_{b}.\ell_{2}p_{b}^{\mu}p_{b}^{\nu}\non\\
&&\ +\f{1}{2}p_{b}^{2}p_{c}^{2}(\ell_{2}^{\mu}p_{b}^{\nu}+\ell_{2}^{\nu}p_{b}^{\mu})\Big{\rbrace}
\ee
Corresponding to this subleading energy-momentum tensor for b'th particle, the subleading metric fluctuation becomes,
\be
\Delta_{(1)}h^{(b)\mu\nu}(x)\ &=&\ -8\pi G\int\f{d^{4}\ell_{1}}{(2\pi)^{4}}\ e^{i\ell_{1}.x}G_{r}(\ell_{1})\ \Big[\Delta_{(1)}\widehat{T}^{(b)\mu\nu}(\ell_{1})-\f{1}{2}\eta^{\mu\nu}\Delta_{(1)}\widehat{T}^{(b)\rho} _{\ \rho}(\ell_{1})\Big]
\ee
The order $\mathcal{O}(G^{2})$ correction to the straight line trajectory of particle-$a$ satisfies the following equation,
\be
\f{d^{2}\Delta_{(2)}Y_{a}^{\mu}(\sigma)}{d\sigma^{2}}\ &=&\ -\Big[\Delta_{(0)}\Gamma^{\mu}_{\nu\rho}\big(r_{a}+v_{a}\sigma+\Delta_{(1)}Y_{a}(\sigma)\big)+\Delta_{(1)}\Gamma^{\mu}_{\nu\rho}(r_{a}+v_{a}\sigma)\Big]\non\\
&&\times \Bigg( v_{a}^{\nu}+\f{d\Delta_{(1)}Y_{a}^{\nu}(\sigma)}{d\sigma}\Bigg)\Bigg(v_{a}^{\rho}+\f{d\Delta_{(1)}Y_{a}^{\rho}(\sigma)}{d\sigma}\Bigg)\Bigg{|}_{\mathcal{O}(G^{2})}
\ee
Now after extracting the order $\mathcal{O}(G^{2})$ contribution from the RHS above equation and using the boundary conditions follows from eq.\eqref{BoundaryGR}:
\be
\Delta_{(2)}Y_{a}^{\mu}(\sigma)\Big{|}_{\sigma=0}\ =\ 0 \hspace{5mm} ,\ \hspace{5mm}\ {d\Delta_{(2)}Y_a^\mu(\sigma)\over d\sigma}\Bigg{|}_{\sigma\rightarrow \infty}\ =\ 0
\ee
we get,
\be
\f{d\Delta_{(2)}Y_{a}^{\mu}(\sigma)}{d\sigma}\ &=&\ \sum_{\substack{b=1\\b\neq a}}^{M+N}\int_{\sigma}^{\infty}d\sigma'\ \Delta_{(0)}\Gamma^{(b)\mu}_{\ \nu\rho}(r_{a}+v_{a}\sigma')\Bigg{\lbrace}v_{a}^{\nu}\f{d\Delta_{(1)}Y_{a}^{\rho}(\sigma')}{d\sigma'}+\f{d\Delta_{(1)}Y_{a}^{\nu}(\sigma')}{d\sigma'}v_{a}^{\rho}\Bigg{\rbrace}\non\\
&&\ -8\pi G\ \sum_{\substack{b=1\\ b\neq a}}^{M+N}\int\f{d^{4}\ell_{1}}{(2\pi)^{4}}\ e^{i\ell_{1}.(r_{a}-r_{b})}G_{r}(\ell_{1})\f{1}{p_{b}.\ell_{1} -i\epsilon}\Big[2\ell_{1} .v_{a}p_{b}.v_{a}p_{b}^{\mu}-\ell_{1}^{\mu}(p_{b}.v_{a})^{2}-p_{b}^{2}\ell_{1}.v_{a}v_{a}^{\mu}\non\\
&&\ +\f{1}{2}p_{b}^{2}v_{a}^{2}\ell_{1}^{\mu}\Big]\times \int_{\sigma}^{\infty}d\sigma' e^{i\ell_{1} .v_{a}\sigma'}\Big{\lbrace} i\ell_{1} \cdot\Delta_{(1)}Y_{a}(\sigma')\Big{\rbrace}\non\\
&&\ +\int_{\sigma}^{\infty}d\sigma'\ \Delta_{(1)}\Gamma^{\mu}_{\nu\rho}(r_{a}+v_{a}\sigma')v_{a}^{\nu}v_{a}^{\rho}
\ee
where
\begingroup
\allowdisplaybreaks
\be
\Delta_{(1)}\Gamma^{\mu}_{\nu\rho}(r_{a}+v_{a}\sigma')&=&\ i(8\pi G)^{2}\sum_{\substack{b=1\\ b\neq a}}^{M+N}\sum_{\substack{c=1\\ c\neq b}}^{M+N}\int \f{d^{4}\ell_{1}}{(2\pi)^{4}}\ e^{i\ell_{1}.(r_{a}-r_{b})}e^{i\ell_{1}.v_{a}\sigma'}G_{r}(\ell_{1})\int\f{d^{4}\ell_{2}}{(2\pi)^{4}}\ e^{-i\ell_{2}.(r_{c}-r_{b})}\non\\
&&\ G_{r}(\ell_{2})\f{1}{p_{b}.\ell_{2}+i\epsilon}\f{1}{p_{b}.(\ell_{2}-\ell_{1})+i\epsilon}\f{1}{p_{c}.\ell_{2}-i\epsilon}\Big[\ell_{1\nu}\lbrace \mathcal{E}_{\rho}^{\ \mu}(p_{b},p_{c},\ell_{1},\ell_{2})\non\\
&&\ -\f{1}{2}\delta_{\rho}^{\mu}\mathcal{E}_{\alpha}^{\ \alpha}(p_{b},p_{c},\ell_{1},\ell_{2})\rbrace +\ell_{1\rho}\lbrace \mathcal{E}_{\nu}^{\ \mu}(p_{b},p_{c},\ell_{1},\ell_{2})-\f{1}{2}\delta_{\nu}^{\mu}\mathcal{E}_{\alpha}^{\ \alpha}(p_{b},p_{c},\ell_{1},\ell_{2})\rbrace\non\\
&&\ -\ell_{1}^{\mu}\lbrace \mathcal{E}_{\nu\rho}(p_{b},p_{c},\ell_{1},\ell_{2})\ -\f{1}{2}\eta_{\nu\rho}\mathcal{E}_{\alpha}^{\ \alpha}(p_{b},p_{c},\ell_{1},\ell_{2})\rbrace \Big]\non\\
&&\ +\ i(8\pi G)^{2}\sum_{b=1}^{M+N}\sum_{c=1}^{M+N}\int\f{d^{4}\ell_{1}}{(2\pi)^{4}}\ e^{i\ell_{1}.(r_{a}-r_{c})}e^{i\ell_{1}.v_{a}\sigma'}G_{r}(\ell_{1})\int\f{d^{4}\ell_{2}}{(2\pi)^{4}}\ e^{i\ell_{2}.(r_{c}-r_{b})}\non\\
&&\ \times G_{r}(\ell_{1}-\ell_{2})G_{r}(\ell_{2})\ \f{1}{p_{b}.\ell_{2}-i\epsilon}\f{1}{p_{c}.(\ell_{1}-\ell_{2})-i\epsilon}\Big{\lbrace}p_{b\alpha}p_{b\beta}-\f{1}{2}p_{b}^{2}\eta_{\alpha\beta}\Big{\rbrace}\non\\
&&\ \Big{\lbrace} p_{c\gamma}p_{c\sigma}-\f{1}{2}p_{c}^{2}\eta_{\gamma\sigma}\Big{\rbrace}\Big[\ell_{1\nu}\Big(\mathcal{F}_{\rho}^{\ \mu,\alpha\beta,\gamma\sigma}(\ell_{1},\ell_{2})-\f{1}{2}\delta_{\rho}^{\mu}\mathcal{F}_{\delta}^{\ \delta,\alpha\beta,\gamma\sigma}(\ell_{1},\ell_{2})\Big)\non\\
&&\ +\ell_{1\rho}\Big(\mathcal{F}_{\nu}^{\ \mu,\alpha\beta,\gamma\sigma}(\ell_{1},\ell_{2})-\f{1}{2}\delta_{\nu}^{\mu}\mathcal{F}_{\delta}^{\ \delta,\alpha\beta,\gamma\sigma}(\ell_{1},\ell_{2})\Big)-\ell_{1}^{\mu}\Big(\mathcal{F}_{\nu\rho }\ ^{,\alpha\beta,\gamma\sigma}(\ell_{1},\ell_{2})\non\\
&&\ -\f{1}{2}\eta_{\nu\rho}\mathcal{F}_{\delta}^{\ \delta,\alpha\beta,\gamma\sigma}(\ell_{1},\ell_{2})\Big)\Big]
\ee
\endgroup
The order $\mathcal{O}\big(\omega(\ln\omega)^{2}\big)$ contribution from the fifth term of  eq.\eqref{sub2leadingTX} in the integration region $L^{-1}>>|\ell_{2}^{\mu}|>>|\ell_{1}^{\mu}|>>\omega$ turns out,
\begingroup
\allowdisplaybreaks
\be
&&\Delta_{(2)}^{(5)}\widehat{T}^{X\mu\nu}(k)\non\\ &=&\ \sum_{a=1}^{M+N}m_{a}e^{-ik.r_{a}}\int_{0}^{\infty}d\sigma\ e^{-i(k.v_{a}-i\epsilon)\sigma}\ \Big{\lbrace} -ik\cdot \Delta_{(2)}Y_{a}(\sigma)\Big{\rbrace}v_{a}^{\mu}v_{a}^{\nu}\non\\
&\simeq &\ -\sum_{a=1}^{M+N}\f{p_{a}^{\mu}p_{a}^{\nu}}{p_{a}.k}\ k_{\alpha}\Bigg[ i(8\pi G)^{2} (p_{a}.k)\ \sum_{\substack{b=1\\ b\neq a}}^{M+N}\sum_{\substack{c=1\\ c\neq a}}^{M+N}\int_{\omega}^{L^{-1}}\f{d^{4}\ell_{2}}{(2\pi)^{4}}G_{r}(\ell_{2})\ \f{1}{(p_{a}.\ell_{2}+i\epsilon)^{2}} \f{1}{p_{c}.\ell_{2}-i\epsilon}\non\\
&& \Big[2p_{a}.\ell_{2}\Big{\lbrace}p_{a}.p_{c}p_{c}^{\beta}-\f{1}{2}p_{c}^{2}p_{a}^{\beta}\Big{\rbrace}-\ell_{2}^{\beta}\Big{\lbrace}(p_{a}.p_{c})^{2}-\f{1}{2}p_{a}^{2}p_{c}^{2}\Big{\rbrace}\Big]\int_{\omega}^{|\vec{\ell}_{2}|}\f{d^{4}\ell_{1}}{(2\pi)^{4}}\ G_{r}(\ell_{1})\non\\
&&\times\f{1}{(p_{a}.\ell_{1}+i\epsilon)^{3}}\f{1}{p_{b}.\ell_{1}-i\epsilon}\ \ell_{1\beta}\ \Big[2p_{a}.\ell_{1}\Big{\lbrace} p_{a}.p_{b}p_{b}^{\alpha}-\f{1}{2}p_{b}^{2}p_{a}^{\alpha}\Big{\rbrace} -\ell_{1}^{\alpha}\Big{\lbrace}(p_{a}.p_{b})^{2}-\f{1}{2}p_{a}^{2}p_{b}^{2}\Big{\rbrace}\Big]\non\\
&&\ -i(8\pi G)^{2}(p_{a}.k)\sum_{\substack{b=1\\ b\neq a}}^{M+N}\sum_{\substack{c=1\\ c\neq b}}^{M+N}\int_{\omega}^{L^{-1}}\f{d^{4}\ell_{2}}{(2\pi)^{4}}G_{r}(\ell_{2})\ \f{1}{(p_{b}.\ell_{2}+i\epsilon)^{2}} \f{1}{p_{c}.\ell_{2}-i\epsilon}\non\\
&& \Big[2p_{b}.\ell_{2}\Big{\lbrace}p_{b}.p_{c}p_{c}^{\beta}-\f{1}{2}p_{c}^{2}p_{b}^{\beta}\Big{\rbrace}-\ell_{2}^{\beta}\Big{\lbrace}(p_{b}.p_{c})^{2}-\f{1}{2}p_{b}^{2}p_{c}^{2}\Big{\rbrace}\Big]\int_{\omega}^{|\vec{\ell}_{2}|}\f{d^{4}\ell_{1}}{(2\pi)^{4}}\ G_{r}(\ell_{1})\non\\
&&\times\f{1}{(p_{a}.\ell_{1}+i\epsilon)^{3}}\f{1}{p_{b}.\ell_{1}-i\epsilon}\ \ell_{1\beta}\ \Big[2p_{a}.\ell_{1}\Big{\lbrace} p_{a}.p_{b}p_{b}^{\alpha}-\f{1}{2}p_{b}^{2}p_{a}^{\alpha}\Big{\rbrace} -\ell_{1}^{\alpha}\Big{\lbrace}(p_{a}.p_{b})^{2}-\f{1}{2}p_{a}^{2}p_{b}^{2}\Big{\rbrace}\Big]\Bigg]
\ee
\endgroup
After evaluation of the two momenta integrals we find,
\begingroup
\allowdisplaybreaks
\be
&&\Delta_{(2)}^{(5)}\widehat{T}^{X\mu\nu}(k)\non\\
&\simeq &\ -iG^{2}\ \sum_{a=1}^{M+N}\Big{\lbrace}\ln\Big((\omega+i\epsilon\eta_{a})L\Big)\Big{\rbrace}^{2}p_{a}^{\mu}p_{a}^{\nu}k_{\alpha}\ \Bigg[\sum_{\substack{b=1\\ b\neq a\\ \eta_{a}\eta_{b}=1}}^{M+N}\sum_{\substack{c=1\\ c\neq a\\ \eta_{a}\eta_{c}=1}}^{M+N}  \ \f{1}{[(p_{a}.p_{c})^{2}-p_{a}^{2}p_{c}^{2}]^{3/2}} \f{1}{[(p_{a}.p_{b})^{2}-p_{a}^{2}p_{b}^{2}]^{5/2}}\non\\
&& \Big{\lbrace}(p_{a}.p_{c})^{3}p_{c\beta}-\f{3}{2}p_{a}.p_{c}p_{a}^{2}p_{c}^{2}p_{c\beta}+\f{1}{2}p_{a}^{2}(p_{c}^{2})^{2}p_{a\beta}\Big{\rbrace} \Bigg{\lbrace}\lbrace(p_{a}.p_{b})^{2}-p_{a}^{2}p_{b}^{2}\rbrace\Big[2p_{b}^{2}p_{a}.p_{b}p_{b}^{\alpha}p_{a}^{\beta}-2(p_{a}.p_{b})^{2}p_{b}^{\alpha}p_{b}^{\beta}\non\\
&&\ -(p_{b}^{2})^{2}p_{a}^{\alpha}p_{a}^{\beta}+p_{b}^{2}p_{a}.p_{b}p_{a}^{\alpha}p_{b}^{\beta} -\Big{\lbrace}(p_{a}.p_{b})^{2}-\f{1}{2}p_{a}^{2}p_{b}^{2}\Big{\rbrace}\lbrace p_{b}^{2}\eta^{\alpha\beta}-p_{b}^{\alpha}p_{b}^{\beta}\rbrace\Big]\non\\
&&\ +3\lbrace p_{a}.p_{b}p_{b}^{\beta}-p_{b}^{2}p_{a}^{\beta}\rbrace\Big[(p_{a}.p_{b})^{3}p_{b}^{\alpha}-\f{3}{2}p_{a}.p_{b}p_{a}^{2}p_{b}^{2}p_{b}^{\alpha}\ +\f{1}{2}p_{a}^{2}(p_{b}^{2})^{2}p_{a}^{\alpha}\Big]\Bigg{\rbrace}\non\\
&& -\ \sum_{\substack{b=1\\ b\neq a\\ \eta_{a}\eta_{b}=1}}^{M+N}\sum_{\substack{c=1\\ c\neq b\\ \eta_{b}\eta_{c}=1}}^{M+N}  \ \f{1}{[(p_{b}.p_{c})^{2}-p_{b}^{2}p_{c}^{2}]^{3/2}}\ \f{1}{[(p_{a}.p_{b})^{2}-p_{a}^{2}p_{b}^{2}]^{5/2}} \Big{\lbrace}(p_{b}.p_{c})^{3}p_{c\beta}-\f{3}{2}p_{b}.p_{c}p_{b}^{2}p_{c}^{2}p_{c\beta}+\f{1}{2}p_{b}^{2}(p_{c}^{2})^{2}p_{b\beta}\Big{\rbrace} \non\\
&&\times \Bigg{\lbrace}\lbrace(p_{a}.p_{b})^{2}-p_{a}^{2}p_{b}^{2}\rbrace\Big[2p_{b}^{2}p_{a}.p_{b}p_{b}^{\alpha}p_{a}^{\beta}-2(p_{a}.p_{b})^{2}p_{b}^{\alpha}p_{b}^{\beta}-(p_{b}^{2})^{2}p_{a}^{\alpha}p_{a}^{\beta}+p_{b}^{2}p_{a}.p_{b}p_{a}^{\alpha}p_{b}^{\beta} -\Big{\lbrace}(p_{a}.p_{b})^{2}-\f{1}{2}p_{a}^{2}p_{b}^{2}\Big{\rbrace}\non\\
&&\times\lbrace p_{b}^{2}\eta^{\alpha\beta}-p_{b}^{\alpha}p_{b}^{\beta}\rbrace\Big] +3\lbrace p_{a}.p_{b}p_{b}^{\beta}-p_{b}^{2}p_{a}^{\beta}\rbrace\Big[(p_{a}.p_{b})^{3}p_{b}^{\alpha}-\f{3}{2}p_{a}.p_{b}p_{a}^{2}p_{b}^{2}p_{b}^{\alpha} +\f{1}{2}p_{a}^{2}(p_{b}^{2})^{2}p_{a}^{\alpha}\Big]\Bigg{\rbrace} \Bigg]\label{TX25}
\ee
\endgroup
The order $\mathcal{O}\big(\omega(\ln\omega)^{2}\big)$ contribution from the sixth term of  eq.\eqref{sub2leadingTX} turns out,
\begingroup
\allowdisplaybreaks
\be
&&\Delta_{(2)}^{(6)}\widehat{T}^{X\mu\nu}(k)\non\\
 &=&\ \sum_{a=1}^{M+N}m_{a}e^{-ik.r_{a}}\int_{0}^{\infty}d\sigma\ e^{-i(k.v_{a}-i\epsilon)\sigma}\ v_{a}^{\mu}\f{d\Delta_{(2)}Y_{a}^{\nu}(\sigma)}{d\sigma}\non\\
&\simeq &\ i(8\pi G)^{2}\sum_{a=1}^{M+N}p_{a}^{\mu} (p_{a}.k)\ \sum_{\substack{b=1\\ b\neq a}}^{M+N}\sum_{\substack{c=1\\ c\neq a}}^{M+N}\int_{\omega}^{L^{-1}}\f{d^{4}\ell_{2}}{(2\pi)^{4}}G_{r}(\ell_{2})\ \f{1}{(p_{a}.\ell_{2}+i\epsilon)^{2}} \f{1}{p_{c}.\ell_{2}-i\epsilon}\non\\
&& \Big[2p_{a}.\ell_{2}\Big{\lbrace}p_{a}.p_{c}p_{c}^{\beta}-\f{1}{2}p_{c}^{2}p_{a}^{\beta}\Big{\rbrace}-\ell_{2}^{\beta}\Big{\lbrace}(p_{a}.p_{c})^{2}-\f{1}{2}p_{a}^{2}p_{c}^{2}\Big{\rbrace}\Big]\int_{\omega}^{|\vec{\ell}_{2}|}\f{d^{4}\ell_{1}}{(2\pi)^{4}}\ G_{r}(\ell_{1})\non\\
&&\times\f{1}{(p_{a}.\ell_{1}+i\epsilon)^{3}}\f{1}{p_{b}.\ell_{1}-i\epsilon}\ \ell_{1\beta}\ \Big[2p_{a}.\ell_{1}\Big{\lbrace} p_{a}.p_{b}p_{b}^{\nu}-\f{1}{2}p_{b}^{2}p_{a}^{\nu}\Big{\rbrace} -\ell_{1}^{\nu}\Big{\lbrace}(p_{a}.p_{b})^{2}-\f{1}{2}p_{a}^{2}p_{b}^{2}\Big{\rbrace}\Big]\non\\
&&\ -i(8\pi G)^{2}\sum_{a=1}^{M+N}(p_{a}.k)\ p_{a}^{\mu}\ \sum_{\substack{b=1\\ b\neq a}}^{M+N}\sum_{\substack{c=1\\ c\neq b}}^{M+N}\int_{\omega}^{L^{-1}}\f{d^{4}\ell_{2}}{(2\pi)^{4}}G_{r}(\ell_{2})\ \f{1}{(p_{b}.\ell_{2}+i\epsilon)^{2}} \f{1}{p_{c}.\ell_{2}-i\epsilon}\non\\
&&\ \Big[2p_{b}.\ell_{2}\Big{\lbrace}p_{b}.p_{c}p_{c}^{\beta}-\f{1}{2}p_{c}^{2}p_{b}^{\beta}\Big{\rbrace}-\ell_{2}^{\beta}\Big{\lbrace}(p_{b}.p_{c})^{2}-\f{1}{2}p_{b}^{2}p_{c}^{2}\Big{\rbrace}\Big]\int_{\omega}^{|\vec{\ell}_{2}|}\f{d^{4}\ell_{1}}{(2\pi)^{4}}\ G_{r}(\ell_{1})\non\\
&&\times\f{1}{(p_{a}.\ell_{1}+i\epsilon)^{3}}\f{1}{p_{b}.\ell_{1}-i\epsilon}\ \ell_{1\beta}\ \Big[2p_{a}.\ell_{1}\Big{\lbrace} p_{a}.p_{b}p_{b}^{\nu}-\f{1}{2}p_{b}^{2}p_{a}^{\nu}\Big{\rbrace} -\ell_{1}^{\nu}\Big{\lbrace}(p_{a}.p_{b})^{2}-\f{1}{2}p_{a}^{2}p_{b}^{2}\Big{\rbrace}\Big]
\ee
\endgroup
After evaluation of the two momenta integrals analogous to $\Delta_{(2)}^{(5)}\widehat{T}^{X\mu\nu}(k)$ we find,
\begingroup
\allowdisplaybreaks
\be
&&\Delta_{(2)}^{(6)}\widehat{T}^{X\mu\nu}(k)\non\\
&\simeq &\ iG^{2}\ \sum_{a=1}^{M+N}\Big{\lbrace}\ln\Big((\omega+i\epsilon\eta_{a})L\Big)\Big{\rbrace}^{2} (p_{a}.k)p_{a}^{\mu}\ \Bigg[ \sum_{\substack{b=1\\ b\neq a\\ \eta_{a}\eta_{b}=1}}^{M+N}\sum_{\substack{c=1\\ c\neq a\\ \eta_{a}\eta_{c}=1}}^{M+N}  \ \f{1}{[(p_{a}.p_{c})^{2}-p_{a}^{2}p_{c}^{2}]^{3/2}}\ \f{1}{[(p_{a}.p_{b})^{2}-p_{a}^{2}p_{b}^{2}]^{5/2}}\non\\
&& \Big{\lbrace}(p_{a}.p_{c})^{3}p_{c\beta}-\f{3}{2}p_{a}.p_{c}p_{a}^{2}p_{c}^{2}p_{c\beta}+\f{1}{2}p_{a}^{2}(p_{c}^{2})^{2}p_{a\beta}\Big{\rbrace}  \Bigg{\lbrace}\lbrace(p_{a}.p_{b})^{2}-p_{a}^{2}p_{b}^{2}\rbrace\Big[2p_{b}^{2}p_{a}.p_{b}p_{b}^{\nu}p_{a}^{\beta}-2(p_{a}.p_{b})^{2}p_{b}^{\nu}p_{b}^{\beta}\non\\
&&\ -(p_{b}^{2})^{2}p_{a}^{\nu}p_{a}^{\beta}+p_{b}^{2}p_{a}.p_{b}p_{a}^{\nu}p_{b}^{\beta} -\Big{\lbrace}(p_{a}.p_{b})^{2}-\f{1}{2}p_{a}^{2}p_{b}^{2}\Big{\rbrace}\lbrace p_{b}^{2}\eta^{\nu\beta}-p_{b}^{\nu}p_{b}^{\beta}\rbrace\Big] +3\lbrace p_{a}.p_{b}p_{b}^{\beta}-p_{b}^{2}p_{a}^{\beta}\rbrace\non\\
&&\times\Big[(p_{a}.p_{b})^{3}p_{b}^{\nu}-\f{3}{2}p_{a}.p_{b}p_{a}^{2}p_{b}^{2}p_{b}^{\nu}\ +\f{1}{2}p_{a}^{2}(p_{b}^{2})^{2}p_{a}^{\nu}\Big]\Bigg{\rbrace}\non\\
&&\ -\sum_{\substack{b=1\\ b\neq a\\ \eta_{a}\eta_{b}=1}}^{M+N}\sum_{\substack{c=1\\ c\neq b\\ \eta_{b}\eta_{c}=1}}^{M+N} \ \f{1}{[(p_{b}.p_{c})^{2}-p_{b}^{2}p_{c}^{2}]^{3/2}}\ \f{1}{[(p_{a}.p_{b})^{2}-p_{a}^{2}p_{b}^{2}]^{5/2}}\ \Big{\lbrace}(p_{b}.p_{c})^{3}p_{c\beta}-\f{3}{2}p_{b}.p_{c}p_{b}^{2}p_{c}^{2}p_{c\beta}+\f{1}{2}p_{b}^{2}(p_{c}^{2})^{2}p_{b\beta}\Big{\rbrace} \non\\
&&\times \Bigg{\lbrace}\lbrace(p_{a}.p_{b})^{2}-p_{a}^{2}p_{b}^{2}\rbrace\Big[2p_{b}^{2}p_{a}.p_{b}p_{b}^{\nu}p_{a}^{\beta}-2(p_{a}.p_{b})^{2}p_{b}^{\nu}p_{b}^{\beta}-(p_{b}^{2})^{2}p_{a}^{\nu}p_{a}^{\beta}+p_{b}^{2}p_{a}.p_{b}p_{a}^{\nu}p_{b}^{\beta} -\Big{\lbrace}(p_{a}.p_{b})^{2}-\f{1}{2}p_{a}^{2}p_{b}^{2}\Big{\rbrace}\non\\
&&\times\lbrace p_{b}^{2}\eta^{\nu\beta}-p_{b}^{\nu}p_{b}^{\beta}\rbrace\Big] +3\lbrace p_{a}.p_{b}p_{b}^{\beta}-p_{b}^{2}p_{a}^{\beta}\rbrace\Big[(p_{a}.p_{b})^{3}p_{b}^{\nu}-\f{3}{2}p_{a}.p_{b}p_{a}^{2}p_{b}^{2}p_{b}^{\nu} +\f{1}{2}p_{a}^{2}(p_{b}^{2})^{2}p_{a}^{\nu}\Big]\Bigg{\rbrace}  \Bigg]\label{TX26}
\ee
\endgroup
The order $\mathcal{O}\big(\omega(\ln\omega)^{2}\big)$ contribution from the seventh term of eq.\eqref{sub2leadingTX} we can directly get by interchanging $\mu\leftrightarrow\nu$ in the expression\eqref{TX26},
\begingroup
\allowdisplaybreaks
\be
&&\Delta_{(2)}^{(7)}\widehat{T}^{X\mu\nu}(k)\non\\
 &=&\ \sum_{a=1}^{M+N}m_{a}e^{-ik.r_{a}}\int_{0}^{\infty}d\sigma\ e^{-i(k.v_{a}-i\epsilon)\sigma}\ v_{a}^{\nu}\f{d\Delta_{(2)}Y_{a}^{\mu}(\sigma)}{d\sigma}\non\\
&\simeq &\ iG^{2}\ \sum_{a=1}^{M+N}\Big{\lbrace}\ln\Big((\omega+i\epsilon\eta_{a})L\Big)\Big{\rbrace}^{2} (p_{a}.k)p_{a}^{\nu}\ \Bigg[ \sum_{\substack{b=1\\ b\neq a\\ \eta_{a}\eta_{b}=1}}^{M+N}\sum_{\substack{c=1\\ c\neq a\\ \eta_{a}\eta_{c}=1}}^{M+N}  \ \f{1}{[(p_{a}.p_{c})^{2}-p_{a}^{2}p_{c}^{2}]^{3/2}} \f{1}{[(p_{a}.p_{b})^{2}-p_{a}^{2}p_{b}^{2}]^{5/2}}\non\\
&&\times \Big{\lbrace}(p_{a}.p_{c})^{3}p_{c\beta}-\f{3}{2}p_{a}.p_{c}p_{a}^{2}p_{c}^{2}p_{c\beta}+\f{1}{2}p_{a}^{2}(p_{c}^{2})^{2}p_{a\beta}\Big{\rbrace} \Bigg{\lbrace}\lbrace(p_{a}.p_{b})^{2}-p_{a}^{2}p_{b}^{2}\rbrace\Big[2p_{b}^{2}p_{a}.p_{b}p_{b}^{\mu}p_{a}^{\beta}-2(p_{a}.p_{b})^{2}p_{b}^{\mu}p_{b}^{\beta}\non\\
&&\ -(p_{b}^{2})^{2}p_{a}^{\mu}p_{a}^{\beta}+p_{b}^{2}p_{a}.p_{b}p_{a}^{\mu}p_{b}^{\beta} -\Big{\lbrace}(p_{a}.p_{b})^{2}-\f{1}{2}p_{a}^{2}p_{b}^{2}\Big{\rbrace}\lbrace p_{b}^{2}\eta^{\mu\beta}-p_{b}^{\mu}p_{b}^{\beta}\rbrace\Big] +3\lbrace p_{a}.p_{b}p_{b}^{\beta}-p_{b}^{2}p_{a}^{\beta}\rbrace\non\\
&&\times\Big[(p_{a}.p_{b})^{3}p_{b}^{\mu}-\f{3}{2}p_{a}.p_{b}p_{a}^{2}p_{b}^{2}p_{b}^{\mu} +\f{1}{2}p_{a}^{2}(p_{b}^{2})^{2}p_{a}^{\mu}\Big]\Bigg{\rbrace}\non\\
&&\ -\sum_{\substack{b=1\\ b\neq a\\ \eta_{a}\eta_{b}=1}}^{M+N}\sum_{\substack{c=1\\ c\neq b\\ \eta_{b}\eta_{c}=1}}^{M+N} \ \f{1}{[(p_{b}.p_{c})^{2}-p_{b}^{2}p_{c}^{2}]^{3/2}}\ \f{1}{[(p_{a}.p_{b})^{2}-p_{a}^{2}p_{b}^{2}]^{5/2}} \Big{\lbrace}(p_{b}.p_{c})^{3}p_{c\beta}-\f{3}{2}p_{b}.p_{c}p_{b}^{2}p_{c}^{2}p_{c\beta}+\f{1}{2}p_{b}^{2}(p_{c}^{2})^{2}p_{b\beta}\Big{\rbrace} \non\\
&&\times \Bigg{\lbrace}\lbrace(p_{a}.p_{b})^{2}-p_{a}^{2}p_{b}^{2}\rbrace\Big[2p_{b}^{2}p_{a}.p_{b}p_{b}^{\mu}p_{a}^{\beta}-2(p_{a}.p_{b})^{2}p_{b}^{\mu}p_{b}^{\beta}-(p_{b}^{2})^{2}p_{a}^{\mu}p_{a}^{\beta}+p_{b}^{2}p_{a}.p_{b}p_{a}^{\mu}p_{b}^{\beta} -\Big{\lbrace}(p_{a}.p_{b})^{2}-\f{1}{2}p_{a}^{2}p_{b}^{2}\Big{\rbrace}\non\\
&&\times\lbrace p_{b}^{2}\eta^{\mu\beta}-p_{b}^{\mu}p_{b}^{\beta}\rbrace\Big] +3\lbrace p_{a}.p_{b}p_{b}^{\beta}-p_{b}^{2}p_{a}^{\beta}\rbrace\Big[(p_{a}.p_{b})^{3}p_{b}^{\mu}-\f{3}{2}p_{a}.p_{b}p_{a}^{2}p_{b}^{2}p_{b}^{\mu} +\f{1}{2}p_{a}^{2}(p_{b}^{2})^{2}p_{a}^{\mu}\Big]\Bigg{\rbrace}  \Bigg]\label{TX27}
\ee
\endgroup

\section{Fourier transforms for deriving early and late time waveforms}
Here we are interested to determine the Fourier transformation of following three functions,
\be
\widetilde{F}_{n,0}(\omega ,\vec{x})\ &=&\ Ce^{i\omega \phi}\ f(\omega)\ \omega^{n-1}\Big{\lbrace}\ln(\omega +i\epsilon)\Big{\rbrace}^{n}\\
\widetilde{F}_{0,n}(\omega ,\vec{x})\ &=&\ Ce^{i\omega \phi}\ f(\omega)\ \omega^{n-1}\Big{\lbrace}\ln(\omega -i\epsilon)\Big{\rbrace}^{n}\\
\widetilde{F}_{n-r,r}(\omega ,\vec{x})\ &=&\ Ce^{i\omega \phi}\ f(\omega)\ \omega^{n-1}\Big{\lbrace}\ln(\omega +i\epsilon)\Big{\rbrace}^{n-r}\ \Big{\lbrace}\ln(\omega -i\epsilon)\Big{\rbrace}^{r}
\ee
where $C$ and $\phi$ are functions of $\vec{x}$ and $f(\omega)$ is a function of $\omega$ which is smooth at $\omega=0$ with $f(0)=1$ and falls off sufficiently as $\omega\rightarrow\pm\infty$ for the convergence of the Fourier transforms. Now the Fourier transformations are given by,
\be
F_{n,0}(t,\vec{x})\ &=&\ C\int_{-\infty}^{\infty}\f{d\omega}{2\pi}\ e^{-i\omega u}\ f(\omega)\ \omega^{n-1}\Big{\lbrace}\ln(\omega +i\epsilon)\Big{\rbrace}^{n}\label{Fn0}\\
F_{0,n}(t,\vec{x})\ &=&\ C\int_{-\infty}^{\infty}\f{d\omega}{2\pi}\ e^{-i\omega u}\ f(\omega)\ \omega^{n-1}\Big{\lbrace}\ln(\omega -i\epsilon)\Big{\rbrace}^{n}\label{F0n}\\
F_{n-r,r}(t,\vec{x})\ &=&\ C\int_{-\infty}^{\infty}\f{d\omega}{2\pi}\ e^{-i\omega u}\ f(\omega)\ \omega^{n-1}\Big{\lbrace}\ln(\omega +i\epsilon)\Big{\rbrace}^{n-r}\ \Big{\lbrace}\ln(\omega -i\epsilon)\Big{\rbrace}^{r}\label{Fn-rr}
\ee
where $u\equiv t-\phi$. For $u>0$ from the convergence of the Fourier integrals we have to close the contour in the lower half plane of complex $\omega$. Similarly for $u<0$ we have to to close the contour in the upper half plane of complex $\omega$. Now for standard principle value of complex logarithms $\ln(\omega\pm i\epsilon)$, the branch cut singularities start at $\omega=\mp i\epsilon$ and extend to $\omega=-\infty \mp i\epsilon$. This implies  $F_{n,0}(t,\vec{x})=0$ for $u<0$ and $F_{0,n}(t,\vec{x})=0$ for $u>0$. Now using the discontinuity of principal valued logarithm, defined as $\ln z =\ln|z| +iArg(z)$ for $-\pi<Arg(z)\leq \pi$, we get $\ln(\omega+i\epsilon)=\ln(\omega-i\epsilon)+2\pi i H(-\omega)$, where $H$ is the Heaviside step function. Substituting $\ln(\omega+i\epsilon)=\ln(\omega-i\epsilon)+2\pi i H(-\omega)$ in eq.\eqref{Fn0} we get,
\be
F_{n,0}(t,\vec{x})\ &=&\ F_{0,n}(t,\vec{x})\ +\ C\int_{-\infty}^{\infty}\f{d\omega}{2\pi}\ e^{-i\omega u} f(\omega)\ \omega^{n-1}\Bigg[ n\Big{\lbrace}\ln(\omega-i\epsilon)\Big{\rbrace}^{n-1}\ 2\pi iH(-\omega)\non\\
&& +\ \f{n(n-1)}{2}\Big{\lbrace}\ln(\omega-i\epsilon)\Big{\rbrace}^{n-2}\ (2\pi i)^{2}H(-\omega)+\cdots +(2\pi i)^{n}H(-\omega)\Bigg]\non\\
\ee
The most singular term in the integrand contributes,
\be
F_{n,0}(t,\vec{x})\ -\ F_{0,n}(t,\vec{x})\ &\simeq &\  inC\ \Bigg(i\f{d}{du}\Bigg)^{n-1}\ \int_{-\infty}^{0}d\omega\ e^{-i\omega u} f(\omega)\ \Big{\lbrace}\ln(\omega-i\epsilon)\Big{\rbrace}^{n-1}\ 
\ee
Now substituting $v=\omega u$ in the above integral we get,
\be
F_{n,0}(t,\vec{x})\ -\ F_{0,n}(t,\vec{x}) &\simeq &  inC\ \Bigg(i\f{d}{du}\Bigg)^{n-1}\ \int_{-\infty\times sgn(u)}^{0}\f{dv}{u}\ e^{-iv}\ f\Big(\f{v}{u}\Big)\ \Big{\lbrace}\ln(v-i\epsilon\times sgn(u))-\ln u\Big{\rbrace}^{n-1}\non\\
&=& -C n!\ i^{n-1}\ \f{(\ln |u|)^{n-1}}{u^{n}}\ +\ \mathcal{O}\Big(\f{(\ln |u|)^{n-2}}{u^{n}}\Big)
\ee
This gives,
\be
F_{n,0}(t,\vec{x})\ &=&\ \begin{cases} -C n!\ i^{n-1}\ \f{(\ln |u|)^{n-1}}{u^{n}} \hspace{1cm}\hbox{for\ }\ u\rightarrow +\infty\\
\\
0\hspace{2cm}\hbox{for\ }\ u\rightarrow -\infty
\end{cases}\label{FourierFn0}\\
F_{0,n}(t,\vec{x})\ &=&\ \begin{cases} 0\hspace{2cm}\hbox{for\ }\  u\rightarrow +\infty\\
\\
+C n!\ i^{n-1}\ \f{(\ln |u|)^{n-1}}{u^{n}} \hspace{1cm}\hbox{for\ }\  u\rightarrow -\infty\label{FourierF0n}
\end{cases}
\ee
Now in the expression \eqref{Fn-rr} substituting $\ln(\omega+i\epsilon)=\ln(\omega-i\epsilon)+2\pi i H(-\omega)$ and keeping the terms up to subleading singular pieces of the integrand we get,
\be
F_{n-r,r}(t,\vec{x})\ &\simeq &\ F_{0,n}(t,\vec{x})\ +\ iC(n-r)\ \int_{-\infty}^{0}d\omega\ e^{-i\omega u}\ f(\omega)\omega^{n-1}\ \Big{\lbrace}\ln(\omega-i\epsilon)\Big{\rbrace}^{n-1}
\ee
Similarly, in the expression \eqref{Fn-rr} substituting $\ln(\omega-i\epsilon)=\ln(\omega+i\epsilon)-2\pi i H(-\omega)$ and keeping the terms up to subleading singular pieces of the integrand we get,
\be
F_{n-r,r}(t,\vec{x})\ &\simeq &\ F_{n,0}(t,\vec{x})\ -\ iCr\ \int_{-\infty}^{0}d\omega\ e^{-i\omega u}\ f(\omega)\omega^{n-1}\ \Big{\lbrace}\ln(\omega+i\epsilon)\Big{\rbrace}^{n-1}
\ee
Then from the last two expressions we find the following relation,
\be
F_{n,0}(t,\vec{x})+F_{0,n}(t,\vec{x})-2F_{n-r,r}(t,\vec{x})\ &\simeq &  iCr\ \int_{-\infty}^{0}d\omega\ e^{-i\omega u}\ f(\omega)\omega^{n-1}\ \Big{\lbrace}\ln(\omega+i\epsilon)\Big{\rbrace}^{n-1}\non\\
&&\ -\  iC(n-r)\ \int_{-\infty}^{0}d\omega\ e^{-i\omega u}\ f(\omega)\omega^{n-1}\ \Big{\lbrace}\ln(\omega-i\epsilon)\Big{\rbrace}^{n-1}\non\\
&\simeq &\ iC(2r-n)\ \ \int_{-\infty}^{0}d\omega\ e^{-i\omega u}\ f(\omega)\omega^{n-1}\ \Big{\lbrace}\ln(\omega+i\epsilon)\Big{\rbrace}^{n-1}\non\\
\ee
where in the last line of RHS we have substituted $\ln(\omega-i\epsilon)=\ln(\omega+i\epsilon)-2\pi i H(-\omega)$ and kept only the leading non-analytic piece of the integrand. Now substituting $v=\omega u$ we get,
\be
&&F_{n,0}(t,\vec{x})+F_{0,n}(t,\vec{x})-2F_{n-r,r}(t,\vec{x})\non\\
 &\simeq & \  iC(2r-n)\ \Big(i\f{d}{du}\Big)^{n-1}\ \int_{-\infty \times sgn(u)}^{0}\f{dv}{u}\ e^{-iv} \ f\Big(\f{v}{u}\Big)\ \Big{\lbrace}\ln(v+i\epsilon\times sgn(u))-\ln u\Big{\rbrace}^{n-1}\non\\
 &=&\ -C(2r-n)\ i^{n-1}(n-1)!\ \f{(\ln |u|)^{n-1}}{u^{n}}\ +\ \mathcal{O}\Big(\f{(\ln |u|)^{n-2}}{u^{n}}\Big)\label{FFF}
\ee
Now combining the results of eq.\eqref{FourierFn0},\eqref{FourierF0n} and \eqref{FFF} we get,
\be
F_{n-r,r}(t,\vec{x})\ &=&\ \begin{cases} -C(n-r)\ (n-1)!\ i^{n-1}\ \f{(\ln |u|)^{n-1}}{u^{n}}\hspace{1cm}\hbox{for\ }\ u\rightarrow +\infty\\
\\
+Cr\ (n-1)!\ i^{n-1}\ \f{(\ln |u|)^{n-1}}{u^{n}}\ \hspace{1cm}\hbox{for\ }\ u\rightarrow -\infty\label{FourierFn-rr}
\end{cases}
\ee

\section{Gravitational energy-momentum tensor}\label{appC}
We consider deviation of the metric $h_{\mu\nu}$ from Minkowski metric  given as $g_{\mu\nu}=\eta_{\mu\nu}+2h_{\mu\nu}$ and $h_{\mu\nu}$ satisfies harmonic gauge $\p_{\mu}h^{\mu\nu}-\f{1}{2}\p^{\nu}h\ =\ 0$. The definitions for Christoffel symbol, Riemann tensor and Ricci scalar are following:
\be
\Gamma^{\alpha}_{\beta\gamma}\ &=&\ \f{1}{2}g^{\alpha\sigma}\big(\p_{\beta}g_{\sigma\gamma}+\p_{\gamma}g_{\sigma\beta}-\p_{\sigma}g_{\beta\gamma}\big)\\
R_{\mu\nu}\ &=&\ R^{\rho}\ _{\mu\rho\nu}\ =\ \p_{\rho}\Gamma^{\rho}_{\mu\nu}-\p_{\nu}\Gamma^{\rho}_{\mu\rho}+\Gamma^{\tau}_{\mu\nu}\Gamma^{\rho}_{\rho\tau}-\Gamma^{\tau}_{\mu\rho}\Gamma^{\rho}_{\tau\nu}\\
R\ &=&\ g^{\mu\nu}R_{\mu\nu}
\ee
To compute various quantities below in terms of linear perturbation we follow the references \cite{DeWitt, Grisaru, Berends}.
\be
g^{\mu\nu}\ &=&\ \eta^{\mu\nu}-2h^{\mu\nu}+4h^{\mu\rho}h_{\rho}^{\ \nu}+\mathcal{O}(h^{3})\\
\sqrt{-det\ g}\ &=&\ 1+h^{\rho}_{\ \rho}-h^{\rho\sigma}h_{\rho\sigma}+\f{1}{2}h_{\rho}^{\ \rho}h_{\sigma}^{\ \sigma}+\mathcal{O}(h^{3})\\
\Gamma^{\alpha}_{\beta\gamma}\ &=&\ (\eta^{\alpha\sigma}-2h^{\alpha\sigma}+4h^{\alpha\kappa}h_{\kappa}^{\ \sigma})\ \big(\p_{\beta}h_{\gamma\sigma}+\p_{\gamma}h_{\beta\sigma}-\p_{\sigma}h_{\beta\gamma}\big)\ +\mathcal{O}(h^{4})
\ee
Ricci tensor components,
\be
R^{(1)}_{\mu\nu}\ &=&\ -\p_{\rho}\p^{\rho}h_{\mu\nu}\\
R^{(2)}_{\mu\nu}\ &=&\ \p_{\mu}h_{\alpha\beta}\p_{\nu}h^{\alpha\beta}+2h^{\alpha\beta}\p_{\mu}\p_{\nu}h_{\alpha\beta}-2h^{\alpha\beta}\p_{\nu}\p_{\beta}h_{\alpha\mu}-2h^{\alpha\beta}\p_{\mu}\p_{\beta}h_{\alpha\nu}+2h^{\alpha\beta}\p_{\alpha}\p_{\beta}h_{\mu\nu}\non\\
&&\ +2\p^{\beta}h_{\nu\alpha}\p_{\beta}h_{\mu}^{\ \alpha}-2\p_{\beta}h_{\nu\alpha}\p^{\alpha}h_{\mu}^{\ \beta}\\
R^{(3)}_{\mu\nu}\ &=&\ 4h^{\rho}_{\ \alpha}\p_{\rho}h^{\alpha\sigma}\p_{\mu}h_{\nu\sigma}+4h^{\rho}_{\ \alpha}\p_{\rho}h^{\alpha\sigma}\p_{\nu}h_{\mu\sigma}-4h^{\rho}_{\ \alpha}\p_{\rho}h^{\alpha\sigma}\p_{\sigma}h_{\mu\nu}-2\p_{\mu}h_{\nu}^{\ \sigma}h^{\rho\alpha}\p_{\sigma}h_{\rho\alpha}\non\\
&&\ -2\p_{\nu}h_{\mu}^{\ \sigma}h^{\rho\alpha}\p_{\sigma}h_{\rho\alpha}+2\p^{\sigma}h_{\mu\nu}h^{\rho\alpha}\p_{\sigma}h_{\rho\alpha}-4h^{\alpha\sigma}\p_{\mu}h_{\rho\sigma}\p_{\nu}h_{\alpha}^{\ \rho}+4h^{\alpha\sigma}\p_{\rho}h_{\mu\sigma}\p_{\alpha}h^{\rho}_{\ \nu}\non\\
&&\ -4h^{\alpha\sigma}\p_{\rho}h_{\mu\sigma}\p^{\rho}h_{\nu\alpha}-4h^{\alpha\sigma}\p_{\sigma}h_{\mu\rho}\p_{\alpha}h^{\rho}_{\ \nu}+4h^{\alpha\sigma}\p_{\sigma}h_{\mu\rho}\p^{\rho}h_{\nu\alpha}+4h^{\rho}_{\ \alpha}h^{\alpha\sigma}\p_{\rho}\p_{\mu}h_{\nu\sigma}\non\\
&&\ +4h^{\rho}_{\ \alpha}h^{\alpha\sigma}\p_{\rho}\p_{\nu}h_{\mu\sigma}-4h^{\rho}_{\ \alpha}h^{\alpha\sigma}\p_{\rho}\p_{\sigma}h_{\mu\nu}-4h^{\rho}_{\ \alpha}h^{\alpha\sigma}\p_{\mu}\p_{\nu}h_{\rho\sigma}
\ee
Ricci scalar components,
\be
R^{(1)}\ &=&\ \eta^{\mu\nu}R_{\mu\nu}^{(1)}\\
R^{(2)}\ &=&\ \eta^{\mu\nu}R^{(2)}_{\mu\nu}-2h^{\mu\nu}R^{(1)}_{\mu\nu}\\
R^{(3)}\ &=&\ \eta^{\mu\nu}R^{(3)}_{\mu\nu}-2h^{\mu\nu}R^{(2)}_{\mu\nu}+4h^{\mu\tau}h_{\tau}^{\ \nu}R^{(1)}_{\mu\nu}
\ee
Now using the definition of gravitational energy-momentum tensor given in eq.\eqref{gravEMtensor}, it's components in expansion of linear metric perturbation,
\begingroup
\allowdisplaybreaks
\be
8\pi G\ T^{(2)h\mu\nu}(x)\  &=&\ -\big[R^{(2)}_{\rho\sigma}\eta^{\mu\rho}\eta^{\nu\sigma}-2\eta^{\nu\sigma}h^{\mu\rho}R^{(1)}_{\rho\sigma}-2\eta^{\mu\rho}h^{\nu\sigma}R^{(1)}_{\rho\sigma}+\eta^{\mu\rho}\eta^{\nu\sigma}hR^{(1)}_{\rho\sigma}-\f{1}{2}\eta^{\mu\nu}\eta^{\rho\sigma}R^{(2)}_{\rho\sigma}+\eta^{\mu\nu}h^{\rho\sigma}R^{(1)}_{\rho\sigma}\non\\
&&\ +R^{(1)}h^{\mu\nu}-\f{1}{2}hR^{(1)}\eta^{\mu\nu}\big]\non\\
&=&\ -2\Big[\f{1}{2}\p^{\mu}h_{\alpha\beta}\p^{\nu}h^{\alpha\beta}+h^{\alpha\beta}\p^{\mu}\p^{\nu}h_{\alpha\beta}-h^{\alpha\beta}\p^{\nu}\p_{\beta}h_{\alpha}^{\mu}-h^{\alpha\beta}\p^{\mu}\p_{\beta}h_{\alpha}^{\nu}+h^{\alpha\beta}\p_{\alpha}\p_{\beta}h^{\mu\nu}\non\\
&&+\p^{\beta}h^{\nu\alpha}\p_{\beta}h_{\alpha}^{\mu}-\p^{\beta}h^{\alpha\nu}\p_{\alpha}h_{\beta}^{\mu}-\f{1}{2}h^{\mu\nu}\p_{\rho}\p^{\rho}h+h^{\mu\rho}\p^{\sigma}\p_{\sigma}h_{\rho}^{\nu}+h^{\nu\rho}\p^{\sigma}\p_{\sigma}h_{\rho}^{\mu}\Big]\non\\
&&+\eta^{\mu\nu}\Big[ \f{3}{2}\p^{\rho}h_{\alpha\beta}\p_{\rho}h^{\alpha\beta}+2h^{\alpha\beta}\p^{\rho}\p_{\rho}h_{\alpha\beta}-\p^{\beta}h^{\alpha\rho}\p_{\alpha}h_{\beta\rho}\Big]\non\\
&&+h\Big[ \p^{\rho}\p_{\rho}h^{\mu\nu}-\f{1}{2}\p^{\rho}\p_{\rho}h\eta^{\mu\nu}\Big]\label{T2h}
\ee

\be
8\pi G\ T^{(3)h\mu\nu}(x)\ &=&\ -\ \Bigg[ \eta^{\mu\rho}\eta^{\nu\sigma}R^{(3)}_{\rho\sigma}-2h^{\mu\rho}\eta^{\nu\sigma}R^{(2)}_{\rho\sigma}-2\eta^{\mu\rho}h^{\nu\sigma}R^{(2)}_{\rho\sigma}+4h^{\mu\rho}h^{\nu\sigma}R^{(1)}_{\rho\sigma}+4h^{\mu\tau}h_{\tau}^{\ \rho}\eta^{\nu\sigma}R^{(1)}_{\rho\sigma}\non\\
&&\ +4h^{\nu\tau}h_{\tau}^{\ \sigma}\eta^{\mu\rho}R^{(1)}_{\rho\sigma}+h\eta^{\mu\rho}\eta^{\nu\sigma}R^{(2)}_{\rho\sigma}-2hh^{\mu\rho}\eta^{\nu\sigma}R^{(1)}_{\rho\sigma}-2h\eta^{\mu\rho}h^{\nu\sigma}R^{(1)}_{\rho\sigma}-h_{\alpha\beta}h^{\alpha\beta}\eta^{\mu\rho}\eta^{\nu\sigma}R^{(1)}_{\rho\sigma}\non\\
&&\ +\f{1}{2}h^{2}\eta^{\mu\rho}\eta^{\nu\sigma}R^{(1)}_{\rho\sigma}-\f{1}{2}\eta^{\mu\nu}R^{(3)}+h^{\mu\nu}R^{(2)}-2h^{\mu\tau}h_{\tau}^{\ \nu}R^{(1)}-\f{1}{2}\eta^{\mu\nu}hR^{(2)}+hh^{\mu\nu}R^{(1)}\non\\
&&\ +\f{1}{2}\eta^{\mu\nu}h^{\alpha\beta}h_{\alpha\beta}R^{(1)}-\f{1}{4}\eta^{\mu\nu}h^{2}R^{(1)}\Bigg]\non\\
&=&\ -\Big[ 4h^{\rho}_{\ \alpha}\p_{\rho}h^{\alpha\sigma}\p^{\mu}h^{\nu}_{\ \sigma}+4h^{\rho}_{\ \alpha}\p_{\rho}h^{\alpha\sigma}\p^{\nu}h^{\mu}_{\ \sigma}-4h^{\rho}_{\ \alpha}\p_{\rho}h^{\alpha\sigma}\p_{\sigma}h^{\mu\nu}-2\p^{\mu}h^{\nu\sigma}h^{\rho\alpha}\p_{\sigma}h_{\rho\alpha}\non\\
&&\ -2\p^{\nu}h^{\mu\sigma}h^{\rho\alpha}\p_{\sigma}h_{\rho\alpha}+2\p^{\sigma}h^{\mu\nu}h^{\rho\alpha}\p_{\sigma}h_{\rho\alpha}-4h^{\alpha\sigma}\p^{\mu}h_{\rho\sigma}\p^{\nu}h_{\alpha}^{\ \rho}+4h^{\alpha\sigma}\p_{\rho}h^{\mu}_{\ \sigma}\p_{\alpha}h^{\rho\nu}\non\\
&&\ -4h^{\alpha\sigma}\p_{\rho}h^{\mu}_{\ \sigma}\p^{\rho}h^{\nu}_{\ \alpha}-4h^{\alpha\sigma}\p_{\sigma}h^{\mu}_{\ \rho}\p_{\alpha}h^{\rho\nu}+4h^{\alpha\sigma}\p_{\sigma}h^{\mu\rho}\p_{\rho}h^{\nu}_{\ \alpha}+4h^{\rho}_{\ \alpha}h^{\alpha\sigma}\p_{\rho}\p^{\mu}h^{\nu}_{\ \sigma}\non\\
&&\ +4h^{\rho}_{\ \alpha}h^{\alpha\sigma}\p_{\rho}\p^{\nu}h^{\mu}_{\ \sigma}-4h^{\rho}_{\ \alpha}h^{\alpha\sigma}\p_{\rho}\p_{\sigma}h^{\mu\nu}-4h^{\rho}_{\ \alpha}h^{\alpha\sigma}\p^{\mu}\p^{\nu}h_{\rho\sigma}\Big]\non\\
&& \ +2h^{\mu\rho}\Big[ \p_{\rho}h_{\alpha\beta}\p^{\nu}h^{\alpha\beta}+2h^{\alpha\beta}\p_{\rho}\p^{\nu}h_{\alpha\beta}-2h^{\alpha\beta}\p^{\nu}\p_{\beta}h_{\alpha\rho}-2h^{\alpha\beta}\p_{\rho}\p_{\beta}h_{\alpha}^{\nu}\non\\
&&\ +2h^{\alpha\beta}\p_{\alpha}\p_{\beta}h_{\rho}^{\nu}\ +2\p^{\beta}h^{\nu}_{\ \alpha}\p_{\beta}h_{\rho}^{\ \alpha}-2\p_{\beta}h^{\nu\alpha}\p_{\alpha}h_{\rho}^{\ \beta}\Big]\non\\
&&\ +2h^{\nu\sigma}\Big[\p^{\mu}h_{\alpha\beta}\p_{\sigma}h^{\alpha\beta}+2h^{\alpha\beta}\p^{\mu}\p_{\sigma}h_{\alpha\beta}-2h^{\alpha\beta}\p_{\sigma}\p_{\beta}h_{\alpha}^{\ \mu}-2h^{\alpha\beta}\p^{\mu}\p_{\beta}h_{\alpha\sigma}\non\\
&&\ +2h^{\alpha\beta}\p_{\alpha}\p_{\beta}h^{\mu}_{\ \sigma}\ +2\p^{\beta}h_{\sigma\alpha}\p_{\beta}h^{\mu\alpha}-2\p_{\beta}h_{\sigma\alpha}\p^{\alpha}h^{\mu\beta}\Big]\non\\
&&\ +4h^{\mu\rho}h^{\nu\sigma}\p_{\alpha}\p^{\alpha}h_{\rho\sigma}+4h^{\mu\alpha}h_{\alpha\rho}\p_{\beta}\p^{\beta}h^{\rho\nu}+4h^{\nu\alpha}h_{\alpha\sigma}\p_{\beta}\p^{\beta}h^{\mu\sigma}\non\\
&&\ -h\ \Big[\p^{\mu}h_{\alpha\beta}\p^{\nu}h^{\alpha\beta}+2h^{\alpha\beta}\p^{\mu}\p^{\nu}h_{\alpha\beta}-2h^{\alpha\beta}\p^{\nu}\p_{\beta}h_{\alpha}^{\ \mu}-2h^{\alpha\beta}\p^{\mu}\p_{\beta}h_{\alpha}^{\ \nu}\non\\
&&\ +2h^{\alpha\beta}\p_{\alpha}\p_{\beta}h^{\mu\nu} +2\p^{\beta}h^{\nu\alpha}\p_{\beta}h^{\mu}_{\ \alpha}-2\p_{\beta}h^{\nu\alpha}\p_{\alpha}h^{\mu\beta}\Big]\non\\
&&\ -2hh^{\mu\rho}\p^{\beta}\p_{\beta}h_{\rho}^{\ \nu}-2hh^{\nu\sigma}\p_{\beta}\p^{\beta}h^{\mu}_{\ \sigma}-h^{\alpha\beta}h_{\alpha\beta}\p_{\rho}\p^{\rho}h^{\mu\nu}+\f{1}{2}h^{2}\p^{\rho}\p_{\rho}h^{\mu\nu}\non\\
&&\ -6\eta^{\mu\nu}h^{\alpha\sigma}\p_{\beta}h_{\rho\sigma}\p^{\beta}h_{\alpha}^{\ \rho}+4\eta^{\mu\nu}h^{\alpha\sigma}\p_{\rho}h_{\beta\sigma}\p_{\alpha}h^{\beta\rho}-3\eta^{\mu\nu}h^{\rho\sigma}\p_{\rho}h_{\alpha\beta}\p_{\sigma}h^{\alpha\beta}\non\\
&&\ -2\eta^{\mu\nu}h^{\rho\sigma}h^{\alpha\beta}\p_{\rho}\p_{\sigma}h_{\alpha\beta}+2\eta^{\mu\nu}h^{\rho\sigma}h^{\alpha\beta}\p_{\sigma}\p_{\beta}h_{\alpha\rho}+2\eta^{\mu\nu}h^{\rho\sigma}h^{\alpha\beta}\p_{\rho}\p_{\beta}h_{\alpha\sigma}\non\\
&&\ -2\eta^{\mu\nu}h^{\rho\sigma}h^{\alpha\beta}\p_{\alpha}\p_{\beta}h_{\rho\sigma}+2\eta^{\mu\nu}h^{\rho\sigma}\p_{\beta}h_{\sigma\alpha}\p^{\alpha}h_{\rho}^{\ \beta}-4\eta^{\mu\nu}h^{\rho\alpha}h_{\alpha}^{\ \sigma}\p_{\beta}\p^{\beta}h_{\rho\sigma}\non\\
&&\ -3h^{\mu\nu}\p_{\rho}h_{\alpha\beta}\p^{\rho}h^{\alpha\beta}-2h^{\mu\nu}h^{\alpha\beta}\p_{\rho}\p^{\rho}h_{\alpha\beta}+2h^{\mu\nu}\p_{\beta}h_{\rho\alpha}\p^{\alpha}h^{\rho\beta}-2h^{\mu\nu}h^{\rho\sigma}\p_{\alpha}\p^{\alpha}h_{\rho\sigma}\non\\
&&\ -2h^{\mu\alpha}h_{\alpha}^{\ \nu}\p_{\beta}\p^{\beta}h+\f{3}{2}\eta^{\mu\nu}h\p_{\rho}h_{\alpha\beta}\p^{\rho}h^{\alpha\beta}+2\eta^{\mu\nu}hh^{\alpha\beta}\p_{\rho}\p^{\rho}h_{\alpha\beta}-\eta^{\mu\nu}h\p_{\beta}h_{\rho\alpha}\p^{\alpha}h^{\rho\beta}\non\\
&&\ +hh^{\mu\nu}\p_{\alpha}\p^{\alpha}h\ +\f{1}{2}\eta^{\mu\nu}h^{\alpha\beta}h_{\alpha\beta}\p_{\rho}\p^{\rho}h-\f{1}{4}\eta^{\mu\nu}hh\p_{\alpha}\p^{\alpha}h\label{T3h}
\ee
\endgroup
The Fourier transform of $T^{(3)h\mu\nu}(x)$ turns out to be,
\be
\widehat{T}^{(3)h\mu\nu}(k)\ &=&\ \f{1}{8\pi G}\int \f{d^{4}\ell_{1}}{(2\pi)^{4}}\int \f{d^{4}\ell_{2}}{(2\pi)^{4}}\ \widehat{h}^{\alpha\beta}(\ell_{1})\widehat{h}^{\rho\sigma}(\ell_{2})\widehat{h}^{\gamma\delta}(k-\ell_{1}-\ell_{2})\ G^{\mu\nu}_{\alpha\beta ,\rho\sigma ,\gamma\delta}(\ell_{1},\ell_{2},k)\label{T3}
\ee
where to extract the expression of $ G^{\mu\nu}_{\alpha\beta ,\rho\sigma ,\gamma\delta}(\ell_{1},\ell_{2},k)$ from eq.\eqref{T3h} we can always choose $h^{\gamma\delta}(x)$ to be the metric on which no derivative operates. This is always possible since in the expression \eqref{T3h} there are two derivatives operating but the expression contains three metric fluctuations. So choosing $h^{\gamma\delta}(x)$ this way, the momentum dependence of $ G^{\mu\nu}_{\alpha\beta ,\rho\sigma ,\gamma\delta}(\ell_{1},\ell_{2},k)$  have the following structure,
\be
 G^{\mu\nu}_{\alpha\beta ,\rho\sigma ,\gamma\delta}(\ell_{1},\ell_{2},k)\sim \# \ell_{1}\ell_{1}+\# \ell_{1}\ell_{2}+\# \ell_{2}\ell_{2} 
\ee
We don't need the explicit expression for $ G^{\mu\nu}_{\alpha\beta ,\rho\sigma ,\gamma\delta}(\ell_{1},\ell_{2},k)$ for the purpose of this paper.

\section{Brief sketch of the derivation of the results in \S\ref{subngrem}}\label{partialderivation}
Here in this appendix we briefly discuss the origin of different terms in eq.\eqref{Jn+Jnh} and eq.\eqref{TnX+Tnh} for the scattering event described in \S\ref{subngrem} in presence of long range electromagnetic and gravitational interactions. 

In presence of both long range electromagnetic and gravitational interactions, the Fourier transformation of (sub)$^{n}$-leading order matter current density given in eq.\eqref{Jnfull} contributes to order $\mathcal{O}\big(\omega^{n-1}(\ln\omega)^{n}\big)$ in the integration regions where all the integrating momenta are larger than $\omega$. Following the analysis of \S\ref{SsubnEM}, we find
\be
\Delta_{(n)}\widehat{J}^{\mu}(k)&\simeq &-i\sum_{a=1}^{M+N}\Bigg[q_{a}\f{k_{\rho}}{p_{a}\cdot k}\ \Bigg(p_{a}^{\mu}\f{\p}{\p p_{a\rho}}\ -\ p_{a}^{\rho}\f{\p}{\p p_{a\mu}} \Bigg)\Big{\lbrace}K_{em}^{cl}+K_{gr}^{cl}\Big{\rbrace}\Bigg] \times \f{1}{n!}\Bigg{[}k_{\sigma}\f{\p}{\p p_{a\sigma}}\Big{\lbrace}K_{em}^{cl}+K_{gr}^{cl}\Big{\rbrace}\Bigg{]}^{n-1} \non\\
&&\  -\ i\omega^{n-1}\ \sum_{a=1}^{M+N}\Big{\lbrace}\ln\lbrace(\omega +i\epsilon\eta_{a})L\rbrace\Big{\rbrace}^{n}\ \mathbf{n}_{\alpha_{1}}\mathbf{n}_{\alpha_{2}}\cdots\mathbf{n}_{\alpha_{n-1}}\mathbf{B}^{(n),\alpha_{1}\alpha_{2}\cdots\alpha_{n-1}\mu}(q_{a},p_{a})\label{UsualJ}
\ee
where the undetermined function $\mathbf{B}^{(n),\alpha_{1}\alpha_{2}\cdots\alpha_{r-1}\mu}\big(q_{a},p_{a}\big)$ is antisymmetric under $\mu$ and any $\alpha_{\ell}$ exchange for $\ell=1,2,...,n-1$. The last line of the above expression follows from the analysis of $\Delta_{(n)}^{rest}\widehat{J}_{\mu}(k)$, which contains terms involving $\Delta_{(r)}Y_{a}(\sigma)$ for $r\geq 2$ within the $\sigma$ integral. Now we also need the analyze the Fourier transform of the gravitational contribution to current density as given in eq.\eqref{currentGR}. It turns out that only the following part of (sub)$^{n}$-leading $\widehat{J}^{\mu}_{h}(k)$ can contribute to order $\mathcal{O}\big(\omega^{n-1}(\ln\omega)^{n}\big)$:
\be
\Delta_{(n)}\widehat{J}_{h}^{\mu}(k)&\simeq &\ \int \f{d^{4}\ell}{(2\pi)^{4}}\ \Delta_{(0)}\widehat{h}^{\alpha\beta}(\ell)\ \Delta_{(n-1)}\widehat{A}^{\rho}(k-\ell)\ \mathcal{F}^{\mu} _{\ \alpha\beta ,\rho}(k,\ell)
\ee
where,
\be
\mathcal{F}^{\mu} _{\ \alpha\beta ,\rho}(k,\ell)&=&\ -k.(k-\ell)\eta_{\alpha\beta}\delta_{\rho}^{\mu}+k_{\rho}\eta_{\alpha\beta}(k-\ell)^{\mu}+2k.(k-\ell)\delta_{\alpha}^{\mu}\eta_{\beta\rho}-2k_{\rho}\delta_{\alpha}^{\mu}(k-\ell)_{\beta}\non\\
&&\ +2k_{\alpha}(k-\ell)_{\beta}\delta_{\rho}^{\mu}-2k_{\alpha}\eta_{\beta\rho}(k-\ell)^{\mu}
\ee
Now in an inductive sense knowing the expression for (sub)$^{(n-1)}$-leading gauge field, if we analyze the expression for $\Delta_{(n)}\widehat{J}_{h}^{\mu}(k)$ in the integration region $R^{-1}<<|\ell^{\mu}|<<\omega$, we get,
\begingroup
\allowdisplaybreaks
\be
\Delta_{(n)}\widehat{J}_{h}^{\mu}(k)&\simeq &\ -i\ \f{1}{n!}\Big{\lbrace}-2iG\ln\lbrace(\omega+i\epsilon)R\rbrace\sum_{b=1}^{N}p_{b}\cdot k\Big{\rbrace}^{n}\ \sum_{a=1}^{M+N}q_{a}\f{p_{a}^{\mu}}{p_{a}.k}\non\\
&&\ -i\sum_{r=1}^{n-1}\f{1}{(n-r)!}\ \Big{\lbrace}-2iG\ln\lbrace(\omega+i\epsilon)R\rbrace\sum_{b=1}^{N}p_{b}\cdot k\Big{\rbrace}^{n-r}\ \sum_{a=1}^{M+N}\ \Bigg[q_{a}\f{k_{\rho}}{p_{a}.k}\non\\
&&\ \times\Bigg(p_{a}^{\mu}\f{\p}{\p p_{a\rho}}\ -\ p_{a}^{\rho}\f{\p}{\p p_{a\mu}} \Bigg)\Big{\lbrace}K_{em}^{cl}+K_{gr}^{cl}\Big{\rbrace}\Bigg] \times \f{1}{r!}\Bigg{[}k_{\sigma}\f{\p}{\p p_{a\sigma}}\Big{\lbrace}K_{em}^{cl}+K_{gr}^{cl}\Big{\rbrace}\Bigg{]}^{r-1}\non\\
&&\  -i \ \omega^{n-1}\sum_{r=1}^{n-1}\f{1}{(n-r)!}\ \Big{\lbrace}-2iG\ln\lbrace(\omega+i\epsilon)R\rbrace\sum_{b=1}^{N}p_{b}\cdot \mathbf{n}\Big{\rbrace}^{n-r}\non\\
&&\times \sum_{a=1}^{M+N}\Big{\lbrace}\ln\lbrace(\omega +i\epsilon\eta_{a})L\rbrace\Big{\rbrace}^{r}\ \mathbf{n}_{\alpha_{1}}\mathbf{n}_{\alpha_{2}}\cdots\mathbf{n}_{\alpha_{r-1}}\mathbf{B}^{(r),\alpha_{1}\alpha_{2}\cdots\alpha_{r-1}\mu}(q_{a},p_{a})\label{Jhextra}
\ee
\endgroup
Now sum over the contributions of eq.\eqref{UsualJ} and eq.\eqref{Jhextra} reproduces the expression of eq.\eqref{Jn+Jnh}.

The Fourier transform of (sub)$^{n}$-leading matter energy-momentum tensor follows from eq.\eqref{FourierTX},
\begingroup
\allowdisplaybreaks
\be
\Delta_{(n)}\widehat{T}^{X\mu\nu}(k)\ &\simeq &\ \  \sum_{a=1}^{M+N}m_{a}e^{-ik.r_{a}}\ \int_{0}^{\infty}d\sigma \ e^{-i(k\cdot v_{a}-i\epsilon)\sigma}\non\\
&&\ \times \Bigg[\ \f{1}{n!}\Big{\lbrace}-ik\cdot \Delta_{(1)}Y_{a}(\sigma)\Big{\rbrace}^{n}\ v_{a}^{\mu}v_{a}^{\nu}\ +\ \f{1}{(n-1)!}\ \Big{\lbrace}-ik\cdot \Delta_{(1)}Y_{a}(\sigma)\Big{\rbrace}^{n-1}v_{a}^{\mu}\f{d\Delta_{(1)}Y_{a}^{\nu}(\sigma)}{d\sigma}\non\\
&&\ +\ \f{1}{(n-1)!}\ \Big{\lbrace}-ik\cdot \Delta_{(1)}Y_{a}(\sigma)\Big{\rbrace}^{n-1}v_{a}^{\nu}\f{d\Delta_{(1)}Y_{a}^{\mu}(\sigma)}{d\sigma}\ +\ \f{1}{(n-2)!}\ \Big{\lbrace}-ik\cdot \Delta_{(1)}Y_{a}(\sigma)\Big{\rbrace}^{n-2}\non\\
&&\ \times \f{d\Delta_{(1)}Y_{a}^{\mu}(\sigma)}{d\sigma}\f{d\Delta_{(1)}Y_{a}^{\nu}(\sigma)}{d\sigma}\ \Bigg]\ +\ \Delta_{(n)}^{(rest)}\widehat{T}^{X\mu\nu}(k)\label{subnleadingTX}
\ee
\endgroup
where $\Delta_{(n)}^{(rest)}\widehat{T}^{X\mu\nu}(k)$ 
 contains terms having one or more $\Delta_{(r)}Y_{a}(\sigma)$ for $r\geq 2$ within the $\sigma$ integral. Now $\Delta_{(n)}\widehat{T}^{X\mu\nu}(k) -\Delta_{(n)}^{(rest)}\widehat{T}^{X\mu\nu}(k)$ can be evaluated analogous to \S\ref{SsubnEM} and contribution at order $\mathcal{O}(\omega^{n-1}(\ln\omega)^{n})$ becomes,
 \begingroup
\allowdisplaybreaks
\be
&&\Delta_{(n)}\widehat{T}^{X\mu\nu}(k) -\Delta_{(n)}^{(rest)}\widehat{T}^{X\mu\nu}(k)\non\\
&\simeq &\ (-i)\sum_{a=1}^{M+N}\f{k_{\rho}k_{\sigma}}{p_{a}.k}\ \Bigg[\Bigg(p_{a}^{\mu}\f{\p}{\p p_{a\rho}}-p_{a}^{\rho}\f{\p}{\p p_{a\mu}}\Bigg)\Big{\lbrace}K_{gr}^{cl}+K_{em}^{cl}\Big{\rbrace}\Bigg]\non\\
&&\times \Bigg[\Bigg(p_{a}^{\nu}\f{\p}{\p p_{a\sigma}}-p_{a}^{\sigma}\f{\p}{\p p_{a\nu}}\Bigg)\ \Big{\lbrace}K_{gr}^{cl}+K_{em}^{cl}\Big{\rbrace}\Bigg]\times\ \f{1}{n!}\ \Bigg(k^{\alpha}\f{\p}{\p p_{a}^{\alpha}}\Big{\lbrace}K_{gr}^{cl}+K_{em}^{cl}\Big{\rbrace}\Bigg)^{n-2}\non\\
&& + \f{i}{n!}\sum_{a=1}^{M+N}(p_{a}.k)\ \Bigg(\f{\p}{\p p_{a\mu}}\Big{\lbrace}K_{gr}^{cl}+K_{em}^{cl}\Big{\rbrace}\Bigg) \Bigg(\f{\p}{\p p_{a\nu}}\Big{\lbrace}K_{gr}^{cl}+K_{em}^{cl}\Big{\rbrace}\Bigg)\Bigg(k^{\alpha}\f{\p}{\p p_{a}^{\alpha}}\Big{\lbrace}K_{gr}^{cl}+K_{em}^{cl}\Big{\rbrace}\Bigg)^{n-2}\label{TX-TXrest}
\ee
\endgroup
Now we also need to analyze the (sub)$^{n}$-leading gravitational energy-momentum tensor having structure of eq.\eqref{Thsketch}. In the gravitational energy-momentum tensor expression we also have to include the following electromagnetic energy-momentum tensor contribution,
\be
\widehat{T}^{(em)\mu\nu}(k)&=&\ \int d^{4}x\ e^{-ik.x}\ \Big[g^{\mu\rho}g^{\nu\sigma}g^{\alpha\beta}F_{\rho\alpha}F_{\sigma\beta}-\f{1}{4}g^{\mu\nu}g^{\rho\sigma}g^{\alpha\beta}F_{\rho\alpha}F_{\sigma\beta}\Big]
\ee
But from our experience of the analysis of sub-subleading gravitational energy-momentum tensor, we expect that only the following part of the gravitational energy-momentum tensor needs to analyze to extract order $\mathcal{O}(\omega^{n-1}(\ln\omega)^{n})$ contribution:
\be
&&\Delta_{(n)}\widehat{T}_{2}^{h\mu\nu}(k)\non\\
&\equiv & \f{1}{8\pi G}\ \int\f{d^{4}\ell_{1}}{(2\pi)^{4}}\ \Bigg[ \Delta_{(0)}\widehat{h}^{\alpha\beta}(\ell_{1})\mathcal{F}^{\mu\nu }\ _{\alpha\beta ,\rho\sigma }(k,\ell_{1})\Delta_{(n-1)}\widehat{h}^{\rho\sigma}(k-\ell_{1})+\Delta_{(n-1)}\widehat{h}^{\alpha\beta}(\ell_{1})\mathcal{F}^{\mu\nu}\ _{ \alpha\beta ,\rho\sigma }(k,\ell_{1})\non\\
&&\times \Delta_{(0)}\widehat{h}^{\rho\sigma}(k-\ell_{1})\Bigg]\label{gravnemten}
\ee
So in an inductive sense substituting the expression of $\Delta_{(n-1)}\widehat{h}^{\rho
\sigma}(\ell)$ in the above expression and analyzing $\Delta_{(n)}\widehat{T}_{2}^{h\mu\nu}(k)$ in the integration region $R^{-1}<<|\ell_{1}^{\mu}|<<\omega$ we get,
\begingroup
\allowdisplaybreaks
\be
&&\Delta_{(n)}\widehat{T}_{2}^{h\mu\nu}(k)\non\\
&\simeq &\ (-i) \f{1}{n!}\Big{\lbrace}-2iG\ln\lbrace(\omega +i\epsilon)R\rbrace\ \sum_{\substack{b=1}}^{N}p_{b}\cdot k\Big{\rbrace}^{n}\ \sum_{a=1}^{M+N}\f{p_{a}^{\mu}p_{a}^{\nu}}{p_{a}.k}\non\\
&& -i\ \f{1}{(n-1)!}\Big{\lbrace}-2iG\ln\lbrace(\omega +i\epsilon)R\rbrace\ \sum_{\substack{b=1}}^{N}p_{b}\cdot k\Big{\rbrace}^{n-1}\sum_{a=1}^{M+N}\f{p_{a}^{\mu}k_{\rho}}{p_{a}.k}\Bigg[\Bigg(p_{a}^{\nu}\f{\p}{\p p_{a\rho}}-p_{a}^{\rho}\f{\p}{\p p_{a\nu}}\Bigg)\Big{\lbrace}K_{gr}^{cl}+K_{em}^{cl}\Big{\rbrace}\Bigg]\non\\
&&\ -i\ \sum_{r=2}^{n-1}\f{1}{(n-r)!}\ \Big{\lbrace}-2iG\ln(\omega +i\epsilon)\ \sum_{\substack{b=1}}^{N}p_{b}\cdot k\Big{\rbrace}^{n-r}\ \sum_{a=1}^{M+N}\f{k_{\rho}k_{\sigma}}{p_{a}.k}\non\\
&&\ \times \ \Bigg[\Bigg(p_{a}^{\mu}\f{\p}{\p p_{a\rho}}-p_{a}^{\rho}\f{\p}{\p p_{a\mu}}\Bigg)\Big{\lbrace}K_{gr}^{cl}+K_{em}^{cl}\Big{\rbrace}\Bigg]\ \Bigg[\Bigg(p_{a}^{\nu}\f{\p}{\p p_{a\sigma}}-p_{a}^{\sigma}\f{\p}{\p p_{a\nu}}\Bigg)\Big{\lbrace}K_{gr}^{cl}+K_{em}^{cl}\Big{\rbrace}\Bigg]\non\\
&&\ \times \f{1}{r!}\Bigg[k^{\alpha}\f{\p}{\p p_{a}^{\alpha}}\Big{\lbrace}K_{gr}^{cl}+K_{em}^{cl}\Big{\rbrace}\Bigg]^{r-2}\non\\
&&\ -i\omega^{n-1}\sum_{r=3}^{n-1}\Big{\lbrace}\ln\lbrace(\omega +i\epsilon)R\rbrace\Big{\rbrace}^{n-r}\f{1}{(n-r)!}\Big{\lbrace}-2iG\sum_{b=1}^{N}p_{b}\cdot \mathbf{n}\Big{\rbrace}^{n-r}\sum_{a=1}^{M+N}\Big{\lbrace}\ln\lbrace(\omega+i\epsilon\eta_{a})L\rbrace\Big{\rbrace}^{r}\non\\
&&\ \times\ \mathbf{n}_{\alpha_{1}}\mathbf{n}_{\alpha_{2}}\cdots \mathbf{n}_{\alpha_{r-1}}\mathbf{C}^{(r),\ \mu\nu\alpha_{1}\cdots\alpha_{r-1}}(q_{a},p_{a})
\label{hatThsubnemgr}
\ee
\endgroup
 where $\mathbf{C}^{(r),\ \mu\nu\alpha_{1}\cdots\alpha_{r-1}}(q_{a},p_{a})$ is anti-symmetric under $\mu \leftrightarrow\alpha_{i}$ as well as $\nu \leftrightarrow\alpha_{j}$ exchange for $i,j=1,2,\cdots,r-1$. The functional behaviour of $\mathbf{C}^{(r)}$ has to  determine by explicit computations, but we expect these functions will only depend on the scattering data. On the other hand in the integration region $L^{-1}>>|\ell_{1}^{\mu}|>>\omega$, analysis of the (sub)$^{n}$-leading gravitational energy-momentum tensor of eq.\eqref{gravnemten}  is very much involved. But from the gauge invariance requirement of the total (sub)$^{n}$-leading energy-momentum tensor, we expect the following structure of contribution at order $\mathcal{O}(\omega^{n-1}(\ln\omega)^{n})$ in the integration region where the integrating momenta are large compare to $\omega$,
 \begingroup
\allowdisplaybreaks
\be
&&\Delta_{(n)}^{(rest)}\widehat{T}^{X\mu\nu}(k)+\Delta_{(n)}\widehat{T}_{2}^{h\mu\nu}(k)\ +\ \Delta_{(n)}\widehat{T}^{(em)\mu\nu}(k)\non\\
&\simeq &- \f{i}{n!}\sum_{a=1}^{M+N}(p_{a}.k)\ \Bigg(\f{\p}{\p p_{a\mu}}\Big{\lbrace}K_{gr}^{cl}+K_{em}^{cl}\Big{\rbrace}\Bigg) \Bigg(\f{\p}{\p p_{a\nu}}\Big{\lbrace}K_{gr}^{cl}+K_{em}^{cl}\Big{\rbrace}\Bigg)\Bigg(k^{\alpha}\f{\p}{\p p_{a}^{\alpha}}\Big{\lbrace}K_{gr}^{cl}+K_{em}^{cl}\Big{\rbrace}\Bigg)^{n-2}\non\\
&& -i\ \omega^{n-1} \sum_{a=1}^{M+N}\Big{\lbrace}\ln\lbrace(\omega+i\epsilon\eta_{a})L\rbrace\Big{\rbrace}^{n}\ \mathbf{n}_{\alpha_{1}}\mathbf{n}_{\alpha_{2}}\cdots \mathbf{n}_{\alpha_{n-1}} \ \mathbf{C}^{(n),\ \mu\nu\alpha_{1}\cdots\alpha_{n-1}}(q_{a},p_{a})\label{TXrest+Th2+Tem}
\ee
\endgroup
 By analyzing the LHS of the above expression for $n\geq 3$ in the specified integration region we can determine the unknown function $\mathbf{C}^{(n)}(q_{a},p_{a})$. Now summing over the contributions of eq.\eqref{TX-TXrest}, eq.\eqref{hatThsubnemgr} and eq.\eqref{TXrest+Th2+Tem} we get the expression of eq.\eqref{TnX+Tnh}.

\bibliography{classicalref}
\bibliographystyle{utphys}

\end{document}